\documentclass[11pt,a4paper]{amsart}

\usepackage{amssymb}
\usepackage{mathtools}
\usepackage{enumitem}
\usepackage[T1]{fontenc}
\usepackage[american]{babel}
\usepackage[numbers]{natbib}

\theoremstyle{plain}
\newtheorem{theorem}{Theorem}
\numberwithin{theorem}{section}

\newtheorem{lemma}[theorem]{Lemma}
\newtheorem{proposition}[theorem]{Proposition}
\newtheorem{corollary}[theorem]{Corollary}

\theoremstyle{definition}
\newtheorem{definition}[theorem]{Definition}

\theoremstyle{remark}
\newtheorem{remark}[theorem]{Remark}

\usepackage[margin=1in]{geometry}
\numberwithin{equation}{subsection}

\usepackage[colorlinks,citecolor=blue,urlcolor=blue,hypertexnames=false]{hyperref}
\providecommand{\doi}[1]{%
  \href{https://doi.org/#1}{\nolinkurl{https://doi.org/#1}}%
}

\newcommand{\D}{\mathbb{D}}
\newcommand{\R}{\mathbb{R}}
\newcommand{\C}{\mathbb{C}}
\newcommand{\E}{\mathbb{E}}
\newcommand{\Prob}{\mathbb{P}}

\newcommand{\1}{\mathbf{1}}

\newcommand{\M}{\mathcal{M}}
\newcommand{\Z}{\mathcal{Z}}

\newcommand{\p}{\mathsf{p}}

\renewcommand{\Re}{\mathrm{Re}}
\renewcommand{\Im}{\mathrm{Im}}
\renewcommand{\epsilon}{\varepsilon}

\newcommand{\gff}{\mathrm{GFF}}
\newcommand{\gffe}[1]{\left\langle#1\right\rangle_{\gff(\varepsilon)}}
\newcommand{\gffcf}[1]{\left\langle#1\right\rangle_{\gff}}
\newcommand{\phie}{\varphi_\varepsilon}
\newcommand{\phit}{\varphi_{\sqrt{t}}}

\DeclareFontEncoding{LS1}{}{}
\DeclareFontSubstitution{LS1}{stix}{m}{n}
\DeclareSymbolFont{symbols2}{LS1}{stixfrak}{m}{n}
\DeclareMathSymbol{\typecolon}{\mathbin}{symbols2}{"25}
\newcommand{\optypecolon}{\mathop{\typecolon}}

\newcommand{\opcolon}{\mathop{:}}

\newcommand{\isingccf}[1]{\left\langle#1\right\rangle_{0}}

\author{S. C. Park}
\address{S. C. Park. Department of Mathematics and Statistics, University of Helsinki. Helsinki, Finland}
\email{sc.m.park@helsinki.fi}

\author{Tuomas Virtanen}
\address{Tuomas Virtanen. Department of Mathematics and Statistics, University of Helsinki. Helsinki, Finland}
\address{Department of Natural and Health Sciences, Åbo Akademi University. Turku, Finland}
\email{tuomas.virtanen@abo.fi, tuomas.virtanen@helsinki.fi}
\urladdr{https://orcid.org/0009-0004-2132-2479}

\author{Christian Webb}
\address{Christian Webb. Department of Mathematics and Statistics, University of Helsinki. Helsinki, Finland}
\email{christian.webb@helsinki.fi}

\date{\today}

\title[Near-critical Ising, sine-Gordon, and bosonization]{Near-critical Ising, sine-Gordon at the free fermion point, and bosonization}

\begin{document}

\begin{abstract} 
    In this article, we study the continuous correlations of the near-critical Ising model in two dimensions with plus boundary conditions, and prove that doubled correlation functions of primary fields (spin, disorder, fermions, energy) in the Ising model are given by correlation functions of the sine-Gordon model at the free fermion point. This is an instance of bosonization. 

    The main ideas involve analyticity of correlation functions in a mass parameter in finite volume and proving that in a perturbative regime, the Taylor coefficients of the correlation functions match due to known bosonization results for the critical Ising model in terms of the Gaussian free field.

    The main techniques on the Ising side involve construction and precise estimates of certain massive holomorphic functions while on the sine-Gordon side, we control an iterated Mayer expansion with techniques going back to Brydges and Kennedy.
\end{abstract}

\subjclass[2020]{Primary 82B20, 81T40, 60G60; Secondary 30E25, 81T08}
\keywords{Bosonization; sine-Gordon model; Ising model}

\maketitle

\setcounter{tocdepth}{1}
{
  \hypersetup{linkcolor=black}
  \tableofcontents
}

\section{Introduction and main results}

\subsection{Background}  Scaling limits of two-dimensional models of statistical physics have long been understood by physicists to have a very rich structure. For many models, the scaling limit at precisely the critical point of the model is conjectured to be described by a \emph{conformal field theory}, and many aspects of the scaling limit become exactly solvable (see e.g. \cite{BPZ,DFMS}). Mathematically, such statements remain conjectures in general, but there has been great progress in understanding them in the setting of the \emph{critical Ising model} \cite{CHI,CHI2,HonKytVik} and the \emph{Liouville model} \cite{KRV}.

Physicists also predict that for very special perturbations of certain critical models (and conformal field theories), the scaling limits still remain exactly solvable (see e.g. \cite{Mussardo}). Some of the prime examples of such models are the \emph{near-critical Ising model} (either perturbed in the thermal or magnetic direction) and the \emph{sine-Gordon model}. Again, mathematically rigorous results about such statements are somewhat more limited, but for the near-critical Ising model (perturbed in the thermal direction) see \cite{WMTB} and more recently e.g. \cite{P, CIM23} and for the sine-Gordon model (at the free fermion point) see e.g. \cite{BMW,BaWe}.

Another fascinating phenomenon physicists have predicted to occur in two-dimensional models is \emph{bosonization/fermionization}. This is a duality that relates the correlation functions of certain very special bosonic models to certain equally special fermionic models. Perhaps the original mathematically rigorous work on such a phenomenon goes back to \cite{ML} which was done in a Hilbert space setting. Recently, some results of this flavor have been given rigorous mathematical proofs in a probabilistic setting: in \cite{BIVW}, it was proven that correlation functions of primary fields in the critical double Ising model can be expressed as Gaussian free field correlation functions, while in \cite{BMW,BaWe}, it was proven that correlation functions related to massive free fermions can be expressed in terms of sine-Gordon correlation functions at the free fermion point. In \cite{Mason}, this was used to show that the scaling limit of the height function in a massive perturbation of the full plane \emph{dimer model} is the sine-Gordon field at the free fermion point, and in \cite{BMR}, such a result was proven for suitable domains with boundaries. See also \cite{BFM} for an instance of bosonization that relates the sine-Gordon model away from the free fermion point to an interacting fermionic model known as the \emph{massive Thirring model} -- this duality is sometimes known as the Coleman correspondence due to it being conjectured by Coleman in \cite{Coleman}. In fact, the main results of this article can be seen as an extension of the Coleman correspondence.

The goal of this article is to extend the bosonization results of \cite{BIVW} from the critical Ising model to the near-critical Ising model (perturbed in the thermal direction). The relevant bosonic model is now the sine-Gordon model at the free fermion point instead of the Gaussian free field. Conceptually, this may not be too surprising to the reader, as the Ising model is in a sense a model of massive Majorana fermions; the double Ising model is then a model of two massive Majorana fermions. This is equivalent to one massive Dirac fermion which by \cite{BaWe} is known to be equivalent to the sine-Gordon model at the free fermion point. Indeed, this bosonization result was already predicted (for spin correlation functions) in \cite{IZ}.

\subsection{The Ising model and the double Ising model}

The Ising model \cite{Lenz, Ising} is one of the most fundamental models of statistical mechanics. It is complicated enough to describe interesting phenomena, namely ferromagnetic phase transitions, but tractable enough to be studied in a mathematically rigorous way. 

In this article, we are concerned with the two-dimensional nearest neighbor Ising model with zero external field and plus boundary conditions. While we will mainly focus on the continuum model, we briefly review as motivation the discrete model. For this purpose, let $\Omega\subset \R^2$ be a nice domain (say bounded and simply connected with smooth boundary), $\delta>0$, and $\Omega_\delta=\Omega\cap \delta \mathbb Z^2$. The set of \emph{spin configurations} on $\Omega_\delta$ is $\{-1,1\}^{\Omega_\delta}$ and the Ising model is a family of probability measures on $\{-1,1\}^{\Omega_\delta}$. More precisely, for each \emph{inverse temperature} $\beta>0$, the probability of a spin configuration $\sigma^\delta\in \{-1,1\}^{\Omega_\delta}$ is 
\begin{equation}\label{eq:isingprob}
\mathbb P_\beta(\sigma^\delta)\propto e^{\beta \sum_{x\sim y}\sigma_x^\delta\sigma_y^\delta},
\end{equation}
where $x\sim y$ means that $x$ and $y$ are nearest neighbors in $\Omega_\delta$ (each pair summed over only once), and we allow also $x$ (or $y$) to be on the boundary (one lattice step outside) of $\Omega_\delta$, in which case the boundary condition is that $\sigma_x^\delta=1$ (or $\sigma_y^\delta=1$).

Since \cite{Onsager}, it has been understood that this model has a phase transition at $\beta=\beta_c=\frac{1}{2}\log (1+\sqrt{2})$. This exact solution of the model spurred a number of celebrated rigorous computations of various statistics of the model, see \cite{mccoy-wu} for a comprehensive treatment. More recently, in \cite{CHI} (see also \cite{mccoy-wu, Pinson, Dubedat} for infinite volume results), it was proven that \emph{spin correlation functions} have scaling limits that enjoy conformal covariance properties, as predicted in \cite{BPZ, Burkhardt-Guim}. More precisely, there exist limiting correlation functions $\langle \sigma_{x_1}\cdots \sigma_{x_n}\rangle_{0;\Omega}$, which transform covariantly under conformal transformations of $\Omega$ (see Remark \ref{rem:confcov}) such that as $\delta\to 0$,
\begin{equation}\label{eq:critspinconv}
\delta^{-n/8}\sum_{\sigma\in \{-1,1\}^{\Omega_\delta}} \sigma_{x_1}^\delta\cdots \sigma_{x_n}^\delta\mathbb P_{\beta_c}(\sigma)=\mathcal C_\sigma^n  \langle \sigma_{x_1}\cdots \sigma_{x_n}\rangle_{0;\Omega}+o(1),
\end{equation}
where $\mathcal C_\sigma = 2^{\frac{5}{48}}e^{\frac32\zeta'(-1)}$, see e.g. \cite[(1.8)]{CHI2} (note they work on the rotated square lattice). The subscript $0$ refers here to the fact that we are exactly at the critical, or massless, point.

To see the full conformal field theory structure predicted by physicists, one needs to also consider other correlation functions in addition to pure spin correlation functions. The relevant \emph{primary} fields in the model are the \emph{disorder field} $\mu$, \emph{fermion fields} $\psi,\psi^*$, and the \emph{energy field} $\epsilon$. These all have lattice level definitions that need to be renormalized by a suitable power of $\delta$ to lead to meaningful correlation functions in the scaling limit. We refer the reader to \cite[Section 1 and Section 2]{CHI2} for details of the lattice level definitions of these fields and conformal invariance results on their critical scaling limit. It also bears mentioning that conformal invariance of the critical model has been seen in the limit in the form of a \emph{Schramm-Loewner Evolution}, or SLE of its {interface curves} \cite{Smirnov, CDHKS}.

At the heart of the aforementioned demonstrations of the conformal invariance of the discrete model has been a robust theory of discrete complex analysis which passes to the continuum limit, following the approach of \cite{Smirnov} (see also \cite{Dubedat, Perk} for related discrete context). Indeed, the key insight for our analysis is the characterization of these limiting correlation functions in terms of objects arising from boundary value problems of complex analysis. We now briefly outline this (purely continuous) characterization of these functions, in order to generalize it later.

The starting point is that if we define for $\eta\in \C\setminus \{0\}$, $\psi^{[\eta]}=\frac{1}{2}(\bar \eta\psi+\eta\psi^*)$, then \emph{spin-weighted fermionic correlation functions} of the form 
\begin{equation}\label{eq:spinweightedfermcrit}
    f(z)=\frac{\langle \sigma_{a_1}\cdots \sigma_{a_n}\psi_z\psi^{[\eta]}_w\rangle_{0;\Omega}}{\langle \sigma_{a_1}\cdots \sigma_{a_n}\rangle_{0;\Omega}} 
\end{equation}
as functions of $z$, are meromorphic on the double cover of $\Omega\setminus \{a_1,\dots,a_n\}$ and uniquely characterized by suitable boundary values on $\partial \Omega$ and $\{a_1,\dots,a_n,w\}$ (see \cite[Section 3]{CHI2}). General spin-weighted fermionic correlation functions (with more insertions of $\psi$ and/or $\psi^{[\eta]}$) can be recovered from two insertions due to a Pfaffian identity fermionic correlation functions satisfy (see \cite[Proposition 2.24]{CHI2}). General correlation functions can then be extracted from the operator product expansions (OPEs) the fields satisfy, proven in \cite[Section 6]{CHI2}, and decorrelation estimates for pure spin correlations (see \cite[Section 2.9]{CHI} or alternatively \cite[Proposition 5.3]{CHI2}).

We now turn to the near-critical Ising model. Near-critical here refers to the fact that instead of choosing that $\beta$ is exactly $\beta_c$, we fix some $m\in \R$, nonzero if not exactly at criticality, and take $\beta=\beta_c-m\delta/2$. It turns out that this is exactly the correct scaling in $\delta$ for perturbing the inverse temperature to see a non-trivial (and different from the critical case, being \emph{massive}) regime in the scaling limit $\delta\to 0$. More precisely, in \cite{P}, it is proven that there exists a family of non-critical limiting correlation functions $\langle \sigma_{a_1}\cdots \sigma_{a_n}\rangle_{m;\Omega}$ for $m<0$ such that instead of \eqref{eq:critspinconv} we have
\begin{equation}\label{eq:nearcritspinconv}
\delta^{-n/8}\sum_{\sigma\in \{-1,1\}^{\Omega_\delta}} \sigma_{a_1}^\delta\cdots \sigma_{a_n}^\delta\mathbb P_{\beta_c-m\delta/2}(\sigma)=\mathcal C_\sigma^n\langle \sigma_{a_1}\cdots \sigma_{a_n}\rangle_{m;\Omega}+o(1).
\end{equation}

The proof again goes through analyzing spin-weighted correlation functions similar to \eqref{eq:spinweightedfermcrit}, which also for $m\neq 0$ satisfy the same boundary conditions as in the critical case, but instead of meromorphicity ($\bar\partial f=0$) $f$ is $m$-\emph{massive holomorphic}, in the sense that it satisfies the equation $\bar\partial f=-i m \bar f$, an instance of the \emph{Vekua equation} \cite{Vekua, Bers}.

For our proof of bosonization of the near-critical Ising model, it turns out to be convenient to generalize this setting slightly. Instead of $m$ a constant (as above), we replace it by a non-constant deterministic function $\alpha$. We construct an $\alpha$-massive counterpart to fermion-fermion correlations \eqref{eq:spinweightedfermcrit} (and disorder-fermion correlations, for technical reasons) directly in the continuum as the unique solutions to $\bar\partial f=-i\alpha \bar f$ with analogous branching and boundary conditions as in the critical case. More general correlation functions can then be constructed from Pfaffian identities, operator product expansions, and decorrelation assumptions on pure spin correlations. To simplify notation slightly, we will from now on write e.g. $\sigma_V:=\prod_{i\in V}\sigma_{v_i}$,\footnote{While for spin and energy observables, the order of the product does not matter, for fermions and disorder observables it does. We will always assume that our index sets are ordered, say $Z=\{1,...,n\}$, in which case we understand that $\psi_Z=\prod_{i\in Z}\psi_{z_i}=\psi_{z_1}\cdots \psi_{z_n}$. We use the same convention for $\psi_W^*$ and $\mu_U$.} within correlation functions, so that a general primary field correlation can be written compactly as 
\begin{equation*}
    \langle \sigma_V \mu_U\psi_Z\psi_W^*\epsilon_A\rangle_{\alpha;\Omega},
\end{equation*}
with the convention that the insertion points are all distinct unless otherwise stated. We will do this construction in detail in Sections \ref{sec:ising-analysis} and \ref{subsec:energy}. 

To our knowledge, this general setup with non-constant mass does not yet have full rigorous scaling limit results available, and as we work purely in the continuum, we do not claim here that continuous correlations constructed in this paper arise as a limit of discrete correlations (though see the closely related setting of the dimer model with a spatially varying mass studied in \cite{BMR}). Nonetheless, when $\alpha$ is a negative constant $m$, the pure spin correlations constructed in this paper coincide with the scaling limit established in \cite{P} (see Proposition \ref{prop:spin}), which may be generalized to any primary field correlation with any $m\in \mathbb R$ \cite{CIP}. In addition, the core assumptions in the continuous construction (Vekua equation, boundary condition, decorrelation near the boundary) have been shown to appear and be valid in far more general settings than in the constant mass case, see \cite{Che, Mah}.

The final fact related to the Ising model we need to discuss is the \emph{double Ising model}. On the lattice level, this simply amounts to taking two independent copies of the probabilistic model and looking at suitable observables constructed from these two independent copies. While we do not go into it further, it is known that there is an exact combinatorial correspondence between the (critical) {dimer} model and the critical double Ising model \cite{Dubedat}, which is expected to generalize to massive versions of both models and should be the discrete counterpart for our continuous correspondence (see e.g. \cite{BHS}).

Working in the continuum, we also define the double Ising model correlation functions directly as deterministic functions, but one should view this definition as being motivated by $(\sigma,\mu,\psi,\psi^*,\epsilon)$ and $(\widetilde \sigma,\widetilde\mu,\widetilde \psi,\widetilde \psi^*,\widetilde \epsilon)$ being independent random objects so that correlation functions factorize suitably. The definition is as follows
\begin{align}\label{eq:doublecf}
    &\left\langle (\sigma\widetilde \sigma)_{V} (\mu\widetilde \mu)_{U}(\psi\widetilde \psi)_{Z} (\psi^* \widetilde \psi^*)_{W}(\epsilon+\widetilde\epsilon)_{A}\right\rangle_{\alpha;\Omega}\\ \nonumber
    &\quad=\sum_{S\subset A} 
    \left\langle \sigma_{V} \mu_{U}\psi_{Z} \psi_{W}^*\epsilon_{S}\right\rangle_{\alpha;\Omega}\left\langle \sigma_{V} \mu_{U}\psi_{Z} \psi_{W}^*\epsilon_{A\setminus S}\right\rangle_{\alpha;\Omega}.
\end{align}
Here we interpret $(\sigma\widetilde \sigma)_V=\prod_{i\in V}\sigma_{v_i}\widetilde \sigma_{v_i}$ and $(\epsilon+\widetilde \epsilon)_A=\prod_{m\in A}(\epsilon_{a_m}+\widetilde \epsilon_{a_m})$.

Note that for the energy field, we are considering $\epsilon+\widetilde\epsilon$, while for the other fields we consider products. One could also consider the product $\epsilon\widetilde\epsilon$, but the sum turns out to be slightly more convenient when discussing the sine-Gordon model. Note that, apart from a choice of branches of square roots, one can recover inductively the Ising correlation functions from the double Ising correlation functions.

\subsection{The sine-Gordon model}

The (Euclidean massless) sine-Gordon model is a fascinating model of two-dimensional quantum field theory. It is predicted (and in some cases proven) to have connections to various types of interesting phenomena: bosonization \cite{Coleman,BaWe,BFM}, solitons \cite[Chapter 6]{Coleman2}, spontaneous symmetry breaking and dynamical mass generation, connections to the classical 2d Coulomb gas  \cite{Frohlich,LRV2}, coupling dependent structure of renormalization counter terms \cite{NRS}, and the Kosterlitz-Thouless transition \cite{Falco}. As already mentioned, it is a fundamental example of a near-critical model where there are many conjectures about exact solvability of certain central quantities -- see e.g. \cite{LZ} and \cite[Part IV]{Mussardo}. The model has been studied mathematically rigorously in various ways -- in addition to the references above, see e.g. \cite{Barashkov,DH}.

While some of the properties described above arise only in the infinite volume limit of the model, we will mainly be focused on the model in finite volume (though see Corollary \ref{cor:infvol} for results about the infinite volume model). Formally, given a nice domain $\Omega\subset \R^2$ and parameters $\mu\in\R$ and $\beta\in(0,8\pi)$, the finite volume model on $\Omega$ is for us a family of probability measures on a suitable space of generalized functions (e.g. $\mathcal D '(\Omega)$, but this will not be important in what follows) of the form \begin{equation}
    \mathbb P_{\mathrm{sG}(\beta,\mu|\Omega)}(d\varphi)=\frac{1}{Z_{\mathrm{sG}(\beta,\mu|\Omega)}}e^{\mu\int_\Omega \opcolon \cos(\sqrt{\beta}\varphi)\opcolon }\mathbb P_{\mathrm{GFF}_\Omega}(d\varphi),
\end{equation}
where $\mathbb P_{\mathrm{GFF}_\Omega}(d\varphi)$ is the law of the Gaussian free field on $\Omega$ with zero Dirichlet boundary conditions on $\partial \Omega$,  $\opcolon \cdot\opcolon $ denotes Wick ordering, and $Z_{\mathrm{sG}(\beta,\mu|\Omega)}$ is a normalization constant (the partition function of the model). One could also study other types of boundary conditions (for example, the ones in \cite{BaWe,BMW} are different), but for the connection to the Ising model (in a simply connected domain with plus boundary conditions), these turn out to be the relevant ones.

The most interesting objects in the model (that have meaningful infinite volume limits, connections to the Coulomb gas, and which are related to exact solvability and bosonization) are correlation functions of the form
\begin{align}
\left\langle \prod_{i\in Z}\partial \varphi(z_i)\prod_{j\in W}\bar\partial \varphi(w_j)\prod_{k\in U}\opcolon e^{i\gamma_k\varphi(u_k)}\opcolon \right\rangle_{\mathrm{sG}(\beta,\mu|\Omega)},
\end{align}
where $\gamma_k\in \R$ for each $k$.

We again use similar shorthand notation as we did for the Ising model and write this as 
\begin{align}
\left\langle (\partial \varphi)_Z(\bar\partial \varphi)_W(\opcolon e^{i\gamma\varphi}\opcolon )_U\right\rangle_{\mathrm{sG}(\beta,\mu|\Omega)}.
\end{align}
Note that we are abusing notation here as $\gamma$ also depends on the index in the product over $U$.

In fact, for the connection to the Ising model, we will consider Wick ordered trigonometric functions of the field rather than the Wick ordered complex exponentials. These are related as follows:
\begin{align*}
    \opcolon \cos(\gamma \varphi(x))\opcolon  =\frac{1}{2}(\opcolon e^{i\gamma\varphi(x)}\opcolon +\opcolon e^{-i\gamma\varphi(x)}\opcolon ) \qquad \text{and} \qquad     \opcolon \sin(\gamma \varphi(x))\opcolon  =\frac{1}{2i}(\opcolon e^{i\gamma\varphi(x)}\opcolon -\opcolon e^{-i\gamma\varphi(x)}\opcolon ). 
\end{align*}

For $\beta\geq 4\pi$, it turns out that such a direct definition of the model or its correlation functions is not possible, but one has to define the model through a limit of models where the Gaussian free field has been regularized. The particular choice of regularization should not matter too much, but the one we use is detailed in Section \ref{sec:gff}. Also to avoid dealing with complications coming from the boundary of the domain, we find it convenient to replace the constant $\mu$ with a spatially varying function $\rho\in C_c^\infty(\Omega)$.  Moreover, while the sine-Gordon model is an interesting model for all values of $\beta\in(0,8\pi)$ ($8\pi$ is the point where the Kosterlitz-Thouless transition happens), the connection to the Ising model occurs only at $\beta=4\pi$, so we focus on this case. We will also drop the $\beta$ from the notation $\langle\cdot \rangle_{\mathrm{sG}(\beta,\mu|\Omega)}$ and simply write $\langle\cdot \rangle_{\mathrm{sG}(\mu|\Omega)}=\langle\cdot \rangle_{\mathrm{sG}(4\pi,\mu|\Omega)}$. Similarly not all observables of the form $\opcolon \cos(\gamma\varphi)\opcolon $ or $\opcolon \sin(\gamma\varphi)\opcolon $ correspond to Ising primary fields -- we only study those with $\gamma\in \{\sqrt{\pi}, \sqrt{4\pi}\}$ (though as discussed in \cite{BMW}, also other values of $\gamma$ have meaningful fermionic interpretations for the complex exponentials). As a final comment about these correlation functions, we mention that for $\beta=4\pi$, the correlation functions of $\opcolon \cos( \sqrt{4\pi}\varphi)\opcolon $ are not well defined, but require further renormalization in addition to Wick ordering. To summarize this discussion, the correlation functions we study in this article are 
\begin{multline}\label{eq:sgcfdef}
\left\langle (\opcolon \cos(\sqrt{\pi}\varphi)\opcolon )_V (\opcolon  \sin(\sqrt{\pi}\varphi)\opcolon )_U (\partial \varphi)_Z(\bar\partial \varphi)_W(\optypecolon \cos(\sqrt{4\pi}\varphi)\optypecolon )_A\right\rangle_{\mathrm{sG}(\rho|\Omega)}\\
\shoveleft\quad=\lim_{\epsilon\to 0}\left\langle (\opcolon \cos(\sqrt{\pi}\varphi_\epsilon)\opcolon )_V (\opcolon  \sin(\sqrt{\pi}\varphi_\epsilon)\opcolon )_U (\partial \varphi_\epsilon)_Z(\bar\partial \varphi_\epsilon)_W(\optypecolon \cos(\sqrt{4\pi}\varphi_\epsilon)\optypecolon )_A\right\rangle_{\mathrm{sG}(\rho|\epsilon,\Omega)}\\
\shoveleft\quad:=\lim_{\epsilon\to 0}\frac{1}{\left\langle e^{\int_\Omega d^2v \rho(v)\opcolon \cos(\sqrt{4\pi}\varphi_\epsilon(v))\opcolon }\right\rangle_{\mathrm{GFF}_\Omega(\epsilon)}}\Big\langle(\opcolon \cos(\sqrt{\pi}\varphi_\epsilon)\opcolon )_V(\opcolon \sin(\sqrt{\pi}\varphi_\epsilon)\opcolon )_U\\
\times(\partial \varphi_\epsilon)_Z(\bar\partial \varphi_\epsilon)_W(\optypecolon \cos(\sqrt{4\pi}\varphi_\epsilon)\optypecolon )_A e^{\int_\Omega d^2v \rho(v)\opcolon \cos(\sqrt{4\pi}\varphi_\epsilon(v))\opcolon }\Big\rangle_{\mathrm{GFF}_\Omega(\epsilon)},
\end{multline}
where $\varphi_\epsilon$ is the regularized Gaussian free field,  $\langle \cdot\rangle_{\mathrm{GFF}_\Omega(\epsilon)}$ denotes integration with respect to its law (see Section \ref{sec:gff}), 
\begin{equation*}
\opcolon \cos(\alpha \varphi_\epsilon(x))\opcolon =c_\alpha \epsilon^{-\frac{\alpha^2}{4\pi}}\cos(\alpha \varphi_\epsilon(x)) \qquad \text{and} \qquad 
\opcolon \sin(\alpha \varphi_\epsilon(x))\opcolon =c_\alpha \epsilon^{-\frac{\alpha^2}{4\pi}}\sin(\alpha \varphi_\epsilon(x)),
\end{equation*}
where $c_\alpha$ is defined in \eqref{eq:wick}, and 
\begin{align}
\optypecolon \cos(\sqrt{4\pi}\varphi_\epsilon(x))\optypecolon =\opcolon \cos( \sqrt{4\pi}\varphi_\epsilon(x))\opcolon -\int_\Omega d^2 y\,\rho(y)\mathcal A(x,y|\epsilon),
\end{align}
where 
\begin{multline}
    \mathcal A(x,y|\epsilon)=\left\langle\opcolon \cos (\sqrt{4\pi}\varphi_\epsilon(x))\opcolon  \opcolon \cos (\sqrt{4\pi}\varphi_\epsilon(y))\opcolon  \right\rangle_{\mathrm{GFF}_\Omega(\epsilon)}\\
    -\left\langle\opcolon \cos (\sqrt{4\pi}\varphi_\epsilon(x))\opcolon   \right\rangle_{\mathrm{GFF}_\Omega(\epsilon)}\left\langle \opcolon \cos (\sqrt{4\pi}\varphi_\epsilon(y))\opcolon  \right\rangle_{\mathrm{GFF}_\Omega(\epsilon)}.
\end{multline}
As we will later argue, $\mathcal A(x,y|\epsilon)$ is not integrable in the $\epsilon\to 0$ limit, so $\optypecolon \cdot\optypecolon $ amounts to subtracting an ``infinite constant'' when compared to the usual Wick ordering $\opcolon \cdot\opcolon $, and this is necessary to construct non-trivial correlation functions.

Existence of such limiting correlation functions is one of the main results of our analysis of the sine-Gordon model in this article as presented in Theorem \ref{th:sgmain}, though note that there we prove existence of a limit in the sense of generalized functions. In the proof of Theorem \ref{th:main}, we then argue that these limiting correlation functions are actually functions and not just generalized functions. 

\subsection{Main results}

Having given an informal definition of the correlation functions in both models (with precise definitions in Section \ref{sec:sgcorr} and Section \ref{sec:ising-analysis}), we can now state our main bosonization result. Briefly put, it states that the correlation functions of the two models agree when we use the dictionary given in Table \ref{tab:cfcorr}.

\begin{table}[ht]
    \centering
\begin{tabular}{|c|c|}
\hline
Double Ising & Sine-Gordon \\
\hline
$\alpha$ & $\rho=-\frac{2}{\pi}\alpha$\\
$\sigma\tilde \sigma$&$\sqrt{2}\opcolon \cos(\sqrt{\pi}\varphi)\opcolon $\\
$\mu\tilde\mu$&$\sqrt{2}\opcolon \sin(\sqrt{\pi}\varphi)\opcolon $\\
$\epsilon+\tilde\epsilon$&$ 2\optypecolon \cos(\sqrt{4\pi}\varphi)\optypecolon $\\
$\psi\tilde\psi$&$4i \sqrt{\pi}\partial \varphi$\\
$\psi^\star\tilde \psi^\star$&$-4i\sqrt{\pi}\bar\partial \varphi$\\
\hline
\end{tabular}
    \caption{The double Ising -- sine-Gordon correspondence.}
    \label{tab:cfcorr}
\end{table}

The precise statement is as follows. 
\begin{theorem}\label{th:main}
For $\Omega\subset \R^2$ a bounded, open, simply connected set with smooth boundary and $\alpha\in C_c^\infty(\Omega)$, the correlation functions of the double Ising model \eqref{eq:doublecf} and sine-Gordon model \eqref{eq:sgcfdef} exist and satisfy 
\begin{align*}
    &\left\langle (\sigma\widetilde \sigma)_{V} (\mu\widetilde \mu)_{U}(\psi\widetilde \psi)_{Z} (\psi^* \widetilde \psi^*)_{W}(\epsilon+\widetilde\epsilon)_{A}\right\rangle_{\alpha;\Omega}\\
    &=\left\langle (\sqrt{2}\opcolon \cos\sqrt{\pi}\varphi\opcolon )_V (\sqrt{2}\opcolon  \sin\sqrt{\pi}\varphi\opcolon )_U (4i \sqrt{\pi}\partial \varphi)_Z(-4i\sqrt{\pi}\bar\partial \varphi)_W(2\optypecolon \cos\sqrt{4\pi}\varphi\optypecolon )_A\right\rangle_{\mathrm{sG}(-\frac{2\alpha}{\pi}|\Omega)}.
\end{align*}
\end{theorem}
The proof of this theorem relies on two key results: one concerning (double) Ising correlation functions and one sine-Gordon correlation functions. We will state these shortly, and then, using them, give a brief proof of Theorem \ref{th:main}. First however, we state two corollaries of Theorem \ref{th:main}.

Our first corollary concerns the situation where $\alpha$ tends to a constant. In this situation, we can identify our (double) Ising correlation functions with scaling limits of correlation functions of the discrete model (constructed in \cite{P}).

\begin{corollary}\label{cor:constantm}
For $m\in \R$, the following holds:
\begin{equation*}
    \langle \sigma_V\mu_U\psi_Z\psi^*_W\epsilon_A\rangle_{m;\Omega}=\lim_{\alpha\to m}\langle \sigma_V\mu_U\psi_Z\psi^*_W\epsilon_A\rangle_{\alpha;\Omega},
\end{equation*}
 where $\alpha\to m$ means pointwise convergence of functions $\alpha$ to $m$, with the condition that $\alpha\in C_c^\infty(\Omega)$ such that $|\alpha| \leq \kappa_\infty$ and $\operatorname{dist}(z,\partial\Omega)^2|\nabla \alpha(z)|\leq \kappa_b$ for some fixed $\kappa_\infty, \kappa_b>0$.

Consequently, under the same assumptions on $\Omega$ as in Theorem \ref{th:main}, and the above assumptions on $\alpha$, the limit
\begin{align*}
    &\left\langle (\sqrt{2}\opcolon \cos\sqrt{\pi}\varphi\opcolon )_V (\sqrt{2}\opcolon  \sin\sqrt{\pi}\varphi\opcolon )_U (4i \sqrt{\pi}\partial \varphi)_Z(-4i\sqrt{\pi}\bar\partial \varphi)_W(2\optypecolon \cos\sqrt{4\pi}\varphi\optypecolon )_A\right\rangle_{\mathrm{sG}(-\frac{2m}{\pi}|\Omega)}\\
    &=\!\lim_{\alpha\to m}\!\Big\langle (\sqrt{2}\opcolon \cos\sqrt{\pi}\varphi\opcolon )_V (\sqrt{2}\opcolon  \sin\sqrt{\pi}\varphi\opcolon )_U (4i\sqrt{\pi}\partial\varphi)_Z (-4i\sqrt{\pi}\bar\partial\varphi)_W (2\optypecolon \cos\sqrt{4\pi}\varphi\optypecolon )_A \Big\rangle_{\mathrm{sG}(-\frac{2\alpha}{\pi}|\Omega)}
\end{align*}
exists, and in the specific case $m<0$:
\begin{align*}
    \left\langle \big(\sqrt{2}\opcolon \cos\sqrt{\pi}\varphi\opcolon \big)_V\right\rangle_{\mathrm{sG}(-\frac{2m}{\pi}|\Omega)}=\left\langle \sigma_V\right\rangle_{m;\Omega}^2,
\end{align*}
with the spin correlation functions agreeing with those constructed as scaling limits in \cite{P, CIM23}.
\end{corollary}
\begin{proof}
The convergence under $\alpha \to m$ is essentially identical to the proof of Lemma \ref{lem:integral-convergence} (basically, using dominated convergence on the summands involving $\alpha$), after series expansion of general correlations have been justified (see also Remark \ref{rem:usual-conditions} and Section \ref{sec:proof-isingmain}, that the radius of convergence is uniform as long as the assumptions on $\alpha$ are kept). For the constant mass case, see Proposition \ref{prop:spin}.
\end{proof}

Our second result concerns the infinite volume limit $\Omega\to \R^2$. While there should be a similar closed system of equations obeyed by sine-Gordon correlations corresponding to $n$-point pure spin or disorder correlations at any $m$ (see \cite{P3} and references therein), for the sake of complete yet concise presentation, we focus on the two-point pure spin or disorder correlation functions and their connection to Painlevé equations.
\begin{corollary}\label{cor:infvol}
    Suppose $m<0$ and let $\{u_1,u_2\}\subset B_R=\{x\in\R^2:|x|<R\}$ for large enough $R$ with $U=\{1, 2\}$. The following limits
    \begin{align*}
    \mathfrak C(|u_1-u_2|):=\left\langle (\sqrt{2}\opcolon \cos(\sqrt{\pi}\varphi)\opcolon )_U\right\rangle_{\mathrm{sG}(-\frac{2m}{\pi}|\C)}=\lim_{R\to\infty}\left\langle (\sqrt{2}\opcolon \cos(\sqrt{\pi}\varphi)\opcolon )_U\right\rangle_{\mathrm{sG}(-\frac{2m}{\pi}|B_R)},\\
    \mathfrak S(|u_1-u_2|):=\left\langle (\sqrt{2}\opcolon \sin(\sqrt{\pi}\varphi)\opcolon )_U\right\rangle_{\mathrm{sG}(-\frac{2m}{\pi}|\C)}=\lim_{R\to\infty}\left\langle (\sqrt{2}\opcolon \sin(\sqrt{\pi}\varphi)\opcolon )_U\right\rangle_{\mathrm{sG}(-\frac{2m}{\pi}|B_R)},
\end{align*}
exist, and may be written
\begin{align*} 
\mathfrak C(r) &= \mathcal{C}_\sigma^{-4}(8|m|)^{1/2}\cosh^2 h_0(|m|r) \exp \left(2\int_{\infty}^{|m|r} \left(\frac{ (h'_0(s))^2}{4}-4 \sinh^2h_0(s)\right) ds\right),\\
\mathfrak S(r) &=\mathcal{C}_\sigma^{-4}(8|m|)^{1/2}\sinh^2 h_0(|m|r) \exp\left(2 \int_{\infty}^{|m|r} \left(\frac{ (h'_0(s))^2}{4}-4 \sinh^2h_0(s)\right) ds\right),
\end{align*}
where $\eta_0=\exp(-2h_0)$ satisfies the Painlev\'e III equation $\eta_0'' = \eta_0^{-1}(\eta'_0)^2 - r^{-1}\eta'_0+ \eta_0^3 - \eta_0^{-1}$ and has the asymptotic $\eta_0(\theta) \sim 1- \frac{ 2}{\pi}K_0(2\theta)$ as $\theta \to \infty$, where $K_\nu$ is the modified Bessel function of the second kind.
\end{corollary}
\begin{proof}
    For the infinite volume limit, see e.g. \cite[the proof of Corollary 4.21]{CIM23} for the Ising correlation, and the sine-Gordon correlation is simply its square. Although in \cite{CIM23} the result is stated in terms of the spin correlation for $m>0$ with free boundary conditions, it is elementary to state it in the $m<0$ setting with plus boundary conditions (which is the case for our continuous correlations model) using Kramers-Wannier duality, e.g. \cite[Remark 4.10, Theorem 4.20]{CIM23}. Then the expression in terms of the Painlev\'e III transcendent is a celebrated result from \cite{WMTB}, see also \cite[Section 4.2]{P} or \cite{P3} for a recent exposition.
\end{proof}

One may wish to compare these results to the main results of \cite{BMW}, where sine-Gordon correlation functions of $\opcolon e^{i\sqrt{4\pi}\gamma\varphi}\opcolon $ with $|\gamma|<1/2$ were constructed, proven to be related to those of so-called twisted fermions, and to satisfy a differential equation related to Painlevé equations as well. A significant difference is that our correlation functions correspond to the boundary case $\gamma=\pm1/2$ for which the fermionic analysis in \cite{BMW} breaks down. Also our analysis on the Ising side relies on entirely different ideas than the fermionic analysis in \cite{BMW}.

We now turn to stating the main tools we need for proving Theorem \ref{th:main}. We begin with the main statement on the Ising side. Note that this result is valid for more general $\alpha$ than in Theorem \ref{th:main}.
\begin{theorem}\label{th:isingmain}
    \begin{enumerate}
        \item For a smooth, bounded, simply connected domain $\Omega \subset \mathbb C$ and $\alpha:\Omega \to \mathbb R$ smooth up to the boundary, the Ising correlation functions
        \begin{equation*}
            \left\langle \sigma_V\mu_U \psi_Z \psi^*_W\epsilon_A\right\rangle_{\alpha;\Omega}
        \end{equation*}
        exist and are locally uniformly bounded, in the sense that their values are bounded by a constant only depending on $\Omega$, $\alpha$, the number of insertion points, and the minimum distance among them and the boundary. Consequently, the double Ising model correlation functions $$\left\langle (\sigma\widetilde \sigma)_V(\mu\widetilde \mu)_U (\psi\widetilde\psi)_Z (\psi^*\widetilde \psi^*)_W(\epsilon+\widetilde \epsilon)_A\right\rangle_{\alpha;\Omega}$$ exist and are locally uniformly bounded as well.
        \item The correlations $\left\langle \sigma_V\mu_U \psi_Z \psi^*_W\epsilon_A\right\rangle_{m\alpha;\Omega}$ are analytic as functions of $m$ in a neighborhood of $\R$ (for each $m_0\in \R$, there exists a convergent Taylor series expansion for $m$ in some neighborhood of $m_0$) and are continuous functions of the insertion points indexed by $U, V, Z, W, A$, away from the diagonal and the boundary.
        Again, this translates immediately to double Ising correlation functions. 
        \item Around $m=0$, we have the Taylor expansions
        \begin{multline}
        \langle \sigma_V\widetilde \sigma_V \mu_U\widetilde \mu_U (\psi\widetilde \psi)_Z(\psi^*\widetilde\psi^*)_W(\epsilon+\widetilde\epsilon)_A\rangle_{m\alpha;\Omega}\\
        \shoveleft=\sum_{r=0}^\infty \left(-\frac{m}{\pi}\right)^{r}\frac{1}{r!}\int_{\Omega^{r}}d^{2r}x  \prod_{j=1}^r \alpha(x_j)\sum_{I\subset A}\sum_{\substack{K\subset [r]:\\ |K|=|I|}}\sum_{\nu:I\leftrightarrow K}\prod_{i\in I}\left(-\langle \epsilon_{a_i}+\widetilde \epsilon_{a_i};\epsilon_{x_{\nu(i)}}+\widetilde \epsilon_{x_{\nu(i)}}\rangle_0^\mathsf T\right)\\
        \times \left\langle \sigma_V\widetilde \sigma_V\mu_U \widetilde \mu_U(\psi\widetilde \psi)_Z(\psi^*\widetilde\psi^*)_W(\epsilon+\widetilde\epsilon)_{A\setminus I};(\epsilon+\widetilde \epsilon)_{x_{b(1)}};\dots;(\epsilon+\widetilde \epsilon)_{x_{b(r-|K|)}}\right\rangle_{0;\Omega}^{\mathsf T}
        \end{multline}
        where $\nu:I\leftrightarrow K$ is a bijection from $I$ to $K$, $b(1),\dots,b(r-|K|)$ is an enumeration of $\{1,\dots,r\}\setminus K$, and $\langle \cdots \rangle_{0;\Omega}^\mathsf T$ denotes a ``cumulant'' in that it is defined through the moments to cumulants map from the Ising correlation functions:
        \begin{equation}\label{eq:mtoc}
            \langle X_1;\dots;X_n\rangle^\mathsf T=\sum_{\lambda\in \Pi_n}(-1)^{|\lambda|-1}(|\lambda|-1)!\prod_{B\in \lambda}\bigg\langle \prod_{b\in B}X_b\bigg\rangle,
        \end{equation}
        where $\Pi_{n}$ denotes the set of partitions of $[n]=\{1,\dots,n\}$, and $|\lambda|$ the number of parts of the partition $\lambda\in \Pi_n$.
    \end{enumerate}
\end{theorem}
A useful way of writing the cumulant that we use in our analysis of these correlation functions (and is readily checked from \eqref{eq:mtoc}) is 
\begin{align}\label{eq:mtoc2}
    &\left\langle \sigma_V\widetilde  \sigma_V\mu_U \widetilde \mu_U\Psi_Z\widetilde \Psi_Z (\epsilon+\widetilde \epsilon)_{W\setminus I};(\epsilon+\widetilde \epsilon)_{x_{b(1)}};\dots;(\epsilon+\widetilde \epsilon)_{x_{b(r-|K|)}}\right\rangle_{0;\Omega}^\mathsf T\\
    \nonumber&\quad =\sum_{C\subset [r]\setminus K}\bigg\langle \sigma_V \widetilde \sigma_V\mu_U\widetilde \mu_U \Psi_Z\widetilde \Psi_Z (\epsilon+\widetilde \epsilon)_{W\setminus I}\prod_{j\in C}(\epsilon+\widetilde \epsilon)_{x_j}
    \bigg\rangle_{0;\Omega}\\
    \nonumber&\qquad\times\sum_{\lambda\in \Pi_{[r]\setminus (K\cup C)}}(-1)^{|\lambda|}|\lambda|!\prod_{B\in \lambda}\bigg\langle \prod_{k\in B}(\epsilon+\widetilde \epsilon)_{x_k}
    \bigg\rangle_{0;\Omega},
\end{align}
where for a set $A$, $\Pi_A$ denotes the set of partitions of $A$ and $\Psi_Z=\prod_{i\in Z}\Psi_{z_i}$, where $\Psi_{z_i}$ stands for $\psi_{z_i}$ or $\psi_{z_i}^*$.

We choose to state the explicit expansion only for the doubled correlation, since it is the one used directly in the proof of bosonization; similar techniques as in the proof of Theorem~\ref{th:isingmain} (namely, a combination of Proposition \ref{pr:spinexp} and Lemma \ref{le:enest}) should also yield an analogous explicit expansion for the correlation $\left\langle \sigma_V\mu_U \psi_Z \psi^*_W\epsilon_A\right\rangle_{m\alpha;\Omega}$.

On the sine-Gordon side, the corresponding result is of the following form. 
\begin{theorem}\label{th:sgmain}
    Let $\Omega$ be as in Theorem \ref{th:main} and  $f_1,\dots,f_p$, $f_1',\dots,f_{p'}'$, $g_1,\dots,g_q$, $g_1',\dots,g_{q'}'$, $h_1,\dots,h_r$ $\in C_c^\infty(\Omega)$ with disjoint supports.
\begin{enumerate}
    \item For $\rho\in C_c^\infty(\Omega)$, the smeared correlation functions 
    \begin{multline*}
        \left\langle \prod_{j=1}^p \partial \varphi(f_j) \prod_{j'=1}^{p'} \bar\partial \varphi(f_{j'}')\prod_{k=1}^q \!\opcolon  \cos(\sqrt{\pi}\varphi)\opcolon (g_k) \prod_{k'=1}^{q'} \!\opcolon \sin(\sqrt{\pi}\varphi)\opcolon (g_{k'}') \prod_{l=1}^r \optypecolon \cos(\sqrt{4\pi}\varphi)\optypecolon (h_l) \right\rangle_{\mathrm{sG}(\rho|\Omega)}\\
        \shoveleft\quad=\lim_{\epsilon\to 0}\int_{\Omega^{p+p'+q+q'+r}}d^{2p}z d^{2p'}w d^{2q}ud^{2q'}v d^{2r}a \prod_{j=1}^p(-\partial f_j(z_j))\prod_{j'=1}^{p'}(-\bar \partial f_{j'}'(w_{j'}))\prod_{k=1}^q g_k(u_k)\\
        \shoveleft\qquad\times \prod_{k'=1}^{q'}g_{k'}'(v_{k'})\prod_{l=1}^r h_l(a_l)\Bigg\langle \prod_{j=1}^p \varphi_\epsilon(z_j)\prod_{j'=1}^{p'}\varphi_\epsilon(w_{j'})\prod_{k=1}^q \opcolon \cos(\sqrt{\pi}\varphi_\epsilon(u_k))\opcolon \\
        \times \prod_{k'=1}^{q'}\opcolon \sin(\sqrt{\pi}\varphi_\epsilon(v_{k'}))\opcolon \prod_{l=1}^r \optypecolon \cos(\sqrt{4\pi}\varphi_\epsilon(a_l))\optypecolon  \Bigg\rangle_{\mathrm{sG}(\rho|\epsilon,\Omega)}
    \end{multline*}
    exist.
    \item For each $\rho\in C_c^\infty(\Omega)$, there exists a complex neighborhood of the real axis into which the function 
    \begin{equation*}
    \mu\!\mapsto\!\Bigg\langle \!\prod_{j=1}^p \!\partial \varphi(f_j)\! \prod_{j'=1}^{p'} \!\bar\partial \varphi(f_{j'}')\!\prod_{k=1}^q \!\opcolon  \cos(\sqrt{\pi}\varphi)\opcolon (g_k)\! \prod_{k'=1}^{q'}\!\opcolon \sin(\sqrt{\pi}\varphi)\opcolon (g_{k'}')\! \prod_{l=1}^r \!\optypecolon \cos(\sqrt{4\pi}\varphi)\optypecolon (h_l)\! \Bigg\rangle_{\mathrm{sG}(\mu\rho|\Omega)}
    \end{equation*}
    has an analytic continuation.
    \item The Taylor coefficients at zero of this analytic function are given by  
    \begin{multline*}
        \frac{\partial^n}{\partial \mu^n}\bigg|_{\mu=0} \Bigg\langle \prod_{j=1}^p \partial \varphi(f_j) \prod_{j'=1}^{p'} \bar\partial \varphi(f_{j'}')\prod_{k=1}^q \opcolon  \cos(\sqrt{\pi}\varphi)\opcolon (g_k)\\
        \shoveright{\times \prod_{k'=1}^{q'}\opcolon \sin(\sqrt{\pi}\varphi)\opcolon (g_{k'}') \prod_{l=1}^r \optypecolon \cos(\sqrt{4\pi}\varphi)\optypecolon (h_l) \Bigg\rangle_{\mathrm{sG}(\mu\rho|\Omega)}}\\
        =\int_{\Omega^{p+p'+q+q'+r+n}} d^{2p}z d^{2p'}w d^{2q}u d^{2q'}v d^{2r}a d^{2n}b\, \prod_{j=1}^p (-\partial f_j(z_j))\prod_{j'=1}^{p'}(-\bar\partial f_{j'}'(w_{j'}))\prod_{k=1}^qg_k(u_k)\\
        \shoveleft\qquad \times \prod_{k'=1}^{q'} g_{k'}'(v_{k'})\prod_{l=1}^r h_l(a_l)\prod_{m=1}^n\rho(b_m)\sum_{S\subset [r]}\sum_{\substack{S'\subset[n]:\\ |S|=|S'|}} \sum_{\pi:S\leftrightarrow S'}\prod_{s\in S}(-\mathcal A(a_{s},b_{\pi(s)}))\\
        \shoveleft\qquad \times \Bigg\langle \prod_{j=1}^p \varphi(z_j)\prod_{j'=1}^{p'}\varphi(w_{j'})\prod_{k=1}^q \opcolon \cos(\sqrt{\pi}\varphi(u_k))\opcolon  \prod_{k'=1}^{q'}\opcolon \sin(\sqrt{\pi}\varphi(v_{k'}))\opcolon  \\
        \prod_{l\notin S} \opcolon \cos(\sqrt{4\pi}\varphi(a_l))\opcolon ; \opcolon \cos(\sqrt{4\pi}\varphi(b_{\alpha_1}))\opcolon ;\dots;\opcolon \cos(\sqrt{4\pi}\varphi(b_{\alpha_{n-|S'|}}))\opcolon \Bigg\rangle_{\mathrm{GFF}_\Omega}^{\mathsf T},
    \end{multline*}
    where $\{1,\dots,n\}\setminus S'=\{\alpha_1,\dots,\alpha_{n-|S'|}\}$, $\mathcal A(x,y)=\lim_{\epsilon\to 0}\mathcal A(x,y|\epsilon)$ (existence of this limit is justified by Lemma~\ref{lem:gffcorr2}), the sum over $\pi$ is over bijections between $S$ and $S'$, and $\langle A_1;\dots;A_n\rangle^{\mathsf T}$ denotes a cumulant, which in this statement is defined through the usual moments to cumulants map and Lemma~\ref{lem:gffcorr2}. Part of the statement is that the integrand is an integrable function.
\end{enumerate}
\end{theorem}

Using these two results, we can now prove Theorem \ref{th:main}. A key fact that we use here is the bosonization dictionary of Table \ref{tab:cfcorr} (with $\optypecolon \cos(\sqrt{4\pi}\varphi)\optypecolon $ replaced by $\opcolon \cos(\sqrt{4\pi}\varphi)\opcolon $) holds for the critical Ising model due to \cite{BIVW}. The setup there allows for multiply connected domains with mixed boundary conditions; see \cite[Proposition 2.15, Remark 2.16]{BIVW} for an explanation that the main theorem is valid for our case, i.e. simply connected domain with plus boundary conditions. The precise fact we need is as follows.\footnote{There is a minor typo in \cite[Remark 2.11]{BIVW}. It is stated there that $\frac{1}{2}(\epsilon+\widetilde \epsilon)$ corresponds to $2\opcolon\cos(\sqrt{2}\Phi)\opcolon$, when it should be just $\opcolon\cos(\sqrt{2}\Phi)\opcolon$, as stated in Proposition \ref{pr:massless}. Note that $\Phi$ in \cite{BIVW} is our $\sqrt{2\pi}\varphi$. To check that this is indeed a typo, we offer here a brief argument (essentially the one of \cite[Remark 2.11]{BIVW}) -- the reader can readily make this precise by viewing these identities as holding within appropriate correlation functions. For $\gamma=\frac{1}{\sqrt{2}}$, \cite[(2.20)]{BIVW} yields \[\opcolon\cos(\tfrac{1}{\sqrt{2}}\Phi(z))\opcolon \opcolon\cos(\tfrac{1}{\sqrt{2}}\Phi(w))\opcolon=\tfrac{1}{2}|z-w|^{-1/2}(1+\opcolon\cos(\sqrt{2}\Phi(w))\opcolon|z-w|+O(|z-w|^2))\]
On the other hand,  the Ising OPE \cite[(6.12)]{CHI2} implies that
\[
\sigma_z\sigma_w=|z-w|^{-1/4}(1+\tfrac{1}{2}\epsilon_w|z-w|+o(|z-w|))
\]
so using the bosonization identity for spin-observables, $\sigma\widetilde\sigma=\sqrt{2}\opcolon\cos(\frac{1}{\sqrt{2}}\Phi)\opcolon$, we have on the one hand
\[
\sigma_z\widetilde \sigma_z \sigma_w\widetilde \sigma_w=|z-w|^{-1/2}(1+\tfrac{1}{2}(\epsilon_w+\widetilde \epsilon_w)|z-w|+o(|z-w|))
\]
and on the other hand 
\[
\sigma_z\widetilde \sigma_z \sigma_w\widetilde \sigma_w=\sqrt{2}\opcolon\cos(\tfrac{1}{\sqrt{2}}\Phi(z))\opcolon \sqrt{2}\opcolon\cos(\tfrac{1}{\sqrt{2}}\Phi(w))\opcolon=|z-w|^{-1/2}(1+\opcolon\cos(\sqrt{2}\Phi(w))\opcolon|z-w|+O(|z-w|^2)).
\]
Equating coefficients gives the correct identity.
}
\begin{proposition}[{\cite[Theorem 1.1, Remark 2.11]{BIVW}}]\label{pr:massless}
We have
\begin{align*}
    &\left\langle \sigma_V\widetilde \sigma_V\mu_U\widetilde \mu_U \psi_Z\widetilde \psi_Z \psi_W^* \widetilde{\psi}_W^* (\epsilon+\widetilde \epsilon)_X\right\rangle_{0;\Omega}\\
    &=\left\langle (\sqrt{2}\opcolon \cos(\sqrt{\pi}\varphi)\opcolon )_V(\sqrt{2}\opcolon \sin(\sqrt{\pi}\varphi)\opcolon )_U(4i \sqrt{\pi}\partial \varphi)_Z(-4i\sqrt{\pi}\bar\partial \varphi)_W(2\opcolon \cos(\sqrt{4\pi}\varphi)\opcolon )_X\right\rangle_{\gff_\Omega}.
\end{align*}
\end{proposition}

\begin{proof}[Proof of Theorem \ref{th:main}]
Existence of the Ising correlation functions follows from item (1) of Theorem \ref{th:isingmain} while the existence of the sine-Gordon correlation functions (in the sense of distributions) follows from item (2) of Theorem \ref{th:sgmain}.
    
By items (1) and (2) of Theorem \ref{th:isingmain}, for $f_1,\dots,f_p$, $f_1',\dots,f_{p'}'$, $g_1,\dots,g_q$, $g_1',\dots,g_{q'}'$, $h_1,\dots,h_r\allowbreak\in C_c^\infty(\Omega)$ with disjoint supports and $\alpha\in C_c^\infty(\Omega)$ (whose support need not be disjoint from the other ones) the functions 
\begin{multline*}
    m\mapsto \int_{\Omega^{p+p'+q+q'+r}}d^{2p}zd^{2p'}w d^{2q}v d^{2q'}ud^{2r}a \prod_{j=1}^p f_j(z_j)\prod_{j'=1}^{p'}f'_{j'}(w_{j'})\prod_{k=1}^q g_k(v_k)\prod_{k'=1}^{q'}g_{k'}'(u_{k'})\\
    \times \prod_{l=1}^r h_l(a_l)\left\langle (\sigma\widetilde \sigma)_{V} (\mu\widetilde \mu)_{U}(\psi\widetilde \psi)_{Z} (\psi^* \widetilde \psi^*)_{W}(\epsilon+\widetilde\epsilon)_{A}\right\rangle_{m\alpha;\Omega}.
\end{multline*}
are well defined for $m\in \R$ and analytic in $m$ in a neighborhood of $\R$. Moreover, by item (3) of Theorem \ref{th:isingmain}, the $n$th Taylor coefficient of this analytic function at zero is 
\begin{multline*}
    \left(-\frac{1}{\pi}\right)^n\int_{\Omega^{p+p'+q+q'+r+n}}d^{2p}zd^{2p'}w d^{2q}v d^{2q'}ud^{2r}a d^{2n}b \prod_{j=1}^p f_j(z_j)\prod_{j'=1}^{p'}f'_{j'}(w_{j'})\prod_{k=1}^q g_k(v_k)\\
    \times \prod_{k'=1}^{q'}g_{k'}'(u_{k'})\prod_{l=1}^r h_l(a_l)\prod_{m=1}^n \alpha(b_m)\sum_{S\subset [r]}\sum_{\substack{S'\subset[n]:\\ |S|=|S'|}} \sum_{\pi:S\leftrightarrow S'}\prod_{s\in S}\left(-\left\langle \epsilon_{a_s}+\widetilde \epsilon_{a_s};\epsilon_{b_{\pi(s)}}+\widetilde \epsilon_{b_{\pi(s)}}\right\rangle_{0;\Omega}^\mathsf T\right)\\
    \times \left\langle (\sigma\widetilde \sigma)_{V} (\mu\widetilde \mu)_{U}(\psi\widetilde \psi)_{Z} (\psi^* \widetilde \psi^*)_{W}(\epsilon+\widetilde\epsilon)_{A\setminus S};\epsilon_{b_{\alpha_1}}+ \widetilde\epsilon_{b_{\alpha_1}};\dots;\epsilon_{b_{\alpha_{n-|S'|}}}+\widetilde\epsilon_{b_{\alpha_{n-|S'|}}}\right\rangle_{0;\Omega}^\mathsf T.
\end{multline*}
 We now make use of Proposition \ref{pr:massless} which says that the critical Ising correlation functions appearing in this series expansion can be expressed in terms of GFF correlation functions using Table \ref{tab:cfcorr} (with $\optypecolon \cos(\sqrt{4\pi}\varphi)\optypecolon $ replaced by $\opcolon \cos(\sqrt{4\pi}\varphi)\opcolon $). A couple of remarks are in order here. First of all, \cite{BIVW} deals with usual correlation functions (or moments) instead of cumulants, but as cumulants can be expressed in terms of moments (by the usual moments to cumulants map \eqref{eq:mtoc}), we can also use the bosonization dictionary for cumulants. Secondly, the GFF correlation functions defined in \cite[Section 2]{BIVW} are a priori defined differently from ours (in Section \ref{sec:gff}), but in Lemma \ref{le:gffequiv}, we argue that they are indeed the same functions.

This means that for $m$ in our neighborhood of the origin, we can write this Taylor coefficient as 
\begin{align*}
    &\left(-\frac{1}{\pi}\right)^n\!\int_{\Omega^{p+p'+q+q'+r+n}}d^{2p}zd^{2p'}w d^{2q}v d^{2q'}ud^{2r}a d^{2n}b \\
    &\qquad\prod_{j=1}^p f_j(z_j)\prod_{j'=1}^{p'}f'_{j'}(w_{j'})\prod_{k=1}^q g_k(v_k)\prod_{k'=1}^{q'}g_{k'}'(u_{k'})\prod_{l=1}^r h_l(a_l)\prod_{m=1}^n \alpha(b_m)\\
    &\qquad\times\sum_{S\subset [r]}\sum_{\substack{S'\subset[n]:\\ |S|=|S'|}} \sum_{\pi:S\leftrightarrow S'}\prod_{s\in S}\left(-\left\langle \big(2\opcolon \cos\sqrt{4\pi}\varphi(a_s)\opcolon \big); \big(2\opcolon \cos\sqrt{4\pi}\varphi(b_{\pi(s)})\opcolon \big)\right\rangle_{\mathrm{GFF}_\Omega}^\mathsf T\right)\\
    &\qquad\times \Big\langle \big(\sqrt{2}\opcolon \cos\sqrt{\pi}\varphi\opcolon \big)_{V}\big(\sqrt{2}\opcolon \sin\sqrt{\pi}\varphi\opcolon \big)_{U}(4i\sqrt{\pi}\partial \varphi)_{Z}(-4i\sqrt{\pi}\bar\partial\varphi)_{W}\\
    &\qquad\qquad\big(2\opcolon \cos\sqrt{4\pi}\varphi\opcolon \big)_{A\setminus S};\big(2\opcolon \cos\sqrt{4\pi}\varphi(b_{\alpha_1})\opcolon \big);\dots;\big(2\opcolon \cos\sqrt{4\pi}\varphi(b_{\alpha_{n-|S'|}})\opcolon \big)\Big\rangle_{\mathrm{GFF}_\Omega}^\mathsf T.
\end{align*}
 By Lemma \ref{le:gffequiv}, we can write 
\begin{align*}
    &\Big\langle \big(\sqrt{2}\opcolon \cos(\sqrt{\pi}\varphi)\opcolon \big)_{V}\big(\sqrt{2}\opcolon \sin(\sqrt{\pi}\varphi)\opcolon \big)_{U}(4i\sqrt{\pi}\partial \varphi)_{Z}(-4i\sqrt{\pi}\bar\partial\varphi)_{W}\\
    &\qquad\big(2\opcolon \cos(\sqrt{4\pi}\varphi)\opcolon \big)_{A\setminus S};\big(2\opcolon \cos(\sqrt{4\pi}\varphi(b_{\alpha_1}))\opcolon \big);\dots;\big(2\opcolon \cos(\sqrt{4\pi}\varphi(b_{\alpha_{n-|S'|}}))\opcolon \big)\Big\rangle_{\mathrm{GFF}_\Omega}^{\mathsf T}\\
    &=\prod_{j=1}^p(4i \sqrt{\pi}\partial_{z_j})\prod_{j'=1}^{p'}(-4i\sqrt{\pi}\bar \partial_{w_{j'}})\Big\langle \big(\sqrt{2}\opcolon \cos(\sqrt{\pi}\varphi)\opcolon \big)_{V}\big(\sqrt{2}\opcolon \sin(\sqrt{\pi}\varphi)\opcolon \big)_{U}\varphi_{Z}\varphi_{W}\\
    &\qquad\big(2\opcolon \cos(\sqrt{4\pi}\varphi)\opcolon \big)_{A\setminus S};\big(\big(2\opcolon \cos(\sqrt{4\pi}\varphi(b_{\alpha_1}))\opcolon \big);\dots;\big(2\opcolon \cos(\sqrt{4\pi}\varphi(b_{\alpha_{n-|S'|}}))\opcolon \big)\Big\rangle_{\mathrm{GFF}_\Omega}^{\mathsf T}.
\end{align*}
Integrating by parts the $z$ and $w$ derivatives (which is allowed by the integrability of both functions due to Theorem \ref{th:isingmain} and Theorem \ref{th:sgmain}), we see by Theorem \ref{th:sgmain} that this is precisely the same as the $n$th Taylor coefficient at zero of the analytic function 
\begin{multline*}
    m\mapsto \bigg\langle \prod_{j=1}^p \big(4i\sqrt{\pi}\partial \varphi(f_j)\big) \prod_{j'=1}^{p'} \big(-4i\sqrt{\pi}\bar\partial \varphi(f_{j'}')\big)\prod_{k=1}^q \big(\sqrt{2}\opcolon  \cos(\sqrt{\pi}\varphi)\opcolon (g_k)\big)\\
    \times \prod_{k'=1}^{q'}\big(\sqrt{2}\opcolon \sin(\sqrt{\pi}\varphi)\opcolon (g_{k'}')\big) \prod_{l=1}^r \big(2\optypecolon \cos(\sqrt{4\pi}\varphi)\optypecolon (h_l)\big) \bigg\rangle_{\mathrm{sG}(-\frac{2}{\pi}m\alpha|\Omega)}
\end{multline*}
Since both this sine-Gordon correlation function and the double Ising correlation function we started with are analytic in a neighborhood of $\R$ and have the same Taylor expansions at the origin, they must agree everywhere on $\R$. 

To conclude, we point out that since we know from Theorem \ref{th:isingmain} that the double Ising correlation functions are in fact continuous functions of the insertion points (away from the diagonal and the boundary) instead of just generalized functions and since the sine-Gordon correlation functions agree with these when integrated against test functions with disjoint supports, we conclude that the sine-Gordon correlation functions can also be understood as being continuous functions instead of just generalized functions.
\end{proof}

\subsection{Structure of the article}

The rest of the article focuses on proving Theorem \ref{th:isingmain} and Theorem \ref{th:sgmain}. The organization is as follows. In Section \ref{sec:gff}, we review the definition and some basic properties of our regularization of the GFF needed in the construction of the sine-Gordon model. In Section \ref{sec:rp}, we introduce and study the renormalized potential of the sine-Gordon model with an iterated Mayer expansion following closely ideas introduced in \cite{BK} and further developed in \cite{BB,BH,BMW,BaWe}. In Section \ref{sec:rpart}, we introduce a renormalized version of the partition function of the sine-Gordon model, and use the results of Section \ref{sec:rp} to control analyticity properties of the renormalized partition function (this follows the analysis of \cite{BaWe} closely). In Section \ref{sec:sgcorr}, we finally use our analysis on the renormalized partition function to construct the sine-Gordon correlation functions and prove Theorem \ref{th:sgmain}.

In Section \ref{sec:ising-analysis}, we turn to analyzing the Ising model and show existence of general correlation functions from the `basic correlation functions', namely spin-weighted two-point fermion-fermion and disorder-fermion correlations. We prove their existence, uniqueness, and various quantitative bounds on them. In Section \ref{sec:ising-combinatorics}, we prove analyticity properties and explicit series expansion in a neighborhood of the origin for the relevant correlation functions. In Section \ref{sec:ising-double}, we study the series expansions of the double Ising correlation functions and finally prove Theorem \ref{th:isingmain} in Section \ref{sec:proof-isingmain}.

\medskip
 
{\bf Acknowledgments:} This work was initiated while S.P. was employed as a Research Fellow at Korea Institute for Advanced Study and supported by a KIAS Individual Grant (MG077202). S.P. thanks Dmitry Chelkak, Kalle Kyt\"ol\"a, and Elliot Blackstone for helpful conversations.

T.V. is grateful for the financial support by the Doctoral Network in Information Technologies and Mathematics at Åbo Akademi University. T.V. was also supported by the ERC grant CONFSTAT funded by the European Union.

C.W. was supported by the Academy of Finland through the grant 348452 and ERC grant CONFSTAT funded by the European Union. Views and opinions expressed are however those of the authors only and do not necessarily reflect those of the European Union or ERC. Neither the European Union nor ERC can be held responsible for them.

S.P. and C.W. wish to thank Kalle Kytölä for putting them in touch -- this article would not have happened without him.

The preprint version of this article is included as an appendix in T.V.'s PhD thesis \cite{Vir}.

\medskip

{\bf AI use:} Gemini 3.1 Pro (Google) and Claude Sonnet 5 (Anthropic) were used to assist in proof reading of the article. In addition Gemini 3.1 Pro (Google) was consulted unsuccessfully for generalizing Lemma \ref{lem:ctest} from the case of a disk to a more general domain. While the approach suggested by AI was flawed, some parts of the discussion have influenced our approach to the proof.

\section{The Gaussian Free Field and its Heat Kernel Regularization}\label{sec:gff}
In this section, we review the regularization for the Gaussian free field that we use in our construction of the correlation functions of the sine-Gordon model. There are various regularizations one could use, but for the analysis of the sine-Gordon model through the renormalized potential, we find it convenient to mimic the construction of \cite{BaWe} and consider a regularization using the heat kernel of the domain $\Omega$ (in \cite{BaWe}, the massive heat kernel of the full plane was used instead).

Much of the material in this section is likely to be familiar to experts at least in some form. We however rely heavily on the heat kernel regularization for our sine-Gordon analysis and to our knowledge, many of the facts we present here do not exist as such in the literature (while very similar results for other regularizations certainly do). 

We begin with a review of some basic estimates involving the heat kernel of a nice domain. We then review the construction of the regularized GFF in terms of this heat kernel as well as some estimates for GFF correlation functions.

\subsection{The heat kernel}
Let us write $\p_\Omega$ for the  transition density of (a temporally rescaled) Brownian motion on $\Omega$ killed at the boundary of $\Omega$. This transition density is of course also the heat kernel on $\Omega$ (with zero boundary conditions). More precisely, if we write $T_\Omega$ for the first exit time from $\Omega$ for (standard planar) Brownian motion,\footnote{If necessary, we will write $T_\Omega^x$ to emphasize that the Brownian motion has been started at $x\in \Omega$.} then there exists a non-negative function $\p_\Omega:(0,\infty)\times\Omega\times \Omega\to [0,\infty)$ such that if $(B(t))_{t\geq 0}$ denotes our Brownian motion and $\Prob_x$ its law when started at $x$, then for each Borel $A\subset \R^2$, each $t>0$ and $x\in \Omega$,
\begin{equation}
\Prob_x\big(\{B(2t)\in A\} \cap \{2t\leq T_\Omega\}\big)=\int_A d^2y \,\p_\Omega(t,x,y).
\end{equation}
As discussed in \cite[Chapter 3.3]{MP}, a basic fact about $\p_\Omega$ is that it can be related to the full plane transition density of Brownian motion. More precisely, if we write 
$$\p(t,x,y)=\p_{\R^2}(t,x,y)=\frac{1}{4\pi t} e^{-\frac{|x-y|^2}{4t}}$$ 
and let
\begin{equation}\label{eq:rt}
R_t(x,y):=\E_x \left[\p(t-\tfrac{1}{2}T_\Omega,B(T_\Omega),y)\1_{\{T_\Omega<2t\}}\right],
\end{equation}
then
\begin{align}\label{eq:tdens}
\p_\Omega(t,x,y)= \p(t,x,y)-R_t(x,y).
\end{align}
A simple but useful consequence is that we have the following estimate
\begin{equation*}
0\leq \p_\Omega(t,x,y)\leq \p(t,x,y).
\end{equation*}
One can readily check from \eqref{eq:rt} and \eqref{eq:tdens} the standard fact that $\p_\Omega$ is really the heat kernel (namely the fundamental solution to the heat equation) of $\Omega$ with zero Dirichlet boundary conditions:
\begin{equation}
\begin{cases}
(\partial_t-\Delta_x)\p_\Omega(t,x,y)&=0, \, \qquad \qquad t>0, x,y\in \Omega\\
\p_\Omega(t,x,y)&=0,  \, \qquad \qquad   t>0, x\in\partial\Omega, y\in \Omega\\
\lim_{t\to 0}\p_\Omega(t,x,y)&=\delta(x,y),  \qquad x,y\in\Omega
\end{cases}.
\end{equation}

We will want to use $\p_\Omega$ to construct a continuous Gaussian process and for this purpose, we will need some regularity estimates for $\p_\Omega$. For some of these, we find it convenient to express $\p_\Omega$ in terms of Laplacian eigenfunctions, but later on, we will also need some regularity estimates for the function $R_t$ that are derived from the probabilistic representation instead of the basis expansion. 

We now discuss how to express $\p_{\Omega}$ in terms of Laplacian eigenfunctions. First of all, it is a basic fact that for $\Omega$ as in Theorem \ref{th:main}, there exists an orthonormal basis $(e_n)_{n=1}^\infty$ of $L^2(\Omega)$ such that, $e_n$ is real, smooth up to the boundary, $-\Delta e_n=\lambda_n e_n$, where $0<\lambda_1\leq \lambda_2\leq \dots$ and $e_n$ vanishes on the boundary. We record here some standard estimates for $\lambda_n$ and $e_n$.

\begin{lemma}\label{lem:evest}
For $\Omega$ as in Theorem \ref{th:main}, there exists a universal $C_\Omega>0$ such that for $n\geq 1$ we have the following estimates 
\begin{equation*}
\lambda_n\geq \frac{2\pi}{|\Omega|}n,
\end{equation*}
where $|\Omega|$ is the Lebesgue measure of $\Omega$,
as well as 
\begin{equation*}
\|e_{n}\|_{L^\infty(\Omega)}\leq C_\Omega\lambda_n^{1/4},  \qquad \|\nabla e_{n}\|_{L^\infty(\Omega)}\leq C_\Omega\lambda_n^{3/4}, \qquad \text{and} \qquad \|\partial_i \partial_j e_n\|_{L^\infty(\Omega)}\leq C_\Omega \lambda_n^{5/4}, 
\end{equation*}
for $i,j\in\{1,2\}$.
\end{lemma}
\begin{proof}
The bound on $\lambda_n$ is \cite[Corollary 1]{LY}.\footnote{Note that part of \cite{LY} is in dimension three or more, but the proof of Theorem 1 and Corollary 1 does not require this.} The estimate for $\|e_{n}\|_{L^\infty(\Omega)}$ is \cite[Theorem 1]{Grieser}, the estimate for $\|\nabla e_n\|_{L^\infty(\Omega)}$ is \cite[Corollary 1.1]{SX}, and the estimate for $\|\partial_i\partial_je_n\|_{L^\infty(\Omega)}$ is \cite[Theorem 2]{Steinerberger}
\end{proof}

In terms of the eigenfunctions and eigenvalues, we can write the heat kernel $\p_{\Omega}$ as
\begin{equation}\label{eq:hk}
\p_{\Omega}(t,x,y)=e^{t\Delta}(x,y)=\sum_{n= 1}^\infty e^{-t \lambda_n}e_{n}(x)e_{n}(y). 
\end{equation}
While it is slightly circular reasoning (some of the estimates of Lemma \ref{lem:evest} are derived from heat kernel estimates), we point out that Lemma \ref{lem:evest} implies that the series above converges (uniformly) and it can be differentiated termwise twice. From this fact, one readily checks that the series truly equals $\p_{\Omega}$ by noting that, if we write $u(t,x,y)$ for this series, then $u$ is a smooth function on $(0,\infty)\times \Omega^2$ which satisfies the heat equation: $(\partial_t-\Delta_x)u(t,x,y)=0$ (with zero boundary conditions) and $\lim_{t\to 0}u(t,x,y)=\delta(x,y)$, and since $\p_{\Omega}$ is also a solution to the same problem, we must have $u=\p_{\Omega}$ by standard uniqueness results for the heat equation. 

We now turn to more detailed estimates for the heat kernel that we will need later in our analysis of the GFF and the renormalized potential for the sine-Gordon model. The relevant results are recorded in the following lemma. 
\begin{lemma}\label{lem:ctest}
Let $\delta>0$, $x_1,x_2,x_3,x_4\in \Omega$ be such that $d(x_i,\partial \Omega)\geq \delta$. Moreover, let $t_0>0$ be arbitrary and $R_t$ be as in \eqref{eq:rt}. Then there exists a $C=C(t_0,\delta,\Omega)$ such that for $t<t_0$
\begin{equation}\label{eq:Rbound1}
|R_t(x_1,x_2)|\leq Ce^{-\frac{\delta^2}{32t}},
\end{equation}
\begin{equation}\label{eq:Rbound2}
\big|R_t(x_1,x_3)-R_t(x_2,x_3)\big|\leq C|x_1-x_2|e^{-\frac{\delta^2}{32t}}
\end{equation}
and
\begin{equation}\label{eq:Rbound3}
\big|R_t(x_1,x_3)-R_t(x_1,x_4)-R_t(x_2,x_3)+R_t(x_2,x_4)\big|\leq C|x_1-x_2||x_3-x_4|e^{-\frac{\delta^2}{200t}}.
\end{equation}
\end{lemma}
\begin{proof}
Before turning to the actual proof, we record the following well known fact: if $d(x,\partial\Omega)\geq \delta$, and $0<t<t_0$, then for a constant $C=C(\delta,t_0)>0$, we have
\begin{equation}\label{eq:etbound}
\Prob_x(T_\Omega<2t)\leq C e^{-\frac{\delta^2}{32t}}.
\end{equation}
For the sake of completeness, we give a short proof. Let $S(x,\delta)$ be the square (aligned with the coordinate axes) of side length $\frac{1}{\sqrt{2}}\delta$ with midpoint $x$. Then (since $S(x,\delta)\subset \overline{\Omega}$)
\[
\Prob_x(T_\Omega<2t)\leq \Prob_x(T_{S(x,\delta)}<2t)=\Prob_0(T_{S(0,\delta)}<2t)\leq 2\Prob_0\left(\sup_{0<s<2t}|W_s|>\frac{1}{2\sqrt{2}}\delta\right),
\]
where $W$ denotes one-dimensional standard Brownian motion and for the last bound, the reasoning is that for 2d-Brownian motion to leave the square, at least one of its coordinates has to leave the interval $\left[-\frac{1}{2\sqrt{2}}\delta,\frac{1}{2\sqrt{2}}\delta\right]$. Next note that since $W\stackrel{d}{=}-W$, we have 
\begin{align*}
\Prob_0\left(\sup_{0<s<2t}|W_s|>\frac{1}{2\sqrt{2}}\delta\right)&=\Prob_0\left(\left\{\sup_{0<s<2t}W_s>\frac{1}{2\sqrt{2}}\delta\right\} \bigcup \left\{\sup_{0<s<2t}(-W_s)>\frac{1}{2\sqrt{2}}\delta\right\}\right)\\
&\leq 2\Prob_0\left(\sup_{0<s<2t}W_s>\frac{1}{2\sqrt{2}}\delta\right).
\end{align*}
Now by the reflection principle, we have 
\begin{align*}
\Prob_0\left(\sup_{0<s<2t}W_s>\frac{1}{2\sqrt{2}}\delta\right)=2\Prob_0\left(W_{2t}>\frac{1}{2\sqrt{2}}\delta\right)&=\frac{2}{\sqrt{2\pi 2t}}\int_{\frac{\delta}{2\sqrt{2}}}^\infty e^{-\frac{u^2}{4t}}du\\
&\leq \frac{1}{\sqrt{\pi t}}\int_{\frac{\delta}{2\sqrt{2}}}^\infty \frac{u}{\delta/(2\sqrt{2})} e^{-\frac{u^2}{4t}}du\\
&=\frac{4\sqrt{2t}}{\delta\sqrt{\pi}}\int_{\frac{\delta}{2\sqrt{2}}}^\infty \frac{u}{2t}e^{-\frac{u^2}{4t}}du\\
&=\frac{4\sqrt{2t}}{\delta\sqrt{\pi}} e^{-\frac{\delta^2}{32 t}}\\
&\leq \frac{4\sqrt{2t_0}}{\delta\sqrt{\pi}}e^{-\frac{\delta^2}{32t}}.
\end{align*}
This shows \eqref{eq:etbound}.

\medskip

Let us now turn to the actual proof. For \eqref{eq:Rbound1}, we recall from \eqref{eq:rt} that 
\begin{align*}
R_t(x,y)&=\E_x\left[\frac{1}{4\pi (t-\frac{1}{2}T_\Omega)}e^{-\frac{|y-B(T_\Omega)|^2}{4(t-\frac{1}{2}T_\Omega)}}\1_{\{T_\Omega<2t\}}\right]
\end{align*}

We then note that since $x\mapsto x e^{-x}$ is bounded on $(0,\infty)$, there exists a universal constant $c>0$ such that 
\[
\frac{1}{4\pi (t-\frac{1}{2}T_\Omega)}e^{-\frac{|y-B(T_\Omega)|^2}{4(t-\frac{1}{2}T_\Omega)}}\leq \frac{c}{|y-B(T_\Omega)|^2}\leq \frac{C}{\delta^2}.
\]
Thus by \eqref{eq:etbound}, there exists some $C=C(\delta,t_0)$ such that
\[
R_t(x,y)\leq \frac{c}{\delta^2}\Prob_x(T_\Omega<2t)\leq \frac{c}{\delta^2}C(\delta,t_0)e^{-\frac{\delta^2}{32t}},
\]
which yields \eqref{eq:Rbound1}.

\medskip

For \eqref{eq:Rbound2}, we first of all note that e.g. by \eqref{eq:hk} and \eqref{eq:tdens}, $R_t(x,y)=R_t(y,x)$. Next we note that by the simple inequality
\[
|e^{-a}-e^{-b}|\leq |a-b|\max(e^{-a},e^{-b})
\]
for $a,b\geq 0$, we have 
\begin{align*}
|\p(s,x,y)-\p(s,x,z)|&\leq \frac{1}{16\pi s^2}\big||x-y|^2-|x-z|^2\big|\max\left(e^{-\frac{|x-y|^2}{4s}},e^{-\frac{|x-z|^2}{4s}}\right)\\
&\leq \frac{\mathrm{diam}(\Omega)}{8\pi s^2}|y-z|\max\left(e^{-\frac{|x-y|^2}{4s}},e^{-\frac{|x-z|^2}{4s}}\right).
\end{align*}
Thus, under our assumption that $d(x_i,\partial\Omega)\geq \delta$ (so that $|x_i-B(T_\Omega)|\geq \delta$)
\begin{align*}
|R_t(x_1,x_3)-R_t(x_2,x_3)|&=|R_t(x_3,x_1)-R_t(x_3,x_2)|\\
&\leq \E_{x_3}\left[|\p(t-\tfrac{1}{2}T_\Omega,B(T_\Omega),x_1)-\p(t-\tfrac{1}{2}T_\Omega,B(T_\Omega),x_2)|\1_{\{T_\Omega<2t\}}\right]\\
&\leq \frac{\mathrm{diam}(\Omega)}{8\pi}|x_1-x_2|\E_{x_3}\left[\frac{1}{(t-\frac{1}{2}T_\Omega)^2}e^{-\frac{\delta^2}{4(t-\frac{1}{2}T_\Omega)}}\1_{\{T_\Omega<2t\}}\right]\\
&=\frac{2\mathrm{diam}(\Omega)}{\pi \delta^4}|x_1-x_2|\E_{x_3}\left[\frac{\delta^4}{(4 (t-\frac{1}{2}T_\Omega))^2}e^{-\frac{\delta^2}{4(t-\frac{1}{2}T_\Omega)}}\1_{\{T_\Omega<2t\}}\right]
\end{align*}
Since $x\mapsto x^2 e^{-x}$ is bounded on $(0,\infty)$, we see that for some universal constant $c>0$,
\[
|R_t(x_1,x_3)-R_t(x_2,x_3)|\leq c\frac{\mathrm{diam}(\Omega)}{\delta^4}|x_1-x_2|\Prob_{x_3}(T_\Omega<2t)
\]
from which \eqref{eq:etbound} concludes the proof of \eqref{eq:Rbound2}.

\medskip

For \eqref{eq:Rbound3}, we take a PDE approach instead of a probabilistic one -- more precisely, we use local interior gradient estimates for solutions of the heat equation. 

Let us write $u(t,x)=R_t(x,x_3)-R_t(x,x_4)$. Our goal is to show that $|u(t,x_1)-u(t,x_2)|\leq C|x_1-x_2||x_3-x_4|e^{-\frac{\delta^2}{200t}}$. For this we note that $u$ satisfies the heat equation in $(0,\infty)\times \Omega$ (since $(t,x)\mapsto R_t(x,x_i)$ does as it is given by the difference of two fundamental solutions to the heat equation). Note though that we are not specifying the boundary conditions of $u$ on $\partial\Omega$. 

The local interior gradient estimate we use is a special case of \cite[Theorem 2.3.9]{Evans}, which we recall here. First of all, we introduce cylinders: for $x\in \R^2$, $t>0$, and $0<r<\sqrt{t}$, let
\begin{equation}
C(x,t;r)=\{(y,s)\in\R^2: |x-y|\leq r, t-r^2\leq s\leq t\}.
\end{equation}
We also write for any $T>0$, $\Omega_T=(0,T]\times \Omega$. Then a special case of \cite[Theorem 2.3.9]{Evans} states that for any $u$ that solves the heat equation in $\Omega_T$ (with any reasonable boundary conditions) and any cylinders $C(x,t;\frac{r}{2})\subset C(x,t;r)\subset \Omega_T$, we have 
\begin{equation}\label{eq:evans}
\|\nabla u\|_{L^\infty(C(x,t;\frac{r}{2}))}\leq \frac{C}{r^5}\|u\|_{L^1(C(x,t;r))}
\end{equation}
for some universal constant $C$.

We apply this to our $u(t,x)=R_t(x,x_3)-R_t(x,x_4)$ with $r=\min(\sqrt{\frac{t}{2}},\frac{\delta}{2})$. We find from \eqref{eq:Rbound2} (applied with $\delta$ replaced with $\frac{\delta}{2}$) that
\begin{align*}
|\nabla u(t,x)|&\leq \|\nabla u\|_{L^\infty(C(x,t;\frac{r}{2}))}\\
&\leq \frac{C}{r^5}\|u\|_{L^1(C(x,t;r))}\\
&\leq  \frac{\tilde C}{r^5}|x_3-x_4|\int_{t-r^2}^t \int_{B(x,r)} e^{-\frac{\delta^2}{128s}} dyds\\
&\leq \frac{\pi \tilde C}{r}|x_3-x_4| e^{-\frac{\delta^2}{128t}}\\
&\leq C(\delta,t_0,\Omega)|x_3-x_4| e^{-\frac{\delta^2}{200t}}
\end{align*}
for suitable constants ($C$ being universal, while $\tilde C, C(\delta,t_0,\Omega)$ depending on $\delta,t_0,\Omega$). Thus $u$ is locally Lipschitz in the spatial variable and by compactness, we recover the claim.
\end{proof}

We now turn to the GFF.

\subsection{The heat kernel regularization of the GFF} 
We can now use the heat kernel to construct our regularized GFF. 
\begin{lemma}\label{lem:GFFreg}
For any $\epsilon>0$, 
\begin{equation}
C_\epsilon(x,y)=\int_{\epsilon^2}^\infty ds\,\p_{\Omega}(s,x,y)
\end{equation}
is the covariance of a continuous Gaussian process $\phie$ on $\Omega$. 
\end{lemma}
\begin{proof}
As this result is rather standard, we will be a bit brief on details and refer to the literature on various facts. 

Let us first prove that $C_\epsilon$ is symmetric and positive definite.  Symmetry, namely that $\p_{\Omega}(s,x,y)\allowbreak=\p_{\Omega}(s,y,x)$ is proven in \cite[Theorem 3.30]{MP}. For positive definiteness, we use the fact that $\p_{\Omega}$ is a transition density (also see \cite[Theorem 3.30]{MP}) and argue that by symmetry, we have for any $s>0$, $x_1,\dots ,x_n\in \Omega$ and $\alpha_1,\dots ,\alpha_n\in \R$
\begin{align*}
\sum_{i,j=1}^n \alpha_i\alpha_j \p_{\Omega}(s,x_i,x_j)&=\sum_{i,j=1}^n \alpha_i\alpha_j \int_{\Omega}d^2z\, \p_{\Omega}(\tfrac{s}{2},x_i,z)\p_{\Omega}(\tfrac{s}{2},z,x_j)\\
&=\sum_{i,j=1}^n \alpha_i\alpha_j \int_{\Omega} d^2z\,\p_{\Omega}(\tfrac{s}{2},x_i,z)\p_{\Omega}(\tfrac{s}{2},x_j,z)\\
&=\int_{\Omega}d^2z\,\left(\sum_{i=1}^n\alpha_i\p_{\Omega}(\tfrac{s}{2},x_i,z)\right)^2\\
&\geq 0.
\end{align*}
Thus for each $s>0$, $\p_{\Omega}(s,\cdot,\cdot)$ is symmetric and positive definite, which implies that also $C_\epsilon$ is as well. Thus there is some Gaussian process $\phie$ with this covariance.

Using standard regularity theory of Gaussian processes (see e.g. \cite[Appendix B]{LRV}), we can choose $\phie$ to be continuous if 
\begin{equation*}
\left|C_\epsilon(x,x)+C_\epsilon(y,y)-2C_\epsilon(x,y)\right|\leq A|x-y|^\gamma
\end{equation*}
for some $A,\gamma>0$. This follows readily from \eqref{eq:hk} and Lemma~\ref{lem:evest}. 
\end{proof}

Next we record some basic facts about GFF-correlation functions in the $\epsilon\to 0$ limit.

We will find it convenient to write $\langle \cdot\rangle_{\gff_{\Omega}(\epsilon)}$ for expectation with respect to the law of $\phie$. The reader may be familiar with the notion that the GFF on a domain $\Omega$ is a Gaussian process whose covariance is the Green's function of $\Omega$. The following lemma connects our definition of $\phie$ as the regularized GFF to this notion.
\begin{lemma}
The function $G_{\Omega}:\Omega\times \Omega\to (-\infty,\infty]$, 
\begin{equation}
G_{\Omega}(x,y)=\lim_{\epsilon\to 0}C_\epsilon(x,y)=\int_0^\infty ds\,\p_{\Omega}(s,x,y)
\end{equation}
is the Green's function of $\Omega$:
\begin{equation}
G_{\Omega}(x,y)=\frac{1}{2\pi}\log |x-y|^{-1}+g_{\Omega}(x,y),
\end{equation}
where $g$ is harmonic in $x\in\Omega$ (and in $y$), symmetric, smooth up to the boundary, and $g(x,y)=-\frac{1}{2\pi}\log |x-y|^{-1}$ for $x\in\partial\Omega$ and $y\in \Omega$.
\end{lemma}
\begin{proof}
This is standard -- see for example \cite[Theorem 3.35]{MP}.
\end{proof}

The previous lemma can also be viewed as a statement about correlation functions of the regularized GFF: it states that for fixed $x,y\in \Omega$ with $x\neq y$
\begin{equation}
\lim_{\epsilon\to 0}\langle \phie(x)\phie(y)\rangle_{\gff_{\Omega}(\epsilon)}=\langle \varphi(x)\varphi(y)\rangle_{\gff_{\Omega}}:=G_{\Omega}(x,y).
\end{equation}

We will need some further information about $\epsilon\to 0$ limits of GFF correlation functions. We begin with uniform covariance estimates.

\begin{lemma}\label{lem:cov}
Uniformly on compact subsets of $\big\{(x,y)\in \Omega\times \Omega: x\neq y\big\}$, as $\epsilon\to 0$,
\begin{align*}
\left\langle\phie(x)\phie(y)\right\rangle_{\gff_{\Omega}(\epsilon)}&\to G_{\Omega}(x,y)
\end{align*}
and uniformly on compact subsets of $\Omega$, as $\epsilon\to 0$,
\begin{align*}
\Big\langle \big(\phie(x)\big)^2\Big\rangle_{\gff_{\Omega}(\epsilon)}+\frac{1}{2\pi}\log(\varepsilon)&\to\frac{1}{4\pi}\big(\gamma-\log 4\big)+g_{\Omega}(x,x),
\end{align*}
where $\gamma$ is the Euler-Mascheroni constant. 
\end{lemma}
\begin{proof}
We begin by noting that 
\begin{align*}
    \left\langle\phie(x)\phie(y)\right\rangle_{\gff_{\Omega}(\epsilon)}&=\int_{\varepsilon^2}^\infty ds\left(\frac{1}{4\pi s}e^{-\frac{|x-y|^2}{4s}}-R_s(x,y)\right)\\
    &=\int_1^\infty ds\left(\frac{1}{4\pi s}e^{-\frac{|x-y|^2}{4s}}-R_s(x,y)\right)+\int_{\varepsilon^2}^1 ds\left(\frac{1}{4\pi s}e^{-\frac{|x-y|^2}{4s}}-R_s(x,y)\right)\\
    &=\int_1^\infty ds\left(\frac{1}{4\pi s}e^{-\frac{|x-y|^2}{4s}}-R_s(x,y)\right)-\int_{\varepsilon^2}^1 ds\,R_s(x,y)+\int_{\varepsilon^2}^1 \frac{ds}{4\pi s}e^{-\frac{|x-y|^2}{4s}},
\end{align*}
where the first two integrals are convergent in the $\varepsilon\to 0$ limit -- the first one since it is independent of $\epsilon$ and converges (with the appropriate uniformity) by Lemma~\ref{lem:GFFreg}, while the second one converges (with the appropriate uniformity) due to the estimate \eqref{eq:Rbound1}.

For the last integral, we note that a change of variables (and an application of dominated convergence) yields
\begin{align*}
\lim_{\varepsilon\to0}\int_{\varepsilon^2}^1 \frac{ds}{4\pi s}e^{-\frac{|x-y|^2}{4s}}&=\int_{|x-y|^2}^\infty \frac{ds}{4\pi s}e^{-\frac{s}{4}}\\
&=\int_{|x-y|^2}^1 \frac{ds}{4\pi s}-\int_{|x-y|^2}^1 ds\frac{1-e^{-\frac{s}{4}}}{4\pi s}+\int_1^\infty \frac{ds}{4\pi s}e^{-\frac{s}{4}}\\
&=\frac{1}{2\pi }\log|x-y|^{-1}-\int_{|x-y|^2}^1 ds\frac{1-e^{-\frac{s}{4}}}{4\pi s}+\int_1^\infty \frac{ds}{4\pi s}e^{-\frac{s}{4}}.
\end{align*}
The integrals converge and we have proved the first part of the lemma (one readily checks also here the required uniformity in the convergence). 

For the second part, we begin by noting that what we have just computed implies that 
\begin{equation*}
g_{\Omega}(x,y)=\int_1^\infty ds\bigg(\frac{1}{4\pi s} e^{-\frac{|x-y|^2}{4s}}-R_s(x,y)\bigg)-\int_0^1 ds\,R_s(x,y)-\int_{|x-y|^2}^1 ds\frac{1-e^{-\frac{s}{4}}}{4\pi s}+\int_1^\infty \frac{ds}{4\pi s}e^{-\frac{s}{4}}.
\end{equation*}
By continuity (and the estimates we have used), this also holds for $x=y$.

This remark allows us to write as $\epsilon\to 0$ (with the required uniformity)
\begin{align*}
\Big\langle\big(\phie(x)\big)^2\Big\rangle_{\gff_{\Omega}(\epsilon)}+\frac{1}{2\pi}\log(\varepsilon)&=\int_{\varepsilon^2}^\infty ds\bigg(\frac{1}{4\pi s}-R_s(x,x)\bigg)-\int_{\varepsilon^2}^1\frac{ds}{4\pi s}\\
&=\int_1^\infty ds\bigg(\frac{1}{4\pi s}-R_s(x,x)\bigg)-\int_{\varepsilon^2}^1 ds\,R_s(x,x)\\
&=g_{\Omega}(x,x)+\int_{0}^1 ds\frac{1-e^{-\frac{s}{4}}}{4\pi s}-\int_1^\infty \frac{ds}{4\pi s}e^{-\frac{s}{4}}+o(1).
\end{align*}
Finally, integration by parts yields
\begin{align*}
\int_{0}^1 ds\frac{1-e^{-\frac{s}{4}}}{4\pi s}-\int_1^\infty \frac{ds}{4\pi s}e^{-\frac{s}{4}}&=\lim_{a\to0}\bigg[\int_{a}^1 \frac{ds}{4\pi s}-\frac{1}{4\pi}\int_{a/4}^\infty \frac{ds}{s}e^{-s}\bigg]\\
&=\notag \lim_{a\to0}\bigg[-\frac{1}{4\pi}\log a-\frac{1}{4\pi}\bigg(-\frac{\log \frac{a}{4}}{e^{a/4}}+\int_{a/4}^\infty ds\,\frac{\log s}{e^s}\bigg)\bigg]\\
&=\notag \lim_{a\to0}\frac{1}{4\pi}\bigg[-\log a\Big(1-e^{-a/4}\Big)-\frac{\log4}{e^{a/4}}-\int_{a/4}^\infty ds\,\frac{\log s}{e^s}\bigg]\\
&=\notag \frac{1}{4\pi}\big(\gamma-\log 4\big),
\end{align*}
which concludes the proof.
\end{proof}

For the sine-Gordon model, we will also need Wick ordered exponentials of the GFF and estimates for correlation functions involving them. We define the Wick ordered exponentials of the GFF as follows: for each $\alpha\in \R$, let 
\begin{equation}\label{eq:wick}
\opcolon e^{i\alpha \varphi_\epsilon(x)}\opcolon =e^{\frac{\alpha^2}{8\pi}(\gamma-\log 4)}\epsilon^{-\frac{\alpha^2}{4\pi}}e^{i\alpha \varphi_\epsilon(x)}=:c_\alpha \epsilon^{-\frac{\alpha^2}{4\pi}}e^{i\alpha\varphi_\epsilon(x)},
\end{equation}
where again $\gamma$ denotes the Euler-Mascheroni constant. The reason for including $e^{\frac{\alpha^2}{8\pi}(\gamma-\log 4)}$ in our definition of normal order is that correlation functions become cleaner looking (and consistent with the definition of \cite{BIVW}) with this choice. For example, one readily checks using Lemma \ref{lem:cov} that with this convention,
\begin{equation*}
\lim_{\epsilon\to 0}\left\langle \opcolon e^{i\alpha \varphi_\epsilon(x)}\opcolon  \opcolon e^{i\alpha' \varphi_\epsilon(y)}\opcolon \right\rangle_{\gff_{\Omega}(\epsilon)}= e^{-\alpha\alpha' G_{\Omega}(x,y)-\frac{\alpha^2}{2}g_{\Omega}(x,x)-\frac{(\alpha')^2}{2}g_{\Omega}(y,y)}.
\end{equation*}
Without adding this constant to our normal order convention, this would get multiplied by $c_\alpha^{-1}c_{\alpha'}^{-1}$.

Using this normal order notation, we can also define Wick ordered trigonometric functions as expected,
\begin{equation}
\opcolon \cos(\alpha\varphi_\epsilon(x))\opcolon =\frac{1}{2}\Big(\opcolon e^{i\alpha \varphi_\epsilon(x)}\opcolon +\opcolon e^{-i\alpha \varphi_\epsilon(x)}\opcolon \Big)
\end{equation}
and
\begin{equation}
\opcolon \sin(\alpha\varphi_\epsilon(x))\opcolon =\frac{1}{2i}\Big(\opcolon e^{i\alpha \varphi_\epsilon(x)}\opcolon -\opcolon e^{-i\alpha \varphi_\epsilon(x)}\opcolon \Big).
\end{equation}

Estimates for the correlation functions involving Wick ordered exponentials are recorded in the following lemma.
\begin{lemma}\label{lem:gffcorr2}
For $\gamma_1,\dots,\gamma_n\in \R$ and distinct points $x_1,\dots,x_n$, $y_1,\dots,y_m$ $\in \Omega$, the limit
\begin{align*}
&\Bigg\langle \prod_{j=1}^n \opcolon e^{i\gamma_j\varphi(x_j)}\opcolon  \prod_{k=1}^m \varphi(y_k)\Bigg\rangle_{\gff_{\Omega}} :=\lim_{\epsilon\to 0}\Bigg\langle \prod_{j=1}^n \opcolon e^{i\gamma_j\phie(x_j)}\opcolon  \prod_{k=1}^m \phie(y_k)\Bigg\rangle_{\gff_{\Omega}(\epsilon)}
\end{align*} 
exists. 

Moreover, the convergence is uniform on compact subsets of $\big\{u\in \Omega^{n+m}: u_i\neq u_j \ \text{for} \ i\neq j\big\}$.
\end{lemma}
\begin{proof}
We begin by noting that by the Leibniz integral rule, 
\begin{align*}
&\Bigg\langle \prod_{j=1}^n \opcolon e^{i\gamma_j\phie(x_j)}\opcolon  \prod_{k=1}^m \phie(y_k)\Bigg\rangle_{\gff_{\Omega}(\epsilon)}=\prod_{k=1}^m \frac{\partial}{\partial t_k}\Bigg|_{t=0}\Bigg\langle \prod_{j=1}^n \opcolon e^{i\gamma_j\phie(x_j)}\opcolon e^{\sum_{k=1}^m t_k  \phie(y_k)} \Bigg\rangle_{\gff_{\Omega}(\epsilon)}.
\end{align*}
By Girsanov's theorem and a routine Gaussian calculation, we find that
\begin{multline*}
\Bigg\langle \prod_{j=1}^n \opcolon e^{i\gamma_j\phie(x_j)}\opcolon e^{\sum_{k=1}^m t_k \phie(y_k)} \Bigg\rangle_{\gff_{\Omega}(\epsilon)}\\
\shoveleft\quad =e^{\frac{1}{2}\langle (\sum_{k=1}^m t_k \phie(y_k))^2\rangle_{\gff_{\Omega}(\epsilon)}} \prod_{j=1}^n e^{i\gamma_j \sum_{k=1}^m t_k \langle \phie(x_j)\phie(y_k)\rangle_{\gff_{\Omega}(\epsilon)}}  \Bigg\langle \prod_{j=1}^n \opcolon e^{i\gamma_j\phie(x_j)}\opcolon \Bigg\rangle_{\gff_{\Omega}(\epsilon)}\\
\shoveleft\quad =e^{\frac{1}{2}\langle (\sum_{k=1}^m t_k \phie(y_k))^2\rangle_{\gff_{\Omega}(\epsilon)}}  \prod_{j=1}^n e^{i\gamma_j \sum_{k=1}^m t_k \langle \phie(x_j)\phie(y_k)\rangle_{\gff_{\Omega}(\epsilon)}} \\
\times e^{-\sum_{1\leq j<j'\leq n}\gamma_j\gamma_{j'}\langle \varphi_\epsilon(x_j)\varphi_\epsilon(x_{j'})\rangle_{\gff_{\Omega}(\epsilon)}} e^{-\sum_{j=1}^n \frac{\gamma_j^2}{2}\big(\langle (\varphi_\epsilon(x_j))^2\rangle_{\gff_{\Omega}(\epsilon)}+\frac{1}{2\pi}\log(\epsilon)-\frac{1}{4\pi}(\gamma-\log 4)\big)}.
\end{multline*} 
Carrying out the $t$-differentiations and setting $t=0$, we see that the question of (uniform) convergence boils down to convergence of terms handled by Lemma~\ref{lem:cov} (note that since we differentiate only once with respect to each $t_k$ and then set $t=0$, we never get terms of the type $\langle (\varphi_\epsilon(y_k))^2\rangle_{\gff_{\Omega}(\epsilon)}$).
\end{proof}

We record a further GFF estimate we shall need in the proof of Theorem~\ref{th:sgmain}. \begin{lemma}\label{lem:smcov}
For $f\in C_c^\infty(\Omega)$ and $x\in \Omega$, we have
\begin{equation}
\lim_{\epsilon\to 0}\left\langle\phie(x)\phie(f)\right\rangle_{\gff_{\Omega}(\epsilon)}=\left\langle \varphi(x)\varphi(f)\right\rangle_{\gff_{\Omega}}:=\int_{\Omega}d^2y\, f(y) G_{\Omega}(x,y)
\end{equation}
where the convergence is uniform on compact subsets of $\Omega$ and the limiting function is a continuously differentiable function of $x\in\Omega$. Moreover, for any compact $K\subset \Omega$, there exists $C_{K,f}>0$ such that 
\begin{equation*}
\sup_{0<\epsilon<c_K}\left|\left\langle \varphi_\epsilon(x)\varphi_\epsilon(f)\right\rangle_{\gff_{\Omega}(\epsilon)}-\left\langle \varphi_\epsilon(y)\varphi_\epsilon(f)\right\rangle_{\gff_{\Omega}(\epsilon)}\right|<C_{K,f}|x-y| 
\end{equation*}
for $x,y\in K$.
\end{lemma}
\begin{proof}
We begin by noting that by Fubini, we can write 
\begin{equation*}
\left\langle\phie(x)\phie(f)\right\rangle_{\gff_{\Omega}(\epsilon)}=\int_{\Omega}d^2y\, f(y)\left\langle\phie(x)\phie(y)\right\rangle_{\gff_{\Omega}(\epsilon)}=\int_{\Omega}d^2y\, f(y)\int_{\epsilon^2}^\infty ds\, \p_{\Omega}(s,x,y).
\end{equation*}
Pointwise convergence of this follows readily from the dominated convergence theorem and noting that 
\begin{equation*}
0\leq \int_{\epsilon^2}^\infty ds\,\p_{\Omega}(s,x,y)\leq G_{\Omega}(x,y)
\end{equation*}
which is integrable in $y$ for each $x\in \Omega$.

For local uniform convergence, it thus suffices to show that for each compact $K\subset \Omega$, 
\begin{equation*}
\sup_{x\in K}\lim_{\epsilon\to 0}\int_{K}d^2 y \int_0^{\epsilon^2}ds\,\p_{\Omega}(s,x,y)=0.
\end{equation*}
For this, we recall that $\p_{\Omega}(s,x,y)\leq \p(s,x,y)=\frac{1}{4\pi s}e^{-\frac{|x-y|^2}{4 s}}$, so (by shifting our integration variable) it is sufficient to prove that for each $R>0$
\begin{equation*}
\lim_{\epsilon\to 0}\int_{B(0,R)}d^2 u\, \int_0^{\epsilon^2}\frac{ds}{4\pi s}e^{-\frac{|u|^2}{4s}}=0.
\end{equation*}
Using Fubini and writing $u=\sqrt{s}v$, we have
\begin{align*}
\int_{B(0,R)}d^2 u\, \int_0^{\epsilon^2}\frac{ds}{4\pi s}e^{-\frac{|u|^2}{4s}}\leq \int_0^{\epsilon^2}ds\int_{\R^2} \frac{d^2v}{4\pi} e^{-\frac{|v|^2}{4}}=\epsilon^2,
\end{align*}
which concludes the proof of local uniform convergence. 

For the differentiability claim, recall that $G_{\Omega}(x,y)=\frac{1}{2\pi}\log |x-y|^{-1}+g_{\Omega}(x,y)$, where $g_{\Omega}(x,y)$ is smooth in $\Omega\times \Omega$. Thus we need to prove that 
\begin{equation*}
x\mapsto \int_{\Omega}d^2 y\, f(y)\log |x-y|^{-1}
\end{equation*}
is continuously differentiable for $f\in C_c^\infty(\Omega)$. This follows readily from the Leibniz integral rule since $\big|\nabla_x \log |x-y|^{-1}\big|=\frac{1}{|x-y|}$ is integrable in $2d$. 

For the final claim, we use \eqref{eq:tdens} and Lemma~\ref{lem:GFFreg} to write 
\begin{align*}
&\left|\left\langle \varphi_\epsilon(x)\varphi_\epsilon(f)\right\rangle_{\gff_{\Omega}(\epsilon)}-\left\langle \varphi_\epsilon(y)\varphi_\epsilon(f)\right\rangle_{\gff_{\Omega}(\epsilon)}\right|\\
&\quad\leq \int_{\Omega}d^2 u\, f(u) \int_1^\infty ds\, |\p_{\Omega}(s,x,u)-\p_{\Omega}(s,y,u)|  \\
&\qquad +\left| \int_{\Omega}d^2u\, f(u) \int_{\epsilon^2}^1 ds\,  (\p(s,x,u)-\p(s,y,u))\right|\\
&\qquad +\int_{\Omega}d^2u \, f(u)\int_{\epsilon^2}^1 ds\, |R_s(x,u)-R_s(y,u)|.
\end{align*}
Using \eqref{eq:hk} and Lemma~\ref{lem:evest}, one can perform a gradient estimate and readily show that 
\begin{equation*}
\int_{\Omega}d^2 u\, f(u) \int_1^\infty ds\, |\p_{\Omega}(s,x,u)-\p_{\Omega}(s,y,u)|\leq C_{K,f}|x-y|.
\end{equation*}
For the last term, a simple application of \eqref{eq:Rbound2} of Lemma~\ref{lem:ctest} readily shows the existence of a $C_{K,f}>0$ (possibly different from above) such that 
\begin{equation*}
\int_{\Omega}d^2u \, f(u)\int_{\epsilon^2}^1 ds\, |R_s(x,u)-R_s(y,u)|\leq C_{K,f} |x-y|.
\end{equation*}
Finally for the term involving the full plane heat kernel, we write 
\begin{align*}
&\Bigg| \int_{\Omega}d^2u\, f(u) \int_{\epsilon^2}^1 ds\,  (\p(s,x,u)-\p(s,y,u))\Bigg| \\
&\quad\leq \int_{\epsilon^2}^1 ds\Bigg|\int_{\Omega}d^2u\, f(u)\frac{e^{-\frac{|x-u|^2}{4s}}}{4\pi s}-\int_{\Omega}d^2u\, f(u)\frac{e^{-\frac{|y-u|^2}{4s}}}{4\pi s}\Bigg|\\
&\quad\leq \int_{0}^1ds\, \int_{\R^2} \frac{d^2 v}{4\pi}|f(x+\sqrt{s}v)-f(y+\sqrt{s}v)|e^{-\frac{|v|^2}{4}}\\
&\quad\leq \|\nabla f\|_\infty |x-y|\int_{\R^2} \frac{d^2 v}{4\pi}e^{-\frac{|v|^2}{4}}\\
&\quad= \|\nabla f\|_\infty |x-y|.
\end{align*}
Putting these estimates together concludes the proof.
\end{proof}

We will also need limiting correlation functions involving derivatives of the fields. Presumably one could construct these as limits of correlation functions involving derivatives of the regularized fields, but to avoid doing further estimates, we define these correlation functions directly in the $\epsilon=0$ limit.

For this purpose, we first give an explicit formula for functions involving the Wick ordered exponentials and the free field itself.

\begin{lemma}\label{le:gffcfexplicit}
For $\gamma_1,\dots,\gamma_n\in \R$ and distinct points $x_1,\dots,x_n,y_1,\dots,y_m\in \Omega$, we have 
\begin{multline}
\Bigg\langle \prod_{j=1}^n \opcolon e^{i\gamma_j\varphi(x_j)}\opcolon  \prod_{k=1}^m \varphi(y_k)\Bigg\rangle_{\gff_{\Omega}}=e^{-\sum_{1\leq j<j'\leq n}\gamma_j \gamma_{j'}G_{\Omega}(x_j,x_{j'})-\frac{1}{2}\sum_{j=1}^n \gamma_j^2 g_{\Omega}(x_j,x_j)}\\
\times \left\langle \prod_{k=1}^m \bigg(\varphi(y_k)+i\sum_{j=1}^n \gamma_j \langle \varphi(y_k)\varphi(x_j)\rangle_{\gff_{\Omega}} \bigg)\right\rangle_{\gff_{\Omega}},
\end{multline}
where the correlation function on the right is reduced to correlation functions involving only $\varphi(y_k)$ by expanding the product and using linearity of the correlation function.
\end{lemma}
\begin{proof}
This is basically Girsanov's theorem, but as the limiting objects are not real valued Gaussian random variables and we have purely imaginary exponents (instead of real as one usually has in Girsanov's theorem), there is some justification needed. 

We start by considering for $\epsilon>0$ the function 
\begin{align}
F_\epsilon(z_1,\dots,z_n)=\Bigg\langle \prod_{j=1}^n \opcolon e^{z_j\varphi_\epsilon(x_j)}\opcolon  \prod_{k=1}^m \varphi_\epsilon(y_k)\Bigg\rangle_{\gff_{\Omega}(\epsilon)}
\end{align}
For $z_j=i\gamma_j$, we know that as $\epsilon\to 0$, $F_\epsilon$ converges to our desired object. We also note that for general $z_1,\dots,z_n\in \C$, $F_\epsilon(z_1,\dots,z_n)$ (where $\opcolon e^{z\varphi_\epsilon(x)}\opcolon =e^{-\frac{z^2}{8\pi}(\gamma-\log 4)} \epsilon^{\frac{z^2}{4\pi}}e^{z\varphi_\epsilon(x)}$) is well defined by standard Gaussian estimates. Moreover, combining Morera's theorem with Fubini, one readily checks that $F_\epsilon$ is an entire function of $z_1,\dots,z_n$.

Let us then compute $F_\epsilon(z_1,\dots,z_n)$ for $z_1,\dots,z_n\in \R$. For this, we can make use of a standard Girsanov (``complete-the-square'') argument and find 
\begin{align*}
&F_\epsilon(z_1,\dots,z_n)\\
&\qquad=\prod_{j=1}^n e^{-\frac{z_j^2}{8\pi}(\gamma-\log 4)}\epsilon^{\frac{z_j^2}{4\pi}}\left\langle \prod_{j=1}^n e^{z_j \varphi_\epsilon(x_j)}\prod_{k=1}^m \varphi_\epsilon(y_k) \right\rangle_{\gff_{\Omega}(\epsilon)}\\
&\qquad=\prod_{j=1}^n e^{-\frac{z_j^2}{8\pi}(\gamma-\log 4)}\epsilon^{\frac{z_j^2}{4\pi}}\left\langle \prod_{j=1}^n e^{z_j \varphi_\epsilon(x_j)}\right\rangle_{\gff_{\Omega}(\epsilon)}\\
&\qquad\qquad\times\left\langle \prod_{k=1}^m \bigg(\varphi_\epsilon(y_k)+\sum_{j=1}^n z_j \langle \varphi_\epsilon(y_k)\varphi_\epsilon(x_j)\rangle_{\gff_{\Omega}(\epsilon)}\bigg)\right\rangle_{\gff_{\Omega}(\epsilon)}\\
&\qquad=\left\langle \prod_{j=1}^n \opcolon e^{z_j \varphi_\epsilon(x_j)}\opcolon \right\rangle_{\gff_{\Omega}(\epsilon)}\left\langle \prod_{k=1}^m \bigg(\varphi_\epsilon(y_k)+\sum_{j=1}^n z_j \langle \varphi_\epsilon(y_k)\varphi_\epsilon(x_j)\rangle_{\gff_{\Omega}(\epsilon)}\bigg)\right\rangle_{\gff_{\Omega}(\epsilon)}.
\end{align*}
Since the right hand side makes perfect sense for complex values of $z_j$ as well, and is also an entire function of $z_1,\dots,z_n$, this identity must in fact hold for arbitrary $z_1,\dots,z_n\in \C$ -- in particular, for $z_j=i\gamma_j$.

We also know from Lemma \ref{lem:cov} and Lemma \ref{lem:gffcorr2}, that for $z_j=i\gamma_j$ this has a limit as $\epsilon\to 0$:
\begin{equation*}
    \lim_{\epsilon\to 0}F_\epsilon(i\gamma_1,\dots,i\gamma_n)=\left\langle \prod_{j=1}^n \opcolon e^{i\gamma_j \varphi(x_j)}\opcolon \right\rangle_{\gff_{\Omega}} \left\langle \prod_{k=1}^m \bigg(\varphi(y_k)+i\sum_{j=1}^n \gamma_j \langle \varphi(y_k)\varphi(x_j)\rangle_{\gff_{\Omega}}\bigg)\right\rangle_{\gff_{\Omega}}.
\end{equation*}
The second factor is already the same as in the statement of the lemma, so we only need to compute the first factor on the right. This was nearly done in Lemma \ref{lem:gffcorr2}. From the routine Gaussian calculation there (with $t_1,\dots,t_m=0$), we have 
\begin{multline*}
    \left\langle \prod_{j=1}^n \opcolon e^{i\gamma_j \varphi(x_j)}\opcolon \right\rangle_{\gff_{\Omega}}=\lim_{\epsilon\to 0}\bigg(e^{-\sum_{1\leq j<j'\leq n}\gamma_j\gamma_{j'}\langle \varphi_\epsilon(x_j)\varphi_\epsilon(x_{j'})\rangle_{\gff_{\Omega}(\epsilon)}}\\
    \times e^{-\sum_{j=1}^n \frac{\gamma_j^2}{2}\big(\langle (\varphi_\epsilon(x_j))^2\rangle_{\gff_{\Omega}(\epsilon)}+\frac{1}{2\pi}\log(\epsilon)-\frac{1}{4\pi}(\gamma-\log 4)\big)}\bigg).
\end{multline*}
The convergence of this to the first factor in the statement of the lemma follows from Lemma~\ref{lem:cov}. This concludes the proof.
\end{proof}

Note that by expanding the product and using Gaussianity, the right hand side of Lemma~\ref{le:gffcfexplicit} can be written entirely in terms of $G_{\Omega}$ and $g_{\Omega}$. In particular, the right hand side is a smooth function of $y_1,\dots,y_m$ (as long as none of the $y_k$ coincide with another $y$ point or an $x$ point). This allows us to make the following definition: if $D_k$ is either $\partial_{y_k}$ or $\bar \partial_{y_k}$ (holomorphic or antiholomorphic derivative with respect to $y_k$), then 
\begin{align}\label{eq:dercordef}
\Bigg\langle \prod_{j=1}^n \opcolon e^{i\gamma_j\varphi(x_j)}\opcolon  \prod_{k=1}^m D_k \varphi(y_k)\Bigg\rangle_{\gff_{\Omega}}&:=\prod_{k=1}^m D_k \Bigg\langle \prod_{j=1}^n \opcolon e^{i\gamma_j\varphi(x_j)}\opcolon  \prod_{k=1}^m \varphi(y_k)\Bigg\rangle_{\gff_{\Omega}}.    
\end{align}
In addition to defining these correlation functions, we will eventually need to identify them with certain critical Ising correlation functions using the results of \cite{BIVW}. The corresponding GFF correlation functions were defined there slightly differently, so we need to show that the definition there (\cite[Definition 2.3]{BIVW}) is equivalent to our definition. This is done in the following lemma. 
\begin{lemma}\label{le:gffequiv}
We have 
\begin{multline}
    \Bigg\langle \prod_{j=1}^n \opcolon e^{i\gamma_j\varphi(x_j)}\opcolon  \prod_{k=1}^m D_k \varphi(y_k)\Bigg\rangle_{\gff_{\Omega}}\\
    \shoveleft\quad =\prod_{k=1}^m D_k \partial_{z_k}|_{z_{k}=0}\Big[ e^{-\sum_{1\leq j<j'\leq n}\gamma_j \gamma_{j'}G_{\Omega}(x_j,x_{j'})-\frac{1}{2}\sum_{j=1}^n \gamma_j^2 g_{\Omega}(x_j,x_j)}\\
    \times e^{i\sum_{j=1}^n \sum_{k=1}^m \gamma_j z_k G_{\Omega}(x_j,y_k)+\sum_{1\leq k<k'\leq m}z_kz_{k'}G_{\Omega}(y_k,y_{k'})+\frac{1}{2}\sum_{k=1}^m z_k^2 g_{\Omega}(y_k,y_k)}\Big].
\end{multline}
\end{lemma}
\begin{proof}
    Note first of all that on the right hand side of the claim, the terms involving only $x_j$ and $x_{j'}$ can be commuted past the differential operators. Also the $z_k^2$ terms do not matter since once we differentiate with respect to $z_k$ and set $z_k=0$, these terms disappear. So by our definition of the correlation functions involved and Lemma \ref{le:gffcfexplicit}, it is enough for us to prove that 
    \begin{align*}
        &\prod_{k=1}^m \partial_{z_k}|_{z_k=0} e^{i\sum_{j=1}^n \sum_{k=1}^m \gamma_j z_k G_{\Omega}(x_j,y_k)+\sum_{1\leq k<k'\leq m}z_kz_{k'}G_{\Omega}(y_k,y_{k'})}\\
        &\qquad\qquad=\left\langle \prod_{k=1}^m \bigg(\varphi(y_k)+i\sum_{j=1}^n \gamma_j \langle \varphi(y_k)\varphi(x_j)\rangle_{\gff_{\Omega}} \bigg)\right\rangle_{\gff_{\Omega}}.
    \end{align*}
    This is of course a rather standard Gaussian fact one could prove with standard Gaussian combinatorial arguments, but let us give a direct analytic proof. Using Lemma \ref{lem:cov}, the Leibniz integral rule, and the formula for the moment generating function of a multivariate Gaussian random variable, we find that 
    \begin{align*}
    &\left\langle \prod_{k=1}^m \bigg(\varphi(y_k)+i\sum_{j=1}^n \gamma_j \langle \varphi(y_k)\varphi(x_j)\rangle_{\gff_{\Omega}} \bigg)\right\rangle_{\gff_{\Omega}}\\
    &\quad =\lim_{\epsilon\to 0}\left\langle \prod_{k=1}^m \bigg(\varphi_\epsilon(y_k)+i\sum_{j=1}^n \gamma_j \langle \varphi_\epsilon(y_k)\varphi_\epsilon(x_j)\rangle_{\gff_{\Omega}(\epsilon)} \bigg)\right\rangle_{\gff_{\Omega}(\epsilon)}\\
    &\quad =\lim_{\epsilon\to 0}\prod_{k=1}^m \partial_{z_k}|_{z_k=0}\left\langle e^{\sum_{k=1}^m z_k(\varphi_\epsilon(y_k)+i\sum_{j=1}^n \gamma_j \langle \varphi_\epsilon(y_k)\varphi_\epsilon(x_j)\rangle_{\gff_{\Omega}(\epsilon)})}\right\rangle_{\gff_{\Omega}(\epsilon)}\\
    &\quad =\lim_{\epsilon\to 0}\prod_{k=1}^m \partial_{z_k}|_{z_k=0} e^{i\sum_{k=1}^m\sum_{j=1}^n \gamma_j z_k \langle \varphi_\epsilon(y_k)\varphi_\epsilon(x_j)\rangle_{\gff_{\Omega}(\epsilon)}} e^{\frac{1}{2}\sum_{k,k'=1}^m z_kz_{k'}\langle \varphi_\epsilon(y_k)\varphi_\epsilon(y_{k'})\rangle_{\gff_{\Omega}(\epsilon)}}.
    \end{align*}
    Since the diagonal terms in the last sum do not contribute to the derivative (since we differentiate only once with respect to $z_k$ and then set it to be zero), we can replace the last exponential by $e^{\sum_{1\leq k<k'\leq m}z_k z_{k'}\langle \varphi_\epsilon(y_k)\varphi_{\epsilon}(y_{k'})\rangle_{\gff_{\Omega}(\epsilon)}}$. Thus by Lemma \ref{lem:cov}, it only remains to justify swapping the order of the $\epsilon\to 0$ limit and the derivatives. This is clear again by Lemma \ref{lem:cov} since the derivatives (after setting $z_k=0$ for all $k$) simply produce a polynomial in the covariances of which we can take a limit by Lemma \ref{lem:cov} and then rewrite as the derivatives acting on the correct quantity. This concludes the proof. 
\end{proof}

\section{The Renormalized Potential of the sine-Gordon Model}\label{sec:rp}
In this section, we begin our analysis of the sine-Gordon model. This will be done through the so-called renormalized potential of the model. The renormalized potential can be viewed as a generating function of the correlation functions featuring in Theorem~\ref{th:sgmain}.  As already mentioned, our approach follows very closely similar analysis in \cite{BB,BH,BMW,BaWe,BK} -- the main difference to previous works is that we use the zero boundary GFF and thus cannot use directly known results (which use different variants of the GFF). While the renormalized potential gives access to the sine-Gordon correlation functions we are interested in, it is a quantity that is computed from the GFF. As our domain is fixed, we will simply write $\langle\cdot\rangle_{\gff(\varepsilon)}$ instead of $\langle\cdot\rangle_{\gff_\Omega(\varepsilon)}$ when convenient.

Let us begin by introducing some notation: first of all we introduce the functional
\begin{equation}
\phi\mapsto v_{\epsilon^2}(\eta,\phi|\varepsilon):=\int d\xi\,\eta(\xi)c_{\sigma\sqrt{\pi}}\varepsilon^{-\frac{\sigma^2}{4}}e^{i\sqrt{\pi}\sigma\phi(x)},
\end{equation}
where we have written $\xi=(x,\sigma)$ and $\int d\xi=\sum_{\sigma\in\{\pm1,\pm2\}}\int_\Omega d^2x$.  The constant $c_{\sigma\sqrt{\pi}}$ is defined in \eqref{eq:wick}. Also here $\eta\in C_c(\Omega\times \{\pm 1,\pm 2\})$ (meaning that for each $\sigma\in \{\pm 1\pm 2\}$, $x\mapsto \eta(x,\sigma)\in C_c(\Omega)$ -- we find it convenient to allow for less smooth functions here than in Theorem~\ref{th:sgmain}) and $\phi\in C(\Omega)$. Further, we define a kind of generalized partition function (which will serve as a generating function for the correlation functions appearing in Theorem~\ref{th:sgmain}) as
\begin{equation}\label{eq:pf}
Z(\eta|\varepsilon)=\gffe{e^{-v_{\epsilon^2}(\eta,\phie|\varepsilon)}}.
\end{equation}

Now, noting that since for each $t>\epsilon^2$ 
\begin{equation*}
(x,y)\mapsto \int_t^\infty ds \,\p_\Omega(s,x,y) \qquad \text{and} \qquad (x,y)\mapsto \int_{\epsilon^2}^t ds \,\p_\Omega(s,x,y)
\end{equation*}
are covariances (see the proof of Lemma~\ref{lem:GFFreg}) and they sum up to $\gffe{\phie(x)\phie(y)}$, we can decompose $\phie=\varphi_{\sqrt{t}}+\psi_{\epsilon,t}$, where $\varphi_{\sqrt{t}}$ and $\psi_{\epsilon,t}$ are independent continuous Gaussian fields (again, see the proof of Lemma~\ref{lem:GFFreg}) where the covariance of $\varphi_{\sqrt{t}}$ is $\int_t^\infty ds \,\p_\Omega(s,x,y)$ and that of $\psi_{\epsilon,t}$ is $\int_{\epsilon^2}^tds \,\p_\Omega(s,x,y)$. This allows introducing the renormalized potential:
\begin{equation}
\phi\mapsto v_t(\eta,\phi|\varepsilon):=-\log\E_{\psi_{\epsilon,t}}\Big(e^{-v_{\epsilon^2}(\eta,\phi+\psi_{\epsilon,t}|\varepsilon)}\Big),
\end{equation}
so that our partition function can be written as 
\begin{equation}\label{eq:pfrp}
Z(\eta|\varepsilon)=\Big\langle e^{-v_t(\eta,\varphi_{\sqrt{t}}|\varepsilon)}\Big\rangle_{\gff(\sqrt{t})}.
\end{equation}
While it seems hard to compute the renormalized potential $v_t$ explicitly, we will show that for small enough $t$, it can be expressed in terms of a suitable series. More precisely, let us introduce for $\eta\in C_c(\Omega\times \{\pm 1,\pm 2\})$ and $\phi\in C(\Omega)$ (real valued) the series
\begin{equation}\label{eq:renpot}
u_t(\eta,\phi|\varepsilon):=\sum_{n=1}^\infty \frac{1}{n!}\int_{(\Omega\times\{\pm1,\pm2\})^n}d\xi_1\dots d\xi_n\,\eta(\xi_1)\dots\eta(\xi_n)\tilde v_t^n(\xi_1,\dots,\xi_n|\varepsilon)e^{i\sqrt{\pi}\sum_{j=1}^n\sigma_j\phi(x_j)}.
\end{equation}
with the coefficients $\tilde v_t^n$ defined recursively by
\begin{equation}\label{eq:vt1}
\tilde v_t^1(\xi_j|\varepsilon):=c_{\sigma_j\sqrt{\pi}}e^{-\frac{\pi}{2}\sigma_j^2\big(\frac{1}{4\pi}\log\varepsilon^2+\int_{\varepsilon^2}^t ds\,\p_\Omega(s,x_j,x_j)\big)}
\end{equation}
and for $I\subset\{1,\dots,n\}$ such that $|I|\geq2$, let
\begin{multline}\label{eq:vti}
\tilde v_t^{|I|}(\xi_I|\varepsilon):=\frac{1}{2}\int_{\varepsilon^2}^t ds\,\sum_{I_1\dot\cup I_2=I}\sum_{i\in I_1, j\in I_2}\pi\sigma_i\sigma_j\p_\Omega(s,x_i,x_j)\tilde v_s^{|I_1|}(\xi_{I_1}|\varepsilon)\tilde v_s^{|I_2|}(\xi_{I_2}|\varepsilon)\\
\times \exp\bigg\{-\frac{1}{2}\sum_{k,l\in I}\pi\sigma_k\sigma_l\int_s^t du\,\p_\Omega(u,x_k,x_l)\bigg\},
\end{multline}
where $\xi_I=(\xi_i)_{i\in I}$ denotes a vector of elements of $\Omega\times \{\pm 1,\pm 2\}$ and $I_1\dot\cup I_2=I$ means that $I_1$ and $I_2$ are disjoint (non-empty) subsets of $I$ so that their union is $I$. We will show that for small enough $t>\epsilon^2$ (``small enough'' depending only on $\eta$ -- not $\epsilon$) this series converges and agrees with the renormalized potential $v_t$.

To state the required estimate, we introduce the following norm 
\begin{equation}
\|f\|_n(\sigma):=
\begin{cases}
\sup_{x\in\Omega}|f(x,\sigma)|, &\text{ if }\quad n=1,\\
\sup_{x_1\in\Omega}\int_{\Omega^{n-1}}dx_2\dots dx_n|f((x_1,\sigma_1),\dots,(x_n,\sigma_n))|, &\text{ if }\quad n\geq 2.
\end{cases}
\end{equation}

\medskip

The main content of this section is the proof of the following proposition.
\begin{proposition}\label{prop:vtest}
For any $t>0$ and $n\geq 1$ there exists a function $h_t^n:(\Omega\times\{\pm1,\pm2\})^{n}\to[0,\infty]$ independent of $\varepsilon$, that is symmetric in its arguments and such that for a fixed $t_0>0$ and $\epsilon^2<t< t_0$ and $n\neq 2$
\begin{equation*}
|\tilde v_t^n(\xi_1,\dots,\xi_n|\varepsilon)|\leq h_t^{n}(\xi_1,\dots,\xi_n).
\end{equation*}
Moreover for each $K\subset \Omega$ compact and $M\geq 1$, there exists a constant $C_{K,M,t_0}$ such that if $\eta\in C_c(\Omega\times \{\pm 1,\pm 2\})$ with $\mathrm{supp}(\eta(\cdot,\sigma))\subset K$ and $\|\eta\|_\infty\leq M$, then the functions $h_t^n$ (with $n\neq 2$) satisfy
\begin{equation}\label{eq:ht-n}
\|\eta^{\otimes n} h_t^n\|_{n}(\sigma)\leq n^{n-2}t^{-1}C_{K,M,t_0}^nt^{\frac{1}{8}\big(8n-\sum_{j=1}^{n}\sigma_j^2\big)}.
\end{equation}
 Here $\eta^{\otimes n}:(\Omega \times \{\pm 1,\pm 2\})^n\to \C$ denotes the function $\eta^{\otimes n}(\xi_1,\dots ,\xi_n)=\prod_{j=1}^n \eta(\xi_j)$.

Furthermore, for $n=2$ and $\sigma_1\sigma_2>-4$, there exists a function $h_t^2$ such that $|\tilde v_t^2|\leq h_t^2$ and \eqref{eq:ht-n} holds, that is 
\begin{equation*}
\|\eta^{\otimes 2} h_t^2\|_{2}(\sigma)\leq C_{K,M,t_0}^2t^{1-\frac{1}{8}(\sigma_1^2+\sigma_2^2)}
\end{equation*}
for some $C_{K,M,t_0}>0$ and all $\eta\in C_c(\Omega\times \{\pm 1,\pm 2\})$ with $\mathrm{supp}(\eta(\cdot,\sigma))\subset K$ and $\|\eta\|_\infty\leq M$.
\end{proposition}

This proposition gives us our main tool for analyzing the renormalized partition function, in that it allows us to control the series expansion \eqref{eq:renpot}. This is recorded in the following corollary.
\begin{corollary}\label{cor:rpot}
For any $K\subset \Omega$ compact, $M\geq 1$, $\eta\in C_c(\Omega\times \{\pm 1,\pm 2\})$ with $\mathrm{supp}(\eta(\cdot,\sigma))\subset K$, $t_0$ as in Proposition~\ref{prop:vtest}, $\|\eta\|_\infty\leq M$, $t_0>0$, and $\mu\in \R\setminus \{0\}$, let 
\begin{equation}
T_\mu=T_\mu(t_0,\eta)=(4 e C_{K,M,t_0}|\mu|)^{-1} \wedge (4 e C_{K,M,t_0}|\mu|)^{-2},
\end{equation}
where $C_{K,M,t_0}$ is the constant from Proposition~\ref{prop:vtest}. Then for $0<\epsilon^2<t< T_\mu$, the series defining $u_t(\mu \eta,\phi|\epsilon)$ converges absolutely, and equals $v_t(\mu \eta,\phi|\epsilon)$ for each $\phi\in C(\Omega)$.
\end{corollary}
\begin{proof}
The proof here is very similar to that of \cite[Corollary 4.3]{BaWe} and we direct the reader there for further details. We begin by pointing out that by Proposition~\ref{prop:vtest}, for say $n\geq 3$ and $\epsilon^2<t< t_0$ and writing for simplicity $C=C_{K,M,t_0}$, we have \begin{align*}
&\frac{1}{n!}\bigg|\mu^n\int_{\Omega^n}d^2x_1\dots d^2x_n\,\eta(\xi_1)\dots\eta(\xi_n)\tilde v_t^n(\xi_1,\dots,\xi_n|\varepsilon)e^{i\sqrt{\pi}\sum_{j=1}^n\sigma_j\phi(x_j)}\bigg|\\
&\quad\leq  \frac{|\mu|^n}{n!} |\Omega| \|\eta^{\otimes n}h_t^n\|_n(\sigma)\\
&\quad\leq |\Omega|  \frac{|\mu|^n}{n!} n^{n-2} t^{-1}  C^n t^{\frac{1}{8}(8n-\sum_{j=1}^n \sigma_j^2)}\\
&\quad\leq |\Omega|  n^{-2} t^{-1} \left(e C|\mu| t^{1-\frac{\sum_{j=1}^n\sigma_j^2}{8n}}\right)^n,
\end{align*} 
where we used the fact that $\frac{n^n}{n!}\leq e^n$. 

If we now have $t<1$ (and $t<T_\mu$) then
\begin{equation*}
t^{1-\frac{\sum_{j=1}^n\sigma_j^2}{8n}}\leq t^{1/2} < T_\mu^{1/2}\leq (4 eC|\mu|)^{-1},
\end{equation*}
so
\begin{align*}
&\sum_{n\geq 3}\frac{1}{n!}\bigg|\mu^n\int_{(\Omega\times\{\pm 1,\pm 2\})^n} d\xi_1\dots d\xi_n\,\eta(\xi_1)\dots\eta(\xi_n)\tilde v_t^n(\xi_1,\dots,\xi_n|\varepsilon)e^{i\sqrt{\pi}\sum_{j=1}^n\sigma_j\phi(x_j)}\bigg|\\
&\quad \leq \sum_{n\geq 3}|\Omega|  n^{-2} t^{-1} \left(4 eC |\mu| t^{1/2}\right)^n ,
\end{align*}
which converges geometrically under our assumption on $t$ (the factor four came from the $\sigma$-sums). If $1<t<T_\mu$ then
\begin{equation*}
t^{1-\frac{\sum_{j=1}^n\sigma_j^2}{8n}}\leq t < T_\mu \leq (4 eC|\mu|)^{-1},
\end{equation*}
so
\begin{align*}
&\sum_{n\geq 3}\frac{1}{n!}\bigg|\mu^n\int_{(\Omega\times\{\pm 1,\pm 2\})^n} d\xi_1\dots d\xi_n\,\eta(\xi_1)\dots\eta(\xi_n)\tilde v_t^n(\xi_1,\dots,\xi_n|\varepsilon)e^{i\sqrt{\pi}\sum_{j=1}^n\sigma_j\phi(x_j)}\bigg|\\
&\quad \leq \sum_{n\geq 3}|\Omega|  n^{-2} t^{-1} \left(4 eC |\mu| t\right)^n ,
\end{align*}
which also converges geometrically under our assumption on $t$.

The $n=1$ and $n=2$ terms do not influence convergence and we conclude that the series defining $u_t(\mu\eta,\phi|\epsilon)$ converges geometrically, uniformly in $\phi\in C(\Omega)$ for (fixed) $\epsilon^2<t<T_\mu$. It remains to show that this series agrees with the renormalized potential.

To simplify notation, let us from now on write $u_t(\phi)=u_t(\mu \eta,\phi|\epsilon)$. Arguing exactly as in \cite[Proof of Corollary 4.3]{BaWe}, one can show that $h_t(\phi)=e^{-u_t(\phi)}$ satisfies the infinite dimensional heat equation 
\begin{equation}\label{eq:hheq}
\partial_t h_t(\phi)=\frac{1}{2}\Delta_{\p_{\Omega}(t,\cdot,\cdot)}h_t(\phi), 
\end{equation}
where 
\begin{equation}
\Delta_{\p_\Omega(t,\cdot,\cdot)}h_t(\phi)=\int_{\Omega^2}d^2x\,d^2y\,\p_{\Omega}(t,x,y)\mathrm{Hess}\, h_t(\phi,x,y),
\end{equation}
and  $\mathrm{Hess}\, h_t$ is the Hessian of $h_t$: if $D^2$ denotes the second Fréchet derivative on $C(\Omega)$,\footnote{Note that while this is not a Banach space -- something one typically assumes for Fréchet analysis -- we can for example restrict to the support of $\eta$ or work on the Banach space of bounded continuous functions on $\Omega$, so this is not an issue.} then for twice Fréchet differentiable functions $F:C(\Omega)\to \C$ (such as $h_t$)
\begin{equation*}
D^2 F(\phi,f_1,f_2)=\int_{\Omega^2}d^2x\,d^2y\,f_1(x)f_2(y)\mathrm{Hess}\, F(\phi,x,y).
\end{equation*}

We will argue that $g_t=e^{-v_t}$ satisfies this same equation and argue that the two solutions must agree, thus $u_t=v_t$. The idea here is to make use of the representation \eqref{eq:hk} to transform the question into a finite dimensional heat equation. Lemma~\ref{lem:evest} ensures that we then have for any nice functional $g$ 
\begin{equation}\label{eq:lapl2}
\Delta_{\p_\Omega(t,\cdot,\cdot)}g(\phi)=\sum_{n=1}^\infty e^{-t \lambda_n}D^2g(\phi,e_n,e_n).
\end{equation}
For each $N\geq 1$, we then introduce the $L^2(\Omega)$-projection $\Pi_N:L^2(\Omega)\to L^2(\Omega)$ which truncates the eigenfunction expansion: if $f=\sum_{n=1}^\infty f_{n}e_{n}$, then 
\begin{equation*}
\Pi_Nf=\sum_{n=1}^Nf_{n}e_{n}.
\end{equation*}
We define $h_t^N(\phi)=h_t(\Pi_N\phi)$, and note that since $h_t$ satisfies the heat equation, \eqref{eq:lapl2} implies that $h_t^N$ satisfies the finite dimensional heat equation
\begin{equation}\label{eq:fdheq}
\partial_t h_t^N(\phi)=\frac{1}{2}\sum_{n=1}^Ne^{-t\lambda_n}D^2h_t^N(\phi,e_n,e_n)
\end{equation}
with initial data $h_{\epsilon^2}^N(\phi)=h_{\epsilon^2}(\Pi_N\phi)$.\footnote{This really is a finite dimensional heat equation in the sense that if we write $\Pi_N\phi=\sum_{n=1}^N\phi_{n}e_n$, then 
\begin{equation*}
\frac{1}{2}\!\sum_{n=1}^N\! e^{-t\lambda_n}D^2h_t^N(\phi,e_n,e_n)=\frac{1}{2}\!\sum_{n=1}^N\! e^{-t\lambda_n}\partial_{\phi_n}^2h_t^N(\phi),
\end{equation*}
which is simply a Laplacian operator with a time-dependent rescaling in the spatial coordinates.}
Moreover, by our controlled series expansion, this solution is bounded in the ``spatial'' $\phi$-coordinates. 

Analogously to \cite{BaWe}, we can realize the field $\psi_{\epsilon,t}$ as the following series
\begin{equation*}
\psi_{\epsilon,t}(x)=\sum_{n=1}^\infty\bigg(\int_{\epsilon^2}^t ds\,e^{-s\lambda_n}\bigg)^{1/2} V_{n}e_{n}(x),
\end{equation*}
where $(V_{n})_{n\geq 1}$ are independent standard  Gaussians. Using this, one can check that $g_t=e^{-v_t}$ satisfies the heat equation \eqref{eq:hheq} and in particular, $g_t^N(\phi)=g_t(\Pi_N\phi)$ satisfies the finite dimensional heat equation \eqref{eq:fdheq} with initial data $g_{\epsilon^2}(\Pi_N\phi)=e^{-v_{\epsilon^2}(\mu \eta,\Pi_N\phi|\epsilon)}$ and by definition, the solution is bounded in $\phi$. 

Thus to argue that $g_t^N=h_t^N$, we must check that the initial data match. For $h_t^N$, the initial data is 
\begin{equation*}
h_{\epsilon^2}(\Pi_N \phi)=e^{-u_{\epsilon^2}(\mu\eta,\Pi_N\phi|\epsilon)}
\end{equation*}
and since 
\begin{equation*}
\tilde v_t^n(\xi|\epsilon)=
\begin{cases}
c_{\sigma\sqrt{\pi}}\epsilon^{-\frac{\sigma^2}{4}}, & n=1\\
0, & \text{else}
\end{cases},
\end{equation*}
we see that 
\begin{equation*}
u_{\epsilon^2}(\mu\eta,\Pi_N\phi|\epsilon)=\int_{\Omega\times \{\pm 1,\pm 2\}}d\xi \,\mu\eta(\xi)c_{\sigma\sqrt{\pi}}\epsilon^{-\frac{\sigma^2}{4}}e^{i\sqrt{\pi}\sigma \Pi_N\phi(x)}=v_{\epsilon^2}(\mu\eta,\Pi_N\phi|\epsilon)
\end{equation*}
so 
\begin{equation*}
h_{\epsilon^2}^N(\phi)=g_{\epsilon^2}^N(\phi)
\end{equation*}
and we must have $h_t^N(\phi)=g_t^N(\phi)$ since we have bounded solutions to finite dimensional heat equation with the same initial data.

To conclude the proof, it is enough to show that $\lim_{N\to \infty}h_t^N(\phi)=h_t(\phi)$ and $\lim_{N\to \infty}g_t^N(\phi)=g_t(\phi)$ for smooth $\phi$. For this, we note that the uniform convergence of the series defining $u_t$ implies that $u_t$ is continuous in $\phi$ and similarly, one can use the definition of $v_t$ to show that $g_t$ is continuous in $\phi$. It thus remains to prove that $\lim_{N\to \infty}\Pi_N\phi=\phi$ uniformly for each compactly supported smooth function (compact support is sufficient since everything is happening in the support of $\eta$). 

For uniform convergence, note that we have (a priori with convergence only in $L^2(\Omega)$), $\phi=\sum_{n=1}^\infty\phi_{n}e_{n}$ where 
\begin{equation*}
\phi_{n}=\int_{\Omega}d^2x\,\phi(x)e_{n}(x)=\big(\lambda_n\big)^{-l}\int_{\Omega}d^2x(-\Delta)^l \phi(x)e_n(x)
\end{equation*}
for each $l\geq 0$. Taking $l=2$, we see by Cauchy-Schwarz that 
\begin{equation*}
|\phi_{n}|\leq C_{\phi} \lambda_n^{-2}.
\end{equation*}
Thus using Lemma~\ref{lem:evest}, we have 
\begin{equation*}
|(\Pi_N \phi)(x)-\phi(x)|\leq C_{\phi}\sum_{n= N+1}^\infty \lambda_n^{-\frac{7}{4}}\leq C_{\phi,\Omega}\sum_{n= N+1}^\infty n^{-7/4}
\end{equation*}
for a suitable constant $C_{\phi,\Omega}$.
This tends to zero as $N\to \infty$, so we conclude that $h_t(\phi)=g_t(\phi)$, implying that $u_t(\phi)=v_t(\phi)$ for all smooth $\phi$. Since $u_t$ and $v_t$ are continuous, we see that $u_t(\phi)=v_t(\phi)$ for all $\phi\in C(\Omega)$ by density.
\end{proof}
Before going into the proof of Proposition~\ref{prop:vtest}, we record some estimates that will be important for handling the cases $n=3$ and $n=4$. We first look at an $n=3$ estimate.
\begin{lemma}\label{lem:3pe}
For $0<\varepsilon^2<t<t_0$ and $\xi_1,\xi_2,\xi_3\in\Omega\times\{\pm1,\pm2\}$, we have
\begin{equation*}
\big|\pi\sigma_1\sigma_3\p_{\Omega}(t,x_1,x_3)+\pi\sigma_2\sigma_3\p_{\Omega}(t,x_2,x_3)\big|\left|1-e^{-\pi\sigma_1\sigma_2\int_{\varepsilon^2}^t ds\,\p_{\Omega}(s,x_1,x_2)}\right|\leq F_t(\xi_1,\xi_2,\xi_3)
\end{equation*}
for some function $F_t$ independent of $\varepsilon$ and symmetric in the arguments. Moreover for any $K\subset \Omega$ compact and $M\geq 1$, we have for each $\eta\in C_c(\Omega \times \{\pm 1,\pm 2\})$ with $\mathrm{supp}(\eta(\cdot,\sigma))\subset K$ and $\|\eta\|_\infty\leq M$, and $\sigma \in \{\pm 1,\pm 2\}^3$ the estimate
\begin{equation*}
\|\eta^{\otimes 3}F_t\|_3(\sigma)\leq C t
\end{equation*}
for some constant $C=C_{K,M,t_0}$.
\end{lemma}
\begin{proof}
Our first remark is that we always have 
\begin{equation*}
\left|1-e^{-\pi\sigma_1\sigma_2\int_{\varepsilon^2}^t ds\,\p_{\Omega}(s,x_1,x_2)}\right|\leq \left|1-e^{-\pi\sigma_1\sigma_2\int_{0}^t ds\,\p_{\Omega}(s,x_1,x_2)}\right|.
\end{equation*}
Thus it is enough to prove the claim for $\epsilon=0$. 

We also note that since (by \eqref{eq:tdens}) we have $\p_{\Omega}(s,x,y)=\p(s,x,y)-R_s(x,y)$ with $0\leq R_s(x,y)\leq \p(s,x,y)$, we have 
\begin{equation*}
\left|1-e^{-\pi\sigma_1\sigma_2\int_{0}^t ds\,\p_{\Omega}(s,x_1,x_2)}\right|\leq \left|1-e^{-\pi\sigma_1\sigma_2\int_{0}^t ds\,\p(s,x_1,x_2)}\right|.
\end{equation*}
Thus (using the triangle inequality), we see that we have 
\begin{align*}
&\big|\pi\sigma_1\sigma_3\p_{\Omega}(t,x_1,x_3)+\pi\sigma_2\sigma_3\p_{\Omega}(t,x_2,x_3)\big|\left|1-e^{-\pi\sigma_1\sigma_2\int_{\varepsilon^2}^t ds\,\p_{\Omega}(s,x_1,x_2)}\right|\\
&\quad\leq \big|\pi\sigma_1\sigma_3\p(t,x_1,x_3)+\pi\sigma_2\sigma_3\p(t,x_2,x_3)\big|\left|1-e^{-\pi\sigma_1\sigma_2\int_{0}^t ds\,\p(s,x_1,x_2)}\right|\\
&\qquad +\big|\pi\sigma_1\sigma_3R_t(x_1,x_3)+\pi\sigma_2\sigma_3R_t(x_2,x_3)\big|\left|1-e^{-\pi\sigma_1\sigma_2\int_{0}^t ds\,\p(s,x_1,x_2)}\right|.
\end{align*}
The first term above is proven to satisfy the required estimate in \cite[Lemma 4.5]{BaWe} (in the notation of \cite{BaWe}, one has to first take the $m\to 0$ limit and choose $\beta$ appropriately to get our claim). It remains to show that the $R_t$-term satisfies the required estimate. We consider separately $\sigma_1\sigma_2<0$ and $\sigma_1\sigma_2>0$.

\medskip

\underline{The case $\sigma_1\sigma_2<0$}: Here our task is to show that 
\begin{equation*}
|R_t(x_1,x_3)-R_t(x_2,x_3)|(e^{\pi|\sigma_1\sigma_2|\int_0^t ds\,\p(s,x_1,x_2)}-1)
\end{equation*}
satisfies the desired estimate. For the exponential term, we rely on \cite[equation (4.41)]{BaWe}, where it is shown that 
\begin{equation}\label{eq:exphk}
e^{\pi|\sigma_1\sigma_2|\int_0^t ds\,\p(s,x_1,x_2)}-1\leq C\int_0^t \frac{dr}{r}e^{-\frac{|x_1-x_2|^2}{4r}}e^{-\frac{|\sigma_1\sigma_2|}{2}\log \big(\frac{|x_1-x_2|}{\sqrt{r}}\wedge 1\big)}.
\end{equation}
for a universal constant $C>0$. For the $R$-term, we recall from \eqref{eq:Rbound2} that if $\delta=d(\mathrm{supp}(\eta(\cdot,\sigma)),\allowbreak\partial \Omega)>0$ (since we assumed that $\eta$ has compact support), then for $x_1,x_2,x_3\in \mathrm{supp}(\eta(\cdot,\sigma))$
\begin{equation*}
|R_t(x_1,x_3)-R_t(x_2,x_3)|\leq C|x_1-x_2|e^{-\frac{\delta^2}{32t}}
\end{equation*}
for $C=C_{K,M,t_0}$.

Thus the relevant estimate is
\begin{align*}
&\sup_{x_1\in \Omega} \int_{\Omega^2}\!\!d^2x_2\,d^2x_3 |\eta(x_1,\sigma_1)\eta(x_2,\sigma_2)\eta(x_3,\sigma_3)|C_{K,t_0}|x_1-x_2|e^{-\frac{\delta^2}{32t}}\!\!\int_0^t \!\frac{dr}{r}e^{-\frac{|x_1-x_2|^2}{4r}}\!e^{-\log\!\big(\frac{|x_1-x_2|^2}{r}\wedge 1\big)}\\
&\quad\leq C_{K,M,t_0}e^{-\frac{\delta^2}{32t}}\int_0^t dr\, r^{1/2} \int_{\R^2}du(|u|+|u|^{-1})e^{-\frac{|u|^2}{4}}\\
&\quad=C_{K,M,t_0}t^{3/2} e^{-\frac{\delta^2}{32t}}\\
&\quad\leq C_{K,M,t_0}t
\end{align*}
for $t<t_0$. Here we made some basic shifts and scalings of the integration variables and the constant $C_{K,M,t_0}$ may change from line to line. By symmetrizing, we find the contribution to $F_t$ from the $R_t$-terms.

\medskip

\underline{The case $\sigma_1\sigma_2>0$}: Here the argument is even simpler. We can simply make use of the bound $\Big|1-e^{-\pi \sigma_1\sigma_2\int_0^t ds\,\p(s,x_1,x_2)}\Big|\leq 1$ and use \eqref{eq:Rbound1} to find that the relevant estimate is 
\begin{align*}
\sup_{x_1\in \Omega}\int_{\Omega^2}d^2x_2\,d^2x_3\, |\eta(x_1,\sigma_1)\eta(x_2,\sigma_2)\eta(x_3,\sigma_3)|R_t(x_1,x_2)&\leq C_{K,M,t_0} e^{-\frac{\delta^2}{32t}}\leq C_{K,M,t_0}t
\end{align*}
for $t<t_0$. Again, $C_{K,M,t_0}$ may change from one instance to another.
\end{proof}

For $n=4$, we have the following estimate which is of a very similar flavor. 
\begin{lemma}\label{lem:4pe}
For $0<\varepsilon^2<t<t_0$ and $\xi_1,\xi_2,\xi_3,\xi_4\in\Omega\times\{\pm1,\pm2\}$, we have
\begin{align*}
&\Bigg|\sum_{i\in \{1,2\}, j\in \{3,4\}}\pi\sigma_i\sigma_j\p_{\Omega}(t,x_i,x_j)\Bigg|\left|1-e^{-\pi\sigma_1\sigma_2\int_{\varepsilon^2}^t ds\,\p_{\Omega}(s,x_1,x_2)}\right|\left|1-e^{-\pi\sigma_3\sigma_4\int_{\varepsilon^2}^t ds\,\p_{\Omega}(s,x_3,x_4)}\right|\\
&\quad\leq G_t(\xi_1,\xi_2,\xi_3,\xi_4),
\end{align*}
for some function $G_t$ independent of $\varepsilon$ and symmetric in its arguments. Moreover, for $K\subset \Omega$ compact, $M\geq 1$, $\eta\in C_c(\Omega \times \{\pm 1,\pm 2\})$ with $\mathrm{supp}(\eta(\cdot,\sigma))\subset K$ and $\|\eta\|_\infty\leq M$, and $\sigma \in \{\pm 1,\pm 2\}^4$ we have
\begin{equation*}
\|\eta^{\otimes 4}G_t\|_4(\sigma)\leq C_{K,M,t_0} t^2
\end{equation*}
for some constant $C=C_{K,M,t_0}$.
\end{lemma}
\begin{proof}
The proof is very similar to that of Lemma~\ref{lem:3pe} and we only sketch the argument leaving the details to the reader. Again, one can use the bound 
\begin{equation*}
\left|1-e^{-\pi\sigma_p\sigma_q\int_{\varepsilon^2}^t ds\,\p_{\Omega}(s,x_p,x_q)}\right|\leq \left|1-e^{-\pi\sigma_p\sigma_q\int_{0}^t ds\,\p(s,x_p,x_q)}\right|
\end{equation*}
as well as write
\begin{align*}
&\Bigg|\sum_{i\in \{1,2\}, j\in \{3,4\}}\pi\sigma_i\sigma_j\p_{\Omega}(t,x_i,x_j)\Bigg|\\
&\quad\leq \Bigg|\sum_{i\in \{1,2\}, j\in \{3,4\}}\pi\sigma_i\sigma_j\p(t,x_i,x_j)\Bigg|+ \Bigg|\sum_{i\in \{1,2\}, j\in \{3,4\}}\pi\sigma_i\sigma_jR_t(x_i,x_j)\Bigg|.
\end{align*}
The bound for the term involving only $\p$ was obtained in \cite[Lemma 4.6]{BaWe}. For the term involving $R_t$, one can, as in the proof of Lemma~\ref{lem:3pe}, make use of \eqref{eq:Rbound1}, \eqref{eq:Rbound2}, and \eqref{eq:Rbound3} to find the required estimate, and the final function $G_t$ is obtained by symmetrizing. We leave further details to the interested reader.
\end{proof}

We now turn to the proof of the main result in this section.
\begin{proof}[Proof of Proposition~\ref{prop:vtest}]
The proof is by induction, though we treat the $n=1,2,3,4$-cases explicitly. 

\medskip

\underline{The case of $n=1$}: Here we begin by recalling from \eqref{eq:vt1} that 
\begin{align*}
\tilde v_t^1(\xi|\varepsilon)&=c_{\sigma\sqrt{\pi}}\exp\left\{-\frac{\pi}{2}\sigma^2\left[\frac{1}{4\pi}\log\varepsilon^{2}+\int_{\varepsilon^2}^tds\left(\frac{1}{4\pi s}-R_s(x,x)\right)\right]\right\}\\
\notag&=c_{\sigma\sqrt{\pi}}\exp\left\{-\frac{1}{8}\sigma^2\log t+\frac{\pi}{2}\sigma^2\int_{\varepsilon^2}^tds\,R_s(x,x)\right\}\end{align*}
Since $0\leq R_s(x,y)$ for all $s,x,y$, \eqref{eq:Rbound1} implies that for any fixed $t_0>0$ there exists a $C=C_{t_0,K}>0$ such that for $x\in K$,  $\int_0^{t_0} ds R_s(x,x)\leq C$. For this $C$, we have for $t<t_0$ and $(x,\sigma)\in K\times \{\pm 1,\pm 2\}$
\begin{equation}\label{eq:ht1}
0\leq \tilde v_t^1(\xi|\epsilon)\leq c_{\sigma\sqrt{\pi}}e^{2\pi C}t^{-\frac{\sigma^2}{8}}=:h_t^{1}(\xi).
\end{equation}
This clearly satisfies the required estimate for Proposition~\ref{prop:vtest}, and we can move on.

\medskip

\underline{The case of $n=2$}: Let us begin the $n=2$ case with an explicit expression for $\tilde v_t^2$. From \eqref{eq:vti} we get
\begin{equation}\label{eq:vt2}
\tilde v_t^2(\xi_1,\xi_2|\varepsilon)=\tilde v_t^1(\xi_1|\varepsilon)\tilde v_t^1(\xi_2|\varepsilon)\left(1-e^{-\pi \sigma_1\sigma_2 \int_{\varepsilon^2}^t ds\,\p_{\Omega}(s,x_1,x_2)}\right).
\end{equation}
Recalling that $0\leq \p_\Omega\leq \p$, using the bound from the $n=1$-case,
\begin{equation}
|\tilde v_t^2(\xi_1,\xi_2|\epsilon)|\leq h_t^1(\xi_1)h_t^1(\xi_2)\left|1-e^{-\pi \sigma_1\sigma_2\int_0^t ds\, \p(s,x_1,x_2)}\right|=:h_t^2(\xi_1,\xi_2). 
\end{equation}
To see that this satisfies the conditions of Proposition~\ref{prop:vtest} (where the relevant situation was $\sigma_1\sigma_2\neq -4$), we make use of \eqref{eq:exphk} (and a similar estimate for $\sigma_1\sigma_2>0$ -- see \cite[equation (4.47)]{BaWe}) and the $n=1$-case  to find
\begin{align*}
\|\eta^{\otimes 2}h_t^2\|_2(\sigma)&=\sup_{x_1\in \Omega}\int_{\Omega}d^2x_2 \,e^{4\pi C}t^{-\frac{\sigma_1^2}{8}-\frac{\sigma_2^2}{8}}|\eta(x_1,\sigma_1)||\eta(x_2,\sigma_2)|\Big(e^{\pi \sigma_1\sigma_2\int_0^t ds\, \p(s,x_1,x_2)}-1\Big)\\
&\leq C\sup_{x_1\in \Omega}\int_{\Omega}d^2x_2\, t^{-\frac{\sigma_1^2}{8}-\frac{\sigma_2^2}{8}}|\eta(x_1,\sigma_1)||\eta(x_2,\sigma_2)|\int_0^t \frac{dr}{r}e^{-\frac{|x_1-x_2|^2}{4r}}e^{\frac{\sigma_1\sigma_2}{2}\log \big(\frac{|x_1-x_2|}{\sqrt{r}}\wedge 1\big)}\\
&\leq C t^{-\frac{\sigma_1^2}{8}-\frac{\sigma_2^2}{8}}\int_0^t dr \int_{\R^2}d^2 u\, e^{-\frac{|u|^2}{4}}e^{\frac{\sigma_1\sigma_2}{2}\log( |u| \wedge 1)},
\end{align*}
where $C=C_{K,M,t_0}$ can change from line to line. For $\sigma_1\sigma_2\neq -4$, this $\R^2$-integral is convergent and we find that for $t<t_0$
\begin{equation*}
\|\eta^{\otimes 2}h_t^2\|_2(\sigma)\leq C t^{1-\frac{\sigma_1^2}{8}-\frac{\sigma_2^2}{8}}
\end{equation*}
for some constant $C=C_{K,M,t_0}$ -- namely the claim of Proposition~\ref{prop:vtest} for $n=2$.

\medskip

\underline{The case of $n=3$}: From the recursion \eqref{eq:vti} and the expression \eqref{eq:vt2} for $\tilde v_t^2$ we can calculate the following expression for $\tilde v_t^3$
\begin{multline*}
\tilde v_t^3(\xi_1,\xi_2,\xi_3|\varepsilon)\\
\shoveleft\quad=\pi\int_{\varepsilon^2}^t ds \prod_{j=1}^3\tilde v_s^{1}(\xi_j|\varepsilon)\bigg[\left(\sigma_1\sigma_2  \p_{\Omega}(s,x_1,x_2)+\sigma_1\sigma_3  \p_{\Omega}(s,x_1,x_3)\right)\left(1-e^{-\pi \sigma_2\sigma_3 \int_{\varepsilon^2}^s dr\,\p_{\Omega}(r,x_2,x_3)}\right)\\
\shoveleft\qquad\quad+\left(\sigma_2\sigma_1  \p_{\Omega}(s,x_2,x_1)+\sigma_2\sigma_3  \p_{\Omega}(s,x_2,x_3)\right)\left(1-e^{-\pi \sigma_1\sigma_3 \int_{\varepsilon^2}^s dr\, \p_{\Omega}(r,x_1,x_3)}\right)\\
\shoveleft\qquad\quad+\left(\sigma_3\sigma_1  \p_{\Omega}(s,x_3,x_1)+\sigma_3\sigma_2  \p_{\Omega}(s,x_3,x_2)\right)\left(1-e^{-\pi \sigma_1\sigma_2 \int_{\varepsilon^2}^s dr\, \p_{\Omega}(r,x_1,x_2)}\right)\bigg]\\
\times\exp\bigg\{-\frac{\pi}{2}\sum_{i,j=1}^3 \sigma_i\sigma_j \int_s^t dr\,\p_{\Omega}(r,x_i,x_j)\bigg\}.
\end{multline*}
The last exponential term is bounded by one (since $\int_s^t dr\, \p_\Omega(r,x,y)$ is a covariance), so by Lemma~\ref{lem:3pe} and the estimate for $\tilde v_s^1$ we find
\begin{equation}
|\tilde v_t^3(\xi_1,\xi_2,\xi_3)|\leq C\int_0^t ds\,h_s^1(\xi_1)h_s^1(\xi_2)h_s^1(\xi_3) F_s(\xi_1,\xi_2,\xi_3)=:h_t^3(\xi_1,\xi_2,\xi_3)
\end{equation}
and the norm can be bounded by
\begin{equation*}
\|\eta^{\otimes 3}h_t^3\|_3(\sigma)\leq C\int_0^t ds\, s^{-\frac{1}{8}\sigma_1^2-\frac{1}{8}\sigma_2^2-\frac{1}{8}\sigma_3^2} s\leq C t^{2-\frac{1}{8}(\sigma_1^2+\sigma_2^2+\sigma_3^2)},
\end{equation*}
as claimed in Proposition~\ref{prop:vtest} (again, $C=C_{K,M,t_0}$ may change from one instance to another). Note that here the $s$-integral is convergent since $\frac{1}{8}(\sigma_1^2+\sigma_2^2+\sigma_3^2)\leq \frac{3}{2}$ (if we allowed for slightly larger values of $\sigma$, the argument would need to be modified following \cite[Lemma 4.7]{BaWe}).

\medskip

\underline{The case of $n=4$}: From the definition \eqref{eq:vti} we can write the following expression for $\tilde v_t^4$
\begin{multline*}
\tilde v_t^4(\xi_1,\xi_2,\xi_3,\xi_4|\varepsilon)\\
\shoveleft\quad=\int_{\varepsilon^2}^t ds\Bigg[\sum_{i=1}^4\sum_{j\in[4]\setminus\{i\}}\pi\sigma_i\sigma_j  \p_{\Omega}(s,x_i,x_j)\tilde v_s^1(\xi_i|\varepsilon)\tilde v_s^3(\xi_{[4]\setminus\{i\}}|\varepsilon)\\
\shoveleft\qquad+\sum_{i\in\{1,2\},j\in\{3,4\}}\pi\sigma_i\sigma_j\p_{\Omega}(s,x_i,x_j)\tilde v_s^2(\xi_1,\xi_2|\varepsilon)\tilde v_s^2(\xi_3,\xi_4|\varepsilon)\\
\shoveleft\qquad+\sum_{i\in\{1,3\},j\in\{2,4\}}\pi\sigma_i\sigma_j\p_{\Omega}(s,x_i,x_j)\tilde v_s^2(\xi_1,\xi_3|\varepsilon)\tilde v_s^2(\xi_2,\xi_4|\varepsilon)\\
+\sum_{i\in\{1,4\},j\in\{2,3\}}\pi\sigma_i\sigma_j\p_{\Omega}(s,x_i,x_j)\tilde v_s^2(\xi_1,\xi_4|\varepsilon)\tilde v_s^2(\xi_2,\xi_3|\varepsilon)\Bigg]e^{-\frac{1}{2}\sum_{i,j\in[4]} \pi\sigma_i\sigma_j \int_s^t du\,\p_{\Omega}(u,x_i,x_j)}.
\end{multline*}
We have written here $[4]=\{1,2,3,4\}$. We look separately at the two different kinds of terms, ones with $\tilde v_s^1\tilde v_s^3$ and ones with $\tilde v_s^2\tilde v_s^2$. 

Let us first consider the $1,3$-case. First of all, we note that again by positive definiteness, we can drop the last exponential term $\big(\sum_{i,j\in [4]}\pi\sigma_i\sigma_j\int_s^t du\, \p_{\Omega}(u,x_i,x_j)\geq 0\big)$. Next we note that since $0\leq \p_{\Omega} \leq  \p$, we have from the $n=1$ and $n=3$ cases
\begin{equation*}
\Bigg|\sum_{i=1}^4 \sum_{j\in[4]\setminus \{i\}}\pi \sigma_i \sigma_j \p_{\Omega}(s,x_i,x_j)\tilde v_s^1(\xi_i|\epsilon)\tilde v_s^3(\xi_{[4]\setminus \{i\}}|\epsilon)\Bigg|\leq 4\pi \sum_{i=1}^4 \sum_{j\in [4]\setminus \{i\}} \p(s,x_i,x_j)h_s^1(\xi_i)h_s^3(\xi_{[4]\setminus \{i\}}). 
\end{equation*}
Using the bounds from the $n=1$ and $n=3$-cases (along with the translation invariance of $\p$ and the fact that $\int_{\R^2}du \,\p(s,u,0)=1$ for all $s>0$) we see that the $\|\cdot \|_4$-norm of this can be bounded by 
\begin{equation*}
4\pi \cdot 4\cdot 3\int_{\R^2}d^2u\, \p(s,u,0)\|\eta h_s^1\|_1 \|\eta^{\otimes 3}h_s^3\|_3\leq C s^{2-\frac{1}{8}\sum_{j=1}^4 \sigma_j^2}.
\end{equation*}
Since $\sigma_j^2\leq 4$, this is integrable in $s$, so when we integrate from $0$ to $t$, we get a bound of the form $C t^{3-\frac{1}{8}\sum_{j=1}^4\sigma_j^2}$ -- precisely of the form needed for Proposition~\ref{prop:vtest}. 

It remains to understand the $\tilde v_2\tilde v_2$-terms, and by symmetry, we can focus on the term 
\begin{multline*}
\sum_{i\in\{1,2\},j\in\{3,4\}}\pi\sigma_i\sigma_j\p_{\Omega}(s,x_i,x_j)\tilde v_s^2(\xi_1,\xi_2|\varepsilon)\tilde v_s^2(\xi_3,\xi_4|\varepsilon)\\
=\sum_{i\in\{1,2\},j\in\{3,4\}}\pi\sigma_i\sigma_j\p_{\Omega}(s,x_i,x_j)\tilde v_s^1(\xi_1|\varepsilon)\tilde v_s^1(\xi_2|\varepsilon)\left(1-e^{-\pi \sigma_1\sigma_2 \int_{\varepsilon^2}^s du\,\p_{\Omega}(u,x_1,x_2)}\right)\\
\times\tilde v_s^1(\xi_3|\varepsilon)\tilde v_s^1(\xi_4|\varepsilon)\left(1-e^{-\pi \sigma_3\sigma_4 \int_{\varepsilon^2}^s dr\,\p_{\Omega}(r,x_3,x_4)}\right).
\end{multline*}
By the estimate \eqref{eq:ht1} and Lemma~\ref{lem:4pe} we have
\begin{align*}
&\Bigg|\sum_{i\in\{1,2\},j\in\{3,4\}}\pi\sigma_i\sigma_j\p_\Omega(s,x_i,x_j)\tilde v_s^2(\xi_1,\xi_2|\varepsilon)\tilde v_s^2(\xi_3,\xi_4|\varepsilon)\Bigg|\\
&\quad\leq h_s^1(\xi_1)h_s^1(\xi_2)h_s^1(\xi_3)h_s^1(\xi_4)G_s(\xi_1,\xi_2,\xi_3,\xi_4).
\end{align*}
So the terms in $\tilde v_t^4$ containing $\tilde v_s^2\tilde v_s^2$ we want to estimate the norm of are
\begin{align*}
&\Bigg|\int_{\varepsilon^2}^t ds\sum_{i\in\{1,2\},j\in\{3,4\}}\pi\sigma_i\sigma_j\p_\Omega(s,x_i,x_j)\tilde v_s^2(\xi_1,\xi_2|\varepsilon)\tilde v_s^2(\xi_3,\xi_4|\varepsilon)e^{-\frac{1}{2}\sum_{i,j\in[4]} \pi\sigma_i\sigma_j \int_s^t du\,\p_\Omega(u,x_i,x_j)}\Bigg|\\
&\quad\leq\int_{0}^t ds\,h_s^1(\xi_1)h_s^1(\xi_2)h_s^1(\xi_3)h_s^1(\xi_4)G_s(\xi_1,\xi_2,\xi_3,\xi_4),
\end{align*}
where again we dropped the last exponential factor.

Calculating the norm, we find by \eqref{eq:ht1} and Lemma~\ref{lem:4pe} that
\begin{align*}
&\sup_{x_1\in\Omega}\int_{\Omega^3}d^2x_2\,d^2x_3\,d^2x_4\,\int_{0}^t ds\,h_s^1(\xi_1)h_s^1(\xi_2)h_s^1(\xi_3)h_s^1(\xi_4)G_s(\xi_1,\xi_2,\xi_3,\xi_4)\\
\notag &\quad\leq C\int_{0}^t ds\,s^{-\frac{1}{8}\sigma_1^2}s^{-\frac{1}{8}\sigma_2^2}s^{-\frac{1}{8}\sigma_3^2}s^{-\frac{1}{8}\sigma_4^2}\|G_s\|_4\\
\notag &\quad\leq C\int_{0}^t ds\,s^{-\frac{1}{8}(\sigma_1^2+\sigma_2^2+\sigma_3^2+\sigma_4^2)}s^2\\
\notag &\quad\leq Ct^{3-\frac{1}{8}(\sigma_1^2+\sigma_2^2+\sigma_3^2+\sigma_4^2)},
\end{align*}
where again $C=C_{K,M,t_0}$ may vary from line to line. Also, again, since $\sigma_j^2\leq 4$, the integral is convergent. 

Combining the two cases, we find a $h_t^4$ which satisfies the claim of Proposition~\ref{prop:vtest} in the $n=4$ case.

We can finally move on to the induction step.

\medskip

\underline{The general case}: Above we have showed the existence of $h_t^n$ for $n=1,3,4$ (as well as for $n=2$ with $\sigma_1\sigma_2\neq -4$) and the bounds for the norms agree with the proposition. For $n\geq5$ we follow the induction proof from \cite{BB,BaWe,BK}. 

Assume that for all $k<n$ and $k\neq 2$, it holds that for $t<t_0$, $|\tilde v_t^{|I|}(\xi_I)|\leq h_t^k(\xi_I)$ and
\begin{equation}\label{eq:IH}
\|\eta^{\otimes k}h_t^k\|_{k}(\sigma)\leq k^{k-2}t^{-1}C^kt^{\frac{1}{8}\big(8k-\sum_{j=1}^{k}\sigma_j^2\big)}
\end{equation}
for some $C=C_{K,M,t_0}$.

We can split the definition of $\tilde v_t^n$ from \eqref{eq:vti} in two parts, one function that does not include factors of $\tilde v_t^2$ and another function does. We look at these separately and begin with the one with no factors of $\tilde v_t^2$. Using the induction hypothesis, the fact that $\int_s^t du\,\p_{\Omega}(u,\cdot,\cdot)$ is a covariance, and that $\p_{\Omega}\leq \p$, we find that 
\begin{align*}
&\Bigg|\frac{1}{2}\int_{\varepsilon^2}^t ds \sum_{\substack{I_1\dot \cup I_2=[n]\\ |I_1|,|I_2|\neq2}}\sum_{i\in I_1,j\in I_2}\pi\sigma_i\sigma_j  \p_{\Omega}(s,x_i,x_j)\tilde v_s^{|I_1|}(\xi_{I_1}|\varepsilon)\tilde v_s^{|I_2|}(\xi_{I_2}|\varepsilon)e^{-\frac{1}{2}\sum_{l,m\in I} \pi\sigma_l\sigma_m \int_s^t du\,\p_{\Omega}(u,x_l,x_m)}\Bigg|\\
&\quad\leq2\pi \int_{0}^t ds \sum_{\substack{I_1\dot \cup I_2=[n]\\ |I_1|,|I_2|\neq2}}\sum_{i\in I_1,j\in I_2}\p(s,x_i,x_j)h_s^{|I_1|}(\xi_{I_1})h_s^{|I_2|}(\xi_{I_2})\\
&\qquad=:h_t^{n,1}(\xi_{I}).
\end{align*}
Here we have used the notation $[n]=\{1,\dots ,n\}$.

Estimating the norm using symmetry, the induction hypothesis \eqref{eq:IH}, and the fact that $\int_{\R^2}d^2x\, \p(s,x,0)=1$, we find
\begin{align*}
\|\eta^{\otimes n}h_t^{n,1}\|_{n}&\leq2\pi  \sum_{\substack{I_1\dot \cup I_2=[n]\\ |I_1|,|I_2|\neq2}}|I_1||I_2|\int_{0}^t ds \int_{\R^2}d^2x\,\p(s,x,0)\|\eta^{\otimes |I_1|}h_s^{|I_1|}\|_{|I_1|}\|\eta^{\otimes |I_2|}h_s^{|I_2|}\|_{|I_2|}\\
\notag &\leq2\pi  \sum_{\substack{I_1\dot \cup I_2=[n]\\ |I_1|,|I_2|\neq2}}|I_1||I_2|\int_{0}^tds\,  |I_1|^{|I_1|-2}s^{-1}C^{|I_1|}s^{\frac{1}{8}\big(8|I_1|-\sum_{j\in I_1}\sigma_j^2\big)}\\
\notag &\quad\times|I_2|^{|I_2|-2}s^{-1}C^{|I_2|}s^{\frac{1}{8}\big(8|I_2|-\sum_{k\in I_2}\sigma_k^2\big)}\\
\notag &=2\pi  C^{n} \sum_{\substack{I_1\dot \cup I_2=[n]\\ |I_1|,|I_2|\neq2}} |I_1|^{|I_1|-1}|I_2|^{|I_2|-1}\int_{0}^tds\, s^{-2+\frac{1}{8}\big(8n-\sum_{j=1}^{n}\sigma_j^2\big)}\\
\notag &\leq2\pi  C^n\sum_{k=1}^{n-1}\binom{n}{k} k^{k-1}(n-k)^{n-k-1} \frac{1}{-1+\frac{1}{8}\big(8n-\sum_{j=1}^n\sigma_j^2\big)}t^{-1+\frac{1}{8}\big(8n-\sum_{j=1}^n\sigma_j^2\big)}\\
\notag &\leq 2\pi C^n  n^{n-2}\frac{2(n-1)}{-1+\frac{1}{8}(8n-\sum_{j=1}^n \sigma_j^2)}t^{-1} t^{\frac{1}{8}\big(8n-\sum_{j=1}^{n}\sigma_j^2\big)}.
\end{align*}
Here we made use of the identity $\sum_{k=1}^n \binom{n}{k} k^{k-1}(n-k)^{n-k-1}=2(n-1)n^{n-2}$ (see \cite[Proof of Proposition 4.1]{BaWe} for an explanation of this identity) and noted that since $\sigma_j^2\leq 4$ and $n\geq 5$, the $s$–integral is convergent. The point now is that for $n\geq 5$,  $\frac{2(n-1)}{-1+\frac{1}{8}(8n-\sum_{j=1}^n \sigma_j^2)}$ is bounded by a universal constant. Thus by possibly increasing $C$, we find the correct bound for terms not involving $\tilde v_2$.

Now for the terms with $|I_1|=2$ or $|I_2|=2$, recalling \eqref{eq:vt2}, Lemma~\ref{lem:3pe} and the induction hypothesis, and dropping the exponential term as before, we find
\begin{align*}
&\Bigg|\frac{1}{2}\int_{\varepsilon^2}^t ds\hspace{-9pt}\sum_{\substack{I_1\dot \cup I_2=[n]\\ |I_1|=2\vee|I_2|=2}}\sum_{i\in I_1,j\in I_2}\hspace{-3pt}\pi\sigma_i\sigma_j  \p_{\Omega}(s,x_i,x_j)\tilde v_s^{|I_1|}(\xi_{I_1}|\varepsilon)\tilde v_s^{|I_2|}(\xi_{I_2}|\varepsilon)e^{-\frac{1}{2}\sum_{l,m\in I} \pi\sigma_l\sigma_m \int_s^t du\,\p_{\Omega}(u,x_l,x_m)}\Bigg|\\
\notag &\leq \int_{0}^t ds\! \sum_{1\leq a<b\leq n}\,\sum_{j\in[n]\setminus\{a,b\}}|\pi\sigma_a\sigma_j\p_{\Omega}(s,x_a,x_j)+\pi\sigma_b\sigma_j\p_{\Omega}(s,x_b,x_j)|\\
\notag &\quad\times|\tilde v_s^{2}(\xi_a,\xi_b|\varepsilon)||\tilde v_s^{n-2}(\xi_{[n]\setminus\{a,b\}}|\varepsilon)|\\
\notag &\leq \sum_{1\leq a<b\leq n}\,\sum_{j\in[n]\setminus\{a,b\}}\int_{0}^t ds|\pi\sigma_a\sigma_j\p_{\Omega}(s,x_a,x_j)+\pi\sigma_b\sigma_j\p_{\Omega}(s,x_b,x_j)|\left|1-e^{-\pi\sigma_a\sigma_b\int_{\varepsilon^2}^sdu\,\p_{\Omega}(u,x_a,x_b)}\right|\\
\notag &\quad\times\tilde v_s^1(\xi_a|\varepsilon)\tilde v_s^1(\xi_b|\varepsilon) |\tilde v_s^{n-2}(\xi_{[n]\setminus\{a,b\}}|\varepsilon)|\\
\notag &\leq \sum_{1\leq a<b\leq n}\,\sum_{j\in[n]\setminus\{a,b\}}\int_{0}^t ds\,F_s(\xi_a,\xi_b,\xi_j)h_s^1(\xi_a)h_s^2(\xi_b) h_s^{n-2}(\xi_{[n]\setminus\{a,b\}})\\
\notag &\quad=:h_t^{n,2}(\xi).
\end{align*}
Using the induction hypothesis \eqref{eq:IH}, Lemma~\ref{lem:3pe}, and symmetry,  we get
\begin{align*}
\|\eta^{\otimes n}h_t^{n,2}\|_n&\leq\frac{n(n-1)}{2}(n-2)C^n (n-2)^{n-4} \int_{0}^t ds\,s^{-2}s^{\frac{1}{8}\big(8n-\sum_{j=1}^{n}\sigma_j^2\big)}\\
\notag &\leq \widetilde C \frac{n(n-1)}{2}(n-2)^{n-3}C^{n}\frac{2}{n-2}t^{-1+\frac{1}{8}\big(8n-\sum_{j=1}^{n}\sigma_j^2\big)}\\
\notag &\leq  \widetilde C n^{n-2}t^{-1}C^n t^{\frac{1}{8}\big(8n-\sum_{j=1}^{n}\sigma_j^2\big)},
\end{align*}
where $\widetilde C$ is the constant from Lemma~\ref{lem:3pe}, and again, we note that the integral is convergent as $n\geq 5$. Just as in the case of no $\tilde v_t^2$-terms, we can again possibly increase $C$ to find the correct bound. 

Thus, choosing $h_t^n=h_t^{n,1}+h_t^{n,2}$ (and symmetrizing) it holds that $|\tilde v_t^n(\xi|\varepsilon)|\leq h_t^{n}(\xi)$ and that 
\begin{equation*}
\|\eta^{\otimes n}h_t^n\|_n\leq n^{n-2}t^{-1}C^n t^{\frac{1}{8}\big(8n-\sum_{j=1}^{n}\sigma_j^2\big)},
\end{equation*}
 as required.
\end{proof}

\section{The Renormalized Partition Function}\label{sec:rpart}
In this section, we will study the $\epsilon\to 0$ asymptotics of the partition function \eqref{eq:pf} through our estimates for the renormalized potential from Proposition~\ref{prop:vtest} and Corollary~\ref{cor:rpot}. We will see that to have a non-trivial $\epsilon\to 0$ limit, we must renormalize the partition function with an explicit multiplicative counter term. After understanding these $\epsilon\to 0$ asymptotics, we will be able to analyze the relevant sine-Gordon correlation functions (in Section~\ref{sec:sgcorr}). The discussion here is again very similar to that in \cite[Section 5]{BaWe}.

The idea in our approach is to use \eqref{eq:pfrp}, namely $Z(\eta|\epsilon)=\langle e^{-v_t(\eta,\varphi_{\sqrt{t}}|\epsilon)}\rangle_{\gff(\sqrt{t})}$, and control this with the expansion provided by Corollary~\ref{cor:rpot}. The issue with this is the $n=2$-term in the expansion of $v_t$, namely 
\begin{equation}
\frac{1}{2}\int_{(\Omega\times \{\pm 1,\pm 2\})^2}d\xi_1\,d\xi_2\,\eta(\xi_1)\eta(\xi_2)\tilde v_t^2(\xi_1,\xi_2|\epsilon)e^{i\sqrt{\pi}\sum_{j=1}^2 \sigma_j\varphi_{\sqrt{t}}(x_j)}.
\end{equation}
Recalling from \eqref{eq:vt2} that $\tilde v_t^2(\xi_1,\xi_2|\epsilon)=\tilde v_t^1(\xi_1|\epsilon)\tilde v_t^1(\xi_2|\epsilon)\left(1-e^{-\pi \sigma_1\sigma_2\int_{\epsilon^2}^t ds\,\p_\Omega(s,x_1,x_2)}\right)$, one can show that if $\sigma_1\sigma_2<0$, then in the $\epsilon\to 0$ limit, this behaves like $|x_1-x_2|^{-\frac{|\sigma_1\sigma_2|}{2}}$ as $x_1\to x_2$. In particular, if $\sigma_1\sigma_2=-4$, then this is not an integrable singularity. It follows that the $n=2$-term is divergent. This divergence can be cured with a single counter term and the $n\neq 2$ terms can all be controlled with Proposition~\ref{prop:vtest}.

Let us begin by defining our counter term and renormalized partition function. There could be various choices here (in particular, compared to \cite{BaWe}, we add a non-singular term to the counter term as this is more convenient for identifying with Ising). For $\xi_1,\xi_2\in\Omega\times \{-2,2\}$ (note that there is no counter term needed for $\sigma=\pm 1$), let 
\begin{multline}\label{eq:ctt}
    A(\xi_1,\xi_2|\varepsilon):=\left\langle \opcolon e^{i\sqrt{\pi}\sigma_1\varphi_\epsilon(x_1)}\opcolon  \opcolon e^{i\sqrt{\pi}\sigma_2\varphi_\epsilon(x_2)}\opcolon  \right\rangle_{\gff(\epsilon)}\\
    -\left\langle \opcolon e^{i\sqrt{\pi}\sigma_1\varphi_\epsilon(x_1)}\opcolon  \right\rangle_{\gff(\epsilon)} \left\langle \opcolon e^{i\sqrt{\pi}\sigma_2\varphi_\epsilon(x_2)}\opcolon  \right\rangle_{\gff(\epsilon)}.
\end{multline}
We can then define (for $\eta\in C_c(\Omega\times \{\pm 1, \pm 2\})$) the renormalized partition function as
\begin{equation}\label{eq:rpf}
\Z(\eta|\varepsilon)=Z(\eta|\varepsilon)\exp\bigg\{-\frac{1}{2}\int_{(\Omega\times\{\pm2\})^2}d\xi_1\,d\xi_2\,\eta(\xi_1)\eta(\xi_2)A(x_1,x_2|\varepsilon)\bigg\}.
\end{equation}

The goal of this section is to prove the following theorem.
\begin{theorem}\label{thm:uc}
For $\eta\in C_c(\Omega\times\{\pm1,\pm2\})$ the following claims hold
\begin{enumerate}[label=(\roman*)]
\item The limit
\begin{equation*}
\Z(\eta):=\lim_{\varepsilon\to 0}\Z(\eta|\varepsilon),
\end{equation*}
exists and is finite.
\medskip
\item The function $\mu\to\Z(\mu\eta)$ is an entire function of $\mu\in\C$.
\medskip
\item If $\eta(x,\sigma)=\overline{\eta(x,-\sigma)}$ for $\sigma\in\{1,2\}$ and $x\in\Omega$, then $\Z(\eta)>0$.
\medskip
\item Let $\alpha\in\C^N$ be some complex parameters and $K\subset \C^N$ compact. If $\eta_\alpha(\cdot),\eta_\alpha(\cdot|\varepsilon)\in C_c(\Omega\times \{\pm 1,\pm 2\})$ depend on $\alpha$ so that
\begin{equation*}
\lim_{\varepsilon\to0}\sup_{\alpha\in K}\|\eta_\alpha(\cdot|\varepsilon)-\eta_\alpha(\cdot)\|_{L^\infty(\Omega\times\{\pm1,\pm2\})}=0,
\end{equation*}
\begin{equation}\label{eq:newetaassu}
\sup_{\alpha\in K}\|\eta_\alpha\|_{L^\infty(\Omega\times\{\pm 1,\pm 2\})}<\infty \qquad \text{and} \qquad \overline{\bigcup_{\alpha\in K}\bigcup_{\sigma\in\{\pm 1,\pm 2\}}\mathrm{supp}(\eta_\alpha(\cdot,\sigma))}\subset\Omega \qquad \text{is compact},
\end{equation}
then 
\begin{equation*}
\lim_{\varepsilon\to0}|\Z(\eta_{\alpha}(\cdot|\varepsilon)|\varepsilon)-\Z(\eta_{\alpha}(\cdot))|=0
\end{equation*}
and the convergence is uniform in $\alpha\in K$.
\end{enumerate}
\end{theorem}

The main step in proving the convergence and analyticity statements of Theorem~\ref{thm:uc}, is to prove suitable bounds. More precisely, we prove the following proposition. 
\begin{proposition}\label{prop:ub}
Fix a compact set $K\subset\Omega$. Then
\begin{enumerate}[label=(\roman*)]
\item For any fixed $M>0$ 
\begin{equation*}
\sup_{\substack{\varepsilon\in(0,1), \eta\in C_c(\Omega\times \{\pm 1,\pm 2\}):\\ \|\eta\|_{L^\infty(K\times\{\pm1,\pm2\})}\leq M,\\ \mathrm{supp}(\eta(\cdot,\sigma))\subset K}}|\Z(\eta|\varepsilon)|<\infty.
\end{equation*}

\item There exists a $\delta>0$ independent of $\varepsilon$ such that
\begin{equation*}
\inf_{\substack{\varepsilon\in(0,1), \eta\in C_c(\Omega\times \{\pm 1,\pm 2\}):\\ \|\eta\|_{L^\infty(K\times\{\pm1,\pm2\})}\leq \delta, \\
\mathrm{supp}(\eta(\cdot,\sigma))\subset K}
}|\Z(\eta|\varepsilon)|>0.
\end{equation*}

\item For any fixed $M>0$, if $\eta(\cdot,\sigma)=\overline{\eta(\cdot,-\sigma)}$ for $\sigma\in\{1,2\}$,
\begin{equation*}
\inf_{\substack{\varepsilon\in(0,1), \eta\in C_c(\Omega\times \{\pm 1,\pm 2\})\\ \|\eta\|_{L^\infty(K\times\{\pm1,\pm2\})}\leq M\\ \mathrm{supp}(\eta(\cdot,\sigma))\subset K\\ \eta(\cdot,\sigma)=\overline{\eta(\cdot,-\sigma)}}}\Z(\eta|\varepsilon)>0.
\end{equation*}
\end{enumerate}
\end{proposition} 

The proof is rather lengthy. We begin with some facts that are needed to control the $n=2$-term: first we recall some basic estimates for Gaussian processes and then some estimates for the counter term.
\begin{lemma}\label{lem:b}
For any fixed $K\subset \Omega$ compact, $M>0$, $t>0$ there exists a constant $C_{K,M,t}$ such that if $\eta\in C_c(\Omega)$ satisfies $\mathrm{supp}(\eta(\cdot,\sigma))\subset K$ and $\|\eta\|_\infty\leq M$, then 
\begin{equation*}
\left\langle e^{\|\eta \nabla \varphi_{\sqrt{t}}\|_{L^\infty(\Omega)}}\right\rangle_{\gff(\sqrt{t})} \leq C_{K,M,t}.
\end{equation*}
\end{lemma}

For the proof (which is very similar to the proof of \cite[Lemma 5.2]{BaWe}), we will need a general fact about extrema of Gaussian processes known as the Borell-TIS inequality  \cite[Theorem 2.1.1.]{AT}. To state the theorem, let us assume that $X$ is an almost surely bounded Gaussian process on a topological space $T$ and define
\begin{equation}
\sigma_T^2:=\sup_{x\in T}\E\Big[X(x)^2\Big].
\end{equation}
Then 
\begin{equation*}
\E\bigg[\sup_{x\in T}X(x)\bigg]<\infty 
\end{equation*}
and for all $u>0$,
\begin{equation}\label{eq:tis}
\Prob\bigg(\sup_{x\in T}X(x)-\E\bigg[\sup_{x\in T}X(x)\bigg]>u\bigg)\leq e^{-u^2/(2\sigma_T^2)}.
\end{equation}
We can now turn to the proof of the Gaussian estimate.

\begin{proof}[Proof of Lemma~\ref{lem:b}]
We begin by pointing out that by symmetry, $\partial_1 \varphi_{\sqrt{t}}$ and $\partial_2\varphi_{\sqrt{t}}$ have the same law, so by Cauchy-Schwarz it is sufficient to prove the statement for $\nabla \varphi_{\sqrt{t}}$ replaced by say $\partial_1 \varphi_{\sqrt{t}}$. Let us write $K$ for the support of $\eta$. Noting that since $\varphi_{\sqrt{t}}$ is centered,
\begin{align*}
\left\langle e^{\|\eta \partial_1 \varphi_{\sqrt{t}}\|_\infty}\right\rangle_{\gff(\sqrt{t})}&\leq \left\langle e^{\|\eta\|_\infty \sup_{x\in K}|\partial_1 \varphi_{\sqrt{t}}|}\right\rangle_{\gff(\sqrt{t})}\\
\notag &\leq \left\langle e^{M\sup_{x\in K}\partial_1 \varphi_{\sqrt{t}}(x)}\right\rangle_{\gff(\sqrt{t})}+\left\langle e^{M\sup_{x\in K}(- \partial_1 \varphi_{\sqrt{t}}(x))}\right\rangle_{\gff(\sqrt{t})}\\
\notag &=2\left\langle e^{M\sup_{x\in K}\partial_1 \varphi_{\sqrt{t}}(x)}\right\rangle_{\gff(\sqrt{t})}.
\end{align*}
We will control this by estimating the tail of the distribution of $\partial_1\varphi_{\sqrt{t}}$ with Borell-TIS. 

To apply Borell-TIS, let us compute $\sigma_{t}^2=\sup_{x\in K}\E\Big[\big(\partial_1 \varphi_{\sqrt{t}}(x)\big)^2\Big]$. Using Lemma~\ref{lem:evest}, we have for $x\in K$
\begin{align*}
\E\Big[\big(\partial_1 \varphi_{\sqrt{t}}(x)\big)^2\Big]=\sum_{n= 1}^\infty \frac{1}{\lambda_n}e^{-t\lambda_n}|\partial_1 e_{n}(x)|^2&\leq C_{t,K}
\end{align*}
for a suitable constant $C_{t,K}>0$.

If we now let $\tilde u=u-\E\big[\sup_{x\in K}\partial_1\varphi_{\sqrt{t}}(x)\big]>0$ we get from \eqref{eq:tis}
\begin{equation}\label{eq:BTIS}
\Prob\bigg(\sup_{x\in\Omega}\partial_1\varphi_{\sqrt{t}}(x)>u\bigg)\leq \exp\left\{-\frac{\tilde u^2}{2\sigma_{t}^2}\right\}\leq \exp\bigg\{-C_{t,K}\bigg(u-\E\bigg[\sup_{x\in K}\partial_1\varphi_{\sqrt{t}}(x)\bigg]\bigg)^2\bigg\}
\end{equation}
for some (possibly different from earlier) constant $C_{t,K}>0$.

Writing
\begin{align*}
\left\langle e^{M \sup_{x\in K}\partial_1\varphi_{\sqrt{t}}(x)}\right\rangle_{\gff(\sqrt{t})}&=\left\langle \int_{\R} du\,M e^{ M u }\1_{\{\sup_{x\in K}\partial_1\varphi_{\sqrt{t}}(x)>u\}}\right\rangle_{\gff(\sqrt{t})}\\
\notag &=\int_{\R} du\,M e^{ M u }\Prob\bigg(\sup_{x\in K}\partial_1\varphi_{\sqrt{t}}(x)>u\bigg),
\end{align*}
and letting $\mu_{t,K}=\E\big[\sup_{x\in K}\partial_1\varphi_{\sqrt{t}}(x)\big]$ (which is finite by Borell-TIS), by \eqref{eq:BTIS}
\begin{align*}
\left\langle e^{M \sup_{x\in K}\partial_1\varphi_{\sqrt{t}}(x)}\right\rangle_{\gff(\sqrt{t})}&\leq\int_{-\infty}^{\mu_{t,K}} du\,M e^{M u}+\int^{\infty}_{\mu_{t,K}} du\,M e^{M u }\Prob\bigg(\sup_{x\in K}\partial_1\varphi_{\sqrt{t}}(x)>u\bigg)\\
\notag &\leq\int_{-\infty}^{\mu_{t,K}} du\,M e^{M u } +\int^{\infty}_{\mu_{t,K}} du\,M e^{M u }e^{-C_{t,K}(u-\mu_{t,K})^2}\\
\notag &\leq\int_{-\infty}^{\mu_{t,K}} du\,M e^{M u }+e^{M  \mu_{t,K}}\int^{\infty}_{0} dv\,M e^{M v-C_{t,K} v^2}.
\end{align*}
This is a finite constant depending only on $M,t,K$. Thus
\begin{equation*}
\left\langle e^{\sup_{x\in\Omega}|\eta(x)\nabla\varphi_{\sqrt{t}}(x)|}\right\rangle_{\gff(\sqrt{t})}\leq \tilde C_{K,M,t}<\infty.
\end{equation*}
\end{proof}

We now turn to estimating the counter term. The following lemma says basically that in the renormalized partition function, we can replace $A$ by $\tilde v_t^2$ for any fixed $t>0$ without influencing convergence.

\begin{lemma}\label{lem:a+v2}
For each $t>0$, there exists a function $g_t\in L_{loc}^1((\Omega\times\{-2,2\})^2)$ $($namely $g_t((\cdot,\sigma),(\cdot,\sigma'))\in L^1(K)$ for any $K\subset \Omega^2$ compact and $\sigma,\sigma'\in \{\pm 2\}$$\,)$ which is independent of $\varepsilon$ such that for all $\xi_1,\xi_2\in\Omega\times \{-2,2\}$ and $t>\epsilon^2$
\begin{equation*}
\big|A(\xi_1,\xi_2|\varepsilon)+\tilde v_t^2(\xi_1,\xi_2|\varepsilon)\big|\leq g_t(\xi_1,\xi_2),
\end{equation*}
with $A(\xi_1,\xi_2|\varepsilon)$ from \eqref{eq:ctt} and $\tilde v_t^2(\xi_1,\xi_2|\varepsilon)$ from \eqref{eq:vt2}.
\end{lemma}
\begin{proof}
First of all, note that for $\sigma_1\sigma_2=4$, the left hand side is bounded and there is nothing to prove.  We can thus focus on the $\sigma_1\sigma_2=-4$-case.

From \eqref{eq:vt2} and \eqref{eq:vt1} we get that
\begin{align*}
\tilde v_t^2(\xi_1,\xi_2|\varepsilon)\1_{\{\sigma_1\sigma_2=-4\}}&=\tilde v_t^1(\xi_1|\varepsilon)\tilde v_t^1(\xi_2|\varepsilon)\left(1-e^{-\pi \sigma_1\sigma_2 \int_{\varepsilon^2}^t ds\,\p_{\Omega}(s,x_1,x_2)}\right)\1_{\{\sigma_1\sigma_2=-4\}}\\
&=c_{\sigma_1\sqrt{\pi}}c_{\sigma_2\sqrt{\pi}}\varepsilon^{-\frac{\sigma_1^2+\sigma_2^2}{4}}e^{-\frac{1}{2}\pi\sigma_1^2\int_{\varepsilon^2}^t ds\,\p_{\Omega}(s,x_1,x_1)}\\
&\qquad \times e^{-\frac{1}{2}\pi\sigma_2^2\int_{\varepsilon^2}^t ds\,\p_{\Omega}(s,x_2,x_2)}\left(1-e^{-\pi \sigma_1\sigma_2 \int_{\varepsilon^2}^t ds\,\p_{\Omega}(s,x_1,x_2)}\right)\1_{\{\sigma_1\sigma_2=-4\}}\\
&=c_{\sqrt{4\pi}}^2\varepsilon^{-2}e^{-2\pi\int_{\varepsilon^2}^t ds\,\p_{\Omega}(s,x_1,x_1)}e^{-2\pi\int_{\varepsilon^2}^t ds\,\p_{\Omega}(s,x_2,x_2)}\!\left(\!1-e^{4\pi \int_{\varepsilon^2}^t ds\,\p_{\Omega}(s,x_1,x_2)}\right).
\end{align*}
Calculating $A$, we have (by symmetry we can choose $\sigma_1=2$, $\sigma_2=-2$)
\begin{align*}
&A(\xi_1,\xi_2|\varepsilon)\1_{\{\sigma_1\sigma_2=-4\}}\\
&\quad=\left\langle\opcolon e^{i\sqrt{4\pi}\varphi_\epsilon(x_1)}\opcolon \opcolon e^{-i\sqrt{4\pi}\varphi_\epsilon(x_2)}\opcolon \right\rangle_{\gff(\epsilon)}-\left\langle\opcolon e^{i\sqrt{4\pi}\varphi_\epsilon(x_1)}\opcolon \right\rangle_{\gff(\epsilon)}\left\langle\opcolon e^{-i\sqrt{4\pi}\varphi_\epsilon(x_2)}\opcolon \right\rangle_{\gff(\epsilon)}\\
&\quad=c_{\sqrt{4\pi}}^2\varepsilon^{-2}\left\langle e^{i\sqrt{4\pi}\big(\varphi_\epsilon(x_1)-\varphi_\epsilon(x_2)\big)}\right\rangle_{\gff(\epsilon)}-\left\langle\opcolon e^{i\sqrt{4\pi}\varphi_\epsilon(x_1)}\opcolon \right\rangle_{\gff(\epsilon)}\left\langle\opcolon e^{-i\sqrt{4\pi}\varphi_\epsilon(x_2)}\opcolon \right\rangle_{\gff(\epsilon)}\\
&\quad=c_{\sqrt{4\pi}}^2\varepsilon^{-2}e^{-2\pi\int_{\varepsilon^2}^\infty ds\,\p_{\Omega}(s,x_1,x_1)}e^{-2\pi\int_{\varepsilon^2}^\infty ds\,\p_{\Omega}(s,x_2,x_2)}\left(e^{4\pi\int_{\varepsilon^2}^\infty ds\,\p_{\Omega}(s,x_1,x_2)}-1\right).
\end{align*}
Thus
\begin{multline*}
\big|A(\xi_1,\xi_2|\epsilon)+\tilde v_t^2(\xi_1,\xi_2|\epsilon)\big|\1_{\{\sigma_1\sigma_2=-4\}}\\
\shoveleft\quad\leq c_{\sqrt{4\pi}}^2\epsilon^{-2} e^{-2\pi \int_{\epsilon^2}^t ds\,(\p_{\Omega}(s,x_1,x_1)+\p_{\Omega}(s,x_2,x_2))}e^{4\pi\int_{\epsilon^2}^t ds \,\p_{\Omega}(s,x_1,x_2)}\\
\shoveleft\qquad\times \Big| e^{-2\pi\int_t^\infty ds\,(\p_{\Omega}(s,x_1,x_1)+\p_{\Omega}(s,x_2,x_2))} e^{4\pi \int_t^\infty ds \,\p_{\Omega}(s,x_1,x_2)} -1\Big|\\
+ c_{\sqrt{4\pi}}^2\epsilon^{-2}e^{-2\pi \int_{\epsilon^2}^t ds\,(\p_{\Omega}(s,x_1,x_1)+\p_{\Omega}(s,x_2,x_2))} \Big|e^{-2\pi\int_t^\infty ds\,(\p_{\Omega}(s,x_1,x_1)+\p_{\Omega}(s,x_2,x_2))}-1\Big|.
\end{multline*}
Let us estimate the various quantities here for fixed a compact set $K\subset {\Omega}$.

First of all, from \eqref{eq:ht1}, we have for each $t>0$ a constant $C_{t,K}>0$ such that for $x_1,x_2\in K$ 
\begin{equation*}
c_{\sqrt{4\pi}}^2\epsilon^{-2}e^{-2\pi\int_{\epsilon^2}^t ds\,(\p_{\Omega}(s,x_1,x_1)+\p_{\Omega}(s,x_2,x_2))}\leq C_{t,K}.
\end{equation*}
Next, we note since $\p_{\Omega}\leq \p$, \cite[Lemma 4.4]{BaWe} implies that for each $t>0$ there exists a constant $C_{t,K}>0$ such that for $x_1,x_2\in K$
\begin{equation*}
e^{4\pi\int_{\epsilon^2}^t ds \,\p_{\Omega}(s,x_1,x_2)}\leq e^{4\pi \int_0^t ds \,\p(s,x_1,x_2)}\leq C_{t,K}|x_1-x_2|^{-2}.
\end{equation*}

Next, we note that Lemma~\ref{lem:evest} implies that for some $C_{t,K}>0$ we have for $x_1,x_2\in K$
\begin{align*}
&\bigg|\int_t^\infty ds(2\p_{\Omega}(s,x_1,x_2)-\p_{\Omega}(s,x_1,x_1)-\p_{\Omega}(s,x_2,x_2))\bigg|\\
&\quad\leq \sum_{n= 1}^\infty\int_t^\infty ds\, e^{-s\lambda_n}\Big|e_{n}(x_1)(e_{n}(x_2)-e_{n}(x_1))-(e_{n}(x_2)-e_{n}(x_1))e_{n}(x_2)\Big|\\
&\quad\leq C_{t,K}|x_1-x_2|.
\end{align*}
Thus for some (possibly different) $C_{t,K}>0$, 
\begin{equation*}
\left| e^{-2\pi\int_t^\infty ds\,(\p_{\Omega}(s,x_1,x_1)+\p_{\Omega}(s,x_2,x_2))} e^{4\pi \int_t^\infty ds \,\p_{\Omega}(s,x_1,x_2)} -1\right|\leq C_{t,K}|x_1-x_2|.
\end{equation*}
Finally, noting that since $\p_{\Omega}$ is non-negative,
\begin{equation*}
\left|e^{-2\pi\int_t^\infty ds\,(\p_{\Omega}(s,x_1,x_1)+\p_{\Omega}(s,x_2,x_2))}-1\right|\leq 1.
\end{equation*}
Combining all these estimates, we find that for some constant $C_{t,K}>0$, 
\begin{equation*}
\big|A(\xi_1,\xi_2|\epsilon)+\tilde v_t^2(\xi_1,\xi_2|\epsilon)\big|\1_{\{\sigma_1\sigma_2=-4\}}\\
\leq C_{t,K}|x_1-x_2|^{-1},
\end{equation*}
which concludes the proof.
\end{proof}

The previous lemma says that we can swap the counter term $A$ for $\tilde v_t^2$. Let us now argue that the $\tilde v_t^2$-counter term is sufficient to cure the divergence in the $n=2$ term.

\begin{lemma}\label{lem:vt2}
For each $t>0$ and $K\subset \Omega$ compact, there exists a function $\widehat g_t\in L^1(K^2)$ which is independent of $\varepsilon$ such that for all $\xi_1,\xi_2\in\Omega\times \{-2,2\}$ 
\begin{equation*}
\left|\tilde v_t^{2}(\xi_1,\xi_2|\varepsilon)\Big(e^{i\sqrt{\pi}\big(\sigma_1\phit(x_1)+\sigma_2\phit(x_2)\big)}-1\Big)\right|\leq \widehat g_t(x_1,x_2)\left(1+\|\nabla\phit\|_{L^\infty(K)}\right),
\end{equation*}
with $\tilde v_t^2(\xi_1,\xi_2|\varepsilon)$ from \eqref{eq:vt2}.
\end{lemma}
\begin{proof}
First of all, if $\sigma_1=\sigma_2$, then the left hand side is bounded, and there is nothing to prove. We thus focus on $\sigma_1\sigma_2=-4$. Let us fix a compact set $K\subset \Omega$. Using \eqref{eq:vt2} and \eqref{eq:ht1}, there exists a constant $C_{t,K}$ so that we can write 
\begin{align*}
&\left|\tilde v_t^{2}(\xi_1,\xi_2|\varepsilon)\Big(e^{i\sqrt{\pi}\big(\sigma_1\phit(x_1)+\sigma_2\phit(x_2)\big)}-1\Big)\right|\1_{\{\sigma_1\sigma_2=-4\}}\\
&\quad\leq C_{t,K}\left|1-e^{4\pi \int_{\varepsilon^2}^t ds\,\p_{\Omega}(s,x_1,x_2)}\right|\left|e^{i\sqrt{4\pi}\big(\phit(x_1)-\phit(x_2)\big)}-1\right|.
\end{align*}
Since $x\mapsto e^{ix}$ is $1$-Lipschitz, we see that the last term can be bounded by the quantity $\sqrt{4\pi}\big\|\nabla \varphi_{\sqrt{t}}\big\|_{L^\infty(K)}|x_1- x_2|$. Since $\p_{\Omega}\leq \p$, it is sufficient to prove local integrability of   
\begin{equation*}
|x_1-x_2| e^{4\pi \int_0^t ds \,\p(s,x_1,x_2)}.
\end{equation*}
It follows from \cite[Lemma 4.4]{BaWe} (note that the statement there is stated for the ``massive heat kernel'', but the estimate is uniform in the mass, so it holds also for our $\p$)  that there exists a constant $C_{t,K}$ such that for $x_1,x_2\in K$, this can be bounded by $C_{t,K}|x_1-x_2|^{-1}$, which is integrable, and the proof is complete.
\end{proof}

We can now turn to the proof of Proposition~\ref{prop:ub}. 
\begin{proof}[Proof of Proposition~\ref{prop:ub}]
\textit{(i)} Using Corollary~\ref{cor:rpot}, we can choose $t$ small enough (independent of $\epsilon$) that for each of the relevant $\eta$, the renormalized partition function can be written as
\begin{align*}
\Z(\eta|\varepsilon)&=\bigg\langle e^{-\sum_{n=1}^\infty \frac{1}{n!}\int_{(\Omega\times\{\pm1,\pm2\})^n}d\xi_1\dots d\xi_n\,\eta(\xi_1)\dots\eta(\xi_n)\tilde v_t^n(\xi_1,\dots,\xi_n|\varepsilon)e^{i\sqrt{\pi}\sum_{j=1}^n\sigma_j\phit(x_j)}}\bigg\rangle_{\gff(\sqrt{t})}\\
&\qquad\times\exp\bigg\{-\frac{1}{2}\int_{(\Omega\times\{\pm2\})^2}d\xi_1\,d\xi_2\,\eta(\xi_1)\eta(\xi_2)A(\xi_1,\xi_2|\varepsilon)\bigg\}\\
&=\bigg\langle \exp\bigg\{-\sum_{n=1}^\infty \frac{1}{n!}\int_{(\Omega\times\{\pm1,\pm2\})^n}d\xi_1\dots d\xi_n\,\eta(\xi_1)\dots\eta(\xi_n)\tilde v_t^n(\xi_1,\dots,\xi_n|\varepsilon)\\
&\qquad\times\Big(e^{i\sqrt{\pi}\sum_{j=1}^n\sigma_j\phit(x_j)}-\1_{\{n=2\}}\1_{\{\sigma_1\sigma_2=\pm 4\}}\Big)\bigg\}\bigg\rangle_{\gff(\sqrt{t})}\\
&\quad\qquad\times \exp\bigg\{-\frac{1}{2}\int_{(\Omega\times\{\pm2\})^2}d\xi_1\,d\xi_2\,\eta(\xi_1)\eta(\xi_2)\Big(A(\xi_1,\xi_2|\varepsilon)+\tilde v_t^2(\xi_1,\xi_2|\varepsilon)\Big)\bigg\}.
\end{align*}
For the second factor on the right hand side we have by Lemma~\ref{lem:a+v2} and by our assumption of $\mathrm{supp}(\eta(\cdot,\sigma))\subset K$ and that $\|\eta\|_{L^\infty(\Omega\times\{\pm1,\pm2\})}\leq M$ that there exists a constant $C_{t,K,M}>0$ such that 
\begin{align*}
\exp\bigg\{-\frac{1}{2}\int_{(\Omega\times\{\pm2\})^2}d\xi_1\,d\xi_2\,\eta(\xi_1)\eta(\xi_2)\Big(A(\xi_1,\xi_2|\varepsilon)+\tilde v_t^2(\xi_1,\xi_2|\varepsilon)\Big)\bigg\}\leq C_{t,K,M}.
\end{align*}
For the first factor on the right hand side we use Proposition~\ref{prop:vtest} (we also fix some small enough $t_0$ for the proposition and consider $t<t_0$), so
\begin{align*}
&\bigg| \exp\bigg\{-\sum_{n=1}^\infty \frac{1}{n!}\int_{(\Omega\times\{\pm1,\pm2\})^n}d\xi_1\dots d\xi_n\,\eta(\xi_1)\dots\eta(\xi_n)\tilde v_t^n(\xi_1,\dots,\xi_n|\varepsilon)\\
&\qquad\times \Big(e^{i\sqrt{\pi}\sum_{j=1}^n\sigma_j\phit(x_j)}-\1_{\{n=2\}}\1_{\{\sigma_1\sigma_2=\pm 4\}}\Big)\bigg\}\bigg|\\
&\quad\leq e^{\sum_{n=1}^\infty \frac{n^{n-2}}{n!}|K|t^{-1}C_{K,M,t_0}^n  \sum_{\sigma_1,\dots ,\sigma_n\in \{\pm 1,\pm 2\}} t^{\frac{1}{8}\big(8n-\sum_{j=1}^n \sigma_j^2\big)}}\\
&\qquad \times e^{\frac{1}{2}\int_{(\Omega\times\{\pm1,\pm2\})^2}d\xi_1\, d\xi_2\,|\eta(\xi_1)||\eta(\xi_2)||\tilde v_t^2(\xi_1,\xi_2|\varepsilon)|\big|e^{i\sqrt{\pi}\sum_{j=1}^2\sigma_j\phit(x_j)}-1\big|\1_{\{\sigma_1\sigma_2=\pm4\}}}
\end{align*}
By Stirling, we have for some $\tau=\tau(K,M,t_0)$: if $t<\tau$, then the above series converges geometrically, and 
\begin{align*}
e^{\sum_{n=1}^\infty \frac{n^{n-2}}{n!}|K|t^{-1}C_{K,M,t_0}^n \sum_{\sigma_1,\dots ,\sigma_n\in \{\pm 1,\pm 2\}}t^{\frac{1}{8}\big(8n-\sum_{j=1}^n \sigma_j^2\big)}}\leq e^{C_{K,M,t_0} t^{-1}}
\end{align*}
again for a possibly different $C_{K,M,t_0}$. For the $n=2$-term, we note that by the assumption of $\|\eta\|_{L^\infty(\Omega \times\{\pm1,\pm2\})}\leq M$, $\mathrm{supp}(\eta(\cdot,\sigma))\subset K$, and Lemma~\ref{lem:vt2}
\begin{align*}
&\frac{1}{2}\int_{(\Omega\times\{\pm1,\pm2\})^2}d\xi_1\, d\xi_2\,|\eta(\xi_1)\eta(\xi_2)\tilde v_t^2(\xi_1,\xi_2|\varepsilon)|\left|e^{i\sqrt{\pi}(\sigma_1\phit(x_1)+\sigma_2\phit(x_2))}-1\right|\1_{\{\sigma_1\sigma_2=\pm 4\}}\\
&\quad\leq M^2\int_{K^2}dx_1\,dx_2\,\widehat g_t(x_1,x_2)\left(1+\big\|\nabla\phit\big\|_{L^\infty(K)}\right)\\
&\quad\leq C_{K,M,t}\left(1+\big\|\nabla\phit\big\|_{L^\infty(K)}\right).
\end{align*}
Therefore, for the relevant values of $t$, there exists a (possibly different) constant $C_{K,M,t}$ such that
\begin{align*}
|\Z(\eta|\varepsilon)|&\leq \left\langle e^{C_{K,M,t}\big(1+\|\nabla \varphi_{\sqrt{t}}\|_{L^\infty(K)}\big)}\right\rangle_{\gff(\sqrt{t})}
\end{align*}
Using Lemma~\ref{lem:b} (in the notation of Lemma~\ref{lem:b}, we find some compactly supported $\eta$ such that $\eta=C_{K,M,t}$ on $K$), we see that this can be bounded by some constant depending only on $K,M,t$. We can then fix $t$ suitably depending only on $K,M$ to find that uniformly in $\epsilon$, this quantity is bounded by some constant depending only on $K,M$. This concludes the proof of the first claim.

The arguments for the proofs of the claims (ii) and (iii) are identical to the corresponding claims of \cite[Proposition 5.3]{BaWe}, and we omit the details. 
\end{proof}

We now turn to analyticity properties of the renormalized partition function. These will allow us to deduce convergence of the renormalized partition function from Proposition~\ref{prop:ub}.

We begin with an estimate that will allow controlling derivatives of the renormalized partition function -- eventually we will use it to prove that the (limiting) renormalized partition function is an entire function as claimed in Theorem~\ref{thm:uc}.

\begin{lemma}\label{lem:mexp}
For fixed $\epsilon>0$ and $\eta\in C_c(\Omega\times\{\pm1,\pm2\})$, the function $\mu\to\Z(\mu\eta|\varepsilon)$ is entire and we have
\begin{equation}
\Z(\mu\eta|\varepsilon)=\sum_{n=0}^\infty\frac{\mu^n}{n!}\M_n(\eta|\varepsilon),
\end{equation}
where
\begin{equation}
\M_n(\eta|\varepsilon)=\int_{(\Omega\times\{\pm1,\pm2\})^n}d\xi_1\dots d\xi_n\,\eta(\xi_1)\dots\eta(\xi_n)\widetilde{\M}(\xi_1,\dots,\xi_n|\varepsilon)
\end{equation}
with
\begin{multline}\label{eq:mexp}
\widetilde{\M}(\xi_1,\dots,\xi_n|\varepsilon)=\frac{1}{n!}\sum_{\tau\in S_n}\Bigg[\sum_{j=0}^{\lfloor\frac{n}{2}\rfloor}\frac{n!}{j!(n-2j)!}(-1)^{n-j}2^{-j}\gffe{\prod_{l=2j+1}^n:e^{i\sqrt{\pi}\sigma_{\tau_l}\phie(x_{\tau_l})}:}\\
\times\bigg(\prod_{l'=1}^{j} A(\xi_{\tau_{2l'-1}},\xi_{\tau_{2l'}}|\varepsilon)\1_{\{\sigma_{\tau_{2l'-1}}\sigma_{\tau_{2l'}}=\pm 4\}}\bigg)\Bigg],
\end{multline}
where $S_n$ denotes the group of permutations of the set $[n]$.

Furthermore, for each $K\subset \Omega$ compact, $M>0$ and $R>0$ there exists a constant $C_{K,M,R}$ independent of $\varepsilon,\eta$ and $n$ such that
\begin{equation}\label{eq:mbound}
\sup_{\substack{\varepsilon\in(0,1)\\ \|\eta\|_{L^\infty(\Omega\times\{\pm1,\pm2\})}\leq M\\ \mathrm{supp}(\eta(\cdot,\sigma))\subset K}}|\M_n(\eta|\varepsilon)|\leq C_{K,M,R}\frac{n!}{R^n}.
\end{equation}
\end{lemma}
\begin{proof}
Recall from \eqref{eq:rpf} and \eqref{eq:pf} that the renormalized partition function is
\begin{align*}
\Z(\mu\eta|\varepsilon)&=\bigg\langle\exp\bigg\{-\mu\int_{\Omega\times\{\pm1,\pm2\}} d\xi\,\eta(\xi):e^{i\sqrt{\pi}\sigma\phie(x)}:\bigg\}\bigg\rangle_{\gff(\varepsilon)}\\
&\quad\times\exp\bigg\{-\frac{\mu^2}{2}\int_{(\Omega\times\{\pm2\})^2}d\xi_1\,d\xi_2\,\eta(\xi_1)\eta(\xi_2)A(\xi_1,\xi_2|\varepsilon)\bigg\},
\end{align*}
since $\int_{\Omega\times\{\pm1,\pm2\}} d\xi\,\eta(\xi):e^{i\sqrt{\pi}\sigma\phie(x)}:$ is a bounded random variable, the expectation factor above is an entire function of $\mu$. Furthermore, for any $\varepsilon>0$
\begin{equation*}
\bigg|\int_{(\Omega\times\{\pm2\})^2}d\xi_1\,d\xi_2\,\eta(\xi_1)\eta(\xi_2)A(\xi_1,\xi_2|\varepsilon)\bigg|<\infty,
\end{equation*}
hence the second factor in $\Z(\mu\eta|\varepsilon)$ is an entire function of $\mu$ and therefore $\Z(\mu\eta|\varepsilon)$ is entire. We can then expand both factors and change the order of expectation and integration
\begin{align*}
&\Z(\mu\eta|\varepsilon)=\sum_{k=0}^\infty\frac{(-\mu)^k}{k!}\int_{(\Omega\times\{\pm1,\pm2\})^k} d\xi_1\dots d\xi_k\,\eta(\xi_1)\dots\eta(\xi_k)\gffe{\prod_{l=1}^k:e^{i\sqrt{\pi}\sigma_l\phie(x_l)}:}\\
&\quad\times\sum_{j=0}^\infty\frac{(-\mu^2)^j}{2^j j!}\int_{(\Omega\times\{\pm1,\pm2\})^{2j}} d\xi_1\dots d\xi_{2j}\,\eta(\xi_1)\dots\eta(\xi_{2j})\prod_{l'=1}^j A(\xi_{2l'-1},\xi_{2l'}|\varepsilon)\1_{\{\sigma_{2l'-1}\sigma_{2l'}=\pm 4\}}.
\end{align*}
Finally, relabeling the integration variables and symmetrizing in the arguments gives us the desired representation.

Since $\Z$ is entire and $\M_n(\eta|\varepsilon)=\partial_\mu^n|_{\mu=0}\Z(\mu\eta|\varepsilon)$ then by Cauchy's integral formula
\begin{equation*}
\M_n(\eta|\varepsilon)=\frac{n!}{2\pi i}\oint_{|w|=R}dw\frac{\Z(w\eta|\varepsilon)}{w^{n+1}}.
\end{equation*}
From Proposition~\ref{prop:ub} we then get
\begin{align*}
\sup_{\substack{\varepsilon\in(0,1), \eta\in C_c(\Omega\times \{\pm 1,\pm 2\})\\ \|\eta\|_{L^\infty(\Omega\times\{\pm1,\pm2\})}\leq M\\ \mathrm{supp}(\eta(\cdot,\sigma))\subset K}}|\M_n(\eta|\varepsilon)|\leq\sup_{\substack{\varepsilon\in(0,1), \eta\in C_c(\Omega\times \{\pm 1,\pm 2\})\\ \|\eta\|_{L^\infty(\Omega\times\{\pm1,\pm2\})}\leq M\\ \mathrm{supp}(\eta(\cdot,\sigma))\subset K\\ |w|=R}}|\Z(w\eta|\varepsilon)|\frac{n!}{R^n}\leq C_{K,M,R}\frac{n!}{R^n}.
\end{align*}
\end{proof}
Next, we will need a tool for showing that the kernel $\widetilde \M$ from \eqref{eq:mexp} satisfies a certain uniqueness property.

\begin{lemma}\label{lem:Mtiluniq}
For $\varepsilon>0$, $\widetilde{\M}(\xi_1,\dots,\xi_n|\varepsilon)$ is the unique continuous function of $\xi_1,\dots,\xi_n$ for which  
\begin{align*}
&\frac{1}{n!}\frac{\partial}{\partial\alpha_1}\dots\frac{\partial}{\partial\alpha_n}\M_n(\alpha_1f_1+\dots+\alpha_nf_n|\varepsilon)\\
&\quad=\int_{(\Omega\times\{\pm1,\pm2\})^n}d\xi_1\dots d\xi_n\,f_1(\xi_1)\dots f_n(\xi_n)\widetilde{\M}(\xi_1,\dots,\xi_n|\varepsilon),
\end{align*}
for all $f_1,\dots,f_n\in C_c^\infty(\Omega\times\{\pm1,\pm2\})$,
\end{lemma}
\begin{proof}
The fact that $\widetilde{\M}$ satisfies this condition can be proven exactly as in \cite[Lemma 5.6]{BaWe} and we omit the details of this claim. 

For uniqueness, let $\widetilde{\M}_1$ and $\widetilde{\M}_2$ be two such kernels. Since both are integral kernels for the object in question, for any fixed $\sigma_1,\dots ,\sigma_n\in \{\pm 1,\pm 2\}$ and $f\in C_c^\infty(\Omega^n)$ 
\begin{align*}
\int_{\Omega^n}\!d^{2n}x\,f(x_1,... ,x_n)\widetilde{\M}_1((x_1,\sigma_1),... ,(x_n,\sigma_n))=\int_{\Omega^n}\!d^{2n}x\,f(x_1,... ,x_n)\widetilde{\M}_2((x_1,\sigma_1),... ,(x_n,\sigma_n)).
\end{align*}
Thus $\widetilde{\M}_1=\widetilde{\M}_2$, as elements of $\mathcal D'(\Omega^n)$ and since we are dealing with continuous functions, this means that $\widetilde{\M}_1=\widetilde{\M}_2$.  
\end{proof}

As a final ingredient for proving convergence of the renormalized partition function, we will need an expression for $\widetilde\M$ in terms of the renormalized potential (for which we have good bounds).
\begin{lemma}\label{lem:mrep}
Let $\eta\in C_c(\Omega\times\{\pm1,\pm2\})$ and choose $t$ small enough so that we can use Corollary~\ref{cor:rpot}. Then $\M_n(\eta|\varepsilon)$ defined in Lemma~\ref{lem:mexp} can be expressed as
\begin{multline*}
\int_{(\Omega\times\{\pm1,\pm2\})^{n}}d\xi_1\dots d\xi_{n}\,\eta(\xi_1)\dots\eta(\xi_{n})\\
\shoveleft\quad\times\sum_{k=0}^{\lfloor\frac{n}{2}\rfloor}\frac{n!}{k!}2^{-k}(-1)^k\prod_{j=1}^{k}\Big(A(\xi_{2j-1},\xi_{2j}|\varepsilon)+\tilde v_t^2(\xi_{2j-1},\xi_{2j}|\varepsilon)\Big)\1_{\{\sigma_{2j-1}\sigma_{2j}=\pm 4\}}\\
\times\sum_{l=1}^{n-2k}\frac{(-1)^l}{l!}\sum_{\substack{1\leq n_1,\dots,n_l\leq n-2k:\\ n_1+\dots+n_l=n-2k}}\frac{1}{n_1!\dots n_l!}\prod_{j=1}^l\tilde v_t^{n_j}(\xi_{2k+1+n_1+\dots+n_{j-1}},\dots,\xi_{2k+n_1+\dots+n_{j}}|\varepsilon)\\
\times\bigg\langle \prod_{j=1}^l\left(e^{i\sqrt{\pi}\sum_{m=2k+1+n_1+\dots+n_{j-1}}^{2k+n_1+\dots+n_{j}}\sigma_m\phit(x_m)}-\1_{\{n_j=2\}}\1_{\{\sigma_m\sigma_{m+1}=-4\}}\right)\bigg\rangle_{\gff(\sqrt{t})}.
\end{multline*}
\end{lemma}
\begin{proof}
Noting from Lemma~\ref{lem:mexp} that $\M_n(\eta|\varepsilon)=\partial_\mu^n|_{\mu=0}\Z(\mu \eta|\varepsilon)$ and we can write $\Z(\mu\eta|\varepsilon)=F_1(\mu)F_2(\mu)$, where
\begin{multline*}
F_1(\mu)=\bigg\langle \exp\bigg\{-\sum_{n=1}^\infty \frac{\mu^n}{n!}\int_{(\Omega\times\{\pm1,\pm2\})^n}d\xi_1\dots d\xi_n\,\eta(\xi_1)\dots\eta(\xi_n)\tilde v_t^n(\xi_1,\dots,\xi_n|\varepsilon)\\
\times\left(e^{i\sqrt{\pi}\sum_{j=1}^n\sigma_j\phit(x_j)}-\1_{\{n=2\}}\1_{\{\sigma_1\sigma_2=\pm 4\}}\right)\bigg\}\bigg\rangle_{\gff(\sqrt{t})}
\end{multline*}
and 
$
F_2(\mu)=\exp\{-\frac{\mu^2}{2}\int_{(\Omega\times\{\pm1,\pm2\})^2}d\xi_1\,d\xi_2\,\eta(\xi_1)\eta(\xi_2)\big(A(\xi_1,\xi_2|\varepsilon)+\tilde v_t^2(\xi_1,\xi_2|\varepsilon)\big)\1_{\{\sigma_1\sigma_2=\pm 4\}}\}.
$

So, we need to calculate derivatives of $F_1$ and $F_2$ at $0$. Note that $F_2$ is entire and non-zero, and $\Z$ is entire by Lemma~\ref{lem:mexp}, thus $F_1$ is entire. By Cauchy's integral formula, Fubini, etc. we can change the order of derivatives and expectation, so
\begin{align*}
F_1^{(k)}(0)&=\bigg\langle \frac{\partial^k}{\partial\mu^k}\bigg|_{\mu=0} \exp\bigg\{-\sum_{n=1}^\infty \frac{\mu^n}{n!}\int_{(\Omega\times\{\pm1,\pm2\})^n}d\xi_1\dots d\xi_n\,\eta(\xi_1)\dots\eta(\xi_n)\\
&\qquad\times\tilde v_t^n(\xi_1,\dots,\xi_n|\varepsilon)\left(e^{i\sqrt{\pi}\sum_{j=1}^n\sigma_j\phit(x_j)}-\1_{\{n=2\}}\1_{\{\sigma_1\sigma_2=\pm 4\}}\right)\bigg\}\bigg\rangle_{\gff(\sqrt{t})}\\
&=\sum_{l=1}^k\frac{(-1)^l}{l!}\sum_{\substack{1\leq n_1,\dots,n_l\leq k:\\ n_1+\dots+n_l=k}}\frac{k!}{n_1!\dots n_l!} \bigg\langle \prod_{j=1}^l\bigg(\int_{(\Omega\times\{\pm1,\pm2\})^{n_j}}d\xi_1\dots d\xi_{n_j}\,\eta(\xi_1)\dots\eta(\xi_{n_j})\\
&\qquad\times\tilde v_t^{n_j}(\xi_1,\dots,\xi_{n_j}|\varepsilon)\left(e^{i\sqrt{\pi}\sum_{m=1}^{n_j}\sigma_m\phit(x_m)}-\1_{\{n_j=2\}}\1_{\{\sigma_1\sigma_2=\pm 4\}}\right)\bigg)\bigg\rangle_{\gff(\sqrt{t})}\\
&=\int_{(\Omega\times\{\pm1,\pm2\})^{k}}d\xi_1\dots d\xi_{k}\,\eta(\xi_1)\dots\eta(\xi_{k})\sum_{l=1}^k\frac{(-1)^l}{l!}\\
&\qquad\times\sum_{\substack{1\leq n_1,\dots,n_l\leq k:\\ n_1+\dots+n_l=k}}\frac{k!}{n_1!\dots n_l!}\prod_{j=1}^l\tilde v_t^{n_j}(\xi_{1+n_1+\dots+n_{j-1}},\dots,\xi_{n_1+\dots+n_{j}}|\varepsilon)\\
&\qquad\times\bigg\langle \prod_{j=1}^l\left(e^{i\sqrt{\pi}\sum_{m=1+n_1+\dots+n_{j-1}}^{n_1+\dots+n_{j}}\sigma_m\phit(x_m)}-\1_{\{n_j=2\}}\1_{\{\sigma_m\sigma_{m+1}=\pm 4\}}\right)\bigg\rangle_{\gff(\sqrt{t})}.
\end{align*}
For $F_2$ we have that $F_2(0)=1$. For odd $n$, $F_2^{(n)}(0)=0$ and for even $n=2k$ we get the formula
\begin{align*}
F_2^{(2k)}(0)&=(-1)^k\frac{(2k)!}{2^k k!}\left[\int_{(\Omega\times\{\pm1,\pm2\})^2}d\xi_1\,d\xi_2\,\eta(\xi_1)\eta(\xi_2)\Big(A(\xi_1,\xi_2|\varepsilon)+\tilde v_t^2(\xi_1,\xi_2|\varepsilon)\Big)\1_{\{\sigma_1\sigma_2=\pm 4\}}\right]^k\\
&=(-1)^k\frac{(2k)!}{2^k k!}\int_{(\Omega\times\{\pm1,\pm2\})^{2k}}d\xi_1\dots d\xi_{2k}\,\eta(\xi_1)\dots\eta(\xi_{2k})\\
&\quad\times\prod_{j=1}^{k}\Big(A(\xi_{2j-1},\xi_{2j}|\varepsilon)+\tilde v_t^2(\xi_{2j-1},\xi_{2j}|\varepsilon)\Big)\1_{\{\sigma_{2j-1}\sigma_{2j}=\pm 4\}}.
\end{align*}
Now, since $F_2^{(2k+1)}(0)=0$, we can write
\begin{align*}
\M_n(\eta|\varepsilon)=\partial_\mu^n|_{\mu=0}\Z(\mu\eta|\varepsilon)=\sum_{k=0}^n\binom{n}{k}F_1^{(n-k)}(0)F_2^{(k)}(0)=\sum_{k=0}^{\lfloor\frac{n}{2}\rfloor}\binom{n}{2k}F_1^{(n-2k)}(0)F_2^{(2k)}(0),
\end{align*}
so
\begin{align*}
\M_{n}(\eta|\varepsilon)&=\sum_{k=0}^{\lfloor\frac{n}{2}\rfloor}\frac{n!}{(2k)!(n-2k)!}(-1)^k\frac{(2k)!}{2^k k!}\int_{(\Omega\times\{\pm1,\pm2\})^{2k}}d\xi_1\dots d\xi_{2k}\,\eta(\xi_1)\dots\eta(\xi_{2k})\\
&\quad\times\prod_{j=1}^{k}\Big(A(\xi_{2j-1},\xi_{2j}|\varepsilon)+\tilde v_t^2(\xi_{2j-1},\xi_{2j}|\varepsilon)\Big)\1_{\{\sigma_{2j-1}\sigma_{2j}=\pm 4\}}\\
&\quad\times \int_{(\Omega\times\{\pm1,\pm2\})^{n-2k}}d\xi_1\dots d\xi_{n-2k}\,\eta(\xi_1)\dots\eta(\xi_{n-2k})\sum_{l=1}^{n-2k}\frac{(-1)^l}{l!}\\
&\quad\times\sum_{\substack{1\leq n_1,\dots,n_l\leq n-2k:\\ n_1+\dots+n_l=n-2k}}\frac{(n-2k)!}{n_1!\dots n_l!}\prod_{j=1}^l\tilde v_t^{n_j}(\xi_{1+n_1+\dots+n_{j-1}},\dots,\xi_{n_1+\dots+n_{j}}|\varepsilon)\\
&\quad\times\bigg\langle \prod_{j=1}^l\left(e^{i\sqrt{\pi}\sum_{m=1+n_1+\dots+n_{j-1}}^{n_1+\dots+n_{j}}\sigma_m\phit(x_m)}-\1_{\{n_j=2\}}\1_{\{\sigma_m\sigma_{m+1}=\pm 4\}}\right)\bigg\rangle_{\gff(\sqrt{t})}\\
&=\int_{(\Omega\times\{\pm1,\pm2\})^{n}}d\xi_1\dots d\xi_{n}\,\eta(\xi_1)\dots\eta(\xi_{n})\\
&\quad\times\sum_{k=0}^{\lfloor\frac{n}{2}\rfloor}\frac{n!}{k!}2^{-k}(-1)^k\prod_{j=1}^{k}\Big(A(\xi_{2j-1},\xi_{2j}|\varepsilon)+\tilde v_t^2(\xi_{2j-1},\xi_{2j}|\varepsilon)\Big)\1_{\{\sigma_{2j-1}\sigma_{2j}=\pm 4\}}\\
&\quad\times\sum_{l=1}^{n-2k}\frac{(-1)^l}{l!}\sum_{\substack{1\leq n_1,\dots,n_l\leq n-2k:\\ n_1+\dots+n_l=n-2k}}\frac{1}{n_1!\dots n_l!}\prod_{j=1}^l\tilde v_t^{n_j}(\xi_{2k+1+n_1+\dots+n_{j-1}},\dots,\xi_{2k+n_1+\dots+n_{j}}|\varepsilon)\\
&\quad\times\bigg\langle \prod_{j=1}^l\left(e^{i\sqrt{\pi}\sum_{m=2k+1+n_1+\dots+n_{j-1}}^{2k+n_1+\dots+n_{j}}\sigma_m\phit(x_m)}-\1_{\{n_j=2\}}\1_{\{\sigma_m\sigma_{m+1}=\pm 4\}}\right)\bigg\rangle_{\gff(\sqrt{t})}.
\end{align*}
This concludes the proof.
\end{proof}
To finally prove convergence of the renormalized partition function as well as Theorem~\ref{thm:uc}, we begin by providing a bound on the kernels $\widetilde\M$.
\begin{lemma}\label{lem:mbound}
There exists a $t_0>0$ such that for $n\geq 1$ and $0<t<t_0$ there exists a function $g_t\in L_{loc}^1((\Omega\times\{\pm1,\pm2\})^n)$, independent of $\varepsilon$, such that for $\epsilon^2<t$
\begin{equation*}
|\widetilde{\M}(\xi_1,\dots,\xi_n|\varepsilon)|\leq g_t(\xi_1,\dots,\xi_n).
\end{equation*}
\end{lemma}
\begin{proof}
The representation in Lemma~\ref{lem:mrep} (combined with Lemma~\ref{lem:Mtiluniq}) shows that $|\widetilde{\M}|$ can be bounded by sums of products of factors of the form
\begin{equation*}
G_1(\widehat\xi_1,\widehat\xi_2|\varepsilon):=\big|A(\widehat \xi_1,\widehat \xi_2|\varepsilon)+\tilde v_t^2(\widehat\xi_1,\widehat\xi_2|\varepsilon)\big|\1_{\{\hat\sigma_{1}\hat\sigma_{2}=\pm 4\}},
\end{equation*}
\begin{equation*}
G_2(\xi'_1,\dots, \xi'_{n'}|\varepsilon):=\big|\tilde v_t^{n'}(\xi'_1,\dots,\xi'_{n'}|\varepsilon)\big|,\qquad\text{with $\sigma_1'\sigma_2'\neq\pm 4$ when $n'= 2$},
\end{equation*}
and
\begin{equation*}
G_3(\widetilde\xi_1,\dots,\widetilde\xi_{2k}|\varepsilon):=\bigg\langle\prod_{j=1}^k\left|\tilde v_t^{2}(\widetilde\xi_{2j-1},\widetilde\xi_{2j}|\varepsilon)\right|\Big| e^{i\sqrt{4\pi}\big(\widetilde\sigma_{2j-1}\phit(\widetilde x_{2j-1})+\widetilde\sigma_{2j}\phit(\widetilde x_{2j})\big)}-1\Big|\bigg\rangle_{\gff(\sqrt{t})}
\end{equation*}
where $\{\widehat\xi_1,\widehat\xi_2\},\{\xi'_1,\dots,\xi'_{n'}\}$ and $\{\widetilde\xi_1,\dots,\widetilde\xi_{2k}\}$ are disjoint subsets of the integration variables $\xi_1,\dots,\xi_n$ of $\widetilde{\M}$. Thus it is enough to prove local integrability for each function $G_i$ separately.

For $G_1$, this follows from Lemma~\ref{lem:a+v2}, for $G_2$, this follows from Proposition~\ref{prop:vtest} and for $G_3$, this follows from Lemma~\ref{lem:vt2} and Lemma~\ref{lem:b}.
\end{proof}

We now prove that the kernels $\widetilde \M$ converge as $\epsilon\to 0$. Combining with previous results, this will allow proving convergence of the renormalized partition function.
\begin{lemma}\label{lem:mconv}
For $\xi_i\neq\xi_j$ for $i\neq j$,
\begin{equation*}
\lim_{\varepsilon\to0}\widetilde{\M}(\xi_1,\dots,\xi_n|\varepsilon)=\widetilde{\M}(\xi_1,\dots,\xi_n),
\end{equation*}
with
\begin{align}\label{eq:mlimit}
\widetilde{\M}(\xi_1,\dots,\xi_n)=\frac{1}{n!}\sum_{\tau\in S_n}\Bigg[\sum_{j=0}^{\lfloor\frac{n}{2}\rfloor}\frac{n!}{j!(n-2j)!}(-1)^{n-j}2^{-j}\left\langle\prod_{l=2j+1}^n:e^{i\sqrt{\pi}\sigma_{\tau_l}\varphi(x_{\tau_l})}:\right\rangle_\gff\\
\nonumber\times\bigg(\prod_{l'=1}^{j} A(\xi_{\tau_{2l'-1}},\xi_{\tau_{2l'}})\1_{\{\sigma_{\tau_{2l'-1}}\sigma_{\tau_{2l'}}=\pm 4\}}\bigg)\Bigg],
\end{align}
where $A(\xi_1,\xi_2)=\left\langle \opcolon e^{i\sqrt{\pi}\sigma_1\varphi(x_1)}\opcolon  \opcolon e^{i\sqrt{\pi}\sigma_2\varphi(x_2)}\opcolon  \right\rangle_{\gff} -\left\langle \opcolon e^{i\sqrt{\pi}\sigma_1\varphi(x_1)}\opcolon  \right\rangle_{\gff} \left\langle \opcolon e^{i\sqrt{\pi}\sigma_2\varphi(x_2)}\opcolon \right\rangle_{\gff}=\lim_{\epsilon\to 0}A(\xi_1,\xi_2|\epsilon)$, and the correlation functions are given by Lemma~\ref{lem:gffcorr2}.
\end{lemma}
\begin{proof}
By Lemma~\ref{lem:gffcorr2} 
\begin{equation*}
\lim_{\varepsilon\to 0}\gffe{\prod_{l=2j+1}^n\opcolon e^{i\sqrt{\pi}\sigma_{\tau_l}\phie(x_{\tau_l})}\opcolon }=\left\langle\prod_{l=2j+1}^n\opcolon e^{i\sqrt{\pi}\sigma_{\tau_l}\varphi(x_{\tau_l})}\opcolon \right\rangle_{\gff}.
\end{equation*}
A similar statement holds for the counter term
\begin{align*}
A(\xi_1,\xi_2|\epsilon)&=\left\langle \opcolon e^{i\sqrt{\pi}\sigma_1\varphi_\epsilon(x_1)}\opcolon  \opcolon e^{i\sqrt{\pi}\sigma_2\varphi_\epsilon(x_2)}\opcolon \right\rangle_{\gff(\epsilon)}\\
&\qquad-\left\langle \opcolon e^{i\sqrt{\pi}\sigma_1\varphi_\epsilon(x_1)}\opcolon \right\rangle_{\gff(\epsilon)}\left\langle \opcolon e^{i\sqrt{\pi}\sigma_2\varphi_\epsilon(x_2)}\opcolon \right\rangle_{\gff(\epsilon)}.
\end{align*}
So $\widetilde\M(\cdot|\varepsilon)$ given by \eqref{eq:mexp} converges as $\varepsilon\to 0$ to the function $\widetilde\M(\cdot)$ defined by \eqref{eq:mlimit}.
\end{proof}

We now have all the ingredients needed for the proof.
\begin{proof}[Proof of Theorem~\ref{thm:uc}]
For $\widetilde{\M}(\cdot)$ given by \eqref{eq:mlimit} define 
\begin{equation*}
\M_n(\eta):=\int_{(\Omega\times\{\pm1,\pm2\})^n}d\xi_1\dots d\xi_n\,\eta(\xi_1)\dots\eta(\xi_n)\widetilde{\M}(\xi_1,\dots,\xi_n).
\end{equation*}
It follows from Lemmas~\ref{lem:mbound}--\ref{lem:mconv} that as $\varepsilon\to 0$, $\M_n(\eta|\varepsilon)$ in Lemma~\ref{lem:mexp} converges to $\M_n(\eta)$ and is finite for all $\eta\in C_c(\Omega\times\{\pm1,\pm2\})$. Furthermore, if we write $K=\mathrm{supp}(\eta(\cdot,\sigma))$ and $M=\|\eta\|_\infty$,  \eqref{eq:mbound} implies that for each $R>0$, $|\M_n(\eta|\varepsilon)|\leq C_{K,M,R}\frac{n!}{R^n}$ then
\[
\left|\sum_{n=0}^\infty \frac{\mu^n}{n!}\M_n(\eta|\varepsilon)\right|\leq\sum_{n=0}^\infty \frac{|\mu|^n}{n!}C_{K,M,R}\frac{n!}{R^n}\leq C_{K,M,R}\sum_{n=0}^\infty \left(\frac{|\mu|}{R}\right)^n 
\]
converges for $|\mu|<R$. Thus, by dominated convergence
\[
\lim_{\varepsilon\to 0}\Z(\mu\eta|\varepsilon)=\sum_{n=0}^\infty \frac{\mu^n}{n!}\lim_{\varepsilon\to 0}\M_n(\eta|\varepsilon)=\sum_{n=0}^\infty \frac{\mu^n}{n!}\M_n(\eta)=:\Z(\mu\eta),
\]
which defines an entire function (since the radius of convergence is at least $R$ for any $R>0$) and thus we have proved $(i)$ and $(ii)$.

The proof of $(iii)$ follows from Proposition~\ref{prop:ub} $(iii)$, i.e. when $\eta(\cdot,\sigma)=\overline{\eta(\cdot,-\sigma)}$ for $\sigma\in\{1,2\}$ it holds that as $\varepsilon\to 0$, $\Z(\eta)>0$ .

The proof of $(iv)$ is also similar -- see e.g. \cite[Proof of Theorem 5.1]{BaWe} for details, though note that the assumption corresponding to \eqref{eq:newetaassu} is missing in \cite{BaWe}. This is needed to control \cite[(5.69)]{BaWe}. We thank E. Saksman and S. Vihko for bringing this to our attention.
\end{proof}

\section{The sine-Gordon Correlation Functions -- Proof of Theorem~\ref{th:sgmain}}\label{sec:sgcorr}
We are finally in a position to prove Theorem~\ref{th:sgmain}. While the proof is a rather direct consequence of our control of the renormalized partition function and certain GFF correlation functions, it is still a bit lengthy, and in particular, notationally heavy. We thus split the proof into smaller steps. 

Our first remark is that we can write the Wick ordered trigonometric functions as sums of Wick ordered exponentials. We also find it convenient to simplify notation involving the definition of $\optypecolon \cdot\optypecolon $. To do this, let us write for $f,\rho\in C_c^\infty(\Omega)$ \begin{align}\label{eq:cepsdef}
c_\epsilon(f,\rho)=\frac{1}{4}\int_{(\Omega\times \{\pm 2\})^2}d\xi_1\,d\xi_2\, f(x_1)\rho(x_2)A(\xi_1,\xi_2|\epsilon),
\end{align}
where $A(\xi_1,\xi_2|\epsilon)$ is as in \eqref{eq:ctt}. Note that since $f$ and $\rho$ do not depend on $\sigma_1,\sigma_2$‚ we can actually write this as 
\begin{equation}
c_\epsilon(f,\rho)=\int_{\Omega^2}d^2x\,d^2y\, f(x)\rho(y)\mathcal A(x,y|\epsilon),
\end{equation} 
where 
\begin{align}
\mathcal A(x,y|\epsilon)&= \left\langle \opcolon \cos(\sqrt{4\pi}\varphi_\epsilon(x))\opcolon  \opcolon \cos(\sqrt{4\pi}\varphi_\epsilon(y))\opcolon  \right\rangle_{\mathrm{GFF}(\epsilon)} \\
\nonumber&\qquad-\left\langle \opcolon \cos(\sqrt{4\pi}\varphi_\epsilon(x))\opcolon \right\rangle_{\mathrm{GFF}(\epsilon)} \left\langle \opcolon \cos(\sqrt{4\pi}\varphi_\epsilon(y))\opcolon  \right\rangle_{\mathrm{GFF}(\epsilon)}
\end{align}
Note that by Lemma \ref{lem:gffcorr2}, $\mathcal A(x,y)=\lim_{\epsilon\to 0}\mathcal A(x,y|\epsilon)$ also exists.

This allows writing
\begin{equation}
\optypecolon \cos(\sqrt{4\pi}\varphi_\epsilon)\optypecolon (h)=\frac{1}{2}\sum_{\tau\in \{-2,2\}}\left(\opcolon e^{i\tau\sqrt{\pi}\varphi_\epsilon}\opcolon (h)-\mu c_\epsilon(h,\rho)\right).
\end{equation}

It is thus sufficient to study Wick ordered exponentials for the theorem. Our first step in doing this is to express the correlation functions of such Wick ordered exponentials for $\epsilon>0$ in terms of the renormalized partition function and GFF correlation functions. This is recorded in the following lemma.
\begin{lemma}\label{lem:cftopart}
We have for $\mu\in \R$, $\rho\in C_c^\infty(\Omega)$, $\sigma_1,\dots,\sigma_n\in\{-1,1\}$,  $\tau_1,\dots,\tau_{n'}\in \{-2,2\}$, and $f_1,\dots,f_n$, $\widetilde f_1,\dots,\widetilde f_{n'}$, $g_1,\dots,g_m$, $\widetilde g_1,\dots,\widetilde g_{m'}$ $\in C_c^\infty(\Omega)$ with disjoint supports:
\begin{multline*}
\Bigg\langle\prod_{j=1}^n \opcolon e^{i\sigma_j\sqrt{\pi}\phie}\opcolon (f_j)\prod_{j'=1}^{n'}\left(\opcolon e^{i\tau_{j'}\sqrt{\pi}\phie}\opcolon (\widetilde f_{j'})- \mu c_\epsilon(\widetilde f_{j'},\rho)\right)\prod_{l=1}^m \partial \phie(g_l)\prod_{l'=1}^{m'}\bar\partial \phie(\widetilde g_{l'})\Bigg\rangle_{\mathrm{sG}(\mu\rho|\varepsilon,\Omega)}\\
\shoveleft\enspace=\prod_{j=1}^n \partial_{s_j}|_{s_j=0}\prod_{j'=1}^{n'}\partial_{t_{j'}}|_{t_{j'}=0}\prod_{l=1}^m \partial_{u_l}|_{u_l=0} \prod_{l'=1}^{m'}\partial_{v_{l'}}|_{v_{l'}=0}\bigg[e^{\frac{1}{2}\sum_{l',k'=1}^{m'}v_{l'}v_{k'}\gffe{\bar\partial \phie(\widetilde g_{l'})\bar\partial \phie(\widetilde g_{k'})}}\\
\shoveleft\quad\times e^{\sum_{l=1}^m \sum_{l'=1}^{m'}u_l v_{l'}\gffe{\partial \phie(g_l)\bar\partial \phie(\widetilde g_{l'})}} e^{\frac{1}{2}\sum_{l,k=1}^m u_l u_k \gffe{\partial \phie(g_l)\partial \phie(g_k)}} e^{-\sum_{j'=1}^{n'}t_{j'}\mu c_\epsilon(\widetilde f_{j'},\rho)} \\
\times \frac{\Z(\eta_\epsilon|\epsilon)}{\Z(-\frac{\mu}{2}\rho|\epsilon)} e^{\frac{1}{2}\int_{(\Omega\times \{\pm 2\})^2}d\xi_1d\xi_2\big(\eta_\epsilon(\xi_1)\eta_\epsilon(\xi_2)-\frac{\mu^2}{4}\rho(x_1)\rho(x_2)\big)A(\xi_1,\xi_2|\epsilon)} \bigg],
\end{multline*}
where $A(\xi_1,\xi_2|\epsilon)$ is as in \eqref{eq:ctt} and
\begin{equation}
\begin{split}
\eta_\epsilon(x,\sigma)=&-\sum_{j=1}^n s_j \delta_{\sigma,\sigma_j}e^{i\sigma_j \sqrt{\pi}\big(\sum_{l=1}^m u_l\gffe{\phie(x)\partial \phie(g_l)}+\sum_{l'=1}^{m'}v_{l'}\gffe{\phie(x)\bar\partial\phie(\widetilde g_{l'})}\big)}f_j(x)\\
&-\sum_{j'=1}^{n'}t_{j'}\delta_{\sigma,\tau_{j'}}e^{i\tau_{j'} \sqrt{\pi}\big(\sum_{l=1}^m u_l\gffe{\phie(x)\partial \phie(g_l)}+\sum_{l'=1}^{m'}v_{l'}\gffe{\phie(x)\bar\partial\phie(\widetilde g_{l'})}\big)}\widetilde f_{j'}(x)\\
&-\1_{\{\sigma\in \{-2,2\}\}}\frac{\mu}{2}e^{i\sigma\sqrt{\pi}\big(\sum_{l=1}^m u_l\gffe{\phie(x)\partial \phie(g_l)}+\sum_{l'=1}^{m'}v_{l'}\gffe{\phie(x)\bar\partial\phie(\widetilde g_{l'})}\big)}\rho(x).
\end{split}
\end{equation}
\end{lemma}
While we do not emphasize it in our notation (to avoid horrendous looking formulas), note that $\eta_\epsilon$ depends on essentially all of our parameters in addition to $\epsilon$ -- in particular, when carrying out the differentiations, one may need to differentiate $\eta_\epsilon$ as well. While notationally tedious, this representation has the benefit that we can use Theorem~\ref{thm:uc} to deduce convergence as $\epsilon\to 0$. We turn to this after the proof of this lemma. 
\begin{proof}
Our starting point is to write (using the Leibniz integral rule)
\begin{multline*}
\Bigg\langle\prod_{j=1}^n \opcolon e^{i\sigma_j\sqrt{\pi}\phie}\opcolon (f_j)\prod_{j'=1}^{n'}\left(\opcolon e^{i\tau_{j'}\sqrt{\pi}\phie}\opcolon (\widetilde f_{j'})- \mu c_\epsilon(\widetilde f_{j'},\rho)\right)\prod_{l=1}^m \partial \phie(g_l)\prod_{l'=1}^{m'}\bar\partial \phie(\widetilde g_{l'})\Bigg\rangle_{\mathrm{sG}(\mu\rho|\varepsilon,\Omega)}\\
\shoveleft=\frac{1}{Z_{\mathrm{sG}(\mu\rho|\varepsilon,\Omega)}}\prod_{j=1}^n \partial_{s_j}|_{s_j=0}\prod_{j'=1}^{n'}\partial_{t_{j'}}|_{t_{j'}=0}\prod_{l=1}^m \partial_{u_l}|_{u_l=0} \prod_{l'=1}^{m'}\partial_{v_{l'}}|_{v_{l'}=0}\Bigg[e^{-\sum_{j'=1}^{n'}t_{j'}\mu c_\epsilon(\widetilde f_{j'},\rho)}\\
 \times \bigg\langle \exp\bigg\{\sum_{j=1}^n s_j \opcolon e^{i\sigma_j\sqrt{\pi}\phie}\opcolon (f_j)+\sum_{j'=1}^{n'}t_{j'}\opcolon e^{i\tau_{j'}\sqrt{\pi}\phie}\opcolon (\widetilde f_{j'})+\sum_{l=1}^m u_l \partial\phie(g_l)+\sum_{l'=1}^{m'}v_{l'}\bar\partial \phie(\widetilde g_{l'})\\
+\mu \opcolon \cos(\sqrt{4\pi}\phie)\opcolon (\rho)\bigg\}\bigg\rangle_{\gff(\epsilon)}\Bigg]
\end{multline*}
for $\sigma_1,\dots,\sigma_n\in \{-1,1\}$, $\tau_1,\dots,\tau_{n'}\in \{-2,2\}$. This is justified since for $\epsilon>0$, we are dealing either with bounded random variables, or exponentials of Gaussian random variables which are all integrable. 

As one can guess from the statement of the lemma, the idea is to write this in terms of the renormalized partition function $\Z(\eta|\epsilon)$ for a suitable $\eta$ depending on our parameters. Our first task is to get rid of the derivative terms in the exponential -- without them, our GFF-expectation would already be of the form $Z(\eta|\epsilon)$ for a suitable $\eta$. This can be done with a routine Girsanov transform/completing the square. There is a minor technicality here, namely that $\partial\phie$ and $\bar\partial \phie$ are complex valued Gaussian fields, and the validity of Girsanov's theorem is not immediate. However, there is a straightforward analyticity argument allowing the use of Girsanov here too -- for details, see \cite[Proof of Lemma 2.6]{BaWe}. This means that we can write 
\begin{align*}
&\bigg\langle\! e^{\sum_{j=1}^n s_j \opcolon e^{i\sigma_j\sqrt{\pi}\phie}\!\opcolon (f_j)+\sum_{j'=1}^{n'}t_{j'}\opcolon e^{i\tau_{j'}\sqrt{\pi}\phie}\!\opcolon (\widetilde f_{j'})+\sum_{l=1}^m u_l \partial\phie(g_l)+\sum_{l'=1}^{m'}v_{l'}\bar\partial \phie(\widetilde g_{l'})+\mu \opcolon \cos(\sqrt{4\pi}\phie)\opcolon (\rho)}\!\bigg\rangle_{\gff(\varepsilon)}\\
&\quad=Z(\eta_\epsilon|\epsilon)\, e^{\frac{1}{2}\sum_{l,k=1}^m u_l u_k \gffe{\partial \phie(g_l)\partial \phie(g_k)}+\frac{1}{2}\sum_{l',k'=1}^{m'}v_{l'}v_{k'}\gffe{\bar\partial \phie(\widetilde g_{l'})\bar\partial \phie(\widetilde g_{k'})}}\\
&\qquad \qquad\times e^{\sum_{l=1}^m \sum_{l'=1}^{m'}u_l v_{l'}\gffe{\partial \phie(g_l)\bar\partial \phie(\widetilde g_{l'})}} ,
\end{align*}
where $\eta_\epsilon(x,\sigma)$ is as in the statement of the lemma.

Now it just remains to rewrite the ratio of partition functions in terms of a ratio of renormalized partition functions. This follows directly from the definition of the renormalized partition function \eqref{eq:rpf}:
\begin{equation*}
\frac{Z(\eta_\epsilon|\epsilon)}{Z_{\mathrm{sG}(\mu\rho|\varepsilon,\Omega)}}=\frac{\Z(\eta_\epsilon|\epsilon)}{\Z(-\frac{\mu}{2}\rho|\epsilon)}e^{\frac{1}{2}\int_{(\Omega\times \{\pm 2\})^2}d\xi_1\,d\xi_2\big(\eta_\epsilon(\xi_1)\eta_\epsilon(\xi_2)-\frac{\mu^2}{4}\rho(x_1)\rho(x_2)\big)A(\xi_1,\xi_2|\epsilon)}.
\end{equation*}
Putting everything together yields the claim.
\end{proof}

Thus to prove the first claim of Theorem~\ref{th:sgmain}, it remains to prove that all relevant derivatives of: (i) the exponential of the quadratic term, (ii) the ratio of renormalized partition functions, and (iii) the factors involving the counter term converge as $\epsilon\to 0$. We treat each of these cases separately. 
\begin{lemma}\label{lem:quad}
For any $S\subset \{1,\dots,m\}$ and $S'\subset \{1,\dots,m'\}$, we have 
\begin{multline*}
\lim_{\epsilon\to 0}\prod_{l\in S}\!\partial_{u_l}|_{u_l=0}\prod_{l'\in S'}\!\partial_{v_{l'}}|_{v_{l'}=0}\bigg[e^{\frac{1}{2}\sum_{l,k=1}^m u_l u_k \gffe{\partial \phie(g_l)\partial \phie(g_k)}}\\
\shoveright{\times e^{\frac{1}{2}\sum_{l',k'=1}^{m'}v_{l'}v_{k'}\gffe{\bar\partial \phie(\widetilde g_{l'})\bar\partial \phie(\widetilde g_{k'})}+\sum_{l=1}^m \sum_{l'=1}^{m'}u_l v_{l'}\gffe{\partial \phie(g_l)\bar\partial \phie(\widetilde g_{l'})}}\bigg]}\\
\shoveleft\quad=\prod_{l\in S}\partial_{u_l}|_{u_l=0}\prod_{l'\in S'}\partial_{v_{l'}}|_{v_{l'}=0}\bigg[e^{\frac{1}{2}\sum_{l,k=1}^m u_l u_k \gffcf{\varphi(\partial  g_l) \varphi(\partial g_k)}}\\
\times e^{\frac{1}{2}\sum_{l',k'=1}^{m'}v_{l'}v_{k'}\gffcf{ \varphi(\bar\partial\widetilde g_{l'}) \varphi(\bar\partial\widetilde g_{k'})}+\sum_{l=1}^m \sum_{l'=1}^{m'}u_l v_{l'}\gffcf{ \varphi(\partial g_l) \varphi(\bar\partial\widetilde g_{l'})}}\bigg].
\end{multline*}
In particular, the limit exists and is finite.
\end{lemma}
\begin{proof}
Let us first note that the derivatives in question (when evaluated at zero) are polynomials in the relevant covariances. We also have (integrating by parts) e.g.
\begin{equation*}
\gffe{\partial \phie(g_l)\partial \phie(g_k)}=\gffe{\phie(\partial g_l)\phie(\partial g_k)}. 
\end{equation*}

Moreover, since we differentiate only once with respect to each parameter and then set the parameters to zero, we only ever encounter covariances of the form $\gffe{\phie(f)\phie(g)}$ with $f$ and $g$ having disjoint support (since we are assuming in Theorem~\ref{th:sgmain} that the test functions have disjoint support). Thus convergence of the individual covariances (and consequently of the relevant derivatives) follows from the uniform convergence provided by Lemma~\ref{lem:cov}. This concludes the proof.
\end{proof}

Let us then turn to the ratio of the renormalized partition functions. 
\begin{lemma}\label{lem:ratconv}
For $S_1\subset \{1,\dots,n\}$, $S_2\subset \{1,\dots,n'\}$, $S_3\subset \{1,\dots,m\}$ and $S_4\subset \{1,\dots,m'\}$, 
\begin{align*}
&\lim_{\epsilon\to 0}\prod_{j\in S_1} \partial_{s_j}|_{s_j=0} \prod_{j'\in S_2}\partial_{t_{j'}}|_{t_{j'}=0}\prod_{l\in S_3}\partial_{u_l}|_{u_l=0}\prod_{l'\in S_4}\partial_{v_{l'}}|_{v_{l'}=0}\frac{\Z(\eta_\epsilon|\epsilon)}{\Z(-\frac{\mu}{2}\rho|\epsilon)}\\
&\quad =\prod_{j\in S_1} \partial_{s_j}|_{s_j=0} \prod_{j'\in S_2}\partial_{t_{j'}}|_{t_{j'}=0}\prod_{l\in S_3}\partial_{u_l}|_{u_l=0}\prod_{l'\in S_4}\partial_{v_{l'}}|_{v_{l'}=0}\frac{\Z(\eta)}{\Z(-\frac{\mu}{2}\rho)},
\end{align*} 
where 
\begin{align}
\eta(x,\sigma)=&-\sum_{j=1}^n s_j \delta_{\sigma,\sigma_j}e^{-i\sigma_j \sqrt{\pi}\big(\sum_{l=1}^m u_l\gffcf{\varphi(x) \varphi(\partial g_l)}+\sum_{l'=1}^{m'}v_{l'}\gffcf{\varphi(x)\varphi(\bar\partial\widetilde g_{l'})}\big)}f_j(x)\\
\nonumber&-\sum_{j'=1}^{n'}t_{j'}\delta_{\sigma,\tau_{j'}}e^{-i\tau_{j'} \sqrt{\pi}\big(\sum_{l=1}^m u_l\gffcf{\varphi(x) \varphi(\partial g_l)}+\sum_{l'=1}^{m'}v_{l'}\gffcf{\varphi(x)\varphi(\bar\partial\widetilde g_{l'})}\big)}\widetilde f_{j'}(x)\\
\nonumber&-\1_{\sigma\in \{-2,2\}}\frac{\mu}{2}e^{-i\sigma\sqrt{\pi}\big(\sum_{l=1}^m u_l\gffcf{\varphi(x) \varphi(\partial g_l)}+\sum_{l'=1}^{m'}v_{l'}\gffcf{\varphi(x)\varphi(\bar\partial\widetilde g_{l'})}\big)}\rho(x).
\end{align}
In particular, the limit in question exists and is finite.
\end{lemma}
\begin{proof}
First of all, we point out the denominator does not depend on the parameters we are differentiating with respect to and since $\mu$ and $\rho$ do not depend on $\epsilon$, its convergence follows immediately from Theorem~\ref{thm:uc} part (i). Moreover, since $\rho$ is real and independent of $\sigma$‚ part (iii) of the same theorem implies that the denominator is positive. Thus it is sufficient to focus entirely on the numerator.

Since Lemma~\ref{lem:smcov} implies that $\eta_\epsilon\to \eta$ locally uniformly in everything, the (locally uniform) convergence of $\Z(\eta_\epsilon|\epsilon)$ to $\Z(\eta)$ follows from Theorem~\ref{thm:uc} part (iv). The fact that this extends to derivatives follows from the fact that $\Z(\eta_\epsilon|\epsilon)$ is an entire function of all of the parameters (since we are dealing with either Gaussian or bounded random variables) so by locally uniform convergence, $\Z(\eta)$ is entire and the derivatives converge.
\end{proof}

As a final ingredient for proving convergence, we show that the derivatives of the factors involving the counter term converge.
\begin{lemma}\label{lem:cttconv}
For $S_1\subset \{1,\dots,n\}$, $S_2\subset \{1,\dots,n'\}$, $S_3\subset \{1,\dots,m\}$ and $S_4\subset \{1,\dots,m'\}$, 
\begin{align*}
&\lim_{\epsilon\to 0}\prod_{j\in S_1} \partial_{s_j}|_{s_j=0} \prod_{j'\in S_2}\partial_{t_{j'}}|_{t_{j'}=0}\prod_{l\in S_3}\partial_{u_l}|_{u_l=0}\prod_{l'\in S_4}\partial_{v_{l'}}|_{v_{l'}=0}\bigg[e^{-\sum_{j'=1}^{n'}t_{j'}\mu c_\epsilon(\widetilde f_{j'},\rho)}\\
&\qquad\qquad\qquad\qquad\qquad\times e^{\frac{1}{2}\int_{(\Omega\times \{\pm 2\})^2}d\xi_1\,d\xi_2\big(\eta_\epsilon(\xi_1)\eta_\epsilon(\xi_2)-\frac{\mu^2}{4}\rho(x_1)\rho(x_2)\big)A(\xi_1,\xi_2|\epsilon)} \bigg]\\
&=\prod_{j\in S_1} \partial_{s_j}|_{s_j=0} \prod_{j'\in S_2}\partial_{t_{j'}}|_{t_{j'}=0}\prod_{l\in S_3}\partial_{u_l}|_{u_l=0}\prod_{l'\in S_4}\partial_{v_{l'}}|_{v_{l'}=0}\bigg[\exp\bigg\{\frac{\mu^2}{8}\int_{(\Omega\times \{\pm 2\})^2}d\xi_1\,d\xi_2 \\
&\qquad\qquad\Big(e^{i\nu_1\sqrt{\pi}\big(\sum_{l=1}^m u_l\gffcf{\varphi(x_1) \varphi(\partial g_l)}+\sum_{l'=1}^{m'}v_{l'}\gffcf{\varphi(x_1)\varphi(\bar\partial\widetilde g_{l'})}\big)}\\
&\quad\times e^{i\nu_2\sqrt{\pi}\big(\sum_{l=1}^m u_l\gffcf{\varphi(x_2)\varphi(\partial g_l)}+\sum_{l'=1}^{m'}v_{l'}\gffcf{\varphi(x_2)\varphi(\bar\partial\widetilde g_{l'})}\big)}-1\Big)\rho(x_1)\rho(x_2)A(\xi_1,\xi_2)\\
&\qquad +\frac{\mu}{2} \sum_{j'=1}^{n'}t_{j'}\int_{(\Omega\times \{\pm 2\})^2}d \xi_1\,d \xi_2\, \delta_{\nu_1,\tau_{j'}}\Big(e^{i\tau_{j'} \sqrt{\pi}\sum_{l=1}^m u_l\big(\gffcf{\varphi(x_1) \varphi(\partial g_l)}-\gffcf{\varphi(x_2) \varphi(\partial g_l)}\big)}\\
&\quad\times e^{i\nu_2\sqrt{\pi}\sum_{l'=1}^{m'}v_{l'}\big(\gffcf{\varphi(x_1)\varphi(\bar\partial\widetilde g_{l'})}-\gffcf{\varphi(x_2)\varphi(\bar\partial\widetilde g_{l'})}\big)}-1\Big)\widetilde f_{j'}(x_1)\rho(x_2)A(\xi_1,\xi_2)\\
&\qquad +\frac{1}{2}\sum_{\substack{j',k'=1:\\ j'\neq k'}}^{n'}t_{j'}t_{k'}\int_{(\Omega\times \{\pm 2\})^2}d\xi_1\,d\xi_2\, \delta_{\tau_{j'},\nu_1}\delta_{\tau_{k'},\nu_2}\\
&\quad\qquad \times e^{i\tau_{j'} \sqrt{\pi}\big(\sum_{l=1}^m u_l\gffcf{\varphi(x_1) \varphi(\partial g_l)}+\sum_{l'=1}^{m'}v_{l'}\gffcf{\varphi(x_1)\varphi(\bar\partial\widetilde g_{l'})}\big)}\\
&\quad\times e^{i\tau_{k'} \sqrt{\pi}\big(\sum_{l=1}^m u_l\gffcf{\varphi(x_2) \varphi(\partial g_l)}+\sum_{l'=1}^{m'}v_{l'}\gffcf{\varphi(x_2)\varphi(\bar\partial\widetilde g_{l'})}\big)}\widetilde f_{j'}(x_1)\widetilde f_{k'}(x_2)A(\xi_1,\xi_2)\bigg\}\bigg],
\end{align*}
where $A(\xi_1,\xi_2|\epsilon)$ is as in \eqref{eq:ctt}. In particular, the limit exists and is finite. Above we wrote $\xi_i=(x_i,\nu_i)$ with $x_i\in \Omega$ and $\nu_i\in \{\pm 2\}$.
\end{lemma}

\begin{proof}
Let us begin by writing out in detail the quantity in question using Lemma \ref{lem:cftopart}. Since the symbols $\sigma_1,\dots,\sigma_n$ are now reserved, let us call our integration variables $\xi_i=(x_i,\nu_i)$. Also, let us introduce some notation that will make the quantities slightly less horrendous looking. For this purpose, we write 
\begin{equation*}
F_\epsilon(x)=\sum_{l=1}^m u_l\langle \varphi_\epsilon(x)\partial \varphi_\epsilon(g_l)\rangle_{\gff(\epsilon)}+\sum_{l'=1}^{m'}v_{l'}\langle \varphi_\epsilon(x)\bar\partial \varphi_\epsilon(\widetilde g_{l'})\rangle_{\gff(\epsilon)}.
\end{equation*}
With this notation, we have 
\begin{align*}
\eta_\epsilon(x,\nu)\!=-\!\sum_{j=1}^n\! s_j \delta_{\nu,\sigma_j}e^{i\sigma_j \sqrt{\pi}F_\epsilon(x)}f_j(x)-\!\sum_{j'=1}^{n'}\!t_{j'}\delta_{\nu,\tau_{j'}}e^{i\tau_{j'}\sqrt{\pi}F_\epsilon(x)}\widetilde f_{j'}(x)-\frac{\mu}{2}\1_{\nu\in \{-2,2\}}e^{i\nu \sqrt{\pi}F_\epsilon(x)}\!\rho(x).
\end{align*}
Note that when integrating $\eta_\epsilon(\xi_1)\eta_\epsilon(\xi_2)$ over $(\Omega\times \{\pm 2\})^2$, the $s_j$ terms disappear from $\eta_\epsilon$ since we have $\nu_1,\nu_2\in \{\pm 2\}$ but $\sigma_j\in \{\pm 1\}$ so always $\delta_{\nu_i,\sigma_j}=0$ in this case. 

Before looking at the quantity of interest, let us note that we can rewrite $c_\epsilon(f,\rho)$ in a slightly more convenient way. We claim that for $\tau\in \{\pm 2\}$, we can write 
\begin{align*}
c_\epsilon(f,\rho)=\frac{1}{2}\int_{(\Omega\times \{\pm 2\})^2}d\xi_1\,d\xi_2\, \delta_{\nu_1,\tau}f(x_1)\rho(x_2)A(\xi_1,\xi_2|\epsilon).
\end{align*}
To see why this is the case, note that 
\begin{multline*}
\frac{1}{2}\int_{(\Omega\times \{\pm 2\})^2}d\xi_1\,d\xi_2\, \delta_{\nu_1,\tau}f(x_1)\rho(x_2)A(\xi_1,\xi_2|\epsilon)\\
\shoveleft\quad=\frac{1}{2}\sum_{\nu_2\in \{\pm 2\}}\int_{\Omega^2}d^2x_1\, d^2 x_2\, f(x_1)\rho(x_2)\Big(\big\langle \opcolon e^{i\tau \sqrt{\pi}\varphi(x_1)}\opcolon  \opcolon e^{i\nu_2\sqrt{\pi}\varphi(x_2)}\opcolon \big\rangle_{\gff(\epsilon)}\\
\shoveright{-\big\langle \opcolon e^{i\tau \sqrt{\pi}\varphi(x_1)}\opcolon \big\rangle_{\gff(\epsilon)}\big\langle \opcolon e^{i\nu_2\sqrt{\pi}\varphi(x_2)}\opcolon \big\rangle_{\gff(\epsilon)}\Big)}\\
\shoveleft\quad=\int_{\Omega^2}d^2x_1\, d^2 x_2\, f(x_1)\rho(x_2)\Big(\big\langle \opcolon e^{i\tau \sqrt{\pi}\varphi(x_1)}\opcolon  \opcolon \cos(\sqrt{4\pi}\varphi(x_2))\opcolon \big\rangle_{\gff(\epsilon)}\\
-\big\langle \opcolon e^{i\tau \sqrt{\pi}\varphi(x_1)}\opcolon \big\rangle_{\gff(\epsilon)}\big\langle \opcolon \cos(\sqrt{4\pi}\varphi(x_2))\opcolon \big\rangle_{\gff(\epsilon)}\Big).
\end{multline*}
Writing $\opcolon e^{i\tau \sqrt{\pi}\varphi(x_1)}\opcolon =\opcolon  \cos(\sqrt{4\pi}\varphi(x_1))\opcolon +i \tfrac{\tau}{2}\opcolon \sin(\sqrt{4\pi}\varphi(x_1))\opcolon $ and noting that the sine-terms are odd under $\varphi\mapsto -\varphi$, while the cosine-terms are even and the law of the regularized GFF is centered, so the expectations involving the sine terms vanish and indeed we can write $c_\epsilon(f,\rho)$ this way.

After these preparatory remarks, we find
\begin{align*}
&\frac{1}{2}\int_{(\Omega\times \{\pm 2\})^2}d\xi_1\,d\xi_2\left(\eta_\epsilon(\xi_1)\eta_\epsilon(\xi_2)-\frac{\mu^2}{4}\rho(x_1)\rho(x_2)\right)A(\xi_1,\xi_2|\epsilon)-\sum_{j'=1}^{n'}t_{j'}\mu c_\epsilon(\widetilde f_{j'},\rho)\\
& =\frac{\mu^2}{8}\int_{(\Omega\times \{\pm 2\})^2}d\xi_1\,d\xi_2 \left(e^{i\nu_1\sqrt{\pi}F_\epsilon(x_1)+i\nu_2\sqrt{\pi}F_\epsilon(x_2)}-1\right)\rho(x_1)\rho(x_2)A(\xi_1,\xi_2|\epsilon)\\
&\enspace +\frac{1}{2}\!\sum_{j_1',j_2'=1}^{n'} t_{j_1'} t_{j_2'}\int_{(\Omega\times \{\pm 2\})^2}\!d\xi_1\,d\xi_2\,\delta_{\nu_1,\tau_{j_1'}}\delta_{\nu_2,\tau_{j_2'}}e^{i\tau_{j_1'}\sqrt{\pi}F_\epsilon(x_1)+i\tau_{j_2'}\sqrt{\pi}F_\epsilon(x_2)}\widetilde f_{j_1'}(x_1)\widetilde f_{j_2'}(x_2)A(\xi_1,\xi_2|\epsilon)\\
&\enspace +\frac{\mu}{2} \sum_{j'=1}^{n'}t_{j'}\int_{(\Omega\times \{\pm 2\})^2}d\xi_1\,d\xi_2\, \delta_{\nu_1,\tau_{j'}}\left( e^{i\tau_{j'}\sqrt{\pi}F_\epsilon(x_1)+i\nu_2 \sqrt{\pi}F_\epsilon(x_2)}-1\right)\widetilde f_{j'}(x_1)\rho(x_2)A(\xi_1,\xi_2|\epsilon).
\end{align*}
For $\epsilon>0$, this is clearly an entire function of the parameters $s,t,u,v,\mu$.

Note that by Lemma~\ref{lem:gffcorr2} and Lemma~\ref{lem:smcov} we see that $\lim_{\epsilon\to 0}A(\xi_1,\xi_2|\epsilon)=A(\xi_1,\xi_2)$ and $\lim_{\epsilon\to 0}\gffe{\phie(x)\phie(f)}=\gffcf{\varphi(x)\varphi(f)}$, so the integrand converges pointwise to the integrand in the statement of the lemma. To conclude, we need an estimate allowing the use of dominated convergence theorem. Let us start with an estimate for $A$. If $\nu_1=\nu_2$‚ then $A$ is bounded, so we can focus on the case $\nu_1\neq \nu_2$. By \eqref{eq:ctt}, we can write 
\begin{align*}
A((x_1,\nu_1),(x_2,-\nu_1)|\epsilon)&=c_{\sqrt{4\pi}}^2\epsilon^{-2}e^{-2\pi \gffe{\phie(x_1)^2}\!-2\pi \!\gffe{\phie(x_2)^2}}\left(e^{4\pi \gffe{\phie(x_1)\phie(x_2)}}-1\right)\!.
\end{align*}
Lemma~\ref{lem:cov} implies that for any compact $K\subset \Omega$, there exists a constant $C_K$ such that for small enough $\epsilon$ (``small enough'' depending only on $K$)
\begin{equation*}
\epsilon^{-2}e^{-2\pi \gffe{\phie(x_1)^2}-2\pi \gffe{\phie(x_2)^2}}\leq C_K
\end{equation*}
for $x_1,x_2\in K$.

Moreover, since $\p_\Omega\geq 0$, we have again for any compact $K\subset \Omega$ the existence of a constant $C_K$ (possibly different from above)
\begin{equation*}
0\leq e^{4\pi \gffe{\phie(x_1)\phie(x_2)}}-1\leq e^{4\pi \int_0^\infty ds\, \p_\Omega(s,x_1,x_2)}=e^{4\pi G_\Omega(x_1,x_2)}\leq C_K |x_1-x_2|^{-2}
\end{equation*}
for $x_1,x_2\in K$. Thus for any compact $K\subset \Omega$, there exists a $C_K>0$ such that for small enough $\epsilon>0$ and any $x_1,x_2\in K$
\begin{equation*}
A((x_1,\nu_1),(x_2,-\nu_1)|\epsilon)\leq C_K|x_1-x_2|^{-2}.
\end{equation*}

Let us now turn to arguing that we can use dominated convergence. First of all, for the $t_{j_1'}t_{j_2'}$-terms, note that we never differentiate twice with respect to a $t$-parameter. Since $\widetilde f_{j_1'}$ and $\widetilde f_{j_2'}$ have disjoint supports for $j_{1}'\neq j_2'$, we have no issue with dominated convergence. 

Let us then look at the $\mu^2$-term. Here $A(\xi_1,\xi_2|\epsilon)$ is bounded if $\nu_1=\nu_2$‚ so we only need to worry about the $\nu_1=-\nu_2$ term. For this, the uniform Lipschitz continuity and boundedness of $F_\epsilon$ (from Lemma \ref{lem:smcov}) implies that we have for some constant $C$ (not depending on $\epsilon$)
\begin{equation*} 
\left| e^{i\nu_1\sqrt{\pi}(F_\epsilon(x_1)-F_\epsilon(x_2))}-1\right|\leq C|x_1-x_2|.
\end{equation*}
Thus using our bound on $A$, we can use dominated convergence. The reasoning for the $\mu t_{j'}$ terms is similar (if $\nu_1=\tau_{j'}=\nu_2$, then $A$ is bounded and we can use dominated convergence, while if $\nu_1=\tau_{j'}=-\nu_2$, we use the same bound on $A$ and the uniform boundedness and uniform Hölder continuity of $F_\epsilon$).

Thus the uniform Lipschitz estimate of Lemma~\ref{lem:smcov} implies that in the $\mu^2$-and $\mu t_{j'}$-terms above, we can bound the integrand by a constant (depending on essentially all of our parameters apart from $\epsilon$) times $|x_1-x_2|^{-1}\1_K$. This singularity is integrable in two dimensions so the use of dominated convergence is justified and we see that the $\mu^2$ and $\mu t_{j'}$-terms converge (locally uniformly in the parameters) as $\epsilon\to 0$. For the $t_{j'}t_{k'}$-terms, we note that since we differentiate only once with respect to each parameter, we never encounter the diagonal terms $j'=k'$. Thus since $\widetilde f_{j'}$ and $\widetilde f_{k'}$ have disjoint supports, the integrand is uniformly bounded and again by the dominated convergence theorem, all of the relevant terms converge (locally uniformly). 

Since we have locally uniform convergence of an entire function, we conclude that also its derivatives converge to the derivatives of the limit. This concludes the proof.
\end{proof}

Putting everything together takes care of our first task:
\begin{proof}[Proof of Theorem~\ref{th:sgmain}, part (i)]
The statement follows from decomposing Wick ordered trigonometric functions into Wick ordered exponentials, combining Lemmas~\ref{lem:cftopart}--\ref{lem:cttconv} and using the product rule for differentiation.
\end{proof}

We now turn to the second statement of Theorem~\ref{th:sgmain}, namely analyticity of the correlation functions as a function of $\mu$ in a neighborhood of the real axis. Again, it is sufficient to focus on the Wick ordered exponentials instead of Wick ordered trigonometric functions. We have in fact already done  all of the preliminary work and we can start the actual proof.
\begin{proof}[Proof of Theorem~\ref{th:sgmain}, part (ii)]
From Lemmas~\ref{lem:cftopart}--\ref{lem:cttconv}, we see that it is enough to establish that the limiting objects appearing in Lemmas~\ref{lem:quad}--\ref{lem:cttconv} are analytic in $\mu$ in a neighborhood of the real axis. For Lemma~\ref{lem:quad}, there is no dependence on $\mu$. Also for Lemma~\ref{lem:cttconv}, we see  that as a function of $\mu$, the limiting object is a polynomial times an exponential of a quadratic polynomial -- thus an entire function. So we only need to focus on the (derivatives of the) ratio of the renormalized partition functions. For this, we note first that by Theorem~\ref{thm:uc} parts (ii) and (iii)
\begin{equation*}
\mu\mapsto \frac{1}{\Z(-\frac{\mu}{2}\rho)}
\end{equation*}
is analytic in a neighborhood of the real axis. On the other hand, one readily checks from Theorem~\ref{thm:uc} part (ii) that in the notation of Lemma~\ref{lem:ratconv}, 
\begin{equation*}
\prod_{j\in S_1} \partial_{s_j}|_{s_j=0} \prod_{j'\in S_2}\partial_{t_{j'}}|_{t_{j'}=0}\prod_{l\in S_3}\partial_{u_l}|_{u_l=0}\prod_{l'\in S_4}\partial_{v_{l'}}|_{v_{l'}=0}\,\Z(\eta)
\end{equation*}
(for $\eta$ as in Lemma~\ref{lem:ratconv}) is an entire function of $\mu$. Putting everything together, we see that the limiting correlation functions are analytic in $\mu$ in a neighborhood of the real axis -- this concludes the proof.
\end{proof}

We now turn to the final statement in Theorem~\ref{th:sgmain}. This is slightly involved as it requires some combinatorial arguments and analysis of GFF correlation functions from the renormalized potential.

Two preliminary remarks:
\begin{itemize}
\item From our proof of item (i) of Theorem~\ref{th:sgmain}, we see that the convergence is uniform in $\mu$ in some neighborhood of the origin (this boils down to the uniform convergence of the renormalized partition function from Theorem~\ref{thm:uc} -- the other terms being explicit). This implies in particular that the Taylor coefficients (when expanding at zero) of 
\begin{multline*}
\mu\mapsto\Bigg\langle \prod_{j=1}^p \partial \varphi_\epsilon(f_j) \prod_{j'=1}^{p'} \bar\partial \varphi_\epsilon(f_{j'}')\prod_{k=1}^q \opcolon  \cos(\sqrt{\pi}\varphi_\epsilon)\opcolon (g_k) \prod_{k'=1}^{q'}\opcolon \sin(\sqrt{\pi}\varphi_\epsilon)\opcolon (g_{k'}')\\
\times\prod_{l=1}^r \optypecolon \cos(\sqrt{4\pi}\varphi_\epsilon)\optypecolon (h_l) \Bigg\rangle_{\mathrm{sG}(\mu\rho|\epsilon,\Omega)}
\end{multline*}
converge to those of
\begin{equation*}
\mu\!\mapsto\!\Bigg\langle\prod_{j=1}^p\!\partial \varphi(f_j)\!\prod_{j'=1}^{p'}\!\bar\partial \varphi(f_{j'}')\!\prod_{k=1}^q\!\opcolon \cos(\sqrt{\pi}\varphi)\opcolon (g_k)\!\!\prod_{k'=1}^{q'}\!\!\opcolon \sin(\sqrt{\pi}\varphi)\opcolon (g_{k'}')\!\prod_{l=1}^r\! \optypecolon \cos(\sqrt{4\pi}\varphi)\optypecolon (h_l)\!\Bigg\rangle_{\mathrm{sG}(\mu\rho|\Omega)}.
\end{equation*}
\medskip
\item It is again enough to focus on Wick ordered exponentials instead of Wick ordered trigonometric functions.
\end{itemize}
The point of these remarks is that it is sufficient for us to compute the Taylor coefficients of 
\begin{equation*}
\Bigg\langle\prod_{j=1}^n \opcolon e^{i\sigma_j\sqrt{\pi}\phie}\opcolon (f_j)\prod_{j'=1}^{n'}\left(\opcolon e^{i\tau_{j'}\sqrt{\pi}\phie}\opcolon (\widetilde f_{j'})- \mu c_\epsilon(\widetilde f_{j'},\rho)\right)\prod_{l=1}^m \partial \phie(g_l)\prod_{l'=1}^{m'}\bar\partial \phie(\widetilde g_{l'})\Bigg\rangle_{\mathrm{sG}(\mu\rho|\varepsilon,\Omega)}
\end{equation*}
with $\sigma_1,\dots,\sigma_n\in \{-1,1\}$, $\tau_1,\dots,\tau_{n'}\in \{-2,2\}$, and try to control their $\epsilon\to 0$ asymptotics. For computing the Taylor coefficients, we begin with the following simple but general fact. 
\begin{lemma}\label{lem:taylorgen}
Let $X, Y_1,\dots,Y_n$ and $Z$ be complex random variables for which $Y_1,\dots,Y_n,Z$ are bounded and $X$ satisfies $\E \big[e^{\alpha |X|}\big]<\infty$ for all $\alpha>0$, also let $c_1,\dots,c_n\in \C$ be deterministic. There exists an open disk centered at the origin in which 
\begin{equation*}
\mu \mapsto \frac{\E\Big[X\prod_{j=1}^n (Y_j-\mu c_j)e^{\mu Z}\Big]}{\E[e^{\mu Z}]}
\end{equation*}
is analytic and in this disk has the following series expansion
\begin{align*}
\sum_{p=0}^\infty \frac{\mu^p}{p!}\sum_{S\subset[n]}(-1)^{n-|S|}\binom{p}{n-|S|} (n-|S|)!\Bigg(\prod_{j\notin S}c_j\Bigg)\Bigg\langle X\prod_{k\in S}Y_k;Z;\dots;Z\Bigg\rangle^\mathsf T,
\end{align*} 
where the binomial coefficient is understood as zero if $p<n-|S|$, and
\begin{align*}
\Bigg\langle X\prod_{k\in S}Y_k;Z;\dots;Z\Bigg\rangle^\mathsf T
\end{align*}
is the joint cumulant of $X\prod_{k\in S}Y_k$ and $p-(n-|S|)$ instances of $Z$.
\end{lemma}
\begin{proof}
We begin by simply expanding the product over $j$:
\begin{equation*}
\frac{\E\Big[X\prod_{j=1}^n (Y_j-\mu c_j)e^{\mu Z}\Big]}{\E[e^{\mu Z}]}=\sum_{S\subset [n]}(-1)^{n-|S|}\mu^{n-|S|}\Bigg(\prod_{j\notin S}c_j\Bigg)\frac{\E\Big[X\prod_{k\in S}Y_k e^{\mu Z}\Big]}{\E[e^{\mu Z}]}.
\end{equation*}
Since the quantities inside the expectations are integrable by our assumptions (for any $\mu\in\C$), one readily checks by Fubini, dominated convergence and Morera's theorem that the expectations are entire functions. Since $\E[e^{\mu Z}]=1$ for $\mu=0$, we conclude that there is a disk surrounding the origin in which $\E[e^{\mu Z}]\neq 0$. In this disk, the ratio of these two entire functions is analytic, so we conclude that the function in question is indeed analytic in some disk.

To compute the series expansion, we first note that 
\begin{align*}
\frac{\E\Big[X\prod_{k\in S}Y_k e^{\mu Z}\Big]}{\E[e^{\mu Z}]}&=\frac{\partial}{\partial t}\bigg|_{t=0}\log \E\Big[e^{tX\prod_{k\in S}Y_k+\mu Z}\Big].
\end{align*} 
We recognize this logarithm as the joint cumulant generating function of the random variables $X\prod_{k\in S}Y_k$ and $Z$. With similar arguments as before, our assumptions ensure that this cumulant generating function is an analytic function of $t$ and $\mu$ in some neighborhood of the origin and in this neighborhood, we have 
\begin{align*}
\log \E\Big[e^{tX\prod_{k\in S}Y_k+\mu Z}\Big]&=\sum_{p,q=0}^\infty \frac{t^p\mu^q}{p!q!}\Bigg\langle X\prod_{k\in S}Y_k;\dots;X\prod_{k\in S}Y_k; Z;\dots;Z\Bigg\rangle^\mathsf T,
\end{align*}
where there are $p$ entries of $X\prod_{k\in S}Y_k$ and $q$ entries of $Z$.

Evaluating the $t$ derivative at zero, we find that there exists some disk in which we have
\begin{align*}
\frac{\E\Big[X\prod_{j=1}^n (Y_j-\mu c_j)e^{\mu Z}\Big]}{\E[e^{\mu Z}]}=\sum_{q=0}^\infty \sum_{S\subset [n]}(-1)^{n-|S|}\frac{\mu^{n-|S|+q}}{q!}\Bigg(\prod_{j\notin S}c_j\Bigg)\Bigg\langle X\prod_{k\in S}Y_k;Z;\dots;Z\Bigg\rangle^{\mathsf T},
\end{align*}
where there are $q$ entries of $Z$ in the cumulant.

Making a change of variables in the sum, where we make $n-|S|+q$ our new summation variable $p$, we write this as 
\begin{align*}
&\frac{\E\Big[X\prod_{j=1}^n (Y_j-\mu c_j)e^{\mu Z}\Big]}{\E[e^{\mu Z}]}\\
&\qquad =\sum_{p=0}^\infty \frac{\mu^p}{p!}\sum_{S\subset[n]}(-1)^{n-|S|}\binom{p}{n-|S|} (n-|S|)!\Bigg(\prod_{j\notin S}c_j\Bigg)\Bigg\langle X\prod_{k\in S}Y_k;Z;\dots;Z\Bigg\rangle^\mathsf T,
\end{align*} 
where there are $p-(n-|S|)$ $Z$ entries in the cumulant and we understand the binomial coefficient as zero if $p<n-|S|$. This concludes the proof.
\end{proof}

We now apply this to the setting relevant to us. 
\begin{lemma}\label{lem:sgtaylor}
For $\epsilon>0$, we have $(\,$for $\sigma_1,\dots,\sigma_n\in \{-1,1\}$, $\tau_1,\dots,\tau_{n'}\in \{-2,2\}\,)$ 
\begin{align*}
&\frac{\partial^p}{\partial \mu^p}\bigg|_{\mu=0} \!\!\Bigg\langle\prod_{j=1}^n \opcolon e^{i\sigma_j\sqrt{\pi}\phie}\opcolon (f_j)\!\prod_{j'=1}^{n'}\!\Big(\opcolon e^{i\tau_{j'}\sqrt{\pi}\phie}\opcolon (\widetilde f_{j'})- \mu c_\epsilon(\widetilde f_{j'},\rho)\Big)\!\prod_{l=1}^m \partial \phie(g_l)\!\prod_{l'=1}^{m'}\bar\partial \phie(\widetilde g_{l'})\Bigg\rangle_{\mathrm{sG}(\mu\rho|\varepsilon,\Omega)}\\
&\enspace=\int_{\Omega^{n+n'+m+m'+p}} d^{2n}x\,d^{2n'}y\,d^{2m}z\,d^{2m'}w\,d^{2p}u\prod_{j=1}^n f_j(x_j)\!\prod_{j'=1}^{n'}\widetilde f_{j'}(y_{j'})\!\prod_{l=1}^m(-\partial g_l(z_l))\!\prod_{l'=1}^{m'} (-\bar\partial\widetilde g_{l'}(w_{l'}))\\
&\quad \times \prod_{k=1}^p \rho(u_k)\sum_{S\subset [n']}\sum_{\substack{S'\subset[p]:\\ |S|=|S'|}} \sum_{\pi:S\leftrightarrow S'}\prod_{s\in S}(-\mathcal A(y_{s},u_{\pi(s)}|\epsilon))\Bigg\langle \prod_{j=1}^n \opcolon e^{i\sigma_j \sqrt{\pi}\varphi_\epsilon(x_j)}\opcolon  \prod_{l=1}^m \varphi_\epsilon(z_l)\prod_{l'=1}^{m'}\varphi_\epsilon(w_{l'})\\
&\qquad\qquad\prod_{k'\notin S}\big(\opcolon e^{i\tau_{k'}\sqrt{\pi}\varphi_\epsilon(y_{k'})}\opcolon\big) ;\big(\opcolon \cos(\sqrt{4\pi}\varphi_\epsilon(u_{\alpha_1}))\opcolon\big) ;\dots;\big(\opcolon \cos(\sqrt{4\pi}\varphi_\epsilon(u_{\alpha_{p-|S'|}}))\opcolon\big) \bigg\rangle_{\mathrm{GFF}(\epsilon)}^\mathsf T,
\end{align*}
where, $\{\alpha_1,\dots,\alpha_{p-|S'|}\}=[p]\setminus S'$ and the sum over $\pi$ is a sum over bijections between $S$ and $S'$.
\end{lemma}
\begin{proof}
We apply Lemma~\ref{lem:taylorgen} with $X=\prod_{j=1}^n \opcolon e^{i\sigma_j \sqrt{\pi}\varphi_\epsilon}\opcolon (f_j)\prod_{l=1}^m \partial\varphi_\epsilon(g_l)\prod_{l'=1}^{m'}\bar\partial\varphi_\epsilon(\widetilde g_{l'})$, $Y_j=\opcolon e^{i\tau_j\sqrt{\pi}\varphi_\epsilon}\opcolon (\widetilde f_j)$, $c_j=c_\epsilon(\widetilde f_j,\rho)$, and $Z=\opcolon \cos(\sqrt{4\pi}\varphi_\epsilon)\opcolon (\rho)$. 

To be precise, $X$ may not satisfy $\E(e^{\alpha |X|})<\infty$, but replacing $X$ by $\widetilde X=X\1_{\{|X|\leq R\}}$, we can make use of this formula and then let $R\to \infty$. The validity of this argument is readily justified with the Cauchy integral formula, Fubini, and dominated convergence.

We thus find that
\begin{align*}
&\frac{\partial^p}{\partial \mu^p}\bigg|_{\mu=0} \Bigg\langle\prod_{j=1}^n \opcolon e^{i\sigma_j\sqrt{\pi}\phie}\opcolon (f_j)\!\prod_{j'=1}^{n'}\!\!\Big(\opcolon e^{i\tau_{j'}\sqrt{\pi}\phie}\opcolon (\widetilde f_{j'})- \mu c_\epsilon(\widetilde f_{j'},\rho)\Big)\!\prod_{l=1}^m \partial \phie(g_l)\!\prod_{l'=1}^{m'}\bar\partial \phie(\widetilde g_{l'})\Bigg\rangle_{\mathrm{sG}(\mu\rho|\varepsilon,\Omega)}\\
&\enspace=\sum_{S\subset[n']}(-1)^{n'-|S|}\binom{p}{n'-|S|}(n'-|S|)!\bigg(\prod_{j'\notin S}c_\epsilon(\widetilde f_{j'},\rho)\bigg)\Bigg\langle  \prod_{j=1}^n \opcolon e^{i\sigma_j \sqrt{\pi}\varphi_\epsilon}\opcolon (f_j)\prod_{l=1}^m \partial\varphi_\epsilon(g_l)\\
&\qquad\prod_{l'=1}^{m'}\bar\partial\varphi_\epsilon(\widetilde g_{l'}) \prod_{k'\in S}\big(\opcolon e^{i\tau_{k'}\sqrt{\pi}\varphi_\epsilon}\opcolon (\widetilde f_{k'})\big);\big(\opcolon \cos(\sqrt{4\pi}\varphi_\epsilon)\opcolon (\rho)\big);\dots;\big(:\cos(\sqrt{4\pi}\varphi_\epsilon)\opcolon (\rho)\big)\Bigg\rangle_{\mathrm{GFF}(\epsilon)}^\mathsf T,
\end{align*}
where we have $p-(n'-|S|)$ instances of $\opcolon \cos(\sqrt{4\pi}\varphi_\epsilon)\opcolon (\rho)$ (and the interpretation that for $p<n'-|S|$, the binomial coefficient is zero).

Next we use multilinearity of the cumulants and Fubini to write this as 
\begin{multline*}
\int_{\Omega^{n+n'+m+m'+p}} d^{2n}x\,d^{2n'}y\,d^{2m}z\,d^{2m'}w\,d^{2p}u\prod_{j=1}^n f_j(x_j)\prod_{j'=1}^{n'}\widetilde f_{j'}(y_{j'})\prod_{l=1}^m (-\partial g_l(z_l))\prod_{l'=1}^{m'}(-\bar\partial \widetilde g_{l'}(w_{l'}))\\
\shoveleft\quad\times \prod_{k=1}^p \rho(u_k)\sum_{S\subset [n']}(-1)^{n'-|S|}\binom{p}{n'-|S|}(n'-|S|)!\mathcal A (y_{\alpha_1},u_1|\epsilon)\dots \mathcal A(y_{\alpha_{n'-|S|}},u_{n'-|S|}|\epsilon)\\
\shoveleft\qquad\times \Bigg\langle \prod_{j=1}^n \opcolon e^{i\sigma_j \sqrt{\pi}\varphi_\epsilon(x_j)}\opcolon  \prod_{l=1}^m \varphi_\epsilon(z_l)\prod_{l'=1}^{m'}\varphi_\epsilon(w_{l'})\prod_{k'\in S}\big(\opcolon e^{i\tau_{k'}\sqrt{\pi}\varphi_\epsilon(y_{k'})}\opcolon\big) ;\\
\big(\opcolon \cos(\sqrt{4\pi}\varphi_\epsilon(u_{n'-|S|+1}))\opcolon\big) ;\dots;\big(\opcolon \cos(\sqrt{4\pi}\varphi_\epsilon(u_p))\opcolon\big) \Bigg\rangle_{\mathrm{GFF}(\epsilon)}^\mathsf T,
\end{multline*}
where we have written $[n']\setminus S=\{\alpha_1,\dots,\alpha_{n'-|S|}\}$. 

Next, we note that if we would replace $u_1,\dots,u_{n'-|S|}$ in the $A$-terms by any subset of $\{u_1,\dots,u_p\}$, this would not change the value of the integral. So if in the $S$ sum we make $[n']\setminus S$ the summation variable (and still call it $S$), we find that this integral can be written as 
\begin{multline*}
\int_{\Omega^{n+n'+m+m'+p}} d^{2n}x\,d^{2n'}y\,d^{2m}z\,d^{2m'}w\,d^{2p}u\prod_{j=1}^n f_j(x_j)\prod_{j'=1}^{n'}\widetilde f_{j'}(y_{j'})\prod_{l=1}^m (-\partial g_l(z_l))\prod_{l'=1}^{m'}(-\bar\partial \widetilde g_{l'}(w_{l'}))\\
\shoveleft\enspace\times \prod_{k=1}^p \rho(u_k)\sum_{S\subset [n']}\sum_{\substack{S'\subset[p]:\\ |S|=|S'|}} \sum_{\pi:S\leftrightarrow S'}\prod_{s\in S}(-\mathcal A(y_{s},u_{\pi(s)}|\epsilon))\Bigg\langle \prod_{j=1}^n \opcolon e^{i\sigma_j \sqrt{\pi}\varphi_\epsilon(x_j)}\opcolon  \prod_{l=1}^m \varphi_\epsilon(z_l)\prod_{l'=1}^{m'}\varphi_\epsilon(w_{l'})\\
\prod_{k'\notin S}\big(\opcolon e^{i\tau_{k'}\sqrt{\pi}\varphi_\epsilon(y_{k'})}\opcolon\big) ; \big(\opcolon \cos(\sqrt{4\pi}\varphi_\epsilon(u_{\alpha_1}))\opcolon\big) ;\dots;\big(\opcolon \cos(\sqrt{4\pi}\varphi_\epsilon(u_{\alpha_{p-|S'|}}))\opcolon\big) \Bigg\rangle_{\mathrm{GFF}(\epsilon)}^\mathsf T,
\end{multline*}
where, as in the statement of the lemma, $\{\alpha_1,\dots,\alpha_{p-|S'|}\}=[p]\setminus S'$ and the sum over $\pi$ is a sum over bijections between $S$ and $S'$. This concludes the proof.
\end{proof}
This representation has two benefits. First of all, we see immediately from Lemma~\ref{lem:gffcorr2} that the integrand converges as $\epsilon\to 0$. Secondly, as we argue now, the integrand is symmetric  in the $u$ variables. The precise statement is as follows. 
\begin{lemma}\label{lem:symend}
The function 
\begin{multline}
C_\epsilon(x,\sigma,y,\tau,z,w,u)\\
\shoveleft= \sum_{S\subset [n']}\sum_{\substack{S'\subset[p]:\\ |S|=|S'|}} \sum_{\pi:S\leftrightarrow S'}\prod_{s\in S}(-\mathcal A(y_{s},u_{\pi(s)}|\epsilon))\Bigg\langle \prod_{j=1}^n \opcolon e^{i\sigma_j \sqrt{\pi}\varphi_\epsilon(x_j)}\opcolon  \prod_{l=1}^m \varphi_\epsilon(z_l)\prod_{l'=1}^{m'}\varphi_\epsilon(w_{l'})\\
\prod_{k'\notin S}\big(\opcolon e^{i\tau_{k'}\sqrt{\pi}\varphi_\epsilon(y_{k'})}\opcolon\big) ;\big(\opcolon \cos(\sqrt{4\pi}\varphi_\epsilon(u_{\alpha_1}))\opcolon\big) ; \dots;\big(\opcolon \cos(\sqrt{4\pi}\varphi_\epsilon(u_{\alpha_{p-|S'|}}))\opcolon\big) \Bigg\rangle_{\mathrm{GFF}(\epsilon)}^{\mathsf T}
\end{multline}
is the unique smooth function which satisfies:
\begin{itemize}
\item $C_\epsilon(\cdot,\nu u)=C_\epsilon(\cdot,u)$ for each $\nu$ which is a permutation of $[p]$ $($here we write $\nu u=(u_{\nu(1)},\dots,u_{\nu(p)}))$.
\medskip
\item $C_{\epsilon}(x,\sigma,y,\tau,z,w,u)\to 0$ if one of the $z$ or $w$ variables tends to $\partial \Omega$.
\medskip
\item For each $\rho\in C_c^\infty(\Omega)$ and $f_1,\dots,f_n$, $\widetilde f_1,\dots,\widetilde f_{n'}$, $g_1,\dots,g_m$, $\widetilde g_1,\dots,\widetilde g_{m'}$ $\in C_c^\infty(\Omega)$ with disjoint supports (not necessarily disjoint from the support of $\rho$), we have 
\begin{align*}
&\frac{\partial^p}{\partial \mu^p}\bigg|_{\mu=0} \Bigg\langle\prod_{j=1}^n \opcolon e^{i\sigma_j\sqrt{\pi}\phie}\opcolon (f_j)\!\prod_{j'=1}^{n'}\!\!\Big(\opcolon e^{i\tau_{j'}\sqrt{\pi}\phie}\opcolon (\widetilde f_{j'})- \mu c_\epsilon(\widetilde f_{j'},\rho)\Big)\!\prod_{l=1}^m \partial \phie(g_l)\!\prod_{l'=1}^{m'}\bar\partial \phie(\widetilde g_{l'})\Bigg\rangle_{\mathrm{sG}(\mu\rho|\varepsilon,\Omega)}\\
&\enspace=\int_{\Omega^{n+n'+m+m'+p}} d^{2n}x\,d^{2n'}y\,d^{2m}z\,d^{2m'}w\,d^{2p}u\prod_{j=1}^n f_j(x_j)\!\prod_{j'=1}^{n'}\widetilde f_{j'}(y_{j'})\!\prod_{l=1}^m(-\partial g_l(z_l))\!\prod_{l'=1}^{m'} (-\bar\partial\widetilde g_{l'}(w_{l'}))\\
&\qquad\qquad\qquad\qquad \times\prod_{k=1}^p \rho(u_k)\, C_\epsilon(x,\sigma,y,\tau,z,w,u).
\end{align*}
\end{itemize}
\end{lemma}
\begin{proof}
Let us first prove uniqueness. Write $\Delta_\epsilon$ for the difference of two such kernels. Arguing exactly as in the proof of Lemma~\ref{lem:Mtiluniq} (writing $\rho=\sum_{j=1}^p \alpha_j \rho_j$ and differentiating once with respect to each $\alpha_i$, one finds that 
\begin{align*}
\int \prod_{j=1}^n f_j(x_j)\prod_{j'=1}^{n'}\widetilde f_{j'}(y_{j'})\prod_{l=1}^m (-\partial g_l(z_l))\prod_{l'=1}^{m'}(-\bar \partial \widetilde g_{l'}(w_{l'}))\prod_{k=1}^p \rho_k(u_k)\Delta_\epsilon(x,y,z,w,u)=0.
\end{align*}
Integrating by parts in $z_1$ and using smoothness, we see that 
\begin{align*}
\partial_{z_1}\int \prod_{j=1}^n f_j(x_j)\prod_{j'=1}^{n'}\widetilde f_{j'}(y_{j'})\prod_{l=2}^m (-\partial g_l(z_l))\prod_{l'=1}^{m'}(-\bar \partial \widetilde g_{l'}(w_{l'}))\prod_{k=1}^p \rho_k(u_k)\Delta_\epsilon(x,y,z,w,u)=0,
\end{align*}
where we integrate over all other variables except $z_1$. Thus this integral is an antiholomorphic function that vanishes at the boundary and must vanish identically as a function of $z_1$. We can continue to argue like this to see that for every $z,w$
\begin{align*}
\int \prod_{j=1}^n f_j(x_j)\prod_{j'=1}^{n'}\widetilde f_{j'}(y_{j'})\prod_{k=1}^p \rho_k(u_k)\Delta_\epsilon(x,y,z,w,u)=0.
\end{align*}
Thus for each fixed $z,w$, $\Delta_\epsilon(x,y,z,w,u)=0$ in the sense of distributions and as we are dealing with a continuous function, it must vanish identically.

It remains to check that $C_\epsilon$ satisfies the conditions of the lemma. The fact that $C_\epsilon$ vanishes if one of the $z$ or $w$ variables tends to the boundary follows from the basic fact that $\varphi_\epsilon$ vanishes almost surely on the boundary (since $\p_\Omega(s,x,x)$ vanishes for $x\in \partial \Omega$).

Since we proved in Lemma~\ref{lem:sgtaylor} that $C_\epsilon$ is a kernel for the Taylor coefficient, it just remains to check that $C_\epsilon$ satisfies the symmetry condition. This follows readily from the fact that cumulants are permutation invariant and by making suitable changes of variables in the $S$ and $\pi$ sums in the definition of $C_\epsilon$.
\end{proof}
The point of this uniqueness result is that to identify the $\epsilon\to 0$ limit of the Taylor coefficient as an integral with a GFF correlation function as its kernel, it is sufficient to find some integrability bounds for $C_\epsilon$, and this we can do by representing and bounding the Taylor coefficient with the results of Lemmas~\ref{lem:cftopart}--\ref{lem:cttconv} (as well as some associated bounds). This reasoning allows us to conclude the proof of Theorem~\ref{th:sgmain}. 

\begin{proof}[Proof of Theorem~\ref{th:sgmain}, part (iii)]
As discussed above Lemma~\ref{lem:taylorgen}, we know that the Taylor coefficients of part (iii) of Theorem~\ref{th:sgmain} are given as the $\epsilon\to 0$ limits of the corresponding regularized Taylor coefficients. Thus it remains to prove that 
\begin{align*}
&\lim_{\epsilon\to 0}\frac{\partial^p}{\partial \mu^p}\bigg|_{\mu=0}\!\!\Bigg\langle\prod_{j=1}^n\!\opcolon e^{i\sigma_j\sqrt{\pi}\phie}\opcolon (f_j)\!\prod_{j'=1}^{n'}\!\!\Big(\!\opcolon e^{i\tau_{j'}\sqrt{\pi}\phie}\opcolon (\widetilde f_{j'}) \!-\! \mu c_\epsilon(\widetilde f_{j'},\rho)\!\Big)\!\prod_{l=1}^m\!\partial \phie(g_l)\!\prod_{l'=1}^{m'}\!\bar\partial \phie(\widetilde g_{l'})\!\!\Bigg\rangle_{\mathrm{sG}(\mu\rho|\varepsilon,\Omega)}\\
&\enspace=\int_{\Omega^{n+n'+m+m'+p}} d^{2n}x\,d^{2n'}y\,d^{2m}z\,d^{2m'}w\,d^{2p}u\prod_{j=1}^n f_j(x_j)\!\prod_{j'=1}^{n'}\widetilde f_{j'}(y_{j'})\!\prod_{l=1}^m(-\partial g_l(z_l))\!\prod_{l'=1}^{m'} (-\bar\partial\widetilde g_{l'}(w_{l'}))\\
&\qquad \times \prod_{k=1}^p \rho(u_k)\sum_{S\subset [n']}\sum_{\substack{S'\subset[p]:\\ |S|=|S'|}} \sum_{\pi:S\leftrightarrow S'}\prod_{s\in S}(-\mathcal A(y_{s},u_{\pi(s)}))\Bigg\langle \prod_{j=1}^n \opcolon e^{i\sigma_j \sqrt{\pi}\varphi(x_j)}\opcolon  \prod_{l=1}^m \varphi(z_l)\prod_{l'=1}^{m'}\varphi(w_{l'})\\
&\qquad\qquad \prod_{k'\notin S}\big(\opcolon e^{i\tau_{k'}\sqrt{\pi}\varphi(y_{k'})}\opcolon\big) ; \big(\opcolon \cos(\sqrt{4\pi}\varphi(u_{\alpha_1}))\opcolon\big) ;\dots;\big(\opcolon \cos(\sqrt{4\pi}\varphi(u_{\alpha_{p-|S'|}}))\opcolon\big) \Bigg\rangle_{\mathrm{GFF}}^\mathsf T,
\end{align*}
where $A(x,y)=\lim_{\epsilon\to 0}A(x,y|\epsilon)$ and the limiting cumulants are defined from the limiting moments through the moments to cumulants map (and Lemma~\ref{lem:gffcorr2}). Part of the statement is that this is an integrable function.

To do this, we note that first of all, we know from Lemma~\ref{lem:gffcorr2} and Lemma~\ref{lem:sgtaylor} that our integrand converges pointwise almost everywhere to the limiting integrand. Thus our only task is to justify interchanging the order of the limit and the integral -- by dominated convergence. To do this, we write the integrand in a different way.

By Lemma~\ref{lem:cftopart}, we can write our Taylor coefficient as 
\begin{multline*}
\frac{\partial^p}{\partial \mu^p}\bigg|_{\mu=0}\prod_{j=1}^n \partial_{s_j}|_{s_j=0}\prod_{j'=1}^{n'}\partial_{t_{j'}}|_{t_{j'}=0}\prod_{l=1}^m \partial_{u_l}|_{u_l=0} \prod_{l'=1}^{m'}\partial_{v_{l'}}|_{v_{l'}=0}\bigg[e^{\frac{1}{2}\sum_{l,k=1}^m u_l u_k \gffe{\partial \phie(g_l)\partial \phie(g_k)}}\\
\shoveleft\qquad\times e^{\frac{1}{2}\sum_{l',k'=1}^{m'}v_{l'}v_{k'}\gffe{\bar\partial \phie(\widetilde g_{l'})\bar\partial \phie(\widetilde g_{k'})}+\sum_{l=1}^m \sum_{l'=1}^{m'}u_l v_{l'}\gffe{\partial \phie(g_l)\bar\partial \phie(\widetilde g_{l'})}}\\
\times \frac{\Z(\eta_\epsilon|\epsilon)}{\Z(-\frac{\mu}{2}\rho|\epsilon)}e^{\frac{1}{2}\int_{(\Omega\times \{\pm 2\})^2}d\xi_1\,d\xi_2\big(\eta_\epsilon(\xi_1)\eta_\epsilon(\xi_2)-\frac{\mu^2}{4}\rho(x_1)\rho(x_2)\big)A(x_1,x_2|\epsilon)} e^{-\sum_{j'=1}^{n'}t_{j'}\mu c_\epsilon(\widetilde f_{j'},\rho)}\bigg]
\end{multline*}
with the same choice of $\eta_\epsilon$ as in Lemma~\ref{lem:cftopart}.

By carrying out all of the derivatives, we find a representation for the Taylor coefficient as an integral 
\begin{multline*}
\int_{\Omega^{n+n'+m+m'+p}} d^{2n}x\,d^{2n'}y\,d^{2m}z\,d^{2m'}w\,d^{2p}u\prod_{j=1}^n f_j(x_j)\prod_{j'=1}^{n'}\widetilde f_{j'}(y_{j'})\prod_{l=1}^m(-\partial g_l(z_l))\prod_{l'=1}^{m'} (-\bar\partial\widetilde g_{l'}(w_{l'}))\\
\times \prod_{k=1}^p \rho(u_k)\,\widetilde C_\epsilon(x,\sigma,y,\tau,z,w,u)
\end{multline*}
where $\widetilde C_\epsilon$ involves products of terms like $\langle \varphi_\epsilon(z_l)\varphi_\epsilon(z_k)\rangle_{\mathrm{GFF}(\epsilon)}$, $\widetilde \M(\xi_1,\dots,\xi_j|\epsilon)$ from \eqref{eq:mexp} (possibly multiplied by factors of continuous functions due to $\eta_\epsilon$) and quantities such as $A(x,y|\epsilon) \allowbreak\times(e^{i(H_\epsilon(x)-H_\epsilon}-1)$ for some $H_\epsilon$ which is Lipschitz continuous uniformly in $\epsilon$. All of these types of quantities are uniformly (in $\epsilon$) integrable separately (with similar arguments as in the proofs of part (i) of Theorem~\ref{th:sgmain} and the proof of Theorem~\ref{thm:uc}) and they do not share any of the integration variables. This means that we have some kernel for the Taylor coefficient that is uniformly integrable. We can then symmetrize this kernel in the $u$-variables (without affecting uniform integrability) and using Lemma~\ref{lem:symend}, we see that $C_\epsilon$ is also uniformly integrable. This justifies the use of the dominated convergence theorem and concludes the proof.
\end{proof}

\section{Basic Correlations of the Ising Model}\label{sec:ising-analysis}

Let $\alpha$ be a real-valued function. Recall that we work with continuously differentiable complex-valued solutions $f$ of the following $\R$-linear equation
\begin{equation}\label{eq:massive-holomorphicity}
	\bar\partial_z f = -i\alpha\bar f,
\end{equation}
in the weak sense. We call a function $f$ satisfying \eqref{eq:massive-holomorphicity} \emph{($\alpha$)massive holomorphic}\footnote{This is more restrictive than the traditional \emph{generalized analytic} or \emph{pseudo-holomorphic} terminology \cite{Bers, Vekua}, referring to general cases where $\alpha$ is not restricted to be real, or has otherwise prescribed phases.}. We also consider \emph{spinors}, i.e., functions with prescribed $-1$ multiplicative monodromies around isolated points. Even so, the coefficient $\alpha$ is always single-valued.

For notational simplicity, we denote any constant only depending on quantities $q_1, q_2, \ldots$ by $C(q_1, q_2, \ldots)$, with the understanding that they are not all the same. We will continue to use the big-O notation for estimates for which no such uniformity is assumed.

We reserve analysis specific to correlations to Section \ref{sec:estimates-section}. For now, we recall how the regularity of $\partial_{\bar z}$ data translates to that of the solution in the form of the following standard result:

\begin{proposition}\label{prop:dzbar-regularity}
	Let a function $g$ be defined in $B_r = B_r(0)$ such that $\left\Vert g \right\Vert_{L^q(B_r)} < \infty $ for some $q \geq 2$. Then its \emph{Cauchy transform} 
    \begin{equation}\label{eq:cauchytransform}[\mathcal{C} g](z) :=[\mathcal{C}_{B_r} g](z) =  \frac{1}{\pi}\int_{B_r}\frac{g(w)}{z-w}d^2w 
    \end{equation}
    has the weak derivative $\partial_{\bar{z}}[\mathcal{C} g] = g$. Furthermore $\mathcal{C} g$ and its weak $\partial_z$-derivative $\mathcal{S} g$ (the \emph{Beurling transform} of $g$) satisfy
	\begin{equation}\label{eq:dzbar-bound1}
		\left\Vert \mathcal{C} g \right\Vert_{L^q(B_r)} \leq 6r\left\Vert g \right\Vert_{L^q(B_r)}, \quad \text{and} \quad 
        \left\Vert \mathcal{S} g \right\Vert_{L^q(B_r)} \leq \mathbf{S}_q \left\Vert g \right\Vert_{L^q(B_r)},
	\end{equation}
    for some constant $\mathbf{S}_q$. If $q>2$, we have the H\"older seminorm bound
    \begin{align}\label{eq:dzbar-holder}
    \left\Vert \mathcal{C} g \right\Vert_{C^{1-\frac{2}{q}}(B_r)} : = \sup_{x,y\in B_r,x\neq y} \frac{|\mathcal{C} g(x)-\mathcal{C} g(y)|}{|x-y|^{1-\frac2q}}\leq C(q)\left\Vert g \right\Vert_{L^q(B_r)}.
    \end{align}
	
    If in addition $g$ is locally $\beta$-H\"older continuous, 
    then $\mathcal{C}g$ is continuously differentiable with derivatives bounded pointwise in the smaller disk $B_{r/4}$, with estimate
	\begin{equation}\label{eq:dzbar-bound2}
		 \left\Vert g \right\Vert_{L^\infty(B_{r/4})} + \left\Vert \mathcal{S} g \right\Vert_{L^\infty(B_{r/4})} \leq C({\beta}) \left(\left\Vert  g \right\Vert_{L^\infty(B_{r})} +  r^\beta \left\Vert g \right\Vert_{C^\beta(B_r)} \right).
	\end{equation}
\end{proposition}
\begin{proof}
	The $L^q$ and $C^\alpha$-estimates are exactly \cite[(4.16), Theorems 4.3.12, 4.3.13, 4.5.3]{AIM}. Differentiability follows from \cite[Theorem 4.7.1]{AIM}.
	
	For the pointwise bound, we need to be mindful of the discontinuity at the boundary. Fix a bump function $\rho$ which is identically $1$ in $B_{3r/4}$, $0$ outside of $B_r$, and differentiable with $|\nabla \rho| \leq \frac{16}{r}$. Then, \cite[(4.123)]{AIM} yields
	\begin{equation*}
		\max_{B_{r/4}} |\mathcal{S}  g| - \min_{B_{r/4}}|\mathcal{S}  g| \leq r^\beta\left\Vert \mathcal{S} [\rho g]\right\Vert_{C^\beta(B_{r})} \leq C({ \beta}) \left({r^{\beta}}\left\Vert g \right\Vert_{C^\beta(B_r)}+ {\left\Vert g \right\Vert_{L^\infty(B_r)}}\right).
	\end{equation*}
	On the other hand, by \eqref{eq:dzbar-bound1},
    $$ \min_{B_{r/4}}\left|\mathcal{S}  g\right|\leq Cr^{-1}\left\Vert \mathcal{S} g \right\Vert_{L^2(B_r)} \leq Cr^{-1}\left\Vert g \right\Vert_{L^2(B_r)} \leq C\left\Vert g \right\Vert_{L^\infty(B_r)}$$
    Then it is straightforward to derive \eqref{eq:dzbar-bound2}.
\end{proof}
\begin{corollary}\label{cor:dzbar-solution-bound}
    Suppose $\partial_{\bar{z}}g_1 = g_2$ almost everywhere in $B_r$ for (single-valued) functions $g_1, g_2$ such that $g_1\in L^2(B_r)$ and $g_2\in L^q(B_r)$ for some $q > 2$. Then
    \begin{align*}
    \Vert g_1 \Vert_{L^\infty(B_{r/2})} &\leq \Big(C(q)r^{1-\frac{2}{q}}\Vert g_2 \Vert_{L^q(B_r)}+C{\Vert g_1\Vert_{L^\infty(\partial B_{3r/4})}}\Big),\qquad\text{and}\\
    \Vert g_1 \Vert_{C^{1-\frac2q}(B_{r/2})}&\leq \Big(C(q) \Vert g_2 \Vert_{L^q(B_r)}+Cr^{-\left(1-\frac{2}{q}\right)}{\Vert g_1\Vert_{L^\infty(\partial B_{3r/4})}}\Big).
    \end{align*}
\end{corollary}
\begin{proof}
Consider $\mathcal{C}g_2^\dagger : = \mathcal{C}g_2 - \mathcal{C}g_2(0)$, which is H\"older continuous by \eqref{eq:dzbar-holder} and in fact $|\mathcal{C}g_2^\dagger|\leq r^{1-\frac{2}{q}}\|\mathcal C g_2\|_{C^{1-\frac{2}{q}}(B_r)}$ in $B_r$. Then the holomorphic function $g_1 -\mathcal{C}g_2^\dagger$ may be estimated separately using Cauchy's formula: showing the second bound without loss of generality,
\begin{align*}
g_1(z)-g_1(z')=\mathcal C g_2^\dagger(z)-\mathcal Cg_2^\dagger(z')+\oint_{\partial B_{3r/4}}\frac{dw}{2\pi i} \frac{z-z'}{(w-z)(w-z')}(g_1(w)- \mathcal Cg_2^\dagger (w)),
\end{align*}
so we have for $z,z'\in B_{r/2}$
\[
\frac{|g_1(z)-g_1(z')|}{|z-z'|^{1-\frac{2}{q}}}\leq \|\mathcal C g_2\|_{C^{1-\frac{2}{q}}(B_r)}+\frac{16|z-z'|^{\frac{2}{q}}}{r}\Big(\|g_1\|_{L^\infty(\partial B_{3r/4})}+r^{1-\frac{2}{q}}\|\mathcal C g_2\|_{C^{1-\frac{2}{q}}(B_r)}\Big).
\]
By Proposition \ref{prop:dzbar-regularity}, 
\[
\|g_1\|_{C^{1-\frac{2}{q}}(B_{r/2})}\leq C(q)\|g_2\|_{L^q(B_r)}+C r^{-(1-\frac{2}{q})}\|g_1\|_{L^\infty(\partial B_{3r/4})}.
\]
\end{proof}

\begin{remark}\label{rem:dzbar-massive-holomorphicity}
    We note two applications of the above theorems directly in the massive holomorphic setup:
    \begin{itemize}
        \item A locally bounded function $f$ satisfying \eqref{eq:massive-holomorphicity} with locally smooth data $\alpha$ is continuously differentiable: this follows from bootstrapping local boundedness to local H\"older continuity, then differentiability, as above. A simple corollary is that a locally bounded function $f$ satisfying \eqref{eq:massive-holomorphicity} classically in a domain except at some isolated points in fact satisfies it at those points as well. 
        \item Moreover, a family $\{f_\iota \}_{\iota \in I}$ of such functions satisfying $|f_\iota| \leq C$ in a domain $\Omega$ is equicontinuous in any compact subset, and therefore by Arzel\`a-Ascoli has a subsequence converging uniformly on compact subsets (in particular, the limit is $\alpha$-massive holomorphic). We may diagonalize to get convergence in all of $\Omega$.
    \end{itemize}
\end{remark}

\subsection{Existence and Uniqueness}
\subsubsection{Definitions of the correlations}
  As mathematical objects, the Ising correlations to be constructed are functions of the position of each inserted field, which are putative limits of suitably renormalized correlations in the corresponding discrete massive model. The $-1$ multiplicative monodromy of fermions around spins and disorders requires some care: we will first fix the positions of spins and disorder insertions $a_1, \ldots, a_n$ in the original domain $\Omega$ then consider the correlations as functions of the fermion insertion(s) $z_j$ which is multiplied by $-1$ whenever $z_j$ moves around any $a_{j'}$. In other words, they are single valued functions as long as $z_j$ are taken on the \emph{double cover} of $\Omega\setminus\{a_1, \ldots, a_n\}$.

  In addition, the sign of the correlation is uniquely determined as soon as the branch of individual square roots $\sqrt{z-a_j}$ are fixed near $a_j$ (i.e. $\sqrt{z-a_j}$ has been defined on the lift of a punctured neighborhood of $a_j$ in $\Omega$ to the double cover).

  We first define a distinguished set of correlations called \emph{basic correlations}.
  
\begin{definition}\label{def:correlations-definition}
	Let $\Omega\subset \mathbb C$ be a bounded simply connected domain and $\alpha$ a real-valued function on $\Omega$ (see Remark \ref{rem:regularity} for regularity assumptions). We define the \emph{basic  correlations of the $\alpha$-massive Ising model on $\Omega$ with plus boundary condition} as follows.
	
	For distinct points $a_1, \ldots, a_n \in \Omega$ and $z, w \in \Omega\setminus \{a_1, a_2, \ldots, a_n\}$, we consider correlations involving spins $\sigma$, disorder $\mu$, and fermion $\psi$, specifically any of the following combinations:
	\begin{itemize}
		\item (disorder-fermion case) where $\hat{k}$ signifies the absence of the $k$-th spin, $$f(z) :=\frac{\langle  \sigma_{a_1}\stackrel{\hat{k}}{\cdots}\sigma_{a_n}\mu_{a_k}\psi_z\rangle_{\alpha;\Omega}}{\langle \sigma_{a_1}\cdots \sigma_{a_n}\rangle_{\alpha;\Omega}}=\frac{\langle \sigma_{a_1}\stackrel{\hat{k}}{\cdots}\sigma_{a_n}\mu_{a_k}\psi_z\rangle_{\alpha}}{\langle \sigma_{a_1}\cdots \sigma_{a_n}\rangle_{\alpha}},$$ or
		\item (fermion-fermion case) where $\eta$ is any complex constant, $$f(z) :=\frac{\langle  \sigma_{a_1}{\cdots}\sigma_{a_n}\psi_z\psi^{[\eta]}_w\rangle_{\alpha;\Omega}}{\langle \sigma_{a_1}\cdots \sigma_{a_n}\rangle_{\alpha;\Omega}}=\frac{\langle  \sigma_{a_1}{\cdots}\sigma_{a_n}\psi_z\psi^{[\eta]}_w\rangle_{\alpha}}{\langle \sigma_{a_1}\cdots \sigma_{a_n}\rangle_{\alpha}}.$$
	\end{itemize}
	
	Then the correlation $f(z)$, as a function of the fermion insertion point $z$, is defined as the unique spinor having $-1$ multiplicative monodromy around each of $a_1, a_2, \ldots, a_n$ (i.e. $f$ is a single-valued function on the double cover of $\Omega\setminus \{a_1, a_2, \ldots, a_n\}$), massive holomorphic in $\Omega$ away from other insertions, having the asymptotics:
	\begin{itemize}
		\item[\emph{(S)}] near a spin insertion at $a\in \Omega$, it has a \emph{spin-type asymptotic}, i.e., there is $b = b(a) = b(a; f) \in \R$ such that $f(z) =b e^{i\pi/4}(z-a)^{-1/2} + o(|z-a|^{-1/2})$;
		\item[\emph{(D)}] near a disorder insertion at $a'\in \Omega$, it has a \emph{disorder-type asymptotic}, i.e., $f(z) = -e^{-i\pi/4}(z-a')^{-1/2} + o(|z-a'|^{-1/2})$;
		\item[\emph{(F)}] near a fermion insertion at $w\in \Omega$, it has a \emph{fermion-type asymptotic}, i.e., $f(z) = \bar\eta (z-w)^{-1} +  o(|z-w|^{-1})$;
	\end{itemize}
	and extending continuously to each $\zeta \in \partial \Omega$ where it satisfies the boundary condition
	\begin{equation}
		\overline{f(\zeta)} = \tau_\zeta f(\zeta),\mbox{ or equivalently }f^2(\zeta) \in \tau_\zeta^{-1}\R_{\geq0}, \tag{RH}
	\end{equation}
	where $\tau_\zeta$ is the unit complex number in the direction of the tangent at $\zeta$ to $\partial\Omega$, oriented counterclockwise.
\end{definition}
\noindent The asymptotics (D), (S), and (F) are the same as expected from the following operator product expansions (OPEs): as $z\to a, w$, we may replace within correlations
\begin{equation}\label{eq:ope}
	\sigma_a\psi_z \sim e^{i\pi/4}\mu_a(z-a)^{-1/2};\quad \mu_a\psi_z \sim -e^{-i\pi/4}\sigma_a(z-a)^{-1/2};\quad \psi_z\psi_w^{[\eta]} \sim \bar{\eta}(z-w)^{-1},
\end{equation}
which is identical to the critical case \cite{CHI2}; the mass affects the correction terms, see \cite{CIP}.
\begin{remark}\label{rem:confcov}
	The defining properties of the basic correlations are compatible with conformal covariance, as physically expected. Since $\varphi'(\zeta)\tau_\zeta = |\varphi'(\zeta)|\tau_{\varphi(\zeta)}$, for any $f$ satisfying (RH) on $\partial\Omega$ and a conformal map $\varphi : \Omega' \to \Omega$, the pullback $\varphi^* f:= (f \circ \varphi)\cdot (\varphi')^{1/2}$ also satisfies (RH) on $\partial\Omega'$. Going further, it is straightforward to verify that the asymptotics (S), (D), and (F) show covariance which is consistent with the physical transformation rules:
	\begin{equation*}
		\sigma_{\Omega'} \mapsto (\sigma_{\Omega} \circ \varphi) \cdot |\varphi'|^{1/8}, \quad 	\mu_{\Omega'} \mapsto (\mu_{\Omega} \circ \varphi) \cdot |\varphi'|^{1/8}, \quad \psi_{\Omega'} \mapsto (\psi_{\Omega} \circ \varphi) \cdot (\varphi')^{1/2}, \quad \psi_{\Omega'}^{[\eta]} \mapsto \psi_{\Omega}^{[\eta {(\varphi')^{1/2}}]} \circ \varphi.
	\end{equation*}
	Under such transformations, the mass itself transforms covariantly:
	\begin{equation*}
		\partial_{\bar z}\left[ \varphi^*f \right] = \left((\partial_{\bar z} f) \circ \varphi\right)\cdot \overline{\varphi'}\sqrt{\varphi'} =\left[ -i(\alpha \circ \varphi)|{\varphi'}|\right] \overline{\varphi^*f}\text{, or }\alpha \mapsto (\alpha\circ\varphi)|\varphi'|,
	\end{equation*}
	i.e., $\varphi^* f$ is $(\alpha\circ\varphi)|\varphi'|$-massive holomorphic. That is, the basic correlations (and all correlations, as we highlight below) remain unchanged under conformal domain transformations if we substitute all terms within correlations according to the above rules.
\end{remark}
	
	\begin{remark}\label{rem:regularity}
	For the full set of Ising correlations, we only need to consider $\alpha$ which is smooth up to the boundary of a smooth domain $\Omega$.
    
    Nonetheless, we may equivalently define the basic correlations in the case of less smooth $\Omega$ using the pull-back on $\D$ and the conformal covariance rules in Remark \ref{rem:confcov}; see also \eqref{eq:sfconfcov} and Remark \ref{rem:h-properties} for more covariance statements. While less straightforward, the advantage of this approach is that we may define (RH) on cases where $\partial \Omega$ does not admit tangents. This is helpful since the basic correlations also serve as concrete massive holomorphic functions or spinors which happen to be basic ingredients in the relevant analysis.
    
    In fact, for basic correlations, we only assume that $\alpha$ is locally smooth and continuous up to the boundary of a domain $\Omega$ with piecewise smooth and Lipschitz boundary, stating any necessary additional regularity in the individual results (see the beginning of Section \ref{sec:estimates-section}).
	\end{remark}

In subsequent sections, we show existence and uniqueness of these basic correlations, as well as their properties and estimates. Specifically:
\begin{itemize}
		\item We show any correlation satisfying the properties outlined above exists and is unique in Section \ref{sec:fermion-existence}.
		\item We give the $L^p$-estimate, crucial in the series expansion process, in Corollary \ref{cor:Lp}.
		\item We show the correlations are smooth with respect to local movements of the insertions in Proposition \ref{prop:spindiff}.
\end{itemize}

The real fermions and complex fermions are related as follows: within correlations,
\begin{equation}\label{eq:fermion-complex}
	\psi^{[\eta]} = \frac12 (\bar{\eta}\psi + \eta \psi^*)\text{ is real, equivalently }\psi =\bar\eta^{-1} \psi^{[\eta]} +i\bar\eta^{-1} \psi^{[i\eta]} \text{ and }\overline{\psi} = \psi^*.
\end{equation}
That is, under complex conjugation of any correlation, complex fermions $\psi, \psi^*$ turn into each other whereas other fields (spin, disorder, energy, and real fermion) stay the same. Any two insertions among fermions and/or disorders anticommute among themselves; otherwise the correlations do not depend on the order of the insertions.

To insert more fermions, we use a Pfaffian decomposition \eqref{eq:pfaffian}, which may be taken either as a definition or a consequence of requiring anticommutation of fermions and the OPE behavior \eqref{eq:ope} whenever two fermion insertions are merged: see Proposition \ref{prop:fermion-antisymmetry}. To insert more disorders, we insert spins at those points, insert as many fermions, and take fermions to spins and use \eqref{eq:ope}. See Proposition \ref{prop:well-def} for compatibility of such a procedure with the above definitions.

Then, the only remaining nontrivial construction at this stage, apart from energy which will be introduced in Section \ref{subsec:energy}, is that of pure spin correlations (for this, we do assume $\alpha$ is smooth up to the boundary). We will show in Lemma \ref{lem:3/2-decomp} (the asymptotic for $f^{\dagger1}$ with $\nu = -1/2$, $o_{-1/2}=1$) that, in fact, the asymptotic (D) in Definition \ref{def:correlations-definition} may be expanded further:
\begin{equation}\label{eq:def-A}
	\frac{\langle\sigma_{a_1}\stackrel{\hat{k}}{\cdots}\sigma_{a_n} \mu_{a_k} \psi_z\rangle_{\alpha}}{\langle \sigma_{a_1}\cdots \sigma_{a_n}\rangle_{\alpha}} = -\frac{e^{-i\pi/4}e^{2\alpha(a_k)|z-a_k|}}{\sqrt{z-a_k}}\left( 1+2\mathcal{A}_\Omega^{(\alpha)}(a_k)(z-a_k) + o(|z-a_k|) \right),
\end{equation}
which also defines the coefficient $\mathcal{A}_\Omega^{(\alpha)}(a_k)$ for the form $\Re\left[\mathcal{A}_\Omega^{(\alpha)}(a_k)da_k\right]$, which is a closed and conformally covariant differential form by Corollary \ref{cor:defA}. Then we recover the massive pure spin correlations by,
\begin{equation}\label{eq:spin-integral}
	 \frac{ \langle \sigma_{a_1}\cdots \sigma_{a_n}\rangle_{\alpha}}{\langle \sigma_{a_1}\cdots \sigma_{a_n}\rangle_{0}} = \frac{ \langle \sigma_{a_2}\cdots \sigma_{a_n}\rangle_{\alpha}}{\langle \sigma_{a_2}\cdots \sigma_{a_n}\rangle_{0}}\exp \bigg\{\Re \bigg[\int_{a_b\in\partial \Omega}^{a_1} \Big(\mathcal{A}_\Omega^{(\alpha)}(a_1') - \mathcal{A}_\Omega^{(0)}(a_1')\Big)da_1'\bigg] \bigg\},
\end{equation}
where ${\langle \sigma_{a_1}\cdots \sigma_{a_n}\rangle_{0}}$ is the scaling limit of the critical Ising spin correlations. We show in Section \ref{sec:well-def-spin} that this ratio is well-defined, i.e. the result is independent of how we do the integration. We state well-definedness of the massive spin correlation $\langle \sigma_{a_1}\cdots \sigma_{a_n}\rangle_{\alpha}$ in Proposition \ref{prop:spin}.

\subsubsection{Analysis of massive holomorphic functions}
A central fact about the local and global properties of massive holomorphic functions, going back to Bers and Vekua \cite{Bers, Vekua}, is the following \emph{similarity principle}.
\begin{theorem}\label{thm:similarity-principle}
	Suppose $f$ satisfies \eqref{eq:massive-holomorphicity}, possibly with isolated points of singularity or monodromy, in $\D$ with $\alpha \in L^q(\D)$ for some $q \geq 2$. Then there exists a (single-valued) function $s$ such that $\underline{f} := e^{-s}f$ is holomorphic at all points where $f$ is regular. Conversely, given a holomorphic $\underline{f}$, possibly with isolated points of singularity or monodromy, there exists an $s$ such that $f = e^s \underline{f}$ satisfies \eqref{eq:massive-holomorphicity} on all points where $f$ is regular.
	
	In both cases, $s$ is $\gamma_q = \left(1-\frac{2}{q}\right)$-H\"older continuous on $\overline \D$ if $q>2$, and $s$ is unique, as long as we require $\Im s \equiv 0$ on $\partial \D$ and $\Re s(z_0) = 0$ at some point $z_0 \in \D$. We have the following quantitative bounds
	\begin{equation}\label{eq:similarity-bound}
		\Vert s \Vert_{W^{1,q}(\D)} \leq C \Vert \alpha \Vert_{L^q(\D)} \qquad \text{ and}, \text{ if } q>2\text{, }\quad \Vert s \Vert_{C^{\gamma_q}(\D)} \leq C \Vert \alpha  \Vert_{L^q(\D)}.
	\end{equation}
	Under this correspondence, $\underline{f}$ is called the \emph{holomorphic part} of $f$.
\end{theorem}
\begin{proof}
	See, e.g., \cite[Lemma 3.1, Theorem 4.1]{BBC}. Note carefully that the proofs of these results only concern the ratios (in their notation) $\bar{w}/w, \bar{F}/F$, which are single-valued and of unit norm even if $w, F$ are double-valued or have a singularity: these results apply to our case as well. The H\"older estimate comes directly from Morrey's inequality.
\end{proof}

We define notions such as the order of zeros and poles and meromorphicity for $f$ identical to those of $\underline{f}$. Since we may pull-back massive holomorphic functions on any simply connected domain to $\D$ as in Remark \ref{rem:confcov}, we also define the notion of \emph{holomorphic part} on general $\Omega$. It is natural to transform the holomorphic part covariantly, which means $s$ is invariant:
\begin{equation}\label{eq:sfconfcov}
	s \mapsto s\circ \varphi; \qquad \text{ and } \qquad \underline{f} \mapsto (\underline f \circ \varphi)\cdot (\varphi')^{1/2}.
\end{equation}

\begin{remark}\label{rem:hol-phase}
	We emphasize already that the holomorphic part coming from a massive fermion correlation $ \left< \mathcal{O} \psi_z \right>_{\alpha}$, etc., is typically \textbf{not}{ the corresponding critical correlation} $ \left< \mathcal{O}\psi_z \right>_{0}$, etc., even up to a real scalar multiple; the reason is that the delicately prescribed phases in Definition \ref{def:correlations-definition} are disrupted by $e^s$. Nevertheless, see also Lemma \ref{lem:onespin}.
\end{remark}

\begin{remark}\label{rem:singularity-order}
	The H\"older estimate for the factor $e^s$ implies that, around any pole or regular point $a \in \C$, there exists a unique $c\in \C\setminus \{0\}$ and $n \in \mathbb Z$ such that
	\begin{equation}\label{eq:mhol-local-expansion}
		f(z) = c(z-a)^n + O(|z-a|^{n+\epsilon})\mbox{ as }z\to a,
	\end{equation}
	for some $\epsilon \in (0,1)$. Still, note that if $f_1, f_2$ are two solutions both satisfying \eqref{eq:mhol-local-expansion} with common $c, n$, their difference is a solution which satisfies as $z\to a$
	\begin{equation}\label{eq:mhol-difference}
		f_1(z) - f_2(z) = O(|z-a|^{n+\epsilon})\mbox{, and thus }f_1(z) - f_2(z) \sim c'(z-a)^{n+1} \mbox{ for some }c' \in \C.
	\end{equation}
	This is the basis for the so-called \emph{formal power series expansion}, cf. \cite{Bers, Vekua}.
	
	Similarly, if $f$ branches around $a$, so does $\underline{f}$ and it may be expressed locally in terms of half-integer powers of $(z-a)$. If there are a finite number of negative powers (as there will be in this article), $f$ satisfies above asymptotics with the integer $n$ replaced by a half-integer $\nu$.
\end{remark}

Using a version of Green's theorem in complex variables (Green-Riemann's theorem) on $f^2$, we see that, around any closed contour $\gamma$ and its interior $D$ where $f$ satisfies \eqref{eq:massive-holomorphicity},
\begin{equation}\label{eq:green-riemann}
	\oint_\gamma f^2 dz = 2i \int_D \bar\partial_z f^2(z) d^2z = 4 \int_D \alpha(z) |f(z)|^2  d^2z \in \R.
\end{equation}
This suggests that locally we may define a function $h(z)= \Im [\int_{z_0}^z f^2 dz]$ henceforth called a \emph{square integral}, where the line integral is done along any contour to $z$ from an arbitrary fixed point $z_0$. The choice of $z_0$ amounts to a choice of a global additive constant, and the resulting integral is invariant under deformations of the integration contour that do not cross a singularity.

Note also that \eqref{eq:green-riemann} should always be applied on \emph{planar} (i.e. not on double cover) subsets $D \subset \Omega$, since $f^2$ is always single-valued. In other words, the definition $h(z) = \Im [\int_{z_0}^{z} f^2 dz] $ does not depend on the branch choice of $f$.

The following comparison principle for $h = \Im [\int f^2 dz] $ is standard for quasi-linear elliptic PDEs such as one obeyed by $h$ (see Remark \ref{rem:h-properties}). We give a `constructive' proof which clearly shows that it is valid as long as the two functions have identical singularities.
\begin{lemma}\label{lem:maximum}
	Suppose $f_1$ and $f_2$ satisfy \eqref{eq:massive-holomorphicity} on a punctured neighborhood of $a$, possibly both branching at $a$ but in any case such that $h_1 = \Im [\int f_1^2 dz]$ and $h_2 = \Im [\int f_2^2 dz]$ are single-valued in the punctured neighborhood. Assume in addition that $f_1^2 - f_2^2$ extends continuously to $a$.
	
	Then $h_1 - h_2$ extends continuously to $a$. If $h_1 - h_2$ attains a local extremum at $a$, then $h_1 -h_2$ is a constant or, equivalently, $f_1^2 \equiv f_2^2$.
\end{lemma}
\begin{proof}
Assume $f_1-f_2$ and $f_1+f_2$ are not identically zero (if either is, $h_1 - h_2$ extends to $a$ trivially). Applying \eqref{eq:mhol-local-expansion}, we may expand near $a$,
\begin{align*}
f_1(z) - f_2(z) &= c_1 (z-a)^{r_1} + O(|z-a|^{r_1 + \epsilon})\quad\text{and}\\
f_1(z) + f_2(z) &= c_2 (z-a)^{r_2} + O(|z-a|^{r_2 + \epsilon}),
\end{align*}
where $c_1, c_2\neq 0$ and, say, $r_1 \geq r_2$.
Since $(f_1^2 - f_2^2)(z) = c_1c_2 (z-a)^{r_1+r_2} + O(|z-a|^{r_1 +r_2 + \epsilon})$ extends to $a$ continuously, we must have $r_1 + r_2 \geq 0$.

Then it is clear that $h_1 - h_2$ extends continuously to $a$: integrate $f_1^2 - f_2^2$ along any line to get the value of $(h_1 - h_2)(a)$, which is well-defined since the oscillation of $h_1 - h_2$ on small circles around $a$ must be small.

Moreover, $a$ cannot be a local extremum. To see this, for simplicity assume $a = 0$ and $c_1c_2 \in \R$ (modifying $\theta_0$ below if necessary), then it is easy to see that for small $t>0$, we have $$(h_1-h_2)(e^{i\theta_0}t) - (h_1 - h_2)(0) = \Im \bigg[\int_0^t e^{i(r_1+r_2+1)\theta_0}x^{r_1+r_2}\big(c_1c_2 + O(t^{\epsilon})\big) dx\bigg],$$ such that, for $t \ll (c_1c_2)^{1/\epsilon}$, the values of $h_1-h_2$ at $z=e^{i\theta_0}t$ are respectively strictly greater or smaller than its value at $z=0$ for $\theta_0 = \frac{\pi}{2(r_1+r_2+1)}, -\frac{\pi}{2(r_1+r_2+1)}$.
\end{proof}
\begin{lemma}\label{lem:uniqueness} Let $f$ satisfy \eqref{eq:massive-holomorphicity} on the double cover of $\Omega \setminus \{ a_1, a_2, \ldots, a_n \}$. Suppose the boundary condition (RH) holds and the asympotic of $f$ is of the type (S) from Definition \ref{def:correlations-definition} near each $a_j$, if any. Then $f \equiv 0$.
\end{lemma}
\begin{proof}
Let $b_j = b(a_j) \in \mathbb R$. Consider $h = \Im [\int f^2 dz]$. Since
\begin{equation*}
f^2(z) = ib_j^2(z-a_j)^{-1} + O(|z-a_j|^{-1 + \epsilon}),
\end{equation*}
we see that $h(z) = b_j^2 \log |z-a_j| + O(1)$ near $z=a_j$.

Integrating $f^2 dz$ counterclockwise along $\partial \Omega$ (i.e. in the direction of $\tau_z$), the boundary condition (RH) implies that $h$ is constant along $\partial \Omega$, so let us fix the additive constant such that $h\equiv 0$ on $\partial \Omega$. In addition, if any $f(\zeta) \neq 0$ on $\zeta \in \partial \Omega$, integrating inwards from a boundary point $\zeta \in \partial \Omega$ perpendicular to $\partial \Omega$ (i.e. in the direction of $i\tau_\zeta$), the same boundary condition (RH) would imply that $h>0$ at points near $\zeta$.

However, by Lemma \ref{lem:maximum}, $h$ can only have a local extremum at one of the $a_j$ with $b_j \neq 0$, where $h \to -\infty$. Therefore, $\max_{\Omega \setminus \{a_1 \ldots, a_n \}} h$ has to be attained on $\partial \Omega$, where $h\equiv 0$.  Therefore, $f \equiv 0$ at all points on $\partial \Omega$. Then the holomorphic part $\underline{f} \equiv 0$ on $\partial \Omega$ as well. So, the meromorphic function $\underline{f}^2$ has to be identically zero since it vanishes on a boundary segment of non-zero length, and therefore $f \equiv 0$ on $\Omega$.
\end{proof}

\begin{remark}\label{rem:h-properties}
	We collect here some properties of the square integral $h = \Im [\int f^2 dz]$, easily verifiable or used in the proof above.
	\begin{itemize}
		\item Under the conformal covariance $f \mapsto (f \circ \varphi)\cdot (\varphi')^{1/2}$, we have invariance $h\mapsto h\circ \varphi$. This is a direct consequence of the change of variables formula for line integrals.
		\item $h$ satisfies the quasi-linear PDE
		\begin{equation*}
			\Delta h = 4 \Im \bigg[\partial_{\bar z}\partial_z \int f^2 dz\bigg] = -4\alpha |f|^2 = -4\alpha |\nabla h|.
		\end{equation*}
		\item In view of Theorem \ref{thm:similarity-principle}, the asymptotics (S) and (D) from Definition \ref{def:correlations-definition} are simply
		\begin{enumerate}
			\item[(S)] $h$ is bounded above in a neighborhood of $a$ (which forces it to satisfy an asymptotic of the form $h(z) = b^2 \log |z-a| + O(1)$ for some $b\in \mathbb R$);
			\item[(D)] $h(z) = - \log |z-a| + O(1)$ near $a$.
		\end{enumerate}
		\item The boundary condition (RH) for $f$ is equivalent to requiring that $h = \Im [\int f^2 dz]$ is constant along the boundary and has \emph{non-positive outer normal derivative}.
	\end{itemize} 
\end{remark}

\subsubsection{The space \texorpdfstring{$V_\Omega^\alpha$}{V_Ω^α}}\label{sec:fermion-existence}
Let distinct points $a_1, a_2, \ldots, a_n; w_1, \ldots, w_{n'} \in \Omega$ be given for $n,n'\geq 0$. Define the vector space $V_\Omega^\alpha = V_\Omega^\alpha(a_1, a_2, \ldots, a_n; w_1, \ldots, w_{n'})$ as the real vector space of spinors $f$ on $\Omega$ whose monodromies are at $a_1, \ldots, a_n$, satisfy \eqref{eq:massive-holomorphicity} at each point away from $a_1, a_2, \ldots, a_n, w_1, \ldots, w_{n'}$ and (RH) on $\partial \Omega$, and have controlled rates of blow-up near those singularities:
\begin{equation}\label{eq:v12}
\begin{split}
&f(z)(z-a_j)^{\frac12}\text{ is bounded near each $a_j$, and }\\
&f(z)(z-w_{j'})\,\text{ is bounded near $w_{j'}$}.
\end{split}
\end{equation}

Then, thanks to Theorem \ref{thm:similarity-principle}, we may define the following quantities for each element $f\in V_\Omega^\alpha$:
\begin{equation}
\beta({a_j})=\beta({a_j};f):=  \lim_{z\to a_j} -e^{i\pi/4}\sqrt{z-a_j}f(z);
\end{equation}
\begin{equation}
\gamma(w_{j'})=\gamma({w_{j'}};f) := \lim_{z\to w} (z-w_{j'})f(z).
\end{equation}

There exists a linear map $V_\Omega^\alpha \to \R^{n+2n'}$ given by
\begin{equation}\label{eq:VtoRd}
	f \mapsto \big(\Re [\beta({a_1})], \ldots, \Re [\beta({a_n})];\, \Re [\gamma({w_1})],\Im [\gamma({w_1})],\ldots,\Re [\gamma({w_{n'}})],\Im [\gamma({w_{n'}})] \big).
\end{equation}
This map is injective by Lemma \ref{lem:uniqueness}: any $f$ in the kernel is bounded near $w_1, \ldots, w_{n'}$ by Remark \ref{rem:singularity-order}, and therefore is massive holomorphic at those points by the first point of Remark \ref{rem:dzbar-massive-holomorphicity}. So we have $\dim_\R V_\Omega^\alpha \leq n+2n'$, and we will show that in fact the dimension is exactly $n+2n'$ in the following.

\begin{proposition}
	The map \eqref{eq:VtoRd} is surjective. In other words, $\dim_\R V_\Omega^\alpha = n+2n'$.
\end{proposition}
\begin{proof}
We already have existence in the critical case, i.e. $\alpha(z) \equiv 0$. This was proven in \cite[Section 7]{CHI2}, but the actual construction will be useful for the massive case as well, so we now recall its idea.

Let us briefly consider the case of the upper half plane $\mathbb{H}$. Given $a_1, \ldots, a_n, w_1, \ldots, w_{n'} \in \mathbb{H}$, consider the real vector space $V_{\mathbb H}^0$ whose elements are holomorphic spinors of the type
	\begin{equation*}
	\underline{f}_\mathbb{H}(z) = \frac{r_{n+2n'-1}z^{n+2n'-1} + \cdots + r_1 z + r_0}{(z-w_1)(z-\overline{w_1})\cdots (z-w_{n'})(z-\overline{w_{n'}})\sqrt{(z-a_1)(z-\overline{a_1})\cdots (z-a_n)(z-\overline{a_n})}}
	\end{equation*}
for real numbers $r_0, \ldots, r_{n+2n'-1}$. Note that, aside from the complication at infinity, $V_\mathbb{H}^0$ is a fitting name: any $\underline{f}_\mathbb{H} \in V_\mathbb{H}^0$ has the necessary bulk asymptotics and satisfies (RH) on $\partial \mathbb{H} = \R$, i.e. $\underline{f}_\mathbb{H} \in \R$ there.

Now, suppose a bounded $\Omega$ and $a_1, \ldots, a_n, w_1, \ldots, w_{n'} \in \Omega$ are given. Then, for a fixed Riemann map $\varphi: \Omega \to \mathbb{H}$, we may consider spinors of the form $ \varphi^* \underline{f}_\mathbb{H} = (\underline{f}_\mathbb{H} \circ \varphi)\cdot \sqrt{\varphi'}$, where $\underline{f}_\mathbb{H} \in V_\mathbb{H}^0\left( \varphi(a_1), \ldots, \varphi(a_n); \varphi(w_1), \ldots, \varphi(w_{n'})\right)$. As in Remark \ref{rem:confcov}, it is simple to check that each $\varphi^* \underline{f}_\mathbb{H}$ belongs to $V_\Omega^0$, except possibly for the validity of (RH) at $\varphi^{-1}(\infty)$.

The space $V_\mathbb{H}^0$ is graded by the order of decay at infinity, i.e. each $\underline{f}_\mathbb{H}$ must decay at one rate among $z^{-1},\ldots,z^{-n-2n'}$. Since $\varphi$ must have a order $1$ boundary pole at $\varphi^{-1}(\infty)$ (easily seen by an explicit M\"obius map $\D\to\mathbb{H}$ or otherwise), the grades respectively correspond to $\varphi^* \underline{f}_\mathbb{H}$ having a zero at $\varphi^{-1}(\infty)$ of order $0, \ldots, n+2n'-1$ (in particular, $\varphi^*\underline{f}_\mathbb{H}$ is continuous at $\varphi^{-1}(\infty)$, and thus satisfies (RH) there). Since each $\varphi^*\underline{f}_\mathbb{H}\in V_\Omega^0$, we have $\dim_\R V_\Omega^0 = n+2n'$, graded by the order of zero at $\varphi^{-1}(\infty)$.

Then, fix a basis $v_1, \ldots, v_{n+2n'} \in V_\Omega^0$ respectively representing each grade, and use Theorem \ref{thm:similarity-principle} to get massive spinors $v'_1=e^{s_1}v_1, \ldots, v'_{n+2n'}=e^{s_{n+2n'}}v_{n+2n'} \in V_\Omega^\alpha$. These $(n+2n')$ spinors are linearly independent and therefore span $V_\Omega^\alpha$, since nonvanishing factors $e^{s_1}, \ldots, e^{s_{n+2n'}}$ cannot modify the order of zero at $\varphi^{-1}(\infty)$. So $\dim_\R V_\Omega^\alpha = n+2n'$.
\end{proof}
\begin{corollary}\label{cor:existence}
	The correlations defined in Definition \ref{def:correlations-definition} uniquely exist.
\end{corollary}
\begin{proof}
Uniqueness was discussed above. We already have the existence of the fermion-fermion correlations, simply by considering the preimage of vectors of the type $(0, \ldots, 0; \Re [\eta], -\Im [\eta])$ under the map \eqref{eq:VtoRd}. For the disorder-fermion correlations, without loss of generality consider the preimage $f$ of the vector $(1, 0, \ldots, 0;)$ with $n'=0$. We claim $f$ is precisely $\frac{\langle  \sigma_{a_2}{\cdots}\sigma_{a_n}\mu_{a_1}\psi_z\rangle_{\alpha}}{\langle \sigma_{a_1}\cdots \sigma_{a_n}\rangle_{\alpha}}$. We only need to check that in this case $\Im [\beta({a_1};f)] = 0$, i.e., $\beta({a_1};f) = 1$.

Cutting out small circles $B_r(a_1) \cup \cdots  \cup B_r(a_n) =: B$ from $\Omega$ and applying \eqref{eq:green-riemann},
\begin{equation*}
	 \oint_{\partial \Omega} f^2 dz - \oint_{\partial B_r(a_1)} f^2 dz- \cdots - \oint_{\partial B_r(a_n)} f^2 dz = 2i \int_{\Omega \setminus B}  \partial_{ \bar z} f^2 d^2 z= 4\int_{\Omega \setminus B} \alpha |f|^2 d^2 z \in \R.
\end{equation*}
But the boundary condition (RH) also implies that $\oint_{\partial \Omega} f^2 dz \in \R$. Taking imaginary parts of both sides, and noting that as $r\to 0$, $\oint_{\partial B_r(a_j)} f^2 dz \to 2\pi \beta({a_j})^2$, we finally have
$$
2\Im[\beta({a_1})]=2\Im[\beta({a_1})]\cdot\Re[\beta({a_1})]=\Im\left[\beta({a_1})^2\right] = \Im[- \beta({a_2})^2 - \cdots - \beta({a_n})^2] = 0,
$$
since $\beta({a_2}), \ldots, \beta({a_n}) \in i\R$.
\end{proof}

We will make use of calculations of this type many more times, so it will be useful to have a notation for it, generally as a bilinear form. For spinors $f, g$ satisfying \eqref{eq:massive-holomorphicity} possibly with distinct masses $\alpha_f, \alpha_g$ on $\Omega \setminus \left(B_\epsilon(a_1) \cup \cdots \cup B_\epsilon(a_n) \cup B_\epsilon(w_1)\cup \cdots \cup B_\epsilon(w_{n'})\right)=: \Omega \setminus B_\epsilon$ and branching around $B_\epsilon(a_1), \cdots, B_\epsilon(a_n)$, define $GR_\epsilon(f, g)$ as the left side of the equality
\begin{equation}\label{eq:gr}
	\oint_{\partial B = \partial B_\epsilon(a_1) \cup \cdots \cup \partial B_\epsilon(w_{n'})} fg dz=	\oint_{\partial \Omega} fg dz  + 2i\int_{\Omega \setminus B_\epsilon} i (\alpha_f\bar{f}g + \alpha_gf\bar{g})d^2z,
\end{equation}
where the RHS comes from applying Green-Riemann's theorem as above. Note that the first term on the RHS is purely real as long as $f, g$ satisfy (RH), and the second term is also real if $\alpha_f = \alpha_g$. Let $GR_0(f, g) := \lim_{\epsilon \downarrow 0} GR_\epsilon(f, g)$. For example, in the absence of $w_j$'s (i.e. $n' = 0$), $GR_0(f,g)={2\pi}\sum_{j=1}^n \beta({a_j};f)\beta({a_j};g)$. We are already able to show the bulk of well-definedness statements.

\begin{proposition}\label{prop:well-def}
    Let $f^{w^\eta}(z) =\frac{\langle\sigma_{a_1}{\cdots}\sigma_{a_n}\psi_z\psi_w^{[\eta]}\rangle_{\alpha}}{\langle \sigma_{a_1}\cdots \sigma_{a_n}\rangle_{\alpha}}$ and $f_k(z) = \frac{\langle \sigma_{a_1}\stackrel{\hat{k}}{\cdots}\sigma_{a_n}\mu_{a_k} \psi_z\rangle_{\alpha}}{\langle \sigma_{a_1}\cdots \sigma_{a_n}\rangle_{\alpha}}$ for all $k = 1, 2, \ldots, n$.
    Then,
    \begin{itemize}
        \item $b(a_k;f^{w^\eta}) =  \Re[\bar\eta f_k(w)]$, or
$$
e^{-\frac{i\pi}{4}}\sqrt{z- a_k}\frac{\langle\sigma_{a_1}{\cdots}\sigma_{a_n}\psi_z\psi_w^{[\eta]}\rangle_{\alpha}}{\langle \sigma_{a_1}\cdots \sigma_{a_n}\rangle_{\alpha}}\xrightarrow{z\to a_k} \frac{\langle  \sigma_{a_1}\stackrel{\hat{k}}{\cdots}\sigma_{a_n}\mu_{a_k}\psi_w^{[\eta]}\rangle_{\alpha}}{\langle \sigma_{a_1}\cdots \sigma_{a_n}\rangle_{\alpha}}.
$$
        \item $\Re[\bar\eta' f^{w^{\eta}}(w')] = -\Re[\bar\eta f^{(w')^{\eta'}}(w)]$, or
$$
\frac{\langle\sigma_{a_1}{\cdots}\sigma_{a_n}\psi_{w'}^{[\eta']}\psi_w^{[\eta]}\rangle_{\alpha}}{\langle \sigma_{a_1}\cdots \sigma_{a_n}\rangle_{\alpha}} = - \frac{\langle\sigma_{a_1}{\cdots}\sigma_{a_n}\psi_w^{[\eta]}\psi_{w'}^{[\eta']}\rangle_{\alpha}}{\langle \sigma_{a_1}\cdots \sigma_{a_n}\rangle_{\alpha}},
$$
        \item $b(a_j; f_k) = -b(a_k;f_j)$ for all $j\neq k$, or        
$$
        \frac{\langle  \sigma_{a_1}\stackrel{\hat{j,k}}{\cdots}\sigma_{a_n}\mu_{a_k}\mu_{a_j}\rangle_{\alpha}}{\langle \sigma_{a_1}\cdots \sigma_{a_n}\rangle_{\alpha}} = -\frac{\langle  \sigma_{a_1}\stackrel{\hat{j,k}}{\cdots}\sigma_{a_n}\mu_{a_j}\mu_{a_k}\rangle_{\alpha}}{\langle \sigma_{a_1}\cdots \sigma_{a_n}\rangle_{\alpha}}.
        $$
    \end{itemize}
\end{proposition}
\begin{proof}
    The first statement follows from setting $\Im \big[GR_0(f^{w^{\eta}}, f_k)\big]$ to zero, making sure to cut out a disk around $w$ whenever $f^{w^{\eta}}$ is involved, etc. That is, writing $b_j^{w^\eta}:=b(a_j;f^{w^\eta})$ and $b_j^k = b(a_j;f_k)$,
     \begin{align*}
    0&=\lim_{\epsilon\to 0}\Im\bigg[\sum_{j}\oint_{\partial B_\epsilon(a_j)}f^{w^\eta}(z)f_k(z)dz+\oint_{\partial B_\epsilon(w)}f^{w^\eta}(z)f_k(z)dz\bigg]\\
    &=\sum_{j\neq k} \lim_{\epsilon\to 0}\Im\bigg[\oint_{\partial B_\epsilon(a_j)}\big(b_j^{w^\eta}b^{k}_j e^{i\pi/2}(z-a_j)^{-1}+o(|z-a_j|^{-1})\big)dz\bigg]\\
    &\quad + \lim_{\epsilon\to 0}\Im\bigg[\oint_{\partial B_\epsilon(a_k)}\big(-b_k^{w^\eta} (z-a_k)^{-1}+o(|z-a_k|^{-1})\big)dz\bigg]\\
    &\quad +\lim_{\epsilon\to 0}\Im \bigg[\oint_{\partial B_\epsilon(w)} \big(\bar \eta(z-w)^{-1}+o(|z-w|^{-1})\big)f_k(z)dz\bigg]\\
    &=\sum_{j\neq k}\Im \big[-2\pi b_j^{w^\eta} b_j^k\big]-\Im \big[2\pi i b_k^{w^\eta}\big]+\Im\big[\bar\eta 2\pi i f_k(w)\big]\\
    &=-2\pi b_k^{w^\eta}+2\pi \Re\big[\bar \eta f_k(w)\big].
    \end{align*}
    The second and third statements are derived using similar arguments for $\Im \big[GR_0(f^{w^{\eta}},f^{(w')^{\eta'}})\big]$ and $\Im \big[GR_0(f_j, f_k)\big]$, respectively.
\end{proof}

Now we are ready to consider multi-point fermion insertions in spin-weighted correlations. For this purpose, we define multi-point correlations by first considering correlations of the form
$$\frac{\langle\sigma_{a_1}{\cdots}\sigma_{a_n}\psi^{[\eta_1]}_{z_1}\cdots\psi^{[\eta_{2n'-1}]}_{z_{2n'-1}}\psi_{z}\rangle_{\alpha}}{\langle \sigma_{a_1}\cdots \sigma_{a_n}\rangle_{\alpha}} \in V_\Omega^\alpha(a_1, a_2, \ldots, a_n; z_1, \ldots, z_{2n'-1}),$$
where $\psi_{z}$ has the behavior \eqref{eq:ope} whenever approaching other insertions. For example,
$$
\frac{\langle\sigma_{a_1}{\cdots}\sigma_{a_n}\psi^{[\eta_1]}_{z_1}\cdots\psi^{[\eta_{2n'-1}]}_{z_{2n'-1}}\psi_{z}\rangle_{\alpha}}{\langle \sigma_{a_1}\cdots \sigma_{a_n}\rangle_{\alpha}} \sim -\frac{\langle\sigma_{a_1}{\cdots}\sigma_{a_n}\psi^{[\eta_2]}_{z_2}\cdots\psi^{[\eta_{2n'-1}]}_{z_{2n'-1}}\rangle_{\alpha}}{\langle \sigma_{a_1}\cdots \sigma_{a_n}\rangle_{\alpha}}\frac{\overline{\eta_1}}{z-z_1}\text{ as }z\to z_1,
$$
the negative sign coming from anticommuting $z$ to the left $2n'-1$ times in order to use \eqref{eq:ope}. Then we may extend to general fermions by \eqref{eq:fermion-complex}.

In any case, as Proposition \ref{prop:fermion-antisymmetry} below states, such a definition is equivalent to defining multi-point correlations using a Pfaffian decomposition, as expected physically. We will write $\Psi$ for either of $\psi,\psi^*$, or equivalently $\psi^{[\eta]}$ with $|\eta|=1$.

\begin{proposition}\label{prop:fermion-antisymmetry}
	Any two fermion insertions among $\psi^{[\eta]}, \psi, \psi^*$ anticommute. We may decompose
	\begin{equation}\label{eq:pfaffian}
		\frac{\langle\sigma_{a_1}{\cdots}\sigma_{a_n}\Psi_{z_1}\cdots\Psi_{z_{2n'}}\rangle_{\alpha}}{\langle \sigma_{a_1}\cdots \sigma_{a_n}\rangle_{\alpha}} = \operatorname{Pf}\left( \frac{\langle\sigma_{a_1}{\cdots}\sigma_{a_n}\Psi_{z_j}\Psi_{z_k}\rangle_{\alpha}}{\langle \sigma_{a_1}\cdots \sigma_{a_n}\rangle_{\alpha}} \right)_{j,k=1}^{2n'}.
	\end{equation}
\end{proposition}
\begin{proof}
	Antisymmetry in two-point correlations is immediate from Proposition \ref{prop:well-def} and linearity in the definition of the complex fermions.
    
    For \eqref{eq:pfaffian} we begin by assuming the first $2n'-1$ $\Psi$'s are of the form $\psi^{[\eta]}$ and the last $\psi$ as above. In this case, the equivalence to the Pfaffian decomposition definition goes through the exact same argument as in the critical case (see e.g. \cite[Section 6.6]{Hon} or \cite[Proposition 2.24]{CHI2}): in short, one shows the Pfaffian recursive relation
    $$
    \frac{\langle\sigma_{a_1}{\cdots}\sigma_{a_n}\psi^{[\eta_1]}_{z_1}\cdots\psi^{[\eta_{2n'-1}]}_{z_{2n'-1}}\psi_{z}\rangle_{\alpha}}{\langle \sigma_{a_1}\cdots \sigma_{a_n}\rangle_{\alpha}} = \sum_{j=1}^{2n'-1}(-1)^j\frac{\langle\sigma_{a_1}{\cdots}\sigma_{a_n}\psi^{[\eta_1]}_{z_1}\stackrel{\hat{j}}{\cdots}\psi^{[\eta_{2n'-1}]}_{z_{2n'-1}}\rangle_{\alpha}}{\langle \sigma_{a_1}\cdots \sigma_{a_n}\rangle_{\alpha}}\frac{\langle\sigma_{a_1}{\cdots}\sigma_{a_n}\psi^{[\eta_{j}]}_{z_j}\psi_z\rangle_{\alpha}}{\langle \sigma_{a_1}\cdots \sigma_{a_n}\rangle_{\alpha}}.
    $$
    by showing that both sides have the same poles, (S) asymptotic at $a_1, \ldots, a_n$, and the (RH) boundary condition. Note that antisymmetry of any two fermions within the $2n'$-point correlation is immediate from the $2$-point correlation case and the right-hand side of \eqref{eq:pfaffian}.

    Then the case of general $\Psi$ follows by linearity.
\end{proof}

\subsection{Estimates of the Basic Correlations}\label{sec:estimates-section}
In order to simplify notation, define
\begin{equation}
    P_{a}(z) := \min \left( {\frac{1}{\sqrt{|z-a|}}},{\frac{\sqrt{\operatorname{dist}(a, \partial\Omega)}}{{|z-a|}}}\right).
\end{equation}
Additionally, in order to give uniform bounds which are valid under small or favorable perturbations, we use the following notation for constants bounding various relevant quantities:
\begin{itemize}
    \item $0 < \delta_{1,k} \leq \frac{1}{16}|a_1 - a_k |$;
    \item $0 < \delta_1 \leq \frac{1}{16}|a_1 - a_k|$ for all $k=2, \ldots, n$;
    \item $0<\delta_0\leq \frac{1}{16}|a_j-a_k|$ for all $j<k$;
    \item $0<\delta_w\leq\min_{1\leq j\leq n}\min(\frac{1}{16}|a_j - w|, \operatorname{dist}(w,\partial\Omega))$;
    \item $ \Vert \alpha \Vert_{L^2(\Omega)} \leq \kappa$;
    \item $\Vert\alpha\Vert_{L^\infty (\Omega)} \leq \kappa_\infty$;
    \item  $\Vert\nabla\alpha\Vert_{L^\infty (\Omega)} \leq \kappa_\infty'$;
    \item $\operatorname{dist}(z,\partial\Omega)^2|\nabla\alpha(z)| \leq \kappa_b$;
    \item $|m_0| \leq M$,
\end{itemize}
along with the assumptions needed to make such constants finite if they are involved in a bound (e.g. if a bound involves $\kappa_\infty$, we assume $\alpha\in L^\infty(\Omega)$). In Section \ref{subsec:series-expansion}, $m_0$ will stand for the point around which power series in $m$ are expanded, after replacing $\alpha$ by $m\alpha$.

\subsubsection{Critical (massless) correlations}
	In this section, we give estimates of the \emph{critical} correlations, which are (massless) holomorphic and admit usual complex analytic techniques and conformal invariance. First, we estimate the spinor in the simplest cases:
    \begin{lemma}\label{lem:1spin-crit}
        Let $\Omega = \D$. Then 
            \begin{equation}\label{eq:1spin}
            f(z) := \frac{\langle \mu_{a}\psi_z \rangle_{0}}{\langle \sigma_{a}\rangle_{0}} = \frac{e^{-i\pi/4}}{\sqrt{\frac{z-a}{1-\bar{a} z}}}\sqrt{\frac{1-|a|^2}{(1-\bar{a} z)^2}} = \frac{e^{-i\pi/4}}{\sqrt{{z-a}}}\sqrt{\frac{1-|a|^2}{1-\bar{a} z}},
        \end{equation}
        and we have $ |f(z)| \leq C\cdot P_a(z)$ in all of $\D$.
    \end{lemma}
    \begin{proof}
        Note that we have an explicit formula in the case of $a=0\in \D$: the expression ${e^{-i\pi/4}}/{\sqrt{z}}$ satisfies (D) and (RH). For general $a \in \D$, seeing that the conformal map $z\mapsto (z-a)/(1-\bar{a}z)$ fixes $\D$ and sends $a$ to $0$, we get the explicit formula using Remark \ref{rem:confcov}. The desired estimate follows from
        \begin{equation*}
\left|\frac{1-|a|^2}{(z-a)(1-\bar az)} \right|  \leq 2\min\left({\frac{1}{{|z-a|}}}, \frac{{\operatorname{dist}(a, \partial\Omega)}}{|z-a|^2}\right),
        \end{equation*}
        using $1-|a| \leq |1-\bar az| \leq 2 $ and $|z-a| \leq |1-\bar{a}z|$ (e.g. consider that the conformal map above maps to the interior of $\D$).
    \end{proof}
    
	\begin{lemma}\label{lem: df-critical-bound}
		Let $\Omega$ be smooth and write $f(z) = \frac{\langle  \sigma_{a_2}\cdots\sigma_{a_n}\mu_{a_1}\psi_z\rangle_{0}}{\langle \sigma_{a_1}\cdots \sigma_{a_n}\rangle_{0}}$.
        Then we have 
        $$
        \left|f(z)\right|\leq C(n, \Omega)\bigg[P_{a_1}(z)+ \sum_{k=2}^n |b_k| P_{a_k}(z)\bigg],
        $$
        for constants $b_k := b(a_k;f)=i\beta(a_k;f)$, which satisfy
        $$
        |b_k|= \bigg| \frac{\langle  \sigma_{a_1}\stackrel{\hat{1,k}}{\cdots}\sigma_{a_n}\mu_{a_1}\mu_{a_k}\rangle_{0}}{\langle \sigma_{a_1}\cdots \sigma_{a_n}\rangle_{0}}\bigg| \leq C(\Omega,\delta_{1,k})\sqrt{\operatorname{dist}(a_1, \partial\Omega)\operatorname{dist}(a_k, \partial\Omega)}.
        $$
	\end{lemma}
\begin{proof}
    First, suppose $\Omega = \D$. 
    Consider $h = \Im [\int f^2 dz]$, which is harmonic away from the insertions $a_1, \ldots, a_n$. In addition, examining the asymptotics and boundary conditions from Remark \ref{rem:h-properties}, we may explicitly verify the decomposition
    $$
    h = G_{a_1} - b_2^2G_{a_2} - \cdots - b_n^2G_{a_n},\qquad\text{or}\qquad f^2/2 = \partial_zG_{a_1} - b_2^2\partial_zG_{a_2} - \cdots - b_n^2\partial_zG_{a_n},
    $$
    where $-G_{a}$ is the Dirichlet Laplace Green's function on $\D$, i.e., $G_a= G_0 \circ\varphi_a$ and $G_0 = -\log|z|$, $\varphi_a(z) = \frac{z-a}{1-\bar az}$, as in the proof of Lemma \ref{lem:1spin-crit}. Since a single $G_a$ happens to be exactly the square integral for $f_a(z):={\langle \mu_{a}\psi_z \rangle_{0}}/{\langle \sigma_{a}\rangle_{0}}$, we may use Lemma \ref{lem:1spin-crit} to estimate each derivative on the right; it remains to show $b_k^2\leq C(\delta_{1,k})\operatorname{dist}(a_1, \partial\D)\operatorname{dist}(a_k, \partial\D)$ for $k\geq 2$, since identification as a disorder correlation will follow from Proposition \ref{prop:well-def}. First, since $\partial_nh = -|f|^2 \leq 0$ by (RH) on the boundary and each $f_a^2$ has a single pole $\frac{-i}{z-a}$ in the bulk,
    \begin{align*}
    0 \geq \int_{\partial \D} \partial_n h |dz| &=- \int_{\partial \D} |f_{a_1}|^2|dz| +b_2^2 \int_{\partial \D} |f_{a_2}|^2|dz|+\cdots +b_n^2 \int_{\partial \D} |f_{a_n}|^2|dz|\\
    &=- \oint_{\partial \D} f_{a_1}^2 dz +b_2^2 \oint_{\partial \D} f_{a_2}^2 dz+\cdots +b_n^2 \oint_{\partial \D} f_{a_n}^2dz \\
    &=-2\pi i\left[-i+b_2^2i+\cdots+b_n^2i\right]=-2\pi(1-b_2^2-\cdots-b_n^2),
    \end{align*}
    where we used (RH) multiple times. So all $b_k$'s are bounded by $1$, and we may assume henceforth $0<8\operatorname{dist}(a_1,\partial \D) \leq \delta_{1,k}<1$, since otherwise we may choose large enough $C(\delta_{1,k})$ to ensure $ C(\delta_{1,k})\operatorname{dist}(a_1, \partial\D)\operatorname{dist}(a_k, \partial\D) \geq 1$.

    In particular, the $(\delta_{1,k}/2)$-ball around $a_1$ intersects with $\partial \D$. We claim that the point $z_0\in \partial \D$ closest to $a_k$ in $\partial\D \setminus B_{\delta_{1,k}/2}(a_1)$ satisfies $|z_0 - a_k| \leq 2\operatorname{dist}(a_k,\partial\D)$. Indeed, if $z_0$ is the minimizer on all of $\partial\D$, $|z_0-a_k|$ is simply $\operatorname{dist}(a_k,\partial\D)$; otherwise, let $z_1$ be the true minimizer, attained in $\partial\D \cap B_{\delta_{1,k}/2}(a_1)$. Considering the triangle $\triangle a_ka_1z_1$, we know that $|a_1 - a_k| \geq \delta_{1,k}$, $|a_k-z_1| = \operatorname{dist}(a_k,\partial\D)$, and $2\pi/3 \leq \angle a_ka_1z_1 \leq \pi$. Then, using the second law of cosines, it is easy to see that $\operatorname{dist}(a_k,\partial\D) \geq \delta_{1,k}$, and we have $|z_0 - a_k| \leq |z_0 - a_1| + |a_1 - a_k| \leq 2\delta_{1,k} \leq 2\operatorname{dist}(a_k,\partial\D)$.

    Then, this bound, along with Lemma \ref{lem:1spin-crit}, yields
    \begin{align*}
        |f_{a_k}(z_0)|^2 =  \frac{1-|a_k|^2}{|z_0-a_k||1-|a_k|^2+\overline{a_k}(a_k-z_0)|}  \geq \frac{1}{8\operatorname{dist}(a_k,\partial\D)},
    \end{align*}
and, $|a_1-z_0| \geq \delta_{1,k}/2$, and the assumptions $|a_1| \geq 7/8$ and $ \operatorname{dist}(a_1,\partial\D) \leq \delta_{1,k}/8$ yield
    \begin{align*}
        |f_{a_1}(z_0)|^2 \!=\! \frac{1-|a_1|^2}{|z_0-a_1||1-|a_1|^2+\overline{a_1}(a_1-z_0)|} \!\leq\! \frac{2\operatorname{dist}(a_1,\partial\D)}{\frac{1}{2}\delta_{1,k}(\frac{7}{16}\delta_{1,k}-2\operatorname{dist}(a_1,\partial\D))}
        \!\leq\! \frac{64\operatorname{dist}(a_1,\partial\D)}{3\delta_{1,k}^2}.
    \end{align*}
        
    Since $h,G_{a_1},\ldots,G_{a_n}$ all have non-positive outer normal derivatives, we have
    $$
    0 \geq \partial_n h(z_0)\geq \partial_nG_{a_1}(z_0)-b_k^2 \partial_nG_{a_k}(z_0)= b_k^2|f_{a_k}(z_0)|^2-|f_{a_1}(z_0)|^2,
    $$
    which yields the desired bound.

    For general $\Omega$, we may fix a Riemann map (whose derivative and its reciprocal being bounded above by $C(\Omega)$) and pull-back to $\D$ as usual, which gives the $\Omega$ dependence in the final statement.
\end{proof}

\begin{lemma}\label{lem:ff-planar-critical}
    Let $\Omega =\D$ and $f(z) = {\langle \psi_z\psi_w^{[\eta]}\rangle_{0}}$. Then we have $$|f(z)|\leq \frac{2|\eta|}{|z-w|}.$$
    Additionally, in $B_{d/8}(w)$, where $d = \operatorname{dist}(w,\partial\D)=1-|w|$, we have $$|f(z)| \geq \frac{|\eta|}{2|z-w|}.$$
\end{lemma}
\begin{proof}
    By dividing by a real scalar if necessary, we may assume $|\eta| = 1$. For $w=0$ and $\eta = e^{i\pi/4}$, we may verify by hand (cf. \cite[Section 1.6]{HoSm}: the $\frac{1}{z-w}$ in the reference becomes $\frac{e^{-\frac{i\pi}{4}}}{z-w}$ due to a constant difference in the definition of (RH)) that
    $$
    {\left\langle \psi_z\psi_0^{\left[\exp{i\frac{\pi}{4}}\right]}\right\rangle_{0}} = e^{-\frac{i\pi}{4}}\frac{z+1}{z}.
    $$
    Then it is straightforward to modify $\eta$ using the covariance rule of $\psi$:
    $$
    {\bigg\langle \psi_z\psi_0^{\left[\exp{i\left(\frac{\theta_0}{2}+\frac{\pi}{4}\right)}\right]}\bigg\rangle_{0}} = e^{-\frac{i\pi}{4}}\frac{e^{i\theta_0}z+1}{e^{\frac{i\theta_0}{2}}z}.
    $$
    Finally, using $\frac{z-w}{1-\bar{w}z}$ as the involution of $\D$ sending $w$ to $0$ with positive derivative at $w$,
    \begin{align*}
    \Bigg|{\bigg\langle \psi_z\psi_w^{\left[\exp{i\left(\frac{\theta_0}{2}+\frac{\pi}{4}\right)}\right]}\bigg\rangle_{0}}\Bigg| = \Bigg|e^{-\frac{i\pi}{4}}\sqrt{\frac{1-|w|^2}{(1-\bar{w}z)^2}}\cdot \sqrt{\frac{1-|w|^2}{(1-|w|^2)^2}}\cdot \frac{e^{i\theta_0}\left(\frac{z-w}{1-\bar{w}z}\right)+1}{e^{\frac{i\theta_0}{2}}\left(\frac{z-w}{1-\bar{w}z}\right)}\Bigg|&\\
    = \Bigg|\sqrt{\frac{1}{(1-\bar{w}z)^2}}\cdot \frac{e^{i\theta_0}\left(\frac{z-w}{1-\bar{w}z}\right)+1}{\left(\frac{z-w}{1-\bar{w}z}\right)}\Bigg|&\leq \frac{2}{|z-w|}.
    \end{align*}
    Finally, the ball $B_{d/8}(w)$ maps into $B_{1/2}(0)$ under the M\"obius transformation $\frac{z-w}{1-\bar wz}$, as seen from direct computation or Koebe. Then the desired lower bound follows.
\end{proof}
\begin{proposition}\label{prop:ff-crit-estimate}
    Let $\Omega$ be smooth and write $f(z) = \frac{\langle \sigma_{a_1} \sigma_{a_2}\cdots\sigma_{a_n}\psi_z\psi_w^{[\eta]}\rangle_{0}}{\langle \sigma_{a_1}\cdots \sigma_{a_n}\rangle_{0}}$.
    Then we have 
    \begin{align}\label{eq:critical-ff-bound}
        |f(z)| \!\leq\! C(n, \Omega) \Bigg[ \sum_{j=1}^n |b_j| P_{a_j}(z) \!+\! \Bigg[\frac{|\eta|}{\sqrt{\operatorname{dist}(w,\partial\Omega))}}+\!\sum_{k=1}^n \frac{|\eta|}{\sqrt{|w-a_k|}}\Bigg] \!P_w(z) \!+ \frac{|\eta|}{|z-w|}\Bigg],
    \end{align}
    for constants $b_j := b(a_j;f)=i\beta(a_j;f)$, which satisfy
    $$
    |b_j|= \bigg| \frac{\langle  \sigma_{a_1}\stackrel{\hat{j}}{\cdots}\sigma_{a_n}\mu_{a_j}\psi_w^{[\eta]}\rangle_{0}}{\langle \sigma_{a_1}\cdots \sigma_{a_n}\rangle_{0}}\bigg|.
    $$
\end{proposition}
\begin{proof}
As above, we only show it on $\D$. Write $f_{\operatorname{flat}}$ for the correlation from Lemma \ref{lem:ff-planar-critical} taken on the same domain and with the same $w$. Define $e(w)$ as the real constant in the expansion
\begin{align*}
    f(z) = \frac{\bar \eta}{z-w} + i\eta e(w) + O(|z-w|)\text{ as }z\to w,
\end{align*}
and similarly for $f_{\operatorname{flat}}$ and $e_{\operatorname{flat}}(w)$. The phase of the constant term is thus fixed, since otherwise we would have nonzero additive monodromy around the logarithmic singularity of $h = \Im [\int f^2 dz]$ at $w$: since (RH) forces $h$ to be constant on $\partial \Omega$ by Remark \ref{rem:h-properties}, this cannot happen.

Again comparing the asymptotics (D), (S), and (F) for $h$ and $h_{\operatorname{flat}} = \Im [\int f_{\operatorname{flat}}^2 dz]$ as in the proof of Lemma \ref{lem: df-critical-bound} (and using the same notation for the Green's functions $G_a = \Im [\int f_a^2 dz]$ and $b_j = b(a_j)$), we have
\begin{equation*}
h = - b_1^2 G_{a_1} - \dots - b_n^2G_{a_n} + h_{\operatorname{flat}} + (e(w) - e_{\operatorname{flat}}(w))G_w.
\end{equation*}
Note that we get two functions corresponding to the singularity at $w$: $h_{\operatorname{flat}}$ which provides the degree 1 singularity, and $G_w$ term calibrates it at logarithmic order. The coefficients $b_j$ are identified by Proposition \ref{prop:well-def}.

Then Lemmas \ref{lem:1spin-crit}, \ref{lem: df-critical-bound}, \ref{lem:ff-planar-critical} would imply the desired results as in the proof of Lemma \ref{lem: df-critical-bound}, given the following bound (still assuming $\Omega = \D$):
\begin{lemma}\label{lem:argument-lemma}
Write $\delta_d := \min(|w-a_1|, \ldots,|w-a_n|, \operatorname{dist}(w,\partial\Omega))$. Then we have $$|e(w) - e_{\operatorname{flat}}(w)| \leq \frac{C(n)|\eta|^2}{\delta_d}.$$
\end{lemma}
\begin{proof}[Proof of Lemma \ref{lem:argument-lemma}]
 Consider the two meromorphic functions $f_r := -b_1^2 f_{a_1}^2 - \cdots -b_n^2 f_{a_n}^2 +f_{\operatorname{flat}}^2$ and $f_w^2$. We know that $f^2$, which is a linear combination of the two functions, is a meromorphic function which is a square of another meromorphic function, single-valued in a $\delta_d$-neighborhood of $w$. Thanks to Lemma \ref{lem:ff-planar-critical} which bounds $|f_{\operatorname{flat}}^2(z)| \geq \frac{|\eta|^2}{2|z-w|^2}$ (for $|z-w| < \frac{\delta_d}{8}$) and Lemma~\ref{lem:1spin-crit} which bounds each $|f_{a_k}^2(z)| \leq \frac{C}{|z-a_k|}$, along with $|b_k| \leq C(n) \max_{1\leq j\leq n} {\frac{|\eta|}{\sqrt{|w-a_j|}}}$, there exists some radius $\delta_d/C(n)$ such that there is no zero of $f_r$ in $B_{\delta_d/C(n)}(w)$, only a double pole at $w$. Then the same Lemmas in turn give the bound $|f_r| \leq \frac{C(n)|\eta|^2}{\delta_d^2}$ on $\partial B_{\delta_d/C(n)}(w)$.

However, $|f_w^2| \geq C(n)/\delta_d $ on $\partial B_{\delta_d/C(n)}(w)$ by \eqref{eq:1spin}. So, by Rouch\'e's theorem, there exists a constant $C(n)$ such that if $|e(w) - e_{\operatorname{flat}}(w)| \geq C(n)|\eta|^2/\delta_d$ then the number of zeros minus the number of poles of $f^2 = f_r + (e(w) - e_{\operatorname{flat}}(w))f_w^2$ in $B_{\delta_d/C(n)}(w)$ is $-1$, i.e. identical to that of $f_w^2$. Since there is already a unique double pole at $w$, $f^2$ must also have a simple zero, contradicting the assumption that $f$ is single valued near $w$.
\end{proof}
This concludes the proof of Proposition~\ref{prop:ff-crit-estimate}.
\end{proof}

\subsubsection{Massive correlations}\label{sec:fermion-analysis}
We start our quantitative estimates by noting the following useful exception to Remark \ref{rem:hol-phase}. 
\begin{lemma}\label{lem:onespin}
	Let $f(z) = \frac{\langle \mu_{a}\psi_z \rangle_{\alpha;\Omega}}{\langle \sigma_{a}\rangle_{\alpha;\Omega}}$. Then its holomorphic part $\underline{f}$, with $s$ normalized so that $\Re [s(a)] = 0$, is precisely the corresponding critical spinor $\underline{f}(z) = \frac{\langle \mu_{a}\psi_z \rangle_{0;\Omega}}{\langle \sigma_{a}\rangle_{0;\Omega}}$. In addition, we have
	\begin{equation}\label{eq:three-line-integrals}
		0< c(\kappa) \leq \oint_{\partial \Omega} f^2 dz=\int_{\partial \Omega} |f|^2 |dz| =- \int_{\partial \Omega} \partial_n h |dz| \leq C(\kappa),
	\end{equation}
	where
    $\partial_n$ refers to the outer normal derivative.

    If $\Omega$ is smooth, for any $\nu \in (0,1)$ and $z\in \Omega$, we have
    \begin{equation*}
        |h(z)| =h(z) \leq {C(\kappa,\Omega, \nu)}\left[\left(\frac{\operatorname{dist}(a,\partial \Omega)}{|z-a|}\right)^{\nu} +1\right] .
    \end{equation*}
\end{lemma}
\begin{proof}
	For the first part, note that $\underline{f}$ is an element of $V^{0}_\Omega(a)$. That also means that under the linear map \eqref{eq:VtoRd} $\underline{f}$ would be taken to $\Re [e^{-s(a)}]$. We have already shown in the proof of Corollary \ref{cor:existence} that such vectors in fact satisfy $\Im [\beta_a(\underline{f})]=\Im [e^{-s(a)}]=0$, i.e., $\Re [e^{-s(a)}] = \exp\{-\Re [s(a)]\} = 1$, i.e., $\underline{f}$ is the desired critical correlation.
	
	The equivalence of the three line integrals in \eqref{eq:three-line-integrals} is a matter of simple calculation from the boundary condition (RH). In addition, they are easily seen to be invariant under conformal pull-back, so we may fix $\varphi: \D \to \Omega$ such that $\varphi(0) = a$ and compute instead $ \int_{\partial \D}  |\varphi^* f|^2 |dz|$. By the above discussion, the holomorphic part is simply the critical correlation, for which we have an explicit formula:
		\begin{equation*}
		\varphi^* f(z) = \frac{\langle \mu_{0}\psi_z \rangle_{(\alpha\circ \varphi)|\varphi'|;\D}}{\langle \sigma_{0}\rangle_{(\alpha\circ \varphi)|\varphi'|;\D}} = e^{s(z)}\frac{\langle \mu_{0}\psi_z \rangle_{0;\D}}{\langle \sigma_{0}\rangle_{0;\D}} = e^{s(z)}\frac{e^{-i\pi/4}}{\sqrt{z}}.
	\end{equation*}
	
	Then the trace, Sobolev, and Cauchy-Schwarz inequalities give (see also \cite[Proposition 8.4]{BBC}, which translates a $W^{1,2}$ for $2s$ to a $W^{1,q}$ bound for $e^{2s}$ for $q<2$)
	$$
	\frac{C}{\Vert s \Vert_{W^{1,2}(\D)}}\leq \frac{\left(\int_{\partial \D}  |dz|\right)^2}{\int_{\partial \D} |e^{-2s}| |dz|} \leq	\int_{\partial \D} |f_\D|^2 |dz| = \int_{\partial \D} |e^{2s}| |dz| \leq C\Vert s \Vert_{W^{1,2}(\D)},
	$$
	so we have \eqref{eq:three-line-integrals}, since from  \eqref{eq:similarity-bound} and by the change of variables formula we have $\Vert s \Vert_{W^{1,2}(\D)} \leq C \Vert (\alpha\circ \varphi)|\varphi'| \Vert_{L^2(\D)} = C \Vert \alpha \Vert_{L^2(\Omega)}$.

    Similarly, integrating on the straight ray $[z/|z|,z]$ from $z/|z| \in\partial \D$ to $z$, we have
    \begin{align*}
    |h\circ\varphi(z)| = \Bigg|\int_{z/|z|}^z \bigg(e^{s(z')}\frac{e^{-i\pi/4}}{\sqrt{z'}}\bigg)^2 dz' \Bigg| &\leq \Vert e^{2s} \Vert_{L^{1/\nu}([z/|z|,z])} \bigg(\int_{{[z/|z|,z]}} \frac{1}{|z'|^{\frac{1}{1-\nu}}}|dz'| \bigg)^{{1-\nu}}\\ 
    &\leq C(\nu)\Vert {s} \Vert_{W^{1,\frac{2}{1+\nu}}(\mathbb D)}\cdot C|z|^{-\nu}\leq C(\kappa, \nu)|z|^{-\nu},
    \end{align*}
    applying \cite[Proposition 8.4]{BBC} and the trace inequality again, on, say, the half of $\D$ cut by the ray $[-z/|z|,z/|z|]$. Then it remains to note that for $z\in \Omega$, such that $|z-a| \leq \frac{1}{2}\operatorname{dist}(a,\partial \Omega)$, we have
    $$
    \left|\varphi^{-1}(z)\right| \geq \frac{|z-a|}{16\operatorname{dist}(a,\partial \Omega)},
    $$
    by, e.g., Koebe. If $|z-a| > \frac{1}{2}\operatorname{dist}(a,\partial \Omega)$ we may simply appeal to the bound for the case $|z-a| = \frac{1}{2}\operatorname{dist}(a,\partial \Omega)$, thanks to the maximum principle, which gives rise to the constant term.
\end{proof}

We now improve the existence statement of Corollary \ref{cor:existence} to a quantitative bound.
\begin{proposition}\label{prop:thebound}
	Let $\Omega$ be smooth and $f(z) =\frac{\langle  \sigma_{a_2}\cdots\sigma_{a_n}\mu_{a_1}\psi_z\rangle_{\alpha}}{\langle \sigma_{a_1}\cdots \sigma_{a_n}\rangle_{\alpha}}$. Then, for all $2\leq j \leq n$ and $\nu\in (0,1)$, we have 
    \begin{equation*}
    |\beta({a_j}; f)|=|b(a_j)| \leq C(\kappa, \Omega,\nu)\left[\left(\frac{\operatorname{dist}(a_1,\partial \Omega)}{\delta_1}\right)^{\nu} +1\right].
    \end{equation*}
\end{proposition}
\begin{proof}
Consider $ f_1(z) = \frac{\langle \mu_{a_1} \psi_z\rangle_{\alpha}}{\langle \sigma_{a_1}\rangle_{\alpha}}$, $h = \Im [\int f^2 dz]$, and $h_1 = \Im [\int f_1^2 dz]$, normalized to be zero on $\partial \Omega$, as usual. We claim that $h\leq h_1$: it suffices to show that $f^2-f_1^2$ extends continuously to $a_1$ and therefore $h-h_1$, which can only blow-up to $-\infty$ at $a_2, \ldots, a_n$, cannot have a bulk maximum, by Lemma~\ref{lem:maximum}. This holds since
\begin{align}\label{eq:singularitypm}
	f + f_1 = -2e^{-i\pi/4}(z-a_1)^{-1/2} + o(|z-a_1|^{-1/2})\quad\text{and}\quad f - f_1 =  O(|z-a_1|^{1/2}),
\end{align}
from \eqref{eq:mhol-local-expansion} and \eqref{eq:mhol-difference}.

Thanks to the continuity of $h$ away from $a_1, \ldots, a_n$ and the maximum principle, we may infer the following topological facts about the subdomains of $\Omega$:
\begin{itemize}
	\item Any sets of type $\{h<r\}, \{h>r\}$ for $r\in \R$ must contain a singularity where $h$ tends to $-\infty, +\infty$, respectively.
	\item $\{h>0\}$ is a connected set: any connected component must include $a_1$. Its boundary includes $\partial \Omega$ since, if there exists a boundary point near which $h\leq 0$, we would have $f \equiv 0$ (see Remark \ref{rem:h-properties} and the proof of Lemma \ref{lem:uniqueness}).
	\item Therefore, the connected components of  $\{h<0\}$ are all simply connected, each containing one or more of $a_2, \ldots, a_n$, since any hole would be a component of $\{h>0\}$.
	\item There is $K= C(\kappa, \Omega, \nu)\left[\left(\frac{\operatorname{dist}(a,\partial \Omega)}{\delta_1}\right)^{\nu} +1\right]>0$, such that $\{h>K\}$ is a simply connected set. To see this, consider a smooth crosscut $\omega \subset \Omega$ dividing $\Omega$ into one component $\Omega_1$ containing $a_1$ and the other $\Omega_2$ containing all others. We may take $\omega$ so that its minimum distance to $a_1$ is at least $\delta_1$
    . Let $K = \max_\omega h_1$, which satisfies the bound of Lemma \ref{lem:onespin}, then $\{h_1 > K\}$ is contained within $\Omega_1$, since $h_1$ must be less than $K$ in $\Omega_2$. Then, because $h \leq h_1$,  we know that $\{h > K\} \subset \{h_1 > K\}\subset \Omega_1$. Since any hole surrounded by $\{h > K\}$ has to contain some other $a_j$, $\{h > K\}$ cannot have any hole.
	\item The boundaries of these simply connected sets, i.e., level sets of the form $\{h = r\}$, are all Lipschitz and piecewise differentiable, since its tangent direction is explicitly given by the direction of $\overline{f^2}$. Zeroes of $f$ are isolated and the level lines bifurcate at angles greater than zero, since $f$ vanishes like a polynomial to the highest order, thanks to Theorem~\ref{thm:similarity-principle}.
\end{itemize}

Given this, define $O_j \subset \Omega$ as the connected component of $\{h<0\}$ containing $a_j$ for $j\geq 2$. If $b(a_j)= 0$ there is nothing to estimate, so assume $b_j:=b(a_j)\neq 0$.

 Define $f_j(z) := \frac{\langle \mu_{a_1} \psi_z\rangle_{-\alpha;O_j}}{\langle \sigma_{a_1}\rangle_{-\alpha;O_j}}$, i.e., the correlation of the $(-\alpha)$-massive model on $O_j$. It is easy to verify that $-if/b_j$ is $(-\alpha)$-massive as well, with $\Im [\int (-if/b_j)^2 dz] = -h/(b_j)^2$, so we may compare $-h/(b_j)^2$ with $h_j := \Im [\int f_j^2 dz]$ within $O_j$. We claim $-h/(b_j)^2 \leq h_j$: both are $0$ on $\partial O_j$ and $-h/(b_j)^2 - h_j$ cannot have a bulk minimum, since the singularities of $-if/b_j, f_j$ at $a_j$ cancel out as in \eqref{eq:singularitypm} and the difference could only blow-up to $+\infty$ at some other $a_{j'}\in O_j$, if any. By Lemma \ref{lem:onespin}, this implies that (note $\Vert \alpha \Vert_{L^2(O_j)} \leq \Vert \alpha \Vert_{L^2(\Omega)} \leq \kappa$):
 
 \begin{equation*}
 	c(\kappa) \leq \int_{\partial O_j} -\partial_n h_j(z) |dz| \leq \int_{\partial O_j} \partial_n h/(b_j)^2 |dz|.
 \end{equation*}

Finally, the divergence theorem on the function $(h+1)^2$ and the domain ${\{0<h<K\}}$ gives (note minus signs in front of inner boundary from $\partial_n$):
\begin{align*}
&\int_{\{0<h<K\}} \Delta (h+1)^2(z) d^2z \\
&\quad= \int_{\partial \Omega} \partial_n (h+1)^2(z) |dz| - \int_{\partial\{h > K\}} \partial_n (h+1)^2(z) |dz| - \int_{\partial\{h < 0\}} \partial_n (h+1)^2(z)| dz|.    
\end{align*}
Let us estimate each term separately.

Since $\Delta (h+1)^2 = 2(h+1)\Delta h + 2|\nabla h|^2 = -8(h+1)\alpha|\nabla h| + 2|\nabla h|^2 \geq -32(h+1)^2\alpha^2$, we have
$$
\int_{\{0<h<K\}} \Delta (h+1)^2(z) d^2z \geq -32(K+1)^2 \kappa^2.
$$
By the boundary condition
$$
\int_{\partial \Omega} \partial_n (h+1)^2(z) |dz| =  \int_{\partial \Omega} 2\partial_n h(z) |dz| \leq 0. 
$$
Note that $f (z) = \frac{\langle \mu_{a_1} \psi_z\rangle_{\alpha;\{h > K\}}}{\langle \sigma_{a_1}\rangle_{\alpha;\{h > K\}}}$ on $\{h > K\}$ ($h\vert_{\{h > K\}}$ satisfies (D) and (RH), cf. Remark \ref{rem:h-properties}). Therefore, by Lemma \ref{lem:onespin}
$$
-\int_{\partial\{h >K \}} \partial_n (h+1)^2 (z)|dz| =- \int_{\partial\{h >K \}} 2\partial_n h (z)|dz| \leq C(\kappa).
$$
Since $\partial_n h \geq 0$ on $\partial\{h < 0\}$,
$$
 \int_{\partial\{h < 0\}} \partial_n (h+1)^2 (z)|dz| = \int_{\partial\{h < 0\}} 2\partial_n h (z)|dz| \geq \int_{\partial O_j} 2\partial_n h(z) |dz| \geq 2b_j^2 c(\kappa, |\Omega|).
$$
Combining these with the divergence theorem above, we get the desired estimate.
\end{proof}
\begin{corollary}\label{cor:dLp}
	In the same setting as Proposition \ref{prop:thebound}, we have
	$$
	|f(z)| \leq C(\kappa_\infty, n, \Omega)\bigg(P_{a_1}(z)+ \sum_{ k=2}^n |b_k|P_{a_k}(z)\bigg),
	$$
    for constants $b_k:= b(a_k;f)=i\beta(a_k;f)$ which satisfy
    $$
    |b_k| = \bigg| \frac{\langle  \sigma_{a_1}\stackrel{\hat{1,k}}{\cdots}\sigma_{a_n}\mu_{a_1}\mu_{a_k}\rangle_{\alpha}}{\langle \sigma_{a_1}\cdots \sigma_{a_n}\rangle_{\alpha}}\bigg|\leq C(\kappa_\infty, n, \Omega, {\delta_0})\min\left(\sqrt{\operatorname{dist}(a_1,\partial\Omega)},\sqrt{\operatorname{dist}(a_k,\partial\Omega)}\right).
    $$
\end{corollary}
\begin{proof}
	Consider the decomposition $f = e^s \underline{f}$ from Theorem \ref{thm:similarity-principle}. The holomorphic part $\underline{f}$ is in the space $V_\Omega^0(a_1, ..., a_n)$, mapped to the vector $(\Re [e^{-s(a_1)}], \Re [ib(a_2)e^{-s(a_2)}], ..., \Re [ib(a_n)e^{-s(a_n)}])$ under the map \eqref{eq:VtoRd}. Note that $e^{-s}$ is globally bounded by $C(\kappa_q,q, \Omega)$ by \eqref{eq:similarity-bound}, \eqref{eq:sfconfcov}.
    
    In other words, we have
    \begin{multline}
    \underline{f}(z) = \Re [e^{-s(a_1)}]\frac{\langle \sigma_{a_2}\cdots\sigma_{a_n} \mu_{a_1}\psi_z\rangle_{0}}{\langle \sigma_{a_1}\cdots \sigma_{a_n}\rangle_{0}} + \Re [ib(a_2)e^{-s(a_2)}]\frac{\langle \sigma_{a_1}\sigma_{a_3}\cdots\sigma_{a_n} \mu_{a_2}\psi_z\rangle_{0}}{\langle \sigma_{a_1}\cdots \sigma_{a_n}\rangle_{0}} \label{eq:massive-decomp} \\ + \dots +\Re [ib(a_n)e^{-s(a_n)}]\frac{\langle \sigma_{a_1} \cdots\sigma_{a_{n-1}}\mu_{a_n}\psi_z\rangle_{0}}{\langle \sigma_{a_1}\cdots \sigma_{a_n}\rangle_{0}},
    \end{multline}
    where the disorder-fermion correlations satisfy the bound coming from Lemma \ref{lem: df-critical-bound} and \eqref{eq:similarity-bound} finishes the proof of the first estimate. In particular, along with Proposition \ref{prop:thebound}, we know that $\Vert f\Vert_{L^{3/2}(\Omega)} \leq C(\kappa_\infty, n, \Omega, {\delta_1})$.

    Consider the \emph{critical} correlation
    $$
    f_{k;0}(z) :=\frac{\langle \sigma_{a_1}\stackrel{\hat{k}}{\cdots}\sigma_{a_n}\mu_{a_k} \psi_z\rangle_{0}}{\langle \sigma_{a_1}\cdots \sigma_{a_n}\rangle_{0}},
    $$
    and compute the imaginary part of the bilinear form $ GR_0(f,f_{k;0})$ (i.e. LHS of \eqref{eq:gr}):
    \begin{align*}
    2\pi(b(a_k; f) + b(a_1;f_{k;0})) = -2 \Im \bigg[\int_\Omega \alpha \bar f f_{k;0} d^2z\bigg] &\leq 2\kappa_\infty\Vert f\Vert_{L^{3/2}(\Omega)}\Vert f_{k;0}\Vert_{L^{3}(\Omega)}\\
    &\leq C(\kappa_\infty, n, \Omega, {\delta_1})\sqrt{\operatorname{dist}(a_k,\partial\Omega)},
    \end{align*}
    where we used Lemma \ref{lem: df-critical-bound} to bound $f_{k;0}$. The critical coefficient $b(a_1;f_{k;0})$ is bounded by $C(\Omega, {\delta_1})\sqrt{\operatorname{dist}(a_k,\partial\Omega)}$ also by Lemma \ref{lem: df-critical-bound}. We may replace $a_k$ by $a_1$ using the (anti)symmetry in $b_k$ (by Proposition \ref{prop:well-def}), at the cost of replacing $\delta_1$ by $\delta_0$.
\end{proof}
    
Using the same decomposition argument, we get the following bound.
\begin{corollary}\label{cor:Lp}
	Let $\Omega$ be smooth and $f(z) =\frac{\langle \sigma_{a_1} \sigma_{a_2}\cdots\sigma_{a_n}\psi_z\psi_w^{[\eta]}\rangle_{\alpha}}{\langle \sigma_{a_1}\cdots \sigma_{a_n}\rangle_{\alpha}}$. Then we have 
        \begin{multline}\label{eq:massive-ff}
           |f(z)| \leq C(\kappa_\infty, n, \Omega, \delta_0) |\eta| \Bigg( \sum_{j=1}^n \left[\sum_{k=1}^n\min\left(\frac{1}{\sqrt{|w-a_k|}}, \frac{\sqrt{\operatorname{dist}(a_j,\partial\Omega)}}{|w-a_k|}\right)\right] P_{a_j}(z) \\ 
           + \Bigg[ \frac{1}{\sqrt{\operatorname{dist}(w,\partial\Omega)}}
           +\sum_{j=1}^n \frac{1}{\sqrt{|w-a_j|}} \Bigg] P_w(z) + \frac{1}{|z-w|} \Bigg).
        \end{multline}
\end{corollary}
\begin{proof}
In this case, we have
\begin{equation*}
    b(a_j;f) = \frac{\langle  \sigma_{a_1}\stackrel{\hat{j}}{\cdots}\sigma_{a_n}\mu_{a_j}\psi_w^{[\eta]}\rangle_{\alpha}}{\langle \sigma_{a_1}\cdots \sigma_{a_n}\rangle_{\alpha}}.
\end{equation*}
Let us consider the analog of \eqref{eq:massive-decomp} in this case. In addition to terms appearing in \eqref{eq:massive-decomp}, we have an additional term in the decomposition
$$
\frac{\langle \sigma_{a_1} \sigma_{a_2}\cdots\sigma_{a_n}\psi_z\psi_w^{[e^{\overline{s(w)}}\eta]}\rangle_{0}}{\langle \sigma_{a_1}\cdots \sigma_{a_n}\rangle_{0}},
$$
which is the only term which gives rise to the two latter terms centered at $w$ in \eqref{eq:massive-ff}, by Proposition \ref{prop:ff-crit-estimate}. Let us carefully collect the contributions to the $P_{a_j}(z)$ term: we have contributions from
\begin{itemize}
    \item $\Re [ib(a_j)e^{-s(a_j)}]\frac{\langle \sigma_{a_1}\stackrel{\hat{j}}{\cdots}\sigma_{a_n} \mu_{a_j}\psi_z\rangle_{0}}{\langle \sigma_{a_1}\cdots \sigma_{a_n}\rangle_{0}}$, which by Theorem \ref{thm:similarity-principle} and Lemma \ref{lem: df-critical-bound} contributes
    $$
    C(\kappa_\infty,n,\Omega)\Bigg| \frac{\langle  \sigma_{a_1}\stackrel{\hat{j}}{\cdots}\sigma_{a_n}\mu_{a_j}\psi_w^{[\eta]}\rangle_{0}}{\langle \sigma_{a_1}\cdots \sigma_{a_n}\rangle_{0}} \Bigg|P_{a_j}(z),
    $$
    which is bounded by the first term in \eqref{eq:massive-ff} by Lemma \ref{lem: df-critical-bound};
    \item $\Re [ib(a_k)e^{-s(a_k)}]\frac{\langle \sigma_{a_1}\stackrel{\hat{k}}{\cdots}\sigma_{a_n} \mu_{a_k}\psi_z\rangle_{0}}{\langle \sigma_{a_1}\cdots \sigma_{a_n}\rangle_{0}}$ for $j\neq k$, which by Lemma \ref{lem: df-critical-bound} contributes
    $$
     C(\kappa_\infty,n,\Omega)\Bigg| \frac{\langle  \sigma_{a_1}\stackrel{\hat{k}}{\cdots}\sigma_{a_n}\mu_{a_k}\psi_w^{[\eta]}\rangle_{0}}{\langle \sigma_{a_1}\cdots \sigma_{a_n}\rangle_{0}} \Bigg|P_{a_j}(z),
    $$
    which is bounded by the first term in \eqref{eq:massive-ff} by Lemma \ref{lem: df-critical-bound};
    \item $\frac{\langle \sigma_{a_1} \sigma_{a_2}\cdots\sigma_{a_n}\psi_z\psi_w^{[e^{\overline{s(w)}}\eta]}\rangle_{0}}{\langle \sigma_{a_1}\cdots \sigma_{a_n}\rangle_{0}}$, which by Proposition \ref{prop:ff-crit-estimate} contributes
    $$
    C(n,\Omega)\Bigg| \frac{\langle  \sigma_{a_1}\stackrel{\hat{j}}{\cdots}\sigma_{a_n}\mu_{a_j}\psi_w^{[e^{\overline{s(w)}}\eta]}\rangle_{0}}{\langle \sigma_{a_1}\cdots \sigma_{a_n}\rangle_{0}}\Bigg|P_{a_j}(z),
    $$
    which is bounded by the first term in \eqref{eq:massive-ff} by Theorem \ref{thm:similarity-principle} and Lemma \ref{lem: df-critical-bound}.
\end{itemize}
\end{proof}

\subsection{Logarithmic derivatives and series expansions}\label{subsec:series-expansion}
\subsubsection{Second order estimates}
Extracting the logarithmic derivative takes more care. To discuss higher order singularities, we define the bigger space $V_{\Omega;\nu}^\alpha \supset V_{\Omega;-\frac12}^\alpha = V_\Omega^\alpha$ by replacing the order $1/2$ in \eqref{eq:v12} by a half-integer $-\nu \geq 1/2$. We have the following explicit decomposition of spinors in this space:
\begin{lemma}\label{lem:3/2-decomp}
    Let $\Omega$ be smooth. Suppose $f\in V_{\Omega;\nu}^\alpha(a_1, \ldots, a_n)$. Then, as $z\to a_j$, for some complex numbers $o_{\nu+l}=o_{\nu+l}(a_j) =o_{\nu+l}(a_j;f)$, $l=0, 1, 2$, we have the asymptotic expansions: defining
    \begin{align*}
    f^{\dagger0}(z) :=&\, f(z),\\ 
    f^{\dagger1}(z) :=&\, f(z)  + e^{-\frac{i\pi}{4}}\frac{\alpha(a_j)\overline{o_{\nu}}}{\nu+1}(\overline{z-a_j})^{\nu+1}+ e^{-\frac{i\pi}{4}}o_{\nu}\cdot(z-a_j)^{\nu}, \\ 
    f^{\dagger2}(z) :=& f^{\dagger1}(z) +  e^{-\frac{i\pi}{4}}    \overline{o_\nu}\cdot\left[\frac{\partial_z\alpha(a_j)}{\nu+1}\cdot(z-a_j)\overline{  ({z-a_j})^{\nu+1}} + \frac{\partial_{\bar z}\alpha(a_j)}{\nu+2}\cdot\overline{  ({z-a_j})^{\nu+2}} \right]\\ 
    &\qquad\quad+\,e^{-\frac{i\pi}{4}}\alpha(a_j)\cdot{  \left(\frac{\alpha(a_j){o_{\nu}}}{\nu+1}({z-a_j})^{\nu+1}\overline{(z-a_j)}+\frac{{\overline{o_{\nu+1}}}}{\nu+2}{(\overline{z-a_j})}^{\nu+2}\right)} \\
    &\qquad\quad+ e^{-\frac{i\pi}{4}} o_{\nu+1}\cdot(z-a_j)^{\nu+1},
    \end{align*}
we have
    \begin{align*}
    f^{\dagger0}(z) =& - e^{-\frac{i\pi}{4}}o_{\nu}\cdot(z-a_j)^{\nu} + O(|z-a_1|^{\nu+\frac34}),\\ 
    f^{\dagger1}(z)=&\, -e^{-\frac{i\pi}{4}}o_{\nu+1}\cdot(z-a_j)^{\nu+1} + O(|z-a_j|^{\nu+\frac{13}{8}}),\\ 
   f^{\dagger2}(z) =&\,  -e^{-\frac{i\pi}{4}}o_{\nu+2}\cdot(z-a_j)^{\nu+2}  + o(|z-a_j|^{\nu+2}).
    \end{align*}
   Note the exponents in the two big $O$ terms may be improved to the respective next smallest integers \emph{a posteriori}.
   
    We also use the following notation for the special cases
    $$o_{-\frac12}(a_j;f)=\beta(a_j;f) = \beta(z=a_j;f(z))\quad\text{ and }\quad o_{\frac12}(a_j;f)=\lambda(a_j;f) = \lambda(z=a_j;f(z)).$$

    In addition, the collection of $o_{\nu+l}(a_j;f)$ for $\nu + l<-1/2$ and $\Re [\beta(a_j;f)]$ for all $a_j$ together uniquely determine $f$.
\end{lemma}
\begin{proof}
    From Theorem \ref{thm:similarity-principle}, we already know that $f(z) = - e^{-\frac{i\pi}{4}}o_{\nu} \cdot (z-a_j)^{\nu} + O(|z-a_j|^{\nu+\frac34})$ for some $o_{\nu}$. 
    
    Then $f^{\dagger1}(z)$ is a spinor which is itself $O(|z-a_j|^{\nu+\frac34})$ and $\partial_{\bar{z}}f^{\dagger1}(z) = -i\alpha(z) \overline{f(z)} + e^{-\frac{i\pi}{4}}\overline{o_\nu}\cdot\alpha(a_j)\cdot (\overline{z-a_j})^{\nu} = O(|z-a_j|^{\nu+\frac34})$ for $z\neq a_j$. Then we see that $\partial_{\bar z}\left[\frac{f^{\dagger1}(z)}{(z-a_j)^{\nu+1}}\right] =\frac{\partial_{\bar z}\left[f^{\dagger1}(z)\right]}{{(z-a_j)^{\nu+1}}} =  O(|z-a_j|^{-1/4})$ and the $L^2$ function $\frac{f^{\dagger1}(z)}{(z-a_j)^{\nu+1}}$ is in fact $\frac{5}{8}$-H\"older continuous in a neighborhood of $a_j$ by Corollary \ref{cor:dzbar-solution-bound}.

    Then, knowing $f^{\dagger1}(z) = - e^{-\frac{i\pi}{4}}o_{\nu+1}\cdot (z-a_j)^{\nu+1} + O(|z-a_j|^{\nu+\frac{13}{8}})$ for some $o_{\nu+1}$, we may play the same game with the spinor $f^{\dagger2}(z)$ and the function $\frac{f^{\dagger2}(z)}{(z-a_j)^{\nu+2}}= O(|z-a_j|^{-\frac{3}{8}})$ (i.e. a priori in $L^2$). For this purpose, we have the sufficient bound
    \begin{align*}
    {\partial_{\bar z}\left[f^{\dagger2}(z)\right]} =&\,{-i\alpha(z) \overline{f(z)} + e^{-\frac{i\pi}{4}}\left[\partial_z\alpha(a_j)\cdot(z-a_j) + \partial_{\bar z}\alpha(a_j)\cdot\overline{(z-a_j)} \right]\cdot\overline{ {(o_\nu}\cdot ({z-a_j})^{\nu})}}\\
    &+{e^{-\frac{i\pi}{4}}\alpha(a_j)\cdot\overline{\left({o_\nu}\cdot ({z-a_j})^{\nu}+\frac{\alpha(a_j)\overline{o_{\nu}}}{\nu+1}(\overline{z-a_j})^{\nu+1}+o_{\nu+1}\cdot(z-a_j)^{\nu+1}\right)}}\\
    =&\, O(|z-a_j|^{\nu+\frac{13}{8}}).
    \end{align*}

    If $f_1, f_2$ have the same collection of $o_{\nu+l}(a_j;f)$ for $\nu + l<-1/2$ and $\Re [\beta(a_j;f)]$, simply note that $f_1 - f_2$ is an element of $V_\Omega^\alpha$ which is taken to zero under the map \eqref{eq:VtoRd} and is thus zero.
\end{proof}

\begin{remark}
    The coefficients $o_\nu, o_{\nu+1}, o_{\nu+2}$ from Lemma \ref{lem:3/2-decomp} can be extracted using a Cauchy-type formula against holomorphic spinors $z^{-\nu-1}, z^{-\nu-2}, z^{-\nu-3}$, respectively: for example, from applying a Green-Riemann's formula in a small annulus $B_r(a_j)\setminus B_t(a_j)$ then sending $t\downarrow 0$, we have for $l=0, 1, 2$
    \begin{align}\label{eq:cauchy-hol}
    o_{\nu+l} = \frac{1}{2\pi i}\oint_{\partial B_r(a_j)} f^{\dagger l}(z)(z-a_j)^{-\nu-l-1}dz-2\int_{B_r(a_j)}\alpha(z)\overline{f^{\dagger l}(z)}(z-a_j)^{-\nu-l-1}d^2z.
    \end{align}
\end{remark}

In fact, similar ideas allow us to give a uniform bound:

\begin{lemma}\label{lem:halforderexpansion-zeroth}
	Suppose $\Omega$ is smooth, and let $f(z)$ be either  $\frac{\langle  \sigma_{a_2}\cdots\sigma_{a_n}\mu_{a_1}\psi_z\rangle_{\alpha}}{\langle \sigma_{a_1}\cdots \sigma_{a_n}\rangle_{\alpha}}$ or $\frac{\langle \sigma_{a_1}\cdots\sigma_{a_n}\psi_z\psi_w^{[\eta]}\rangle_{\alpha}}{\langle \sigma_{a_1}\cdots \sigma_{a_n}\rangle_{\alpha}}$. If $0 < 2d \leq \min(\operatorname{dist}(a_j,\partial\Omega),\delta_0)$ and $d_w = \min_{k=1,2,\ldots,n} |w - a_k|$, there is a constant $C = C(\Omega,\kappa_\infty,n,\delta_0, \kappa_b)$ such that for $z\in B_{d/4}(a_j)$
    \begin{align}\label{eq:uniform-bound}
    &\left|f(z) + e^{-i\pi/4}\beta (a_j)\cdot (z-a_j)^{-1/2} + 2e^{-i\pi/4}\alpha(a_j)\overline{\beta (a_j)}\cdot (\overline{z-a_j})^{1/2}\right|\\ \nonumber
    &\leq C\cdot\big(d^{-1}|\beta (a_j)| + |\eta|d^{-1/2}d_w^{-1/2}\big){|z-a_j|^{1/2} }{|z-w|^{-1}},
    \end{align}
   where we omit the $\eta$ term and $|z-w|^{-1}$ factor in the case of the disorder-fermion correlation.
   
   See also Corollaries \ref{cor:dLp} and \ref{cor:Lp} for estimates of $\beta(a_j)$.
\end{lemma}
\begin{proof}
        Let us handle the case of $f(z) = \frac{\langle \sigma_{a_1}\cdots\sigma_{a_n}\psi_z\psi_w^{[\eta]}\rangle_{\alpha}}{\langle \sigma_{a_1}\cdots \sigma_{a_n}\rangle_{\alpha}}$; the other case is simpler.
        
        Write $f^w(z):=(z-w)f(z)$ and note that in $B_d(a_j)$, by Corollary \ref{cor:Lp},
        \begin{align}\label{eq:fw-near}
            \left|\sqrt{z-a_j}\cdot f^w(z)\right| &\leq C(\Omega,\kappa_\infty,n,\delta_0)\big(|\beta (a_j)||w-z |+ |\eta|d^{1/2}d_w^{-1/2}\big),
        \end{align}
        so
        $$
        \left|\partial_{\bar{z}}(\sqrt{z-a_j}\cdot f^w(z))\right| =\left|\sqrt{z-a_j}\cdot \partial_{\bar{z}}f^w(z)\right| \leq C(\Omega,\kappa_\infty,n,\delta_0)\big(|\beta (a_j)||w-z| + |\eta|d^{1/2}d_w^{-1/2}\big),
        $$
        which yields, with Corollary \ref{cor:dzbar-solution-bound},
        \begin{equation}\label{eq:holder-1}
        \Vert \sqrt{z-a_j}\cdot  f^w(z)\Vert_{C^{3/4}(B_{d/2}(a_j))}\leq C(\Omega,\kappa_\infty,n,\delta_0)\big(|\beta (a_j)| + |\eta|d^{1/2}d_w^{-1/2}\big)d^{-3/4}.
        \end{equation}
        
        Write $f_b(z)=-e^{-\frac{i\pi}{4}}\beta (a_j)\cdot (z-a_j)^{-\frac12} -  2e^{-\frac{i\pi}{4}}\alpha(a_j)\overline{\beta (a_j)}\cdot (\overline{z-a_j})^{\frac12}$ and $f_b^w(z):=(z-w)f_b(z)$. In fact, $\frac{f^w(z)-f^w_b(z)}{\sqrt{z-a_j}}$ is single-valued and is in $L^2(B_{d/2}(a_j))$: compute 
        \begin{align*}
            &\left| \partial_{\bar{z}}\frac{f^w(z)-f^w_b(z)}{\sqrt{z-a_j}}\right| \\ \nonumber
            &= \Bigg| \frac{-i\alpha(z)\overline{f(z)}+e^{-\frac{i\pi}{4}}\alpha(a_j)\overline{\beta (a_j)}\cdot (\overline{z-a_j})^{-\frac12}}{(z-w)^{-1}\sqrt{z-a_j}}\Bigg|\\ \nonumber
            &=\Bigg| \frac{\alpha(z)(\overline{f(z)+e^{-\frac{i\pi}{4}}\beta (a_j)\cdot ({z-a_j})^{-\frac12})}+(\alpha(z)-\alpha(a_j))\overline{e^{-\frac{i\pi}{4}}\beta (a_j)\cdot ({z-a_j})^{-\frac12})}}{(z-w)^{-1}\sqrt{z-a_j}}\Bigg|\\ \nonumber
            &\leq \Bigg| \frac{\alpha(z)(\overline{({z-a_j})^{\frac12}f^w(z)+e^{-\frac{i\pi}{4}}\beta (a_j)\cdot(z-w))}}{|{z-a_j}|}\cdot\frac{z-w}{\overline{z-w}}\Bigg|+\left| \frac{(\alpha(z)-\alpha(a_j))\overline{\beta (a_j))}}{(z-w)^{-1}(z-a_j)}\right|\\ \nonumber
            &\leq C(\Omega,\kappa_\infty,n,\delta_0)\big(|\beta (a_j)| + |\eta|d^{1/2}d_w^{-1/2}\big)\left(d^{-3/4}|z-a_j|^{-1/4} +\Vert \nabla \alpha \Vert_{L^\infty(B_{d/2}(a_j))} \right),
        \end{align*}
        where we used \eqref{eq:holder-1} on the first term (note that $({z-a_j})^{\frac12}f^w(z)+e^{-\frac{i\pi}{4}}\beta (a_j)(z-w) = 0$ at $z=a_j$ and the latter term also trivially satisfies a H\"older bound of the form \eqref{eq:holder-1}).

         Then we may use Corollary \ref{cor:dzbar-solution-bound} with $q=3$ and \eqref{eq:fw-near} to say that, in $B_{d/2}(a_j)$, the modulus of $ \frac{f^w(z)-f^w_b(z)}{\sqrt{z-a_j}}$ in $B_{d/4}(a_j)$ is at most
        \begin{align*}
            C(\Omega,\kappa_\infty,n,\delta_0)\big(|\beta (a_j)| + |\eta|d^{1/2}d_w^{-1/2}\big)\left(d^{-1} + d^{1/3}\big(d^{-3/4+5/12} +d^{2/3}\Vert \nabla \alpha \Vert_{L^\infty(B_{d/2}(a_j))}\big) \right),
        \end{align*}
        which fits into the desired form of estimate.
\end{proof}

\begin{remark}\label{rem:logarithmic-correction}
    A much simplified version of the argument as in the proof of Lemma \ref{lem:halforderexpansion-zeroth} easily yields, for $z \in B_{\delta_w/2}(w)$,
    \begin{align}\label{eq:logarithmic-correction}
        \Bigg|\frac{\langle \sigma_{a_1}\cdots\sigma_{a_n}\psi_z\psi_w^{[\eta]}\rangle_{\alpha}}{\langle \sigma_{a_1}\cdots \sigma_{a_n}\rangle_{\alpha}} - \frac{\bar \eta}{z-w} +2i \eta\cdot \alpha(w)\cdot \log|z-w| \Bigg| \leq C(\Omega,\kappa_\infty,n,\delta_0, \kappa_b, \delta_w).
    \end{align}
That is, for $C=C(\Omega,\kappa_\infty,n,\delta_0, \kappa_b, \delta_w)$, we have that
\begin{align*}
f^{\dagger}(z)=\frac{\langle \sigma_{a_1}\cdots\sigma_{a_n}\psi_z\psi_w^{[\eta]}\rangle_{\alpha}}{\langle \sigma_{a_1}\cdots \sigma_{a_n}\rangle_{\alpha}} - \frac{\bar \eta}{z-w} +2i \eta\cdot \alpha(w)\cdot \log|z-w|
\end{align*}
satisfies
\begin{align*} 
\big|\partial_{\bar z} f^{\dagger}(z)\big|= \Bigg|-i{\alpha (z)\overline{\frac{\langle \sigma_{a_1}\cdots\sigma_{a_n}\psi_z\psi_w^{[\eta]}\rangle_{\alpha}}{\langle \sigma_{a_1}\cdots \sigma_{a_n}\rangle_{\alpha}}}} +i\alpha(w)\overline{ \frac{\bar\eta}{{z-w}}}\Bigg|\leq C\cdot|z-w|^{-1/4},
\end{align*}
with identical bound on $f^{\dagger}(z)$ itself, implies that $f^{\dagger}$ is continuous and satisfies the above bound by Corollary \ref{cor:dzbar-solution-bound}.
\end{remark}

\begin{proposition}\label{prop:spindiff}
	Suppose $\Omega$ is smooth. Then the correlations defined in Definition \ref{def:correlations-definition} are twice differentiable under movements of spin and disorder insertions. The asymptotic expansions justified in Lemma \ref{lem:3/2-decomp} (e.g. \eqref{eq:def-A}) may be differentiated term-by-term to uniquely identify the derivative.
\end{proposition}
\begin{proof}
    First, take the case of $f(a_1;z) := \frac{\langle \sigma_{a_2}{\cdots}\sigma_{a_n}\mu_{a_1}\psi_z\rangle_{\alpha}}{\langle \sigma_{a_1}\cdots \sigma_{a_n}\rangle_{\alpha}}$. For small $t>0$, consider without loss of generality (see points (1) and (2) below for general settings) $d_t(z): = f(a_1 + t;z) - f(a_1;z)$, which is a massive holomorphic spinor defined on a common double cover of $\Omega \setminus \left(\{a_2, \ldots, a_n\} \cup B_{2t}(a_1)\right)$.

    Write $f^{[w^{\eta}]}(z) := \frac{\langle \sigma_{a_1}{\cdots}\sigma_{a_n}\psi_{z}\psi^{[{\eta}]}_w\rangle_{\alpha}}{\langle \sigma_{a_1}\cdots \sigma_{a_n}\rangle_{\alpha}}$ and consider the following variant of $\Im [GR_{2t}(d_t, f^{[w^{\eta}]})]$ (note from \eqref{eq:gr} that the other contributions are zero) for  $0<4t<t_1\ll 1$ and $0<t_2$:
    \begin{equation}\label{eq:fw-im}
    \Im \!\left[\oint_{\partial B_{t_1}(a_1)} \!d_t \cdot f^{[w^{\eta}]} dz + \oint_{\partial B_{t_2}(a_2)\cup\cdots\cup \partial B_{t_2}(a_n)} \!d_t \cdot f^{[w^{\eta}]} dz + \oint_{\partial B_{t_2}(w)}\!d_t\cdot f^{[w^{\eta}]} dz\right] \!= 0,
    \end{equation}
    where, letting $t_2\downarrow 0$, the second term becomes purely real by (D) and the third term tends to $2\pi i \bar{\eta} d_t(w)$. Therefore, taking $\eta = 1, i$, we can bound the imaginary and real parts of $d_t(w)$, respectively, using $\Im [\oint_{B_{t_1}(a_1)} d_t\cdot f^{[w^{\eta}]} dz]$.
    
    In particular, if we divide $d_t$ by a renormalization factor $M_t >0$ such that $\int_{\partial B_{t_1}(a_1)} \frac{|d_t|}{M_t} |dz| \equiv 1$, we get a subsequential $\alpha$-massive limit $g$ to which $d_t/M_t$ converges uniformly on compact subsets of $\Omega\setminus (B_{t_1}(a_1)\cup\{a_2, \ldots, a_n\})$ as $t \downarrow 0$ (see Remark \ref{rem:dzbar-massive-holomorphicity}) with $\int_{\partial B_{t_1}(a_1)} |g| |dz| = 1$. In addition, since $d_t/M_t$ has the (S) asymptotic at every $a_2, \ldots, a_n$ and the (RH) boundary condition, $g$ has (S) asymptotic at those points also (e.g. use dominated convergence on \eqref{eq:cauchy-hol}) and the (RH) boundary condition. Also note that the convergence $f(a_1 + t; \cdot) \to f(a_1; \cdot)$ as $t\downarrow 0$, uniformly on compact subsets of $\Omega\setminus \{a_1, \ldots, a_n\}$, may be done using similar precompactness arguments (i.e. use the uniform-on-compacts bound from Corollary \ref{cor:dLp} to exhibit a subsequential limit; by the same dominated convergence argument, the limit must have a (D) asymptotic at $a_1$, (S) at $a_2, \ldots, a_n$, and the (RH) boundary condition). 

    We first show that $M_t = O(t)$. For contradiction, assume $M_t/t \to \infty$, possibly along a subsequence. For small enough $t_1$, we consider instead
    $$
    d_t^\dagger = \left(f(a_1 + t;z) +e^{-\frac{i\pi}{4}}(z-a_1-t)^{-\frac12}\right)  - \left( f(a_1;z) +e^{-\frac{i\pi}{4}}(z-a_1)^{-\frac12} \right)=: d_t - D_t,
    $$
    which satisfies $M_t^{-1}d_t^\dagger \to g$ on $\Omega \setminus  B_{t_1}(a_1)$, since $M_t^{-1}D_t \to 0$ uniformly on compact subsets under the assumption. Instead of $f^{[w^{\eta}]}(z) = \frac{\langle \sigma_{a_1}{\cdots}\sigma_{a_n}\psi_{z}\psi^{[{\eta}]}_w\rangle_{\alpha}}{\langle \sigma_{a_1}\cdots \sigma_{a_n}\rangle_{\alpha}}$, take 
    a spinor $g^{[w^{\eta}]}$ which is $\alpha$-massive in $B_{2t_1}(a_1)\setminus \{w\}$ and satisfies
    $$
    g^{[w^{\eta}]}(z)\sim \frac{\bar \eta}{z-w} \quad \text{ and } \quad c(t_1, \kappa_\infty)\leq \frac{g^{[w^{\eta}]}(z)}{|\eta||z-a_1|^{1/2}|w-a_1|^{-1/2}|z-w|^{-1}} \leq C(t_1, \kappa_\infty).
    $$
    We will show the existence of such a spinor in Lemma \ref{lem:formalspinors}. Write $B_t := B_{2t_1}(a_1) \setminus B_{4t}(a_1)$, then we may verify the two-sided identity

    \begin{align*}
    &\Im \bigg[  \oint_{\partial B_{2t_1}(a_1)}d_t^\dagger \cdot g^{[w^{\eta}]} dz  - 2\pi i \bar{\eta} d_t^\dagger (w)\bigg] \\ \nonumber
    &=\lim_{t_2\downarrow 0} \,\Im \bigg[  \oint_{\partial B_{2t_1}(a_1)}d_t^\dagger \cdot g^{[w^{\eta}]} dz-\oint_{\partial B_{4t}(a_1)} d_t^\dagger \cdot g^{[w^{\eta}]} dz  - \oint_{\partial B_{t_2}(w)}d_t^\dagger \cdot g^{[w^{\eta}]} dz\bigg] \\ \nonumber
    &= \,\Im\left[2i\int_{B_t} \partial_{\bar{z}} \left(d_t^\dagger\cdot g^{[w^{\eta}]}\right) d^2z\right] \\ \nonumber
    &= \,\Im\left[2i\int_{B_t} \partial_{\bar{z}} \left(( - D_t)\cdot g^{[w^{\eta}]}\right) d^2z\right]\\ \nonumber 
    &=\,\Im\left[2\int_{B_t}\alpha\cdot {D_t}\cdot \overline{g^{[w^{\eta}]}}d^2z\right],
    \end{align*}
    where $0<4t < |w-a_1| < 2t_1$ and $d^\dagger_t(z) = O(|z-a_1|^{1/2})$ by Lemma \ref{lem:halforderexpansion-zeroth}. Since we also have $D_t = O(t|z-a_1|^{-3/2})$ in $B_t$, dividing by $M_t$ and applying (versions of) the dominated convergence theorem 
    (as $t\downarrow 0$), we see in fact by the definition of $g(w)$ that the convergence $M_t^{-1}d_t(w),M_t^{-1}d_t^\dagger(w) \xrightarrow{t\downarrow 0} g(w)$ may be extended to any fixed $w \in B_{2t_1}(a_1)\setminus\{a_1\}$, where we have
    \begin{equation}\label{eq:g-contradiction}
    \Im\left[2\pi i \bar{\eta} g(w)\right] = \Im \bigg[  \oint_{\partial B_{2t_1}(a_1)}g \cdot g^{[w^{\eta}]} dz\bigg]  = O(|w-a_1|^{-1/2}).
    \end{equation}
    Therefore, $g \in V_{\alpha;-\frac12}^\Omega$. Write $o(t):= o_{-\frac32}(a_1+t;f(a_1+t;z))\equiv0$ as integrals using \eqref{eq:cauchy-hol}, then note that 
    \begin{equation}\label{eq:expansion-differentiation}
        0\equiv M_t^{-1}\left( o(t)-o(0) \right) \xrightarrow{t\downarrow 0}o_{-\frac32}(a_1;g) - \tfrac12  o_{-\frac12}(a_1;f(a_1;\cdot)).
    \end{equation} Since $o_{-\frac32}(a_1;g)=0$ and $o_{-\frac12}(a_1;f(a_1;\cdot)) =1$, we have a contradiction, hence $M_t = O(t)$.
    
    Since there is still a subsequential $\alpha$-massive limit $g$ as $t\downarrow 0$ outside of $B_{t_1}(a_1)$, it would remain to show that we can still extend $g$ everywhere and uniquely characterize it by the term-by-term differentiation coming from \eqref{eq:def-A}. Inspecting the previous argument from \eqref{eq:fw-im}, we cannot disregard $t^{-1}D_t$ anymore: $t^{-1}d^\dagger_t(z)$ converges to $g^\dagger(z):=g(z)+ \frac{e^{\frac{i\pi}{4}}}{2(z-a)^{3/2}}$, in particular $g$ may be extended to $B_{2t_1}(a_1)\setminus \{a_1\}$. So, instead of \eqref{eq:g-contradiction}, we have
    $$
    \Im\left[2\pi i \bar{\eta} g^\dagger(w) \right] = \Im \Bigg[  \oint_{\partial B_{2t_1}(a_1)}g^\dagger(z) \cdot g^{[w^{\eta}]}(z) dz +2\int_{B_{2t_1}(a_1)}\alpha(z)\cdot \frac{e^{\frac{i\pi}{4}}\overline{g^{[w^{\eta}]}(z)}}{2(z-a)^{3/2}}d^2z\Bigg],
    $$
    which is still $O(|w-a_1|^{-1})$. In particular, $g^\dagger$ is strictly smaller than the explicit singularity $\frac{e^{\frac{i\pi}{4}}}{2(w-a)^{3/2}}$, and $g \in V_{\alpha;-\frac32}^\Omega$ (and, $g$ is not identically zero). The unique identification of the factors $o_{-\frac32}(a_1;g)$ and $o_{-\frac12}(a_1;g)$ as those coming from term-by-term differentiation can then be performed as in \eqref{eq:expansion-differentiation}.

    Now, we carefully track the ingredients in the previous argument to claim that one may carry out similar proofs on other derivatives. We had:
    \begin{enumerate}
        \item Explicit identification of the singular part $-e^{-\frac{i\pi}{4}}(z-a_1)^{-\frac12}$, for which differentiability can be shown independently;
        \item The kernels $f^{[w^{\eta}]}, g^{[w^{\eta}]}$, the latter having a $\frac12$-order zero at $a_1$, sufficient for dominated convergence against the $\frac32$-order pole coming from differentiation of the singular part.
    \end{enumerate}
    To extend the point (1) for other cases, we note:
    \begin{itemize}
    \item The singular part of $f(a_1;z)$ at $a_j$, $j>1$ is $b(a_j; f(a_1;\cdot ))\cdot e^{-\frac{i\pi}{4}}(z-a_1)^{-\frac12}$, so we need to show that $b(a_j; f(a_1;\cdot ))$ is differentiable as a function of $a_j$. By differentiating \eqref{eq:cauchy-hol} under the integral sign, we see that it is at least differentiable as a function of $a_1$. However, since $$b(a_j; f(a_1;\cdot ))= \frac{\langle  \sigma_{a_1}\stackrel{\hat{1,j}}{\cdots}\sigma_{a_n}\mu_{a_1}\mu_{a_j}\rangle_{\alpha}}{\langle \sigma_{a_1}\cdots \sigma_{a_n}\rangle_{\alpha}}$$
    by Proposition \ref{prop:well-def}, we see that it must be differentiable as a function of $a_j$ by symmetry.
    \item For the case $f(z) = \frac{\langle\sigma_{a_1}{\cdots}\sigma_{a_n}\psi_z\psi_w^{[\eta]}\rangle_{\alpha}}{\langle \sigma_{a_1}\cdots \sigma_{a_n}\rangle_{\alpha}}$, note that $d_t$ in this case does not have a pole at $w$ and we may use the same argument, given that the $b$-coefficients are given by
    $$
    b(a_k; f) = \frac{\langle  \sigma_{a_1}\stackrel{\hat{k}}{\cdots}\sigma_{a_n}\mu_{a_k}\psi_w^{[\eta]}\rangle_{\alpha}}{\langle \sigma_{a_1}\cdots \sigma_{a_n}\rangle_{\alpha}},
    $$
    again by Proposition \ref{prop:well-def}. This is differentiable as a function of $a_k$ by the previous proof.
    \item For the second derivatives, we use the first three terms from Lemma \ref{lem:3/2-decomp} as the explicit singular part. Here, we again start from differentiability under movements of disorder insertions, then establish differentiability of relevant singular coefficients, then carry out the same proof.
    \end{itemize}

    To address the point (2) for second derivatives, we need a counterpart of $g^{[w^{\eta}]}$ with $\frac32$-order zero at a given $a$. Therefore, we state and prove the following lemma for both cases.
    \begin{lemma}\label{lem:formalspinors}
        Let $l=\frac12,\frac32$. Then for any $\eta\in \C$ and $w\in B_{2t_1}(a)$, there exists some $g^{[w^{\eta}]}$ in $B_{2t_1}(a)$ such that 
    $$
    g^{[w^{\eta}]}(z)\sim \frac{\bar \eta}{z-w} \quad \text{ and } \quad c(t_1, \kappa_\infty)\leq \frac{g^{[w^{\eta}]}(z)}{\left|\frac{(z-a)^l\bar\eta}{(w-a)^l(z-w)}\right|} \leq C(t_1, \kappa_\infty).
    $$
    \end{lemma}
    \begin{proof}[Proof of Lemma \ref{lem:formalspinors}]
     Consider the holomorphic function $F(z) := \frac{(z-a)^l\bar\eta}{(w-a)^l(z-w)}$ and note that we need a version of Theorem \ref{thm:similarity-principle} where the boundary condition for $s$ is replaced by the condition $s(w)=0$: such a result is indeed known, see \cite[Theorem 3.13]{Vekua} or \cite[Theorem 3]{Bers-function}.
     \end{proof}
     This concludes the proof of Proposition~\ref{prop:spindiff}.
\end{proof}
A similar but simpler precompactness argument yields the following, noticing that the limit preserves the defining conditions (S), (D), (F), (RH) away from $a_n$. The convergence of the $\beta$, $\lambda$-coefficients is again clear from \eqref{eq:cauchy-hol}. We omit further details.
\begin{proposition}\label{prop:spin-boundary-limit}
Suppose $a_n$ is sent to some $a_b\in \partial \Omega$. Then $$\frac{\langle \sigma_{a_1}\cdots\sigma_{a_n}\psi_z\psi_w^{[\eta]}\rangle_{\alpha}}{\langle \sigma_{a_1}\cdots \sigma_{a_n}\rangle_{\alpha}} \hspace{-1pt}\to\hspace{-1pt} \frac{\langle \sigma_{a_1}\cdots\sigma_{a_{n-1}}\psi_z\psi_w^{[\eta]}\rangle_{\alpha}}{\langle \sigma_{a_1}\cdots \sigma_{a_{n-1}}\rangle_{\alpha}}\text{ and }\frac{\langle \sigma_{a_2}\cdots\sigma_{a_n}\mu_{a_1}\psi_z\rangle_{\alpha}}{\langle \sigma_{a_1}\cdots \sigma_{a_n}\rangle_{\alpha}} \hspace{-1pt}\to\hspace{-1pt} \frac{\langle \sigma_{a_2}\cdots\sigma_{a_{n-1}}\mu_{a_1}\psi_z\rangle_{\alpha}}{\langle \sigma_{a_1}\cdots \sigma_{a_{n-1}}\rangle_{\alpha}},$$
uniformly on compact subsets away from $a_1, \ldots, a_{n-1}$. In addition, any quantity obtained as $\beta$ or $\lambda$ of the above correlations (in the sense of Lemma \ref{lem:3/2-decomp}) converges to the corresponding quantity of the limit as well.
\end{proposition}

\begin{corollary}\label{cor:defA}
	Let $\Omega$ be smooth. Define $\mathcal{A}_\Omega^{(\alpha)}(a_1)$ by
    \begin{equation}
    2\mathcal{A}_\Omega^{(\alpha)}(a_1) =  \lambda\left(z=a_1; \frac{\langle  \sigma_{a_2}\cdots\sigma_{a_n}\mu_{a_1}\psi_z\rangle_{\alpha}}{\langle \sigma_{a_1}\cdots \sigma_{a_n}\rangle_{\alpha}} \right),
    \end{equation}
    in the sense of Lemma~\ref{lem:3/2-decomp}.
    Then $\Re \left[ \mathcal{A}_\Omega^{(\alpha)}(a_1)da_1\right] = \Re\left[ \mathcal{A}_\Omega^{(\alpha)}(a_1)\right] dx_1 -\Im\left[ \mathcal{A}_\Omega^{(\alpha)}(a_1)\right] dy_1 $ is a closed form, where $a_1 = x_1 + iy_1$.

    Under a conformal map $\varphi$ defined in $\Omega$, it satisfies the covariance rule
    \begin{equation}
    \mathcal{A}_\Omega^{(\alpha)}(a_1) = \mathcal{A}_{\varphi(\Omega)}^{(|(\varphi^{-1})'|\cdot(\alpha\circ\varphi^{-1}))}(\varphi(a_1))\cdot \varphi'(a_1) + \frac{\varphi''(a_1)}{8\varphi'(a_1)}.
    \end{equation}
\end{corollary}
\begin{proof}
    Let $f(z) = \frac{\langle  \sigma_{a_2}\cdots\sigma_{a_n}\mu_{a_1}\psi_z\rangle_{\alpha}}{\langle \sigma_{a_1}\cdots \sigma_{a_n}\rangle_{\alpha}}$. Then the asymptotics for the derivatives with respect to $a_1$, by Proposition \ref{prop:spindiff} and only showing the holomorphic terms from Lemma \ref{lem:3/2-decomp} and $o_{\frac32} = o_{\frac32}(a_1;f)$, are given by
    \begin{align*}
        &\partial_{x_{1}}f(z) = -e^{-i\pi/4}\bigg(\frac{1}{2(z-a_1)^{3/2}}  -\frac{\mathcal{A}_\Omega^{(\alpha)}(a_1)}{(z-a_1)^{1/2}}+\bigg[2\partial_{x_{1}}\mathcal{A}_\Omega^{(\alpha)}(a_1)+\frac{3}{2}o_{\frac32}\bigg](z-a_1)^{1/2}+\cdots\bigg)\\
        &\partial_{y_{1}}f(z) =-e^{-i\pi/4}\bigg(\frac{i}{2(z-a_1)^{3/2}}  -\frac{\mathcal{A}_\Omega^{(\alpha)}(a_1)i}{(z-a_1)^{1/2}}+\bigg[2\partial_{y_{1}}\mathcal{A}_\Omega^{(\alpha)}(a_1)+\frac{3i}{2}o_{\frac32}\bigg](z-a_1)^{1/2}+\cdots\bigg).
    \end{align*}
    Then, as in the proof of Proposition \ref{prop:well-def}, $\Im [GR_0(\partial_{x_{a_1}}f,\partial_{x_{a_1}}f)]=0$ and $\Im [GR_0(\partial_{y_{a_1}}f,\partial_{y_{a_1}}f)]\allowbreak=0$ yield (the anti-holomorphic terms not contributing a nonzero term)
    \begin{align*}
        \Im \!\left[2\partial_{x_{a_1}}\mathcal{A}_\Omega^{(\alpha)}(a_1) + \mathcal{A}_\Omega^{(\alpha)}(a_1)^2 +\frac{3}{2}o_{\frac32}\right]= 0 \quad\text{and}\quad
        \Im \!\left[2i\partial_{y_{a_1}}\mathcal{A}_\Omega^{(\alpha)}(a_1) - \mathcal{A}_\Omega^{(\alpha)}(a_1)^2 -\frac{3}{2}o_{\frac32} \right]= 0,
    \end{align*} 
    respectively. Since $\mathcal{A}_\Omega^{(\alpha)}$ is twice differentiable, combining the two identities yields the desired closedness. The calculation for the covariance rule is identical to the critical case, given Lemma~\ref{lem:3/2-decomp}: see, e.g., \cite[(1.10)]{CHI}.
\end{proof}

\subsubsection{Series expansion and continuity under mass perturbation}
We now move to a major analytical goal of this section, which is to show that the Ising correlations with mass $\alpha$ scaled by a constant $m\in \mathbb R$ may be expanded as a power series in $m$ around any fixed $m_0 \in \mathbb R$. Such expansion also allows us to study how correlations behave as we perturb the mass function $\alpha$.

 For fixed $m_0\in \R$, notice that the correlations $f^{[w^{\eta}]}(z):=\frac{\langle  \sigma_{a_1}\cdots\sigma_{a_n}\psi_{z}\psi^{[\eta]}_w\rangle_{m_0\alpha}}{\langle \sigma_{a_1}\cdots \sigma_{a_n}\rangle_{m_0\alpha}}$ serve as the Green's functions solving the $\R$-linear equation $\partial_{\bar{z}}f^{[w^{\eta}]} + im_0\alpha\overline{f^{[w^{\eta}]}} = \pi\bar\eta \delta_w$ on the double cover of $\Omega\setminus\{a_1, \ldots,a_n \}$ with boundary condition (S) and (RH): if $\phi \in L^q(B')$ ($q>1$) is a real function defined in a ball $B'$ in $\Omega$ away from $a_1,\ldots,a_n$ (for example, a branch of some data $\phi$ globally defined on the double cover), for any ball $B$ whose closure is contained in $B'$ we have
 \begin{align*}
 \oint_{\partial{B}}  \int_{B'}\phi(w) \left[ f^{[w^{\eta}]}(z) -\frac{\bar{\eta}}{z-w} \right] d^2wdz =&   \int_{B'}\phi(w)\oint_{\partial{B}} \left[ f^{[w^{\eta}]}(z) -\frac{\bar \eta}{z-w} \right]dz d^2w\\
 =&\int_{B'}\frac{\phi(w)}{2i}\int_B \left[ -im_0\alpha(z)\overline{f^{[w^{\eta}]}(z)} \right]d^2z d^2w\\
 =&\frac{i}{2}\int_B im_0\alpha(z) \Bigg[\overline{\int_{B'} \phi(w)f^{[w^{\eta}]}(z)d^2w }\Bigg]d^2z,
 \end{align*}
 where we have used Green-Riemann's theorem on the function $f^{[w^{\eta}]}(z) -\frac{\bar{\eta}}{z-w} \in L^2(B')$ (see Corollary \ref{cor:Lp}, and \eqref{eq:mhol-local-expansion} for the fact that the next singularity must be of strictly smaller order). We may apply the formula since this function is in e.g. $W^{1,\frac{3}{2}}(B)$ (where Green-Riemann's theorem still holds from elementary density arguments and the trace inequality) with weak $\partial_{\bar z}$-derivative given by $-im_0\alpha\overline{f^{[w^{\eta}]}}$: consider $\mathcal C_{B'}[-im_0\alpha\overline{f^{[w^{\eta}]}}]\in W^{1,\frac32}(B')$ (recall the notation $\mathcal C_B$ from \eqref{eq:cauchytransform}), then $f^{[w^{\eta}]}(z) -\frac{\bar{\eta}}{z-w} -\mathcal C_{B'}[-im_0\alpha\overline{f^{[w^{\eta}]}}](z) \in  L^2(B')$ has zero $\partial_{\bar{z}}$-derivative almost everywhere (i.e. holomorphic) in $B'$, and thus in $W^{1,\frac32}(B)$.
 
 Since $B$ was arbitrary, we must have, almost everywhere, \begin{align*}\partial_{\bar z}\left[\int_{B'}\phi(w)  f^{[w^{\eta}]}(z) d^2w \right] =&\, \partial_{\bar{z}} \left[ \int_{B'} \frac{\bar \eta\phi(w)}{z-w} d^2w\right]-im_0\alpha(z)\overline{\int_{B'} \phi(w)f^{[w^{\eta}]}(z)d^2w}\\ 
 =&\, \pi\bar \eta\phi(z) -im_0\alpha(z)\overline{\int_{B'} \phi(w) f^{[w^{\eta}]}(z)d^2w},\end{align*}as desired.
 
 Accordingly, within correlations, for complex numbers $c = a+bi$ define the linear combination (and often write $\psi^{[\star]}$ for $\psi^{[1]}$ or $\psi^{[i]}$)
    \begin{equation}\label{eq:star}
            c \star \psi^{[\star]} :=a  \psi^{[\star = 1]}- b\psi^{[\star = i]} = \frac{c\psi + \bar{c} \psi^*}{2}.
    \end{equation}
Then we may attempt to solve the equations $$\partial_{\bar{z}}A_{p+1} + im_0\alpha\overline{A_{p+1}} = -i\alpha \overline{A_p}\quad\text{and}\quad\partial_{\bar{z}}B_{p+1} + im_0\alpha\overline{B_{p+1}} = -i\alpha \overline{B_p},$$ for $A_p (z)= A_p(a_1, \ldots, a_n;z)$, etc., with the initial condition
\begin{equation}
A_0(z) = \frac{\langle  \sigma_{a_2}\cdots\sigma_{a_n}\mu_{a_1}\psi_z\rangle_{m_0\alpha}}{\langle \sigma_{a_1}\cdots \sigma_{a_n}\rangle_{m_0\alpha}}\text{ \quad and \quad} B_0(z)=\frac{\langle  \sigma_{a_1}\cdots\sigma_{a_n}\psi_{z}\psi^{[\eta]}_w\rangle_{m_0\alpha}}{\langle \sigma_{a_1}\cdots \sigma_{a_n}\rangle_{m_0\alpha}},
\end{equation}
by
    \begin{equation}\label{eq:ApBp}
    \begin{split}
        &A_{p+1}(z) = \frac{1}{\pi}\int_{\Omega} \left[i\alpha(x)\overline{A_p(x)}\right]\star \frac{\langle  \sigma_{a_1}\cdots\sigma_{a_n}\psi_{z}\psi^{[\star]}_x\rangle_{m_0\alpha}}{\langle \sigma_{a_1}\cdots \sigma_{a_n}\rangle_{m_0\alpha}}d^2x,\quad\text{and}\\
        &B_{p+1}(z) = \frac{1}{\pi}\int_{\Omega} \left[i\alpha(x)\overline{B_p(x)}\right]\star \frac{\langle  \sigma_{a_1}\cdots\sigma_{a_n}\psi_{z}\psi^{[\star]}_x\rangle_{m_0\alpha}}{\langle \sigma_{a_1}\cdots \sigma_{a_n}\rangle_{m_0\alpha}}d^2x.
    \end{split}
    \end{equation}
    We may expect, at least formally, that:
    \begin{equation}\label{eq:series-expansion}
    \begin{split}
        \frac{\langle  \sigma_{a_2}\cdots\sigma_{a_n}\mu_{a_1}\psi_z\rangle_{m\alpha}}{\langle \sigma_{a_1}\cdots \sigma_{a_n}\rangle_{m\alpha}} &= \sum_{p=0}^\infty A_p(z){(-1)^p(m-m_0)^p},\quad\text{and}\\
        \frac{\langle  \sigma_{a_1}\cdots\sigma_{a_n}\psi_{z}\psi^{[\eta]}_w\rangle_{m\alpha}}{\langle \sigma_{a_1}\cdots \sigma_{a_n}\rangle_{m\alpha}} &= \sum_{p=0}^\infty B_p(z){(-1)^p(m-m_0)^p},
    \end{split}
    \end{equation}
    since the right hand sides may be formally differentiated and shown to satisfy the equation $\partial_{\bar{z}}f + im\alpha\bar{f} =0$ away from $a_1, \ldots, a_n, w$: e.g.
    \begin{align*}
    \partial_{\bar z}\sum_{p=0}^\infty A_p\cdot {(-1)^p(m-m_0)^p} &= -im_0\alpha\sum_{p=0}^\infty  \overline{A_p}\cdot {(-1)^p(m-m_0)^p}-i\alpha\sum_{p=1}^\infty  \overline{A_{p-1}}\cdot {(-1)^p(m-m_0)^p}\\
    &=-im\alpha\sum_{p=0}^\infty \overline{A_p}\cdot {(-1)^p(m-m_0)^p}.
    \end{align*}
    We show that this is indeed the case. First note the bounds we already have for the integrands.
    \begin{remark}\label{rem:up-analysis-1}
Note that the definition \eqref{eq:ApBp} implies that (with $\mathfrak{A}_0(z) := A_0(z)$) there is some function $\mathfrak{A}_p(u,z) = \mathfrak{A}_p(a_1, \ldots,a_n;u,z) $ such that the following equality is true at the level of integrands (note $\mathfrak{A}_{p-1}(u) = \mathfrak{A}_{p-1}(u_1,\ldots,u_{p-1},u_p)$ with $u_p$ taking the place of $z$) for $p\geq 1$:
    \begin{equation}
        A_p (z) = \int_{\Omega^p} \mathfrak{A}_p(u,z) d^{2p}u = \int_{\Omega^p}\frac1\pi\left[i\alpha(u_p)\overline{\mathfrak{A}_{p-1}(u)}\right]\star\frac{\langle  \sigma_{a_1}\cdots\sigma_{a_n}\psi_{z}\psi^{[\star]}_{u_p}\rangle_{m_0\alpha}}{\langle \sigma_{a_1}\cdots \sigma_{a_n}\rangle_{m_0\alpha}}d^{2p}u,
    \end{equation}
    for $u=(u_1,u_2,\ldots,u_p)\in \Omega^p$. Then we have for all $p\geq 0$ (we set $z=u_{p+1}$)
    \begin{align}\label{eq:frak-estimate}
        \left|\mathfrak{A}_p(u,u_{p+1})\right|&\leq C^p \left|\frac{\langle  \mu_{a_1}\sigma_{a_2}\cdots\sigma_{a_n}\psi_{u_1}\rangle_{m_0\alpha}}{\langle \sigma_{a_1}\cdots \sigma_{a_n}\rangle_{m_0\alpha}}\right| \prod_{k=1}^p \bigg|\frac{\langle  \sigma_{a_1}\cdots\sigma_{a_n}\psi_{u_{k+1}}\psi^{[\star]}_{u_{k}}\rangle_{m_0\alpha}}{\langle \sigma_{a_1}\cdots \sigma_{a_n}\rangle_{m_0\alpha}}\bigg|\\ 
        \nonumber&\leq \sum_{j=1}^n\frac{C^p}{\sqrt{|u_1-a_j|}} \prod_{k=1}^p \Bigg(\sum_{j,j'=1}^n\frac{1}{|u_k-a_j|}\Bigg(\frac{\sqrt{\operatorname{dist}(a_{j'},\partial\Omega)}}{\sqrt{|u_{k+1}-a_{j'}|}}+1\Bigg)+\frac{1}{|u_k-u_{k+1}|} \Bigg),
    \end{align}
    for $C = C(\kappa_\infty, n,M, \Omega, \delta_0)$, since by Corollaries \ref{cor:dLp} and \ref{cor:Lp} we have
    \begin{equation*}
        \left|\frac{\langle  \mu_{a_1}\sigma_{a_2}\cdots\sigma_{a_n}\psi_{u_1}\rangle_{m_0\alpha}}{\langle \sigma_{a_1}\cdots \sigma_{a_n}\rangle_{m_0\alpha}}\right| \leq C \sum_{j=1}^n \frac{1}{\sqrt{|u_1-a_j|}},
    \end{equation*}
    and
    \begin{multline}\label{eq:frak-kernel-bound}
        \bigg|\frac{\langle  \sigma_{a_1}\cdots\sigma_{a_n}\psi_{u_{k+1}}\psi^{[\star]}_{u_k}\rangle_{m_0\alpha}}{\langle \sigma_{a_1}\cdots \sigma_{a_n}\rangle_{m_0\alpha}}\bigg|\leq C\sum_{j,j'=1}^n\frac{1}{|u_k-a_j|}\frac{\sqrt{\operatorname{dist}(a_{j'},\partial\Omega)}}{\sqrt{|u_{k+1}-a_{j'}|}} \\
        +C\sum_{j=1}^n\frac{1}{\sqrt{|u_k-a_j|}}\frac{1}{\sqrt{|u_k-u_{k+1}|}}+\frac{C}{|u_k-u_{k+1}|},
    \end{multline}
    and we use the AM-GM inequality $    \frac{2}{\sqrt{|u_k-a_j|}\sqrt{|u_k-u_{k+1}|}} \leq \frac{1}{{|u_k-a_j|}}+\frac{1}{{|u_k-u_{k+1}|}}$.

    Similarly, we have
    \begin{equation}
    B_p (z) \!=\! \int_{\Omega^p}\! \mathfrak{B}_p(w;u,z) d^{2p}u =\! \int_{\Omega^p}\! \frac1\pi\left[i\alpha(u_p)\overline{\mathfrak{B}_{p-1}(w;u)}\right]\star\frac{\langle  \sigma_{a_1}\cdots\sigma_{a_n}\psi_{z}\psi^{[\star]}_{u_p}\rangle_{m_0\alpha}}{\langle \sigma_{a_1}\cdots \sigma_{a_n}\rangle_{m_0\alpha}}d^{2p}u,
    \end{equation}
    with (we set $u_0=w$ and again use Corollary \ref{cor:Lp}, but in a slightly different form from \eqref{eq:frak-kernel-bound})
    \begin{multline}\label{eq:frak-B-estimate}
        \left|\mathfrak{B}_p(u_0;u,u_{p+1})\right|\leq {C^p|\eta|}\cdot \prod_{k=0}^p \Bigg(\sum_{j,j'=1}^n\frac{1}{\sqrt{|u_k-a_j|}}\frac{1}{\sqrt{|u_{k+1}-a_{j'}|}} \\ 
        +\sum_{j=1}^n\frac{1}{\sqrt{|u_k-a_j|}}\frac{1}{\sqrt{|u_k-u_{k+1}|}}+\frac{1}{|u_k-u_{k+1}|} \Bigg).
    \end{multline}
\end{remark}
        
\begin{proposition}\label{prop:series-expansion-f}
    Let $\Omega$ be smooth 
    (let $\delta_w =1$ in case of $A_p$). There is a constant $c = c(M, \Omega, \kappa_\infty,n,\delta_0,\delta_w)>0$ such that the two series \eqref{eq:series-expansion} converge absolutely, uniformly on compact subsets of $\Omega\cup \partial \Omega$ away from $a_1, a_2, \ldots, a_n, w$, for $m\in (m_0 - c, m_0 + c)$.
\end{proposition}
\begin{proof}
    We first show that the series converges. The following lemma gives a useful uniform bound.
    \begin{lemma}\label{lem:lp-Ap-Bp}
    For $p \geq 1$ and for $C=C(M,\Omega,\kappa_\infty,n,\delta_0)$, we have
    \begin{align}\label{eq:summand-bound}
    |A_p(z)|\leq \int_{\Omega^p} \left|\mathfrak{A}_p(u,z) \right| d^{2p}u  &\leq C^p\Bigg[1+\sum_{j=1}^n \frac{\sqrt{\operatorname{dist}(a_j,\partial\Omega)}}{\sqrt{|z-a_j|}} \Bigg],\\ 
    \nonumber |B_1(z)|\leq \int_{\Omega} |\mathfrak{B}_1(w;u,z)| d^{2}u &\leq C|\eta|\Bigg[\sum_{j,j'=1}^n \frac{1}{\sqrt{|z-a_j||w-a_{j'}|}}+\left|\log|z-w|\right| \Bigg],\\ 
    \nonumber |B_p(z)| \leq\int_{\Omega^p} |\mathfrak{B}_p(w;u,z)| d^{2p}u &\leq C^p|\eta|\Bigg[\sum_{j,j'=1}^n \frac{1}{\sqrt{|z-a_j||w-a_{j'}|}} \Bigg]\text{ for }p\geq2.
    \end{align}
    \end{lemma}
    \begin{proof}
    These bounds directly follow by integrating \eqref{eq:frak-estimate} and \eqref{eq:frak-B-estimate} sequentially from $u_1$ to $u_p$ (e.g. inductively showing that they integrate to \eqref{eq:summand-bound}), using the uniform integral bound
    $$
    \int_{\Omega}\frac{1}{|u-a|^\rho|u-a'|^{\rho'}}d^2u \leq C(\rho,\rho',\Omega),
    $$
    for $\rho+\rho'<2$. The only time $\rho+\rho'=2$ happens is for $B_1$, where we have the contribution
    $$
    \int_{\Omega}\frac{1}{|u_0-u_1||u_1-u_2|}d^2u_1 \leq C(\Omega)\left(1+|\log|u_0-u_2||\right),
    $$
    giving rise to the logarithmic term.
    \end{proof}
    
Let us keep presenting the $f(z) =\sum_{p=0}^\infty B_p(z) {(-1)^p(m-m_0)^p} $ case without loss of generality. Then, from \eqref{eq:summand-bound} note that, on compact subsets $z\in \Omega' \Subset (\Omega\cup\partial \Omega)\setminus \{a_1, \ldots, a_n,w\}$, we have for $p\geq 1$
\begin{equation}\label{eq:uniform-compact-bound}
|B_p(z)| \leq C(\operatorname{dist}(\Omega',\{a_1,\ldots,a_n,w\}))\cdot C(M,\Omega,\kappa_\infty,n,\delta_0,\delta_w)^p|\eta|.
\end{equation}
So there is a radius $c=c(M,\Omega,\kappa_\infty,n,\delta_0,\delta_w)>0$ where both series converge absolutely, uniformly on compacts.

Now, given any smooth function $\phi$ with compact support away from $a_1, \ldots, a_n, w$, we have
    \begin{align*}
    \int_\Omega  f \phi_{\bar{z}}d^2z &= \sum_{p=0}^\infty  \left[\int_\Omega  B_p\phi_{\bar{z}}d^2z\right] {(-1)^p(m-m_0)^p}= \sum_{p=0}^\infty  \left[\int_\Omega  -\phi\cdot\partial_{\bar{z}} B_p d^2z\right] {(-1)^p(m-m_0)^p}\\
    &= \int_\Omega \phi \Bigg[im\alpha\overline{\sum_{p=0}^\infty B_p\cdot{(-1)^p(m-m_0)^p}} \Bigg] d^2z= \int_\Omega \phi  \left[im\alpha \overline{f}   \right] d^2z,
    \end{align*}
    where we used Fubini and integrated by parts using \eqref{eq:green-riemann}, leveraging the formal calculation. So $f$ solves $\partial_{\bar{z}}f = -im\alpha \bar{f}$ away from $a_1, \ldots, a_n,w$.

    Given the bound \eqref{eq:summand-bound}, we see that $f\in V^{m\alpha}_\Omega(a_1,\ldots,a_n;w)$. We know that the singularity at $w$ is not perturbed in the highest order. In addition, by definition, we have
    \begin{multline}\label{eq:B-frak-phase}
    -e^{i\pi/4}\lim_{z\to a_j}\mathfrak B_p(w;u,z)\cdot(z-a_j)^{1/2}\\
    = \frac1\pi\left[i\alpha(u_p)\overline{\mathfrak{B}_{p-1}(w;u)}\right]\star\beta\bigg(z=a_j;\frac{\langle  \sigma_{a_1}\cdots\sigma_{a_n}\psi_{z}\psi^{[\star]}_{u_p}\rangle_{m_0\alpha}}{\langle \sigma_{a_1}\cdots \sigma_{a_n}\rangle_{m_0\alpha}}\bigg)\in i\mathbb R.
    \end{multline}
    We may exchange the limit $-e^{i\pi/4}\lim_{z\to a_j}f(z)\cdot(z-a_j)^{1/2}$ with the sum and the integral using dominated convergence; so we have $\beta(a_j;f) \in i \mathbb R$. This uniquely identifies the limit (Corollary~\ref{cor:existence}), therefore \eqref{eq:series-expansion} holds.
\end{proof}

\begin{remark}\label{re:continuity}
    In fact, tracking the integration of bounds \eqref{eq:frak-estimate} and \eqref{eq:frak-B-estimate} from the proof of Lemma \ref{lem:lp-Ap-Bp} more carefully, we may easily note that $A_p, B_p$ are continuous functions of the insertion points $a_1, \ldots, a_n$ and (for the $B_p$ case) $w$, away from one another and the boundary.

    For example, if the only insertion point was $a_1$, we may do a change of variable $u_k-a_1 =: u_k'$ and offload the dependence on $a_1$ onto the integration domain; then it is easy to see from dominated convergence (and the fact that the integrands themselves are locally continuous functions of the insertion points, see Proposition \ref{prop:spindiff}) that the value of the integral depends continuously on $a_1$, away from the boundary. Given more insertion points far apart from one another and the boundary, we may partition $\Omega$ into subdomains in each of which there is one insertion point only and apply the dominated convergence argument.
\end{remark}

Now, we specialize to the case of the series expansion for the logarithmic derivatives.

   \begin{corollary}\label{cor:halforderexpansion-p}
	Consider the functions $A_p$ in \eqref{eq:ApBp} for $p\geq1$. Then there are some complex numbers $\beta_{(p)}:=\beta(a_1;A_p) \in i\mathbb R,\, \beta'_{(p)}:=\beta'(a_1;A_p) \in i\mathbb R\,, \lambda_{(p)}=\lambda(a_1;A_p)$ such that as $z\to a_1$
    \begin{equation}\label{eq:ap-coeff-asymp}
     A_p(z) = - e^{-\frac{i\pi}{4}}\left(\beta_{(p)} \cdot (z-a_1)^{-\frac12} + \lambda_{(p)}\cdot({z-a_1})^{\frac12}+ \beta'_{(p)} \cdot (\overline{z-a_1})^{\frac12} \right) + o(|z-a_1|^{\frac12}).
    \end{equation}
    
    In addition, if $0 < 2d \leq \min\big(\operatorname{dist}(a_1,\partial\Omega),\delta_0\big)$, there is a constant $C = C(M,\Omega,\kappa_\infty,n,\delta_0, \kappa_b)$ such that  {in }$B_{d/16}(a_1)$
    \begin{align}\label{eq:ap-coeff-bound}
    \left|A_p(z) +e^{-\frac{i\pi}{4}} \beta_{(p)}\cdot (z-a_1)^{-\frac12} + e^{-\frac{i\pi}{4}}\beta'_{(p)} \cdot (\overline{z-a_1})^{\frac12} \right| \leq C^p d^{-\frac12}|z-a_1|^{\frac12},
    \end{align}
    and in fact
    \begin{equation}\label{eq:coeff-estimate}
        |\beta_{(p)}| \leq C^p d^{\frac12},\qquad |\beta_{(p)}'| \leq C^p d^{\frac12},\qquad\text{and}\qquad  |\lambda_{(p)}| \leq C^p d^{-\frac12}.
    \end{equation}
\end{corollary}
\begin{proof}
        Note that $A_0$ also satisfies \eqref{eq:ap-coeff-asymp} and \eqref{eq:ap-coeff-bound} by Lemmas \ref{lem:3/2-decomp} and \ref{lem:halforderexpansion-zeroth} (the phase condition on $\beta_{(p)}$ and \eqref{eq:coeff-estimate} do not hold for $A_0$).

        Then, we can use an argument similar to the proof of Lemma \ref{lem:halforderexpansion-zeroth}, showing these bounds inductively. Let $p\geq 1$. By \eqref{eq:summand-bound} and the induction hypothesis (or the equivalent bounds for $A_0$), we know that $(z-a_1)^{1/2}A_p(z)$ satisfies, in $B_d(a_1)$, $|(z-a_1)^{1/2}A_p(z)| \leq C^pd^{\frac12}$ and 
        $$
        |\partial_{\bar z}[(z-a_1)^{1/2}A_p(z)]|=|\alpha(z)\cdot (z-a_1)^{1/2}A_{p-1}(z)|\leq C^p.
        $$
        Thus we have by Corollary \ref{cor:dzbar-solution-bound}, with $q=8$, $\Vert \sqrt{z-a_1}\cdot  A_p(z)\Vert_{C^{3/4}(B_{d/2}(a_j))}\leq C^pd^{-\frac14}$, so 
        $$
        \big|\sqrt{z-a_1}\cdot  A_p(z)+e^{-i\pi/4}\beta_{(p)} \big|\leq C^pd^{-\frac14}|z-a_1|^{\frac34},
        $$
        for some $\beta_{(p)}$, as desired. Then \eqref{eq:summand-bound} shows the corresponding bound in \eqref{eq:coeff-estimate} and a calculation akin to \eqref{eq:B-frak-phase} shows that $\beta_{(p)}\in i\mathbb R$.
        
        For the remaining part (i.e. the existence of $\lambda_{(p)}$ and $\beta'_{(p)}$ in \eqref{eq:ap-coeff-asymp} and the corresponding estimates), write, with some $\beta'_{(p)}$ to be specified later, $$A_p^\dagger(z):= A_p(z) + e^{-i\pi/4}\beta_{(p)} \cdot (z-a_1)^{-1/2} + e^{-i\pi/4}\beta'_{(p)} \cdot (\overline{z-a_1})^{1/2}$$ and compute in $B_{d/2}(a_1)$
        \begin{align*}
            &\bigg| \partial_{\bar{z}}\frac{A_p^\dagger(z)}{\sqrt{z-a_1}}\bigg| = \bigg| \frac{-i\alpha(z)(\overline{m_0A_{p}(z)+A_{p-1}(z)}) + \frac{e^{-i\pi/4}}2\beta'_{(p)}\cdot(\overline{z-a_1})^{-1/2}}{\sqrt{z-a_1}}\bigg|.
        \end{align*}
Since $\left|A_{p}(z) +e^{-i\pi/4}\beta_{(p)}\cdot (z-a_1)^{-1/2}\right|\leq  C^pd^{-\frac14} |z-a_1|^{\frac{1}{4}}$ and better bounds hold for $A_{p-1}$ by the induction hypothesis, we may set $\beta'_{(p)} = -2i\alpha(a_1)(m_0\overline{\beta_{(p)}}+\overline{\beta_{(p-1)}}) \in i\mathbb R$ (note that this choice implies the bound for $\beta'_{(p)}$ in \eqref{eq:coeff-estimate}) and carry out the argument as in the proof of Lemma \ref{lem:halforderexpansion-zeroth}, i.e., it is already easy to see that the expansion \eqref{eq:ap-coeff-asymp} holds with \emph{some} $\lambda_{(p)}$.

To give a concrete bound, note that we have
$$
\bigg| \frac{A_p^\dagger(z)}{\sqrt{z-a_1}}\bigg| \leq C^pd^{-\frac14}|z-a_1|^{-\frac14} \qquad\text{and}\qquad\bigg| \partial_{\bar{z}}\frac{A_p^\dagger(z)}{\sqrt{z-a_1}}\bigg| \leq C^pd^{-\frac14}|z-a_1|^{-\frac14}.
$$
Then Corollary \ref{cor:dzbar-solution-bound} (with $q=3$) gives, say, in $B_{d/4}(a_j)$,
$$
\bigg| \frac{A_p^\dagger(z)}{\sqrt{z-a_1}}\bigg| \leq C^p(d^{\frac{1}{3}+\frac{1}{3}\cdot\frac32}+d^{-\frac{1}{2}})
$$
as desired.
\end{proof}

\begin{corollary}\label{cor:series-expansion-a}
    Let $\lambda(a_j;f)$ be defined as in Lemma \ref{lem:3/2-decomp}. Recall the definition
    $$
    2\mathcal{A}_\Omega^{(m\alpha)}(a_1) =  \lambda\left(z=a_1; \frac{\langle  \sigma_{a_2}\cdots\sigma_{a_n}\mu_{a_1}\psi_z\rangle_{m\alpha}}{\langle \sigma_{a_1}\cdots \sigma_{a_n}\rangle_{m\alpha}} \right),
    $$
    then, there is $c = c(M,\Omega,\kappa_\infty,n,\delta_0, \kappa_b)$ such that for $m\in (m_0 - c, m_0 + c)$, we have
    \begin{equation}
    2\mathcal{A}_\Omega^{(m\alpha)}(a_1)= \sum_{p=0}^\infty \lambda\left(a_1;A_p\right){(-1)^p(m-m_0)^p}.
    \end{equation}
    In particular, we have 
    \begin{equation}
        \left|\mathcal{A}_\Omega^{(m\alpha)}(a_1)- \mathcal{A}_\Omega^{(m_0\alpha)}(a_1) \right| \leq \frac{C(M,\Omega,\kappa_\infty,n,\delta_0, \kappa_b)}{d^{\frac12}}.
    \end{equation}
    If $a_{b} \in \partial \Omega$,
    \begin{equation}
    \int_{a_b}^{a_1} \Big(\mathcal{A}_\Omega^{(m\alpha)}(a_1') -\mathcal{A}_\Omega^{(m_0\alpha)}(a_1')\Big) da_1' =    \sum_{p=1}^\infty {(-1)^p(m-m_0)^p}\int_{a_b}^{a_1}\frac{1}{2}\lambda\left(a_1';A_p\right) da_1'.    
    \end{equation}
    In fact, we have
    \begin{equation}\label{eq:A-integrability}
        \left|\mathcal{A}_\Omega^{(\alpha)}(a_1)- \mathcal{A}_\Omega^{(0)}(a_1) \right| \leq \frac{C(\Omega,\kappa_\infty,n,\delta_0, \kappa_b)}{d^{\frac12}}.
    \end{equation}
\end{corollary}
\begin{proof}
    Using Cauchy's formula \eqref{eq:cauchy-hol} and the bounds \eqref{eq:ap-coeff-bound} and \eqref{eq:coeff-estimate}, the first two statements follow from Fubini and dominated convergence.

    The third statement follows since each coefficient with $p\ge1$ in the series expansion for $\mathcal{A}_\Omega^{(m\alpha)}(a_1)$ obeys the $d^{-\frac12}$ bound of the desired form by \eqref{eq:coeff-estimate}, and the fourth follows by another application of Fubini.

    The fifth statement \eqref{eq:A-integrability} follows by setting $M=2$ and adding, say, at most $2/c$ copies of the third estimate from $0$ to $m_0=1$, since the radius of convergence around any point in $[0, 2)$ is bounded below by $c$.
\end{proof}

\begin{lemma}\label{lem:integral-convergence}
Suppose $\alpha_N$ and $\alpha$ all satisfy the following bounds for a fixed $\kappa_b$: $$\operatorname{dist}(z,\partial\Omega)^2|\nabla\alpha_N(z)| \leq \kappa_b \qquad\text{and}\qquad\operatorname{dist}(z,\partial\Omega)^2|\nabla\alpha(z)| \leq \kappa_b.$$

Let $\alpha_N \to \alpha$ pointwise. Then the basic correlations converge: for fixed $a_1,\ldots,a_n,w,z,m,\eta$,
    $$\frac{\langle  \sigma_{a_1}\cdots\sigma_{a_n}\psi_{z}\psi^{[\eta]}_w\rangle_{\alpha_N}}{\langle \sigma_{a_1}\cdots \sigma_{a_n}\rangle_{\alpha_N}} \to \frac{\langle  \sigma_{a_1}\cdots\sigma_{a_n}\psi_{z}\psi^{[\eta]}_w\rangle_{\alpha}}{\langle \sigma_{a_1}\cdots \sigma_{a_n}\rangle_{\alpha}}\text{;}\qquad\frac{\langle  \sigma_{a_2}\cdots\sigma_{a_n}\mu_{a_1}\psi_z\rangle_{\alpha_N}}{\langle \sigma_{a_1}\cdots \sigma_{a_n}\rangle_{\alpha_N}} \to \frac{\langle  \sigma_{a_2}\cdots\sigma_{a_n}\mu_{a_1}\psi_z\rangle_{\alpha}}{\langle \sigma_{a_1}\cdots \sigma_{a_n}\rangle_{\alpha}}.  $$

    As a result, all correlations obtained as coefficients $\beta, \lambda$ of the above correlations (in the sense of Lemma \ref{lem:3/2-decomp}) also converge. 

    As a consequence, we have
    $$
    \langle \sigma_{a_1}\cdots \sigma_{a_n}\rangle_{\alpha_N} \to \langle \sigma_{a_1}\cdots \sigma_{a_n}\rangle_{\alpha},
    $$
    where this quantity is calculated recursively from \eqref{eq:spin-integral} with fixed choices and ordering of paths $\gamma_1, \dots, \gamma_n$, connecting some boundary points $a_{b1},\dots,a_{bn}$ to $a_1,\dots,a_n$, respectively.
\end{lemma}
\begin{proof}
        Replace $\alpha$ by $m\alpha$. The pointwise convergence of basic correlations under $\alpha_N \to \alpha$ within any single radius of convergence of the series expansion provided by Proposition \ref{prop:series-expansion-f} follows from dominated convergence. Then, as in the proof of Corollary \ref{cor:series-expansion-a}, we may find a finite sequence of masses $0=m_0<m_1<\cdots<m_K=1$ such that each $m_{i+1}$ is in the radius of convergence around $m_i$, to inductively extend the convergence to $m=1$.

    The convergence of functions in the bulk implies the coefficient convergence, using dominated convergence on \eqref{eq:cauchy-hol} as usual. Then, we may use the bound in Corollary \ref{cor:series-expansion-a} to again use dominated convergence as $\alpha_N\to\alpha$ and obtain pure spin correlation convergence.
\end{proof}

\subsubsection{Well-Definedness of Pure Spin Correlations}\label{sec:well-def-spin}
We finish this section by removing the implicit dependence on the ordering and paths used for the pure spin correlation in Lemma \ref{lem:integral-convergence}, i.e., showing that the closed forms $\Re [\mathcal A_\Omega^{(\alpha)}(a_1) da_1]$ (see Corollary \ref{cor:defA}) and the recursive definition \eqref{eq:spin-integral} indeed give rise to a well-defined spin correlation. Concretely, we need to check that the correlations do not have a monodromy around any other spin $a_2,\ldots,a_n$, and do not depend on the boundary point chosen as the starting point of the line integral or the order in which we add the spins.

Let us first point out the general strategy for extracting (differences of) forms $\Re[\mathcal A_\Omega^{(\alpha)}(a_1) ]da_1$. Suppose we write, $f_n(z) = \frac{\langle  \sigma_{a_2}\cdots\sigma_{a_n}\mu_{a_1}\psi_z\rangle_{\alpha}}{\langle \sigma_{a_1}\cdots \sigma_{a_n}\rangle_{\alpha}}$, $h_n = \Im [\int f^2_n dz]$, $\mathcal A_n(a_1):=\frac12 \lambda(a_1;f_n)$, etc., for fixed distinct points $a_2, a_3, \ldots$ in a smooth domain $\Omega$. Then it is straightforward to check that
\begin{align}\label{eq:extract-a}
\mathcal A_{n+1}(a_1) - \mathcal A_n(a_1) = \frac i4 \lim_{z\to a_1}\left[ f_{n+1}^2(z)-f_n^2(z)\right],
\end{align}
thanks to Lemma \ref{lem:3/2-decomp} (specifically, the asymptotic for $f^{\dagger1}$ with $\nu=-1/2$ and $o_\nu = 1$).

\begin{lemma}\label{lem:spin-no-monodromy}
    In the above setting, we have $\Re \left[\int_{\partial B_r(a_2)} \mathcal A_n(a_1') da_1'\right] \xrightarrow{r\downarrow 0}0$ for each $n$.
\end{lemma}
\begin{proof}
    Note that for $n=1$ we do not have an insertion at $a_2$, so the integral is identically zero by Corollary \ref{cor:defA}. Then, it suffices to show that for each $n$ we have
    $$\Re\bigg[\int_{\partial B_r(a_2)} \big(\mathcal A_{n+1}(a_1) - \mathcal A_n(a_1)\big) da_1\bigg] \xrightarrow{r\downarrow 0}0.$$
    
    Therefore, let us study the function $(f_{n+1}^2-f_n^2)(z)$ for $a_1\in B_{r_0}(a_2)$ for small enough $r_0$, so that we may take $\delta_1 = \frac{1}{16}|a_1 - a_2|$ and $a_3, \ldots, a_n \notin B_{4r_0}(a_2) \subset \Omega$. We have an a priori bound for these functions for $z\in  B_{2r_0}(a_2)\cup \partial B_{2r_0}(a_2)$ thanks to Corollary \ref{cor:dLp} and Proposition \ref{prop:thebound} (take $\nu=1/4$ in the latter):
    \begin{equation}\label{eq:a-priori-fn}
    \left| f_n (z)\right| \leq C(\kappa_\infty,n,\Omega,r_0)\cdot \bigg({\frac{|a_1-a_2|^{-1/4}}{|z-a_2|^{1/2}}+\frac{1}{|z-a_1|^{1/2}}}\bigg).
    \end{equation}
    
    In addition, since $f_{n+1} - f_n$ is an $\alpha$-holomorphic spinor whose blow-up at $a_1$ is canceled, we may use Theorem \ref{thm:similarity-principle} to write
    $$
    f_{n+1}(z) - f_n(z)=e^{s(z)}\underline{f}(z),
    $$
    for a holomorphic spinor $\underline{f}$ and some $s$ satisfying $|s| \leq C(\kappa_\infty, r_0)$. We know $\underline{f}^2$ is a meromorphic function which has at most a simple pole at $a_2$, zero at $a_1$, and still satisfies a bound coming from \eqref{eq:a-priori-fn} on, say, $ \partial B_{2r_0}(a_2)$. The maximum principle on $\frac{z-a_2}{z-a_1}\cdot\underline{f}^2(z)$ easily gives, for $z\in  B_{2r_0}(a_2)\cup \partial B_{2r_0}(a_2)$,
    $$
    |f_{n+1}(z) - f_n(z)|^2=|e^{s(z)}|^2|\underline{f}(z)|^2 \leq C(\kappa_\infty,n,\Omega,r_0)\cdot |a_1-a_2|^{-1/2}\cdot \frac{|z-a_1|}{|z-a_2|}.
    $$
    Combining this bound with \eqref{eq:a-priori-fn}, we have
    $$
    |f_{n+1}^2(z) - f_n^2(z)| \leq C(\kappa_\infty,n,\Omega,r_0)\cdot\frac{|a_1-a_2|^{-1/2}}{|z-a_2|}.
    $$
    Then the function $(z-a_1)(f_{n+1}^2 - f_n^2)(z)$ satisfies
    $$
    \left|(z-a_2)(f_{n+1}^2 - f_n^2)(z)\right| \leq C(\kappa_\infty,n,\Omega,r_0)\cdot |a_1-a_2|^{-1/2},
    $$ and
    \begin{align*}
    \left|\partial_{\bar z}\left[(z-a_2)(f_{n+1}^2 - f_n^2)(z)\right]\right| &= \left| 2i\alpha(z)\cdot(z-a_1)\cdot(|f_{n}(z)|^2 - |f_{n+1}(z)|^2)\right| \\
    &\leq C(\kappa_\infty,n,\Omega,r_0)\cdot |a_1-a_2|^{-1/2}.
    \end{align*}
    By Corollary \ref{cor:dzbar-solution-bound} (with $q=4$), we have for $z\in B_{r_0}(a_2)$
    $$
    \left|(z-a_2)(f_{n+1}^2 - f_n^2)(z)- ib(a_1)^2 \right| \leq C(\kappa_\infty, n, \Omega, r_0)\cdot |a_1-a_2|^{-1/2} \cdot |z-a_2|^{1/2},
    $$
    where $ib(a_1)^2:=\lim_{z\to a_2}(z-a_2)(f_{n+1}^2 - f_n^2)(z)$ for $b(a_1) \in \mathbb R$ by (S) asymptotic at $a_2$ for $f_n, f_{n+1}$.

    Finally, using \eqref{eq:extract-a}, we have
    \begin{align*}
    \Re \bigg[\int_{\partial B_r(a_2)} \big(\mathcal A_{n+1}(a_1) - \mathcal A_n(a_1)\big) da_1 \bigg]=&\, \Re \bigg[\int_{\partial B_r(a_2)} -\frac12\left(\frac{b(a_1)^2}{2(a_2-a_1)} + O(1)\right)\, da_1\bigg]\\
    =&\,  \Re \bigg[\int_{0}^{2\pi} -\frac{b(a_2+re^{i\theta })^2}{4re^{i\theta}}  ire^{i\theta}d\theta\bigg] + O(r)\\
    =&\,O(r),
    \end{align*}
    as desired.
\end{proof}

\begin{lemma}\label{lem:crit-welldef}
    Suppose $\Omega$ is smooth and $\alpha$ is smooth up to the boundary. Then $$\Re \bigg[\int_{a_b}^{a_1} \Big(\mathcal{A}_\Omega^{(\alpha)}(a_1') - \mathcal{A}_\Omega^{(0)}(a_1')\Big)da_1'\bigg]$$ does not depend on the choice of $a_b\in\partial \Omega$.
\end{lemma}
\begin{proof}
    It suffices to prove that
    $$
    \Re \bigg[\int_{a_b^1}^{a_b^2} \Big(\mathcal{A}_\Omega^{(\alpha)}(a_1') - \mathcal{A}_\Omega^{(0)}(a_1')\Big)da_1'\bigg] = 0
    $$
    for any $a_b^1, a_b^2 \in \partial \Omega$ thanks to closedness.
    
    Let us first assume $\alpha \in C_c^\infty(\Omega)$. Thanks to conformal covariance (again from Corollary \ref{cor:defA}, and note $da_1$ transforms conformally as well), we may assume that $a_1 \in B_{R}\cap \mathbb{H} \subset B_{3R}\cap \mathbb{H} \subset  \Omega \subset \mathbb{H}$ and $a_2,\ldots, a_n\notin B_{3R}$, with segment $[-3R,3R] \subset \partial \Omega \cap \R$. By shrinking $R$ if necessary, we may assume $\alpha \equiv 0$ in $B_{2R}\cap \mathbb{H}$ since $\alpha$ has compact support.
    
    Write $f(z) = \frac{\langle  \sigma_{a_2}\cdots\sigma_{a_n}\mu_{a_1}\psi_z\rangle_{\alpha}}{\langle \sigma_{a_1}\cdots \sigma_{a_n}\rangle_{\alpha}}$, $f_{(0)}(z) = \frac{\langle  \sigma_{a_2}\cdots\sigma_{a_n}\mu_{a_1}\psi_z\rangle_{0}}{\langle \sigma_{a_1}\cdots \sigma_{a_n}\rangle_{0}}$ and set $h = \Im [\int f^2 dz]$, $h_{(0)} =  \Im [\int f^2_{(0)} dz]$ as usual. Then it suffices to show that  $\Re \left[\mathcal{A}_\Omega^{(\alpha)}(a_1) - \mathcal{A}_\Omega^{(0)}(a_1)\right]$ uniformly extends to zero as $a_1$ approaches $[-R,R]$. As in \eqref{eq:extract-a}, extract (where $z=x+iy$)
    \begin{align*}
       \Re\left[ \mathcal{A}_\Omega^{(\alpha)}(a_1) - \mathcal{A}_\Omega^{(0)}(a_1)\right] &=\lim_{z\to a_1}\frac{1}{4}\Im \left[f^2 - f_{(0)}^2 \right](z)=\lim_{z\to a_1}\frac14\partial_x(h(z) - h_{(0)}(z)).
    \end{align*}
    
    Now, since $\Delta (h-h_{(0)}) = 0$ and $h-h_{(0)}$ is continuously differentiable (since the only possible discontinuity of its derivative $f^2 - f_{(0)}^2$ at $a_1$ has been handled), $\partial_x (h-h_{(0)})$ is also harmonic in $B_{2R}\cap \mathbb{H}$. By Corollary \ref{cor:dLp}, the size of $f^2 - f_{(0)}^2$ on $\partial B_{2R}\cap \mathbb{H}$ is bounded above, uniformly in $a_1\in B_R\cap \mathbb{H}$. Due to harmonicity, we have $\partial_x(h(z) - h_{(0)}(z))\vert_{z=a_1} = O(\Im [a_1])$ as $a_1$ approaches $[-R,R]$, as desired.

    Then, one obtains the general statement from Lemma \ref{lem:integral-convergence}, by taking $\alpha_N$ to be the product of $\alpha$ with a suitable sequence of bump functions such that $\alpha_N \in C_c^\infty(\Omega)$ and $\alpha_N \to \alpha$ pointwise.
\end{proof}

Thanks to the previous two lemmas, we are able to freely deform the integration contours. It remains to show the following.

\begin{lemma}\label{lem:spin-no-order}
    The correlation $\langle \sigma_{a_1}\cdots \sigma_{a_n}\rangle_{\alpha}$, as defined by \eqref{eq:spin-integral}, does not depend on the order in which $a_1, \ldots, a_n$ are introduced.
\end{lemma}
\begin{proof}
    Taking the $0$-spin and $1$-spin correlations as the trivial induction base cases, it suffices to check that any ratio
    $$
    \frac{\langle \sigma_{a_1}\cdots \sigma_{a_n}\sigma_{a}\sigma_{a^\dagger}\rangle_{\alpha}}{\langle \sigma_{a_1}\cdots \sigma_{a_n}\rangle_{\alpha}}
    $$
    does not depend on whether $a$ or $a^{\dagger}$ is introduced first. Examining \eqref{eq:spin-integral} and the well-definedness of this ratio for $\alpha = 0$, we need to show that
    \begin{equation*}
    \Re\bigg[\int_{a_b}^a \mathcal A^{(\alpha;0)}_{a'} da'\bigg] + \Re\bigg[\int_{a_b^\dagger }^{a^\dagger}\mathcal A^{(\alpha;0)}_{a'^\dagger;a} da'^\dagger\bigg] = \Re\bigg[\int_{a_b^\dagger }^{a^\dagger} \mathcal A^{(\alpha;0)}_{a'^\dagger} da'^\dagger\bigg] + \Re\bigg[\int_{a_b}^a\mathcal A^{(\alpha;0)}_{a';a^\dagger} da'\bigg],
    \end{equation*}
    where $\mathcal A^{(\alpha;0)}_{a'^\dagger;a}:=\mathcal A^{(\alpha)}_{a'^\dagger;a}-\mathcal A^{(0)}_{a'^\dagger;a}$,
    $$
    \mathcal A^{(\alpha)}_{a'} := \frac12 \lambda\left(z=a;\frac{\langle \sigma_{a_1}\cdots \sigma_{a_n}\mu_{a}\psi_{z}\rangle_{\alpha}}{\langle \sigma_{a_1}\cdots \sigma_{a_n}\sigma_a\rangle_{\alpha}}\right) \quad\text{and}\quad\mathcal A^{(\alpha)}_{a';a}:=\frac12 \lambda\left(z=a';\frac{\langle \sigma_{a_1}\cdots \sigma_{a_n}\sigma_{a}\mu_{a'}\psi_{z}\rangle_{\alpha}}{\langle \sigma_{a_1}\cdots \sigma_{a_n}\sigma_a\sigma_{a'}\rangle_{\alpha}}\right),
    $$
    etc., and we integrate along some $\gamma$ and $\gamma^\dagger$ connecting $a_b$ to $a$ and $a_b^\dagger$ to $a^\dagger$, respectively. We know from Corollary \ref{cor:defA} and Lemma \ref{lem:spin-no-monodromy} that the integrals do not depend on the deformations of $\gamma$ and $\gamma^\dagger$ that fix the endpoints, so let us fix two concrete paths $\gamma$ and $\gamma^\dagger$ henceforth. In fact, we may write $g_1(a,a^\dagger)$ and $g_2(a,a^\dagger)$ for the left and right hand sides, respectively, and extend them to functions $g_1(a',a'^{\dagger})$ and $g_2(a',a'^\dagger)$ for $a'$ and $a'^\dagger$ in small neighborhoods of $\gamma$ and $\gamma^\dagger$, respectively.

    Let us show $g_1 \equiv g_2$ in that neighborhood. Write $a' = x'+iy'$ and $a'^\dagger = x'^\dagger + iy'^\dagger$. It suffices to show that each mixed second derivative coincides, i.e.,
    \begin{equation}\label{eq:twice-diff}
    \begin{split}
     &\partial_{x'x'^\dagger}g_1(a',a'^\dagger) = \partial_{x'x'^\dagger}g_2(a',a'^\dagger), \quad\partial_{x'y'^\dagger}g_1(a',a'^\dagger) = \partial_{x'y'^\dagger}g_2(a',a'^\dagger),\\ 
     &\partial_{y'x'^\dagger}g_1(a',a'^\dagger) = \partial_{y'x'^\dagger}g_2(a',a'^\dagger), \quad\partial_{y'y'^\dagger}g_1(a',a'^\dagger) = \partial_{y'y'^\dagger}g_2(a',a'^\dagger);
     \end{split}
     \end{equation}
     or
     \begin{align*}
     &\partial_{x'}\Re[\mathcal A^{(\alpha;0)}_{a'^\dagger;a'}]=\partial_{x'^\dagger}\Re[\mathcal A^{(\alpha;0)}_{a';a'^\dagger}],\quad -\partial_{x'}\Im[\mathcal A^{(\alpha;0)}_{a'^\dagger;a'}]=\partial_{y'^\dagger}\Re [\mathcal A^{(\alpha;0)}_{a';a'^\dagger}],\\ 
     \nonumber&\partial_{y'}\Re[\mathcal A^{(\alpha;0)}_{a'^\dagger;a'}]=-\partial_{x'^\dagger}\Im [\mathcal A^{(\alpha;0)}_{a';a'^\dagger}],\quad \partial_{y'}\Im[\mathcal A^{(\alpha;0)}_{a'^\dagger;a'}]=\partial_{y'^\dagger}\Im [\mathcal A^{(\alpha;0)}_{a';a'^\dagger}],
    \end{align*}
    since we may integrate them from the boundary. To elaborate, \eqref{eq:twice-diff} would imply that the gradients $\nabla_{a'}$ of $\partial_{x'^\dagger}g_1(a',a'^\dagger)$ (resp. $\partial_{y'^\dagger}g_1(a',a'^\dagger)$) and $\partial_{x'^\dagger}g_2(a',a'^\dagger)$ (resp. $\partial_{y'^\dagger}g_2(a',a'^\dagger)$) are the same, and as $a'\to a_b$, by Proposition \ref{prop:spin-boundary-limit}, we have the same boundary values $$\partial_{x'^\dagger}g_1(a',a'^\dagger),\,\partial_{x'^\dagger}g_2(a',a'^\dagger) \to \Re[\mathcal A^{(\alpha;0)}_{a'^\dagger}]\quad\text{and}\quad\partial_{y'^\dagger}g_1(a',a'^\dagger),\,\partial_{y'^\dagger}g_2(a',a'^\dagger) \to -\Im[\mathcal A^{(\alpha;0)}_{a'^\dagger}].$$
    
    By Proposition \ref{prop:spindiff}, we may extract 
    $$
    \partial_{x'}\mathcal A^{(\alpha)}_{a'^\dagger;a'}:=\frac12 \lambda\left(z=a'^\dagger;\partial_{x'}\left(\frac{\langle \sigma_{a_1}\cdots \sigma_{a_n}\sigma_{a'}\mu_{a'^\dagger}\psi_{z}\rangle_{\alpha}}{\langle \sigma_{a_1}\cdots \sigma_{a_n}\sigma_{a'}\sigma_{a'^\dagger}\rangle_{\alpha}}\right)\right), \text{ etc.}.
    $$
    However, under the same differentiation with respect to $a'=x'+iy'$, note that the fixed singularity at $a'^\dagger$ vanishes: we have, e.g.,
    \begin{equation}\label{eq:canceled-singularity}
    \begin{split}
    \partial_{x'}\left(\frac{\langle \sigma_{a_1}\cdots \sigma_{a_n}\sigma_{a'}\mu_{a'^\dagger}\psi_{z}\rangle_{\alpha}}{\langle \sigma_{a_1}\cdots \sigma_{a_n}\sigma_{a'}\sigma_{a'^\dagger}\rangle_{\alpha}}\right) &= -2\cdot e^{-\frac{i\pi}{4}}\cdot\partial_{x'}\mathcal A^{(\alpha)}_{a'^\dagger;a'}\cdot \sqrt{z-a'^\dagger}+o(|z-a'^\dagger|^{1/2}),\\
    \partial_{y'}\left(\frac{\langle \sigma_{a_1}\cdots \sigma_{a_n}\sigma_{a'}\mu_{a'^\dagger}\psi_{z}\rangle_{\alpha}}{\langle \sigma_{a_1}\cdots \sigma_{a_n}\sigma_{a'}\sigma_{a'^\dagger}\rangle_{\alpha}}\right) &= -2\cdot e^{-\frac{i\pi}{4}}\cdot\partial_{y'}\mathcal A^{(\alpha)}_{a'^\dagger;a'}\cdot \sqrt{z-a'^\dagger}+o(|z-a'^\dagger|^{1/2}),
    \end{split}
    \end{equation}
    whereas
    \begin{equation}\label{eq:generic-derivative}
    \begin{split}
    \partial_{x'}\left(\frac{\langle \sigma_{a_1}\cdots \sigma_{a_n}\sigma_{a'}\mu_{a'^\dagger}\psi_{z}\rangle_{\alpha}}{\langle \sigma_{a_1}\cdots \sigma_{a_n}\sigma_{a'}\sigma_{a'^\dagger}\rangle_{\alpha}}\right) &= -\frac{e^{-\frac{i\pi}{4}}}{2}\cdot\beta^{(\alpha)}(a';a'^\dagger)\cdot ({z-a'})^{-3/2}+o(|z-a'|^{-3/2}),\\
    \partial_{y'}\left(\frac{\langle \sigma_{a_1}\cdots \sigma_{a_n}\sigma_{a'}\mu_{a'^\dagger}\psi_{z}\rangle_{\alpha}}{\langle \sigma_{a_1}\cdots \sigma_{a_n}\sigma_{a'}\sigma_{a'^\dagger}\rangle_{\alpha}}\right) &= -\frac{ie^{-\frac{i\pi}{4}}}{2}\cdot\beta^{(\alpha)}(a';a'^\dagger)\cdot ({z-a'})^{-3/2}+o(|z-a'|^{-3/2}),
    \end{split}
    \end{equation}
    where $\beta^{(\alpha)}(a';a'^\dagger) = \beta\left(z=a';\frac{\langle \sigma_{a_1}\cdots \sigma_{a_n}\sigma_{a'}\mu_{a'^\dagger}\psi_{z}\rangle_{\alpha}}{\langle \sigma_{a_1}\cdots \sigma_{a_n}\sigma_{a'}\sigma_{a'^\dagger}\rangle_{\alpha}}\right)=-i\frac{\langle \sigma_{a_1}\cdots \sigma_{a_n}\mu_{a'}\mu_{a'^\dagger}\rangle_{\alpha}}{\langle \sigma_{a_1}\cdots \sigma_{a_n}\sigma_{a'}\sigma_{a'^\dagger}\rangle_{\alpha}}$ by Proposition \ref{prop:well-def}. Analogous formulas hold for the $x'^\dagger,y'^\dagger$ derivatives of $\frac{\langle \sigma_{a_1}\cdots \sigma_{a_n}\sigma_{a'^\dagger}\mu_{a'}\psi_{z}\rangle_{\alpha}}{\langle \sigma_{a_1}\cdots \sigma_{a_n}\sigma_{a'}\sigma_{a'^\dagger}\rangle_{\alpha}}$, and studying $\Im [GR_0]$ between the four combinations between the derivatives of either function (see \eqref{eq:gr} and the proof of Proposition \ref{prop:well-def}) yields
    \begin{align*}
        &\Im\left[\partial_{x'}\mathcal A^{(\alpha)}_{a'^\dagger;a'}\cdot \beta^{(\alpha)}(a'^\dagger;a')+\partial_{x'^\dagger}\mathcal A^{(\alpha)}_{a';a'^\dagger}\cdot \beta^{(\alpha)}(a';a'^\dagger)\right]=0;\\
        &\Im\left[\partial_{x'}\mathcal A^{(\alpha)}_{a'^\dagger;a'}\cdot i\beta^{(\alpha)}(a'^\dagger;a')+\partial_{y'^\dagger}\mathcal A^{(\alpha)}_{a';a'^\dagger}\cdot \beta^{(\alpha)}(a';a'^\dagger)\right]=0;\\
        &\Im\left[\partial_{y'}\mathcal A^{(\alpha)}_{a'^\dagger;a'}\cdot i\beta^{(\alpha)}(a'^\dagger;a')+\partial_{y'^\dagger}\mathcal A^{(\alpha)}_{a';a'^\dagger}\cdot i\beta^{(\alpha)}(a';a'^\dagger)\right]=0;\\
        &\Im\left[\partial_{y'}\mathcal A^{(\alpha)}_{a'^\dagger;a'}\cdot \beta^{(\alpha)}(a'^\dagger;a')+\partial_{x'^\dagger}\mathcal A^{(\alpha)}_{a';a'^\dagger}\cdot i\beta^{(\alpha)}(a';a'^\dagger)\right]=0.
    \end{align*}
    Since $\beta^{(\alpha)}(a';a'^\dagger) =- \beta^{(\alpha)}(a'^\dagger;a')\in i\mathbb R$, it is easy to verify \eqref{eq:twice-diff} if $\beta^{(\alpha)}(a;a^\dagger) \neq 0$. In fact, thanks to the  (second) differentiability of both $\mathcal {A}_{a';a'^\dagger}^{(\alpha)}$ and $\beta^{(\alpha)}(a';a'^\dagger)$ with respect to $a', a'^\dagger$, it suffices that $\beta^{(\alpha)}(a';a'^\dagger)$ is nonzero for $(a',a'^\dagger)$ in, e.g., a dense subset of $\Omega^2$. 

    For this genericity, suppose that there is some $a'^\dagger$ and an open neighborhood $D\subset \Omega$ such that $\beta^{(\alpha)}(a';a'^\dagger) = 0$ for all $a'\in D$. Then, instead of \eqref{eq:generic-derivative}, we have the behavior, where $\lambda = \lambda \left( z=a';\frac{\langle \sigma_{a_1}\cdots \sigma_{a_n}\sigma_{a'}\mu_{a'^\dagger}\psi_{z}\rangle_{\alpha}}{\langle \sigma_{a_1}\cdots \sigma_{a_n}\sigma_{a'}\sigma_{a'^\dagger}\rangle_{\alpha}}\right)$,
    \begin{align*}
    &\partial_{x'}\left(\frac{\langle \sigma_{a_1}\cdots \sigma_{a_n}\sigma_{a'}\mu_{a'^\dagger}\psi_{z}\rangle_{\alpha}}{\langle \sigma_{a_1}\cdots \sigma_{a_n}\sigma_{a'}\sigma_{a'^\dagger}\rangle_{\alpha}}\right) = \frac{e^{-\frac{i\pi}{4}}}{2}\cdot\lambda\cdot ({z-a'^\dagger})^{-1/2}+o(|z-a'^\dagger|^{-1/2}),\\ 
    &\partial_{y'}\left(\frac{\langle \sigma_{a_1}\cdots \sigma_{a_n}\sigma_{a'}\mu_{a'^\dagger}\psi_{z}\rangle_{\alpha}}{\langle \sigma_{a_1}\cdots \sigma_{a_n}\sigma_{a'}\sigma_{a'^\dagger}\rangle_{\alpha}}\right) = \frac{ie^{-\frac{i\pi}{4}}}{2}\cdot\lambda\cdot ({z-a'^\dagger})^{-1/2}+o(|z-a'^\dagger|^{-1/2}),
    \end{align*}
    with both derivatives having the asymptotic (S) at $a_1, \ldots, a_n$ and also $a'^\dagger$ (since \eqref{eq:canceled-singularity} is trivially (S)), as well as the boundary condition (RH). However, this means that both derivatives should have a real multiple of the (D) asymptotic at $a'$ by Corollary \ref{cor:existence} (see also its proof), which forces $\lambda =0$, setting both derivatives to be identically zero as functions of $z$.

    To see why this is a contradiction, choose some $a_0\in D$ and $e_0>0$ such that $L:=\{a_0+x: x\in (-e_0, e_0)\} \subset D$. We know that for all $a'\in L$, the spinor $z \mapsto \frac{\langle \sigma_{a_1}\cdots \sigma_{a_n}\sigma_{a'}\mu_{a'^\dagger}\psi_{z}\rangle_{\alpha}}{\langle \sigma_{a_1}\cdots \sigma_{a_n}\sigma_{a'}\sigma_{a'^\dagger}\rangle_{\alpha}}$ may be defined on a fixed double cover of $\Omega\setminus (\{a_1, \ldots, a_n, a'^\dagger\}\cup L)$. Then the derivative condition
    $$
    \partial_{x'}\left(\frac{\langle \sigma_{a_1}\cdots \sigma_{a_n}\sigma_{a'}\mu_{a'^\dagger}\psi_{z}\rangle_{\alpha}}{\langle \sigma_{a_1}\cdots \sigma_{a_n}\sigma_{a'}\sigma_{a'^\dagger}\rangle_{\alpha}}\right) \equiv 0
    $$
    implies that there is a fixed spinor $f$ on the double cover of $\Omega\setminus (\{a_1, \ldots, a_n, a'^\dagger\}\cup L)$ such that $f(z) = \frac{\langle \sigma_{a_1}\cdots \sigma_{a_n}\sigma_{a'}\mu_{a'^\dagger}\psi_{z}\rangle_{\alpha}}{\langle \sigma_{a_1}\cdots \sigma_{a_n}\sigma_{a'}\sigma_{a'^\dagger}\rangle_{\alpha}}$ for all $a'\in L$. However, since $\frac{\langle \sigma_{a_1}\cdots \sigma_{a_n}\sigma_{a'}\mu_{a'^\dagger}\psi_{z}\rangle_{\alpha}}{\langle \sigma_{a_1}\cdots \sigma_{a_n}\sigma_{a'}\sigma_{a'^\dagger}\rangle_{\alpha}}$ has a (square-root) zero at $z=a'$ by the assumption that $\beta^{(\alpha)}(a';a'^\dagger) = 0$, we have $f(z)\to 0$ as $z\to L$. This sets $f(z) \equiv 0$, as seen by, e.g. using Theorem \ref{thm:similarity-principle} to write $f = e^s\underline{f}$ and considering the holomorphic function $\underline{f}^2$ which must vanish on $L$. This contradicts the fact that $f$ must have a (D) asymptotic at $z=a'^\dagger$.

    Therefore, $\beta^{(\alpha)}(\cdot ;\cdot)$ is nonzero at least on a dense subset of $\Omega^2$, and the proof is complete.
\end{proof}

\begin{proposition}\label{prop:spin} 
    We have   
    \begin{align}\label{eq:spin-uniform-bound}
        0 < \langle\sigma_{a_1}\cdots\sigma_{a_n}\rangle_\alpha< C(\Omega,\kappa_b,\kappa_\infty,n,\delta_{all}),
    \end{align}
    where $\delta_{all}>0$ is smaller than $\frac{1}{16}|a_j-a_k|$ for any $j\neq k$ and $\frac12\operatorname{dist}(a_j,\partial \Omega)$ for any $j=1,\ldots,n$.

    In addition, if $\alpha \equiv m < 0$, $\langle \sigma_{a_1}\cdots \sigma_{a_n}\rangle_{m}$ is exactly the continuous correlation obtained as a scaling limit in \cite{P, CIM23}.
\end{proposition}
\begin{proof}
    The critical correlation $\langle\sigma_{a_1}\cdots\sigma_{a_n}\rangle_0$ satisfies \eqref{eq:spin-uniform-bound} by \cite[Theorem 7.1]{CHI2} ($q=0$ case). Then the line integral in \eqref{eq:spin-integral} is bounded by explicitly integrating \eqref{eq:A-integrability}, and it is easy to see that we may choose the integration contour for $a_1'$ uniformly away from other points and obtain \eqref{eq:spin-uniform-bound} for the massive correlation.

    For the constant mass case, while \eqref{eq:spin-integral} expresses natural decorrelation near the boundary given by RSW-type estimates \cite{DCGP, P2}, we note a purely continuous argument. To show that our $n$-spin correlation coincides with the correlation identified in the discussion in \cite[Section 1.3]{P}, it suffices to show that
    \begin{equation*}
	 \frac{ \langle \sigma_{a_1}\cdots \sigma_{a_n}\rangle_{0}}{\langle\sigma_{a_1}\rangle_{0}\cdots \langle\sigma_{a_n}\rangle_{0}}\cdot\frac{ \langle \sigma_{a_2}\cdots \sigma_{a_n}\rangle_{\alpha}}{\langle\sigma_{a_2}\cdots \sigma_{a_n}\rangle_{0}}\cdot \exp \left\{\Re \bigg[\int_{a_b}^{a_1} \Big(\mathcal{A}_\Omega^{(\alpha)}(a_1') - \mathcal{A}_\Omega^{(0)}(a_1')\Big)da_1' \bigg]\right\}
\end{equation*}
may be taken as closely to $1$ as desired, by taking $a_1, \ldots, a_n$ closely to $\partial \Omega$. Inductively, we may assume, for the $n-1$ point case, that
$$
\frac{ \langle \sigma_{a_2}\cdots \sigma_{a_n}\rangle_{\alpha}}{\langle\sigma_{a_2}\cdots \sigma_{a_n}\rangle_{0}} =\frac{ \langle \sigma_{a_2}\cdots \sigma_{a_n}\rangle_{\alpha}}{\langle \sigma_{a_2}\rangle_{0}\cdots \langle \sigma_{a_n}\rangle_{0}} \frac{ \langle \sigma_{a_2}\rangle_{0}\cdots \langle \sigma_{a_n}\rangle_{0}}{\langle\sigma_{a_2}\cdots \sigma_{a_n}\rangle_{0}}\to 1
$$
under the same limit. The critical ratio $\frac{ \langle \sigma_{a_1}\cdots \sigma_{a_n}\rangle_{0}}{\langle\sigma_{a_1}\rangle_{0}\cdots \langle\sigma_{a_n}\rangle_{0}} \to 1$ under the same limit due to \cite[(1.3)]{P}, so the discussion below holds for $\beta = \beta_c$. It remains to note that the exponential factor is also taken to $1$ as $a_1 \to a_b$ again by \eqref{eq:A-integrability}.
\end{proof}

\section{Analyticity of Ising Correlations}\label{sec:ising-combinatorics}
In this section, we show that pure spin and spin-weighted Ising primary field correlations are analytic in the mass $m \in \mathbb R$ and that the power series coefficients around $m_0 =0$ happen to be exactly as expected from formally differentiating the following discrete correspondence (at criticality, the \emph{energy} is simply defined $\epsilon_w=\frac{i}{2}\psi_w\psi_w^*$)
$$
\langle  \mathcal O^\delta \rangle_{m\alpha^\delta} =  \frac{\langle  \mathcal O^\delta \exp\{-{m}\int_\Omega \frac{d^2u}{\pi}\alpha^\delta(u)\epsilon^\delta_u \}\rangle_{0}}{\langle \exp\{-{m}\int_\Omega \frac{d^2u}{\pi} \alpha^\delta(u)\epsilon^\delta_u\} \rangle_{0}},
$$
which we will not attempt to make rigorous in the continuum theory but use as a helpful heuristic. Specifically, we would expect that the expansion of a massive correlation around $m=0$ takes the form
$$
\langle  \mathcal O \rangle_{m\alpha} = \sum_{k=0}^\infty \frac{\partial_m^k\langle  \mathcal O \rangle_{0}}{k!}m^k,$$
with, e.g., $$\partial_m\langle  \mathcal O \rangle_{0} = \left\langle  \mathcal O\right\rangle_{0}\left\langle \int_\Omega \frac{d^2u}{\pi} \alpha(u)\epsilon_u \right\rangle_{0}-\left\langle  {\mathcal O}\int_\Omega \frac{d^2u}{\pi} \alpha(u)\epsilon_u\right\rangle_{0},
$$
and so on.

\begin{remark}\label{rem:usual-conditions}
 Recall that a power series is absolutely convergent within its radius of convergence and its composition with an analytic function is still analytic, which allow for rearrangement, expansion, and re-grouping of terms in a product or composition of multiple series. Nonetheless, it will be useful to have a notion of uniformity across field insertion points staying uniformly away from each other. Given a power series
$$
\sum_{p=0}^\infty {(m-m_0)^p}E_{p,X},
$$
for a function depending on a finite set of insertion points in $\Omega$ indexed by $X$, we say the coefficients satisfy a \emph{locally exponential bound} if
\begin{equation}\label{eq:local-uniform-boundedness}
|E_{p,X}| \leq C(\Omega,M,\kappa_\infty,\kappa_b,\delta_{all},|X|)^p,
\end{equation}
where $\delta_{all}>0$ is smaller than $\frac{1}{16}|x_1-x_2|$ and $\frac12\operatorname{dist}(x_1,\partial \Omega)$ for any $x_1,x_2$ indexed by $X$. Note that a uniform nonzero radius of convergence is provided by estimates of type \eqref{eq:local-uniform-boundedness}, depending on only the six parameters therein.

In fact, the coefficients $E_{p,X}$ will frequently be given by an integral of some function $\mathfrak{E}_{p,X}(\cdot)$ over $\Omega^p$, as in Remark \ref{rem:up-analysis-1}, and we will have a bound of type \eqref{eq:local-uniform-boundedness} for the integral of $|\mathfrak{E}_{p,X}(\cdot)|$ (i.e. integral of the absolute value of the integrand). This implies we are free to use Fubini to merge products of integrals over (powers of) $\Omega$ into a single integral and, e.g., symmetrize over the $p$ components.

In the following, any series expansion will be said to satisfy the \emph{usual conditions} if they have such locally exponentially bounded coefficients (given as explicit bounded integrals involving $\alpha$) and depend continuously on the points indexed by $X$, away from one another and the boundary. Note that usual conditions are preserved under compositions of such series with other analytic functions.
\end{remark}

\subsection{Pure spin correlations}
Our goal is to prove the following. Recall the notation $\sigma_A = \sigma_{a_1}\cdots \sigma_{a_n}$, etc. We will also write $\sigma_{A^\dagger} = \sigma_{a_2}\cdots \sigma_{a_n}$.
\begin{proposition}\label{pr:spinexp}
For distinct points $a_1,\dots,a_n$, the correlation $\log{\langle\sigma_A\rangle_{m\alpha}}$ may be expanded around any $m=m_0$ with the usual conditions. Near $m_0=0$, we have 
\begin{multline}\label{eq:log-ratio}
\log \frac{\langle\sigma_A\rangle_{m\alpha}}{\isingccf{\sigma_A}}=\sum_{p=0}^\infty \frac{(-m)^p}{p!}\int_{\Omega^p}\frac{d^{2p}u}{\pi^p}\, \prod_{j=1}^p \alpha(u_j)\sum_{\Lambda \in \Pi_p}(-1)^{|\Lambda|-1}(|\Lambda|-1)!\\ 
\times \Bigg(\prod_{B\in \Lambda}\frac{\isingccf{\sigma_A\prod_{k\in B}\epsilon_{u_k}}}{\isingccf{\sigma_A}}-\prod_{B\in \Lambda}\isingccf{\prod_{k\in B}\epsilon_{u_k}}\Bigg),
\end{multline}
where $\Pi_p$ denotes the set of partitions of $\{1,\dots,p\}$.
\end{proposition}
Note that series expansion with locally exponentially bounded coefficients follows from Corollary \ref{cor:series-expansion-a}, using the fact that we may integrate the bound \eqref{eq:coeff-estimate} avoiding other spins by closedness and Lemma \ref{lem:crit-welldef}. Then, since $\lambda(a_1';A_p)$ depends locally continuously on $a_1', a_2, \ldots, a_n$ by Proposition \ref{prop:spindiff} (and implicitly dominated convergence on \eqref{eq:cauchy-hol}), another application of dominated convergence implies that $\int_{a_b}^{a_1}\frac12 \lambda(a_1';A_p) da_1'$ also depends locally continuously on $a_1, \ldots, a_n$, as desired. That is, the expansion is under usual conditions.

It remains to identify the expansion around $m_0=0$. To this end, we need the following. Recall the functions $A_p(z)$ from \eqref{eq:ApBp}, at $m_0 = 0$.
\begin{proposition}\label{pr:Ap-induction}
For $p\geq 1$, we have
    \begin{align}\label{eq:Ap-combinatorial-hypothesis}
        &A_{p}(z)=\int_{\Omega^p} \mathfrak{A}^{\text{sym}}_p(u,z) d^{2p}u = \int_{\Omega^p}\frac{d^{2p}u}{\pi^p}\, \prod_{j=1}^p \alpha(u_j)\frac{1}{p!}\sum_{\Lambda\in \Pi_p}(|\Lambda|-1)!(-1)^{|\Lambda|-1}\\ 
        \nonumber\times\sum_{S\in\Lambda}&\Bigg[\frac{\big\langle \sigma_{A^\dagger}\mu_{a_1}\psi_{z}\prod_{j\in S}\epsilon_{u_{j}}\big\rangle_0 }{\left\langle \sigma_{A}\right\rangle_0 }-\frac{\left\langle \sigma_{A^\dagger}\mu_{a_1}\psi_{z}\right\rangle_0 \big\langle \sigma_{A}\prod_{j\in S}\epsilon_{u_{j}}\big\rangle_0 }{\left\langle \sigma_{A}\right\rangle_0^{2}}\Bigg]\prod_{B\in\Lambda\setminus\{S\}}\frac{\big\langle\sigma_{A}\prod_{j\in B}\epsilon_{u_{j}}\big\rangle_0 }{\left\langle \sigma_{A}\right\rangle_0 },
    \end{align}
    where $\mathfrak{A}^{\text{sym}}_p$ is defined as the integrand on the right (see Remark \ref{rem:up-analysis-2} for its integrability).
    
\end{proposition}
\begin{remark}\label{rem:up-analysis-2}
    The proof of Proposition \ref{pr:Ap-induction} will only use equalities valid at the level of the integrand and permutations of components of $u$; therefore, the integrand in \eqref{eq:Ap-combinatorial-hypothesis}, which is evidently invariant under the swap between any $u_j$ and $u_{j'}$, is equal to $$\mathfrak{A}^{\text{sym}}_p(u,z) = \frac{1}{p!}\sum_{s \in S_p}\mathfrak{A}_p(u_{s_1},u_{s_2},\ldots,u_{s_p},z),$$ where the function $\mathfrak{A}_p$ is defined in Remark \ref{rem:up-analysis-1}. Then the estimates \eqref{eq:frak-estimate} and \eqref{eq:summand-bound}, also suitably symmetrized, hold for $|\mathfrak{A}^{\text{sym}}_p|$. In particular, the integral in \eqref{eq:Ap-combinatorial-hypothesis} is well-defined.
\end{remark}
\begin{proof}[Proof of Proposition \ref{pr:Ap-induction}]
Since $\epsilon = \frac{i}{2}\psi \psi ^*$ at criticality, the RHS for the $p = 1$ case resolves to
\begin{align*}
    A_{1}(z)=&\frac{1}{\pi}\int_{\Omega}\alpha(u_1)\Bigg[\frac{\left\langle \sigma_{A^\dagger}\mu_{a_1}\psi_{z}\epsilon_{u_{1}}\right\rangle_0 }{\left\langle \sigma_{A}\right\rangle_0 }-\frac{\left\langle \sigma_{A^\dagger}\mu_{a_1}\psi_{z}\right\rangle_0 \left\langle \sigma_{A}\epsilon_{u_{1}}\right\rangle_0 }{\left\langle \sigma_{A}\right\rangle_0 ^{2}}\Bigg]d^2u_1\\
    =& \frac{1}{\pi}\int_{\Omega}\alpha(u_1)\frac{i}{2}\Bigg[\frac{\left\langle \sigma_{A^\dagger}\mu_{a_1}\psi_{z}\psi_{u_1}\psi_{u_1}^*\right\rangle_0 }{\left\langle \sigma_{A}\right\rangle_0 }-\frac{\left\langle \sigma_{A^\dagger}\mu_{a_1}\psi_{z}\right\rangle_0 \left\langle \sigma_{A}\psi_{u_1}\psi_{u_1}^*\right\rangle_0 }{\left\langle \sigma_{A}\right\rangle_0 ^{2}}\Bigg]d^2u_1.
\end{align*}
This is the induction base case, which we will now show.

Let us simplify the correlations in the integrand, carefully dealing with $\mu_{a_1} = \lim_{a'_1\to a_1} e^{-i\frac\pi4}\allowbreak\times(a_1'-a_1)^{\frac12}\sigma_{a_1}\psi_{a_1'}$ (within correlations, see \eqref{eq:ope}). We have
\begin{align*}
    &\frac{\left\langle \sigma_{A^\dagger}\mu_{a_1}\psi_{z}\psi_{u_1}\psi_{u_1}^*\right\rangle_0 }{\left\langle \sigma_{A}\right\rangle_0 } = \lim_{a'_1\to a_1} e^{-i\pi/4}(a_1'-a_1)^{1/2}\frac{\big\langle \sigma_{A}\psi_{a_1'}\psi_{z}\psi_{u_1}\psi_{u_1}^*\big\rangle_0 }{\left\langle \sigma_{A}\right\rangle_0 } = \lim_{a'_1\to a_1} e^{-i\pi/4}(a_1'-a_1)^{1/2}&\\
    &\qquad\times\Bigg[ \frac{\big\langle \sigma_{A}\psi_{a'_1}\psi_{z}\big\rangle_0 }{\left\langle \sigma_{A}\right\rangle_0 } \frac{\left\langle \sigma_{A}\psi_{u_1}\psi_{u_1}^*\right\rangle_0 }{\left\langle \sigma_{A}\right\rangle_0 } - \frac{\big\langle \sigma_{A}\psi_{a_1'}\psi_{u_1}\big\rangle_0 }{\left\langle \sigma_{A}\right\rangle_0 }\frac{\left\langle \sigma_{A}\psi_{z}\psi_{u_1}^*\right\rangle_0 }{\left\langle \sigma_{A}\right\rangle_0 }+\frac{\left\langle \sigma_{A}\psi_{z}\psi_{u_1}\right\rangle_0 }{\left\langle \sigma_{A}\right\rangle_0 }\frac{\big\langle \sigma_{A}\psi_{a_1'}\psi_{u_1}^*\big\rangle_0 }{\left\langle \sigma_{A}\right\rangle_0 }\Bigg]&\\
    &\quad= \frac{\left\langle \sigma_{A^\dagger}\mu_{a_1}\psi_{z}\right\rangle_0 }{\left\langle \sigma_{A}\right\rangle_0 } \frac{\left\langle \sigma_{A}\psi_{u_1}\psi_{u_1}^*\right\rangle_0 }{\left\langle \sigma_{A}\right\rangle_0 } - \frac{\left\langle \sigma_{A^\dagger}\mu_{a_1}\psi_{u_1}\right\rangle_0 }{\left\langle \sigma_{A}\right\rangle_0 }\frac{\left\langle \sigma_{A}\psi_{z}\psi_{u_1}^*\right\rangle_0 }{\left\langle \sigma_{A}\right\rangle_0 }+\frac{\left\langle \sigma_{A}\psi_{z}\psi_{u_1}\right\rangle_0 }{\left\langle \sigma_{A}\right\rangle_0 }\frac{\left\langle \sigma_{A^\dagger}\mu_{a_1}\psi_{u_1}^*\right\rangle_0 }{\left\langle \sigma_{A}\right\rangle_0 }.&
\end{align*}
Now we may simplify the integral above (see also \eqref{eq:fermion-complex} and the remark below for how correlations behave under conjugation):
\begin{align*}
    A_{1}(z)=& \frac{1}{\pi}\int_{\Omega}\alpha(u_1)\frac{i}{2}\Bigg[-\frac{\left\langle \sigma_{A^\dagger}\mu_{a_1}\psi_{u_1}\right\rangle_0 }{\left\langle \sigma_{A}\right\rangle_0 }\frac{\left\langle \sigma_{A}\psi_{z}\psi_{u_1}^*\right\rangle_0 }{\left\langle \sigma_{A}\right\rangle_0 }+\frac{\left\langle \sigma_{A}\psi_{z}\psi_{u_1}\right\rangle_0 }{\left\langle \sigma_{A}\right\rangle_0 }\frac{\left\langle \sigma_{A^\dagger}\mu_{a_1}\psi_{u_1}^*\right\rangle_0 }{\left\langle \sigma_{A}\right\rangle_0}\Bigg]d^2u_1\\
    =&\frac{1}{\pi}\int_{\Omega}\Bigg[i\alpha(u_1)\frac{\left\langle \sigma_{A^\dagger}\mu_{a_1}\psi_{u_1}^*\right\rangle_0 }{\left\langle \sigma_{A}\right\rangle_0}\Bigg] \star \frac{\big\langle \sigma_{A}\psi_{z}\psi_{u_1}^{[\star]}\big\rangle_0 }{\left\langle \sigma_{A}\right\rangle_0 }d^2u_1,
\end{align*}
thanks to \eqref{eq:star}, as desired. We see that a form of Pfaffian expansion is still valid for disorder insertions; for notational simplicity, we will henceforth write $\mu_a$ as $\sigma_a\psi_a$ to do such expansions,  carefully converting the combination back to $\mu_a$ at the end (respectively marked by $\stackrel{\sigma\psi}{=}$ and $\stackrel{\mu}{=}$).

Assuming \eqref{eq:Ap-combinatorial-hypothesis} as the inductive hypothesis for $p$, we have (again using \eqref{eq:star})
\begin{align*}
    A_{p+1}(z)
    =& \frac{1}{\pi}\int_{\Omega}\alpha(u_{p+1})\frac{i}{2}\Bigg[\overline{A_{p}(u_{p+1})}\frac{\left\langle \sigma_{A}\psi_{z}\psi_{u_{p+1}}\right\rangle_0 }{\left\langle \sigma_{A}\right\rangle_0 }-A_{p}(u_{p+1})\frac{\big\langle \sigma_{A}\psi_{z}\psi_{u_{p+1}}^*\big\rangle_0 }{\left\langle \sigma_{A}\right\rangle_0 }\Bigg]d^{2}u_{p+1}.
\end{align*}
If we plug in the inductive hypothesis for $p$, we will be left with the integral over $\Omega^{p+1}$ (with measure $\frac{d^{2(p+1)}u}{\pi^{p+1}}\prod_{j=1}^{p+1}\alpha(u_j)$ as desired) of a linear combination of terms which are all products of $\sigma_A$-weighted Ising correlations. We check that they contribute all types of terms present in \eqref{eq:Ap-combinatorial-hypothesis} for $p+1$ with correct constants in front.

Fix $\Lambda\in \Pi_p$ and $S\in \Lambda$, which fixes all the pre-factors in front of the correlations and the post-factor $\prod_{B\in\Lambda\setminus\{S\}}\frac{\left\langle\sigma_{A}\prod_{j\in B}\epsilon_{u_{j}}\right\rangle_0 }{\left\langle \sigma_{A}\right\rangle_0 }$. Consider the first term in the summand in the second line of \eqref{eq:Ap-combinatorial-hypothesis}. We have the product
\begin{align}\label{eq:spin-inductive-step}
    \nonumber\frac{i}{2}\frac{\big\langle \sigma_{A^\dagger}\mu_{a_1}\psi_{u_{p+1}}^*\prod_{j\in S}\epsilon_{u_{j}}\big\rangle_0 }{\left\langle \sigma_{A}\right\rangle_0 }&\frac{\left\langle \sigma_{A}\psi_{z}\psi_{u_{p+1}}\right\rangle_0 }{\left\langle \sigma_{A}\right\rangle_0 }-\frac{i}{2}\frac{\big\langle \sigma_{A^\dagger}\mu_{a_1}\psi_{u_{p+1}}\prod_{j\in S}\epsilon_{u_{j}}\big\rangle_0 }{\big\langle \sigma_{A}\big\rangle_0 }\frac{\big\langle \sigma_{A}\psi_{z}\psi_{u_{p+1}}^*\big\rangle_0 }{\left\langle \sigma_{A}\right\rangle_0 }\\
    \stackrel{\sigma\psi}{=} \left(\frac{i}{2}\right)^{|S|+1}&\frac{\big\langle \sigma_{A}\psi_{a_1}\psi_{u_{p+1}}^*\prod_{j\in S}\psi_{u_{j}}\psi_{u_{j}}^*\big\rangle_0 }{\left\langle \sigma_{A}\right\rangle_0 }\frac{\left\langle \sigma_{A}\psi_{u_{p+1}}\psi_{z}\right\rangle_0 }{\left\langle \sigma_{A}\right\rangle_0 }\\
    \nonumber &-\left(\frac{i}{2}\right)^{|S|+1}\frac{\big\langle \sigma_{A}\psi_{a_1}\psi_{u_{p+1}}\prod_{j\in S}\psi_{u_{j}}\psi_{u_{j}}^*\big\rangle_0 }{\left\langle \sigma_{A}\right\rangle_0 }\frac{\big\langle \sigma_{A}\psi_{u_{p+1}}^*\psi_{z}\big\rangle_0 }{\left\langle \sigma_{A}\right\rangle_0 }.
\end{align}
Comparing with the recursive expansion for Pfaffians (e.g. see the proof of Proposition \ref{prop:fermion-antisymmetry}), we see that the above two terms come from the expansion of the $2|S|+4$ point correlation $\frac{\left\langle \sigma_{A}\psi_{a_1}\psi_z\prod_{j\in S\cup\{p+1\}}\psi_{u_{j}}\psi_{u_{j}}^*\right\rangle_0 }{\left\langle \sigma_{A}\right\rangle_0 }$ that keeps $\psi_z$ in the $2$-point correlation. In fact, thanks to the fact that $A_{p+1}$ is an integral on all of $\Omega^{p+1}$, we may in total obtain $2|S|+2$ out of the $2|S|+3$ terms in the expansion by adding the terms where the places of each $u_j$ and $u_{p+1}$ are swapped; once integrated, this process would add $|S|$ copies of the same integral, so we would need to divide by $|S|+1$ to keep the same normalization for $A_{p+1}$. Marking this symmetrization for the integrand with $\stackrel{sym}{=}$, we have, where $S' = S\cup \{p+1\}$,
\begin{align*}
   \stackrel{sym}{=} &\frac{1}{|S|+1}\left(\frac{i}{2}\right)^{|S|+1}\sum_{j'\in S'}\Bigg[\frac{\big\langle \sigma_{A}\psi_{u_{j'}}^*\psi_{a_1}\prod_{j\in S'\setminus\{j'\}}\psi_{u_{j}}\psi_{u_{j}}^*\big\rangle_0 }{\big\langle \sigma_{A}\big\rangle_0 }\frac{\big\langle \sigma_{A}\psi_{u_{j'}}\psi_{z}\big\rangle_0 }{\big\langle \sigma_{A}\big\rangle_0 }\\
   &\qquad\qquad\qquad\qquad\qquad\qquad-\frac{\big\langle \sigma_{A}\psi_{u_{j'}}\psi_{a_1}\prod_{j\in S'\setminus\{j'\}}\psi_{u_{j}}\psi_{u_{j}}^*\big\rangle_0 }{\big\langle \sigma_{A}\big\rangle_0 }\frac{\big\langle \sigma_{A}\psi_{u_{j'}}^*\psi_{z}\big\rangle_0 }{\big\langle \sigma_{A}\big\rangle_0 }\Bigg]\\
   =&\frac{1}{|S'|}\left(\frac{i}{2}\right)^{|S|+1}\Bigg[\frac{\big\langle \sigma_{A}\psi_{a_1}\psi_{z}\prod_{j\in S'}\psi_{u_{j}}\psi_{u_{j}}^*\big\rangle_0 }{\big\langle \sigma_{A}\big\rangle_0 }-\frac{\big\langle \sigma_{A}\prod_{j\in S'}\psi_{u_{j}}\psi_{u_{j}}^*\big\rangle_0 }{\big\langle \sigma_{A}\big\rangle_0 }\frac{\big\langle \sigma_{A}\psi_{a_1}\psi_{z}\big\rangle_0 }{\big\langle \sigma_{A}\big\rangle_0 }\Bigg]\\
   \stackrel{\mu}{=}&\frac{1}{|S'|}\Bigg[\frac{\big\langle \sigma_{A^\dagger}\mu_{a_1}\psi_{z}\prod_{j\in S'}\epsilon_{u_{j}}\big\rangle_0 }{\big\langle \sigma_{A}\big\rangle_0 }-\frac{\big\langle \sigma_{A}\prod_{j\in S'}\epsilon_{u_{j}}\big\rangle_0 }{\big\langle \sigma_{A}\big\rangle_0 }\frac{\big\langle \sigma_{A^\dagger}\mu_{a_1}\psi_{z}\big\rangle_0 }{\big\langle \sigma_{A}\big\rangle_0 }\Bigg].
\end{align*}
This is exactly what the induction hypothesis seems to state for $p+1$, apart from the pre-factor $1/|S'|$. This apparent mismatch comes from the fact that $S'$ is fixed to be the element of the partition $\Lambda':=\Lambda\cup\{S'\}\setminus\{S\} \in \Pi_{p+1}$ that contains $p+1$. This suggests another symmetrization, now between $u_{p+1}$ and all other $u_j$, requiring another pre-factor $\frac{1}{p+1}$, correctly transforming $\frac{1}{p!}$ to $\frac{1}{(p+1)!}$. Then any pair $(\Lambda',S')$ where $\Lambda'\in \Pi_{p+1}$ and $|S'|>1$ appears exactly $|S'|$ times, precisely canceling out the mismatching pre-factor.

Now let us look at the second term in the second line of \eqref{eq:Ap-combinatorial-hypothesis}. In much a similar way as in the first term, we have $-\frac{\left\langle \sigma_{A}\prod_{j\in S}\epsilon_{u_{j}}\right\rangle_0  }{\left\langle \sigma_{A}\right\rangle_0}$ times
\begin{align*}
&\,\frac{i}{2}\frac{\big\langle \sigma_{A^\dagger}\mu_{a_1}\psi_{u_{p+1}}^*\big\rangle_0 }{\left\langle \sigma_{A}\right\rangle_0 }\frac{\left\langle \sigma_{A}\psi_{z}\psi_{u_{p+1}}\right\rangle_0 }{\left\langle \sigma_{A}\right\rangle_0 }-\frac{i}{2}\frac{\left\langle \sigma_{A^\dagger}\mu_{a_1}\psi_{u_{p+1}}\right\rangle_0  }{\left\langle \sigma_{A}\right\rangle_0}\frac{\big\langle \sigma_{A}\psi_{z}\psi_{u_{p+1}}^*\big\rangle_0 }{\left\langle \sigma_{A}\right\rangle_0 }\\
\stackrel{\sigma\psi}{=}&\,\frac{i}{2}\frac{\big\langle \sigma_{A}\psi_{a_1}\psi_{u_{p+1}}^*\big\rangle_0 }{\left\langle \sigma_{A}\right\rangle_0 }\frac{\left\langle \sigma_{A}\psi_{z}\psi_{u_{p+1}}\right\rangle_0 }{\left\langle \sigma_{A}\right\rangle_0 }-\frac{i}{2}\frac{\left\langle \sigma_{A}\psi_{a_1}\psi_{u_{p+1}}\right\rangle_0  }{\left\langle \sigma_{A}\right\rangle_0}\frac{\big\langle \sigma_{A}\psi_{z}\psi_{u_{p+1}}^*\big\rangle_0 }{\left\langle \sigma_{A}\right\rangle_0 }\\
=&\,\frac{i}{2}\frac{\big\langle \sigma_{A}\psi_{a_1}\psi_{u_{p+1}}^*\psi_{z}\psi_{u_{p+1}}\big\rangle_0 }{\left\langle \sigma_{A}\right\rangle_0 }+\frac{i}{2}\frac{\left\langle \sigma_{A}\psi_{a_1}\psi_{z}\right\rangle_0  }{\left\langle \sigma_{A}\right\rangle_0}\frac{\big\langle \sigma_{A}\psi_{u_{p+1}}^*\psi_{u_{p+1}}\big\rangle_0 }{\left\langle \sigma_{A}\right\rangle_0 }\\
\stackrel{\mu}{=}&\,\frac{\left\langle \sigma_{A^\dagger}\mu_{a_1}\psi_{z}\epsilon_{u_{p+1}}\right\rangle_0 }{\left\langle \sigma_{A}\right\rangle_0 }-\frac{\left\langle \sigma_{A^\dagger}\mu_{a_1}\psi_{z}\right\rangle_0  }{\left\langle \sigma_{A}\right\rangle_0}\frac{\left\langle \sigma_{A}\epsilon_{u_{p+1}}\right\rangle_0 }{\left\langle \sigma_{A}\right\rangle_0 }.
\end{align*}
This corresponds to $\Lambda'=(\Lambda\cup \{p+1\})\in \Pi_{p+1}$, so we still have a mismatch by the factor $\frac{|\Lambda|}{p+1}$. The fact that $(-1)^{|\Lambda'|} = -(-1)^{|\Lambda|}$ is addressed by the shared factor $-\frac{\left\langle \sigma_{A}\prod_{j\in S}\epsilon_{u_{j}}\right\rangle_0  }{\left\langle \sigma_{A}\right\rangle_0}$. Note that each $S \in \Lambda$ gives rise to the same term, i.e., if $\Lambda'\in \Pi_{p+1}$ with $\{p+1\} \in \Lambda'$, the combination $(\Lambda', \{p+1\})$ would appear $|\Lambda'|-1$ times, so we can multiply by $|\Lambda'|-1$ instead (which exactly transforms $(|\Lambda|-1)! = (|\Lambda'|-2)!$ to $(|\Lambda'|-1)!$ as desired) and have $(\Lambda', \{p+1\})$ appear exactly once. Then, after the same symmetrization from above, we get that any $(\Lambda', S')$, with $|S'|=1$, appears once, as desired, while the normalization $\frac{1}{p+1}$ precisely supplies the remaining factor.

Therefore the inductive hypothesis is true for $p+1$ and we have shown \eqref{eq:Ap-combinatorial-hypothesis}.
\end{proof}

\begin{proof}[Proof of Proposition \ref{pr:spinexp}]
By Corollary \ref{cor:series-expansion-a} and \eqref{eq:spin-integral}, we have
$$
  \log\frac{ \langle \sigma_{A'}\rangle_{m\alpha}}{\langle \sigma_{A'}\rangle_{0}}\bigg|_{a_1'=a_b}^{a_1'=a_1} = \Re\bigg[\int_{a_b}^{a_1} \left(\mathcal{A}_\Omega^{(m\alpha)}-\mathcal{A}_\Omega^{(0)}\right)(a_1')da_1'\bigg]= \sum_{p=1}^\infty (-m)^p \cdot\Re \bigg[\int_{a_b}^{a_1}\frac12\lambda\left(a_1';A_p\right)da_1'\bigg],
$$
where $\sigma_{A'} := \sigma_{a_1'}\sigma_{a_2}\cdots\sigma_{a_n}, A_p = A_p(a_1',a_2,\ldots,a_n;\cdot)$.

We claim that we can extract $\lambda$ from $A_p$ by using the critical OPE \cite[(6.6)]{CHI2} which we reproduce below:
\begin{align}
\nonumber\mu_{a}\psi_z &= -\frac{e^{-\frac{i\pi}{4}}}{\sqrt{z-a}}\left(\sigma_{a}+4\partial_a\sigma_a\cdot(z-a)+O(|z-a|^{2})\right),\\ \label{eq:critical-ope}
\sigma_{a}\psi_z &= \frac{e^{\frac{i\pi}{4}}}{\sqrt{z-a}}\left(\mu_{a}+4\partial_a\mu_a\cdot(z-a)+O(|z-a|^{2})\right),
\end{align}
directly on the integrand $\mathfrak{A}^{\text{sym}}_p = \mathfrak{A}^{\text{sym}}_p(a_1',a_2, \ldots, a_n; \cdot)$; in other words, we have
\begin{align}\label{eq:lambda-inside}
    \lambda\left(a_1';A_{p}\right)&=4\int_{\Omega^{p}}\frac{d^{2p}u}{\pi^p}\, \prod_{j=1}^p \alpha(u_j)\frac{1}{p!}\sum_{\Lambda\in \Pi_p}(|\Lambda|-1)!(-1)^{|\Lambda|-1}\\ 
    \nonumber&\hspace{-25pt}\times\sum_{C\in\Lambda}\Bigg[\frac{\big\langle \sigma_{A^\dagger}\partial_{a_1'}\sigma_{a_1'}\prod_{j\in C}\epsilon_{u_{j}}\big\rangle_0 }{\left\langle \sigma_{A'}\right\rangle_0 }-\frac{\big\langle \sigma_{A^\dagger}\partial_{a_1'}\sigma_{a_1'}\big\rangle_0 \big\langle \sigma_{A'}\prod_{j\in C}\epsilon_{u_{j}}\big\rangle_0 }{\left\langle \sigma_{A'}\right\rangle_0^{2}}\Bigg]\prod_{B\in\Lambda\setminus\{C\}}\frac{\big\langle\sigma_{A'}\prod_{j\in B}\epsilon_{u_{j}}\big\rangle_0 }{\left\langle \sigma_{A'}\right\rangle_0 }\\
    \nonumber&=4\int_{\Omega^{p}}\frac{d^{2p}u}{\pi^p}\, \prod_{j=1}^p \alpha(u_j)\frac{1}{p!}\sum_{\Lambda\in \Pi_p}(|\Lambda|-1)!(-1)^{|\Lambda|-1}\cdot\partial_{a_1'}\Bigg[\prod_{B\in\Lambda}\frac{\big\langle\sigma_{A'}\prod_{j\in B}\epsilon_{u_{j}}\big\rangle_0 }{\left\langle \sigma_{A'}\right\rangle_0}\Bigg].
\end{align}
We also claim that we may integrate with respect to $a_1'$ from the boundary to $a_1$ under the $\int_{\Omega^p}$ sign, so we would have
\begin{multline*}
\log \frac{\langle\sigma_A\rangle_{m\alpha}}{\isingccf{\sigma_A}} - \log \frac{\langle\sigma_{A^\dagger}\rangle_{m\alpha}}{\isingccf{\sigma_{A^\dagger}}}=\sum_{p=0}^\infty \frac{(-m)^p}{p!}\int_{\Omega^p}\frac{d^{2p}u}{\pi^p}\, \prod_{j=1}^p \alpha(u_j)\sum_{\Lambda \in \Pi_p}(-1)^{|\Lambda|-1}(|\Lambda|-1)!\\ 
\times \left(\prod_{B\in \Lambda}\frac{\isingccf{\sigma_A\prod_{k\in B}\epsilon_{u_k}}}{\isingccf{\sigma_A}}-\prod_{B\in \Lambda}\frac{\isingccf{\sigma_{A^\dagger}\prod_{k\in B}\epsilon_{u_k}}}{\isingccf{\sigma_{A^\dagger}}}\right),
\end{multline*}
which yields \eqref{eq:log-ratio} by inducting on $n = |A|$. Note that we also used that
$$
\frac{\big\langle\sigma_{A'}\prod_{j\in B}\epsilon_{u_{j}}\big\rangle_0 }{\left\langle \sigma_{A'}\right\rangle_0} \to \frac{\big\langle\sigma_{A^\dagger}\prod_{j\in B}\epsilon_{u_{j}}\big\rangle_0 }{\left\langle \sigma_{A^\dagger}\right\rangle_0}\text{ as }\sigma_{a_1'}\to\partial \Omega,
$$
by Propositions \ref{prop:spin-boundary-limit} and \ref{prop:fermion-antisymmetry}.

We show these two claims at one go, i.e.,
\begin{equation}\label{eq:line-integral-under}
\Re \bigg[\int_{a_b}^{a_1} \lambda(a_1';A_p) da_1'\bigg] = \int_{\Omega^p}\Re \left[\int_{a_b}^{a_1} \lambda(a_1';\mathfrak{A}_p^{\text{sym}}) da_1'\right]d^{2p}u,
\end{equation}
where $\lambda(a_1';\mathfrak{A}_p^{\text{sym}})$ is the symmetrization of
\begin{align*}
    \lambda(a_1';\mathfrak{A}_p) &=\frac1\pi \left[i\alpha(u_p)\overline{\mathfrak{A}_{p-1}(u)}\right]\star\lambda\Bigg(z = a_1',\frac{\langle  \sigma_{a_1'}\sigma_{a_2}\cdots\sigma_{a_n}\psi_{z}\psi^{[\star]}_{u_p}\rangle_{0}}{\langle \sigma_{a_1'}\sigma_{a_2}\cdots \sigma_{a_n}\rangle_{0}}\Bigg)\\ 
    &= \frac{4}{\pi i}\left[i\alpha(u_p)\overline{\mathfrak{A}_{p-1}(u)}\right]\star\frac{\langle  \partial_{a_1'}\mu_{a_1'}\sigma_{a_2}\cdots\sigma_{a_n}\psi^{[\star]}_{u_p}\rangle_{0}}{\langle \sigma_{a_1'}\sigma_{a_2}\cdots \sigma_{a_n}\rangle_{0}}.
\end{align*}
Let $d = \frac{1}{4}\min(\operatorname{dist}(a_1',\partial \Omega), \delta_0)$. We extract
$$
\int_{a_b}^{a_1} \lambda(a_1';A_p) da_1' = -e^{i\pi/4}\int_{a_b}^{a_1} \lim_{\epsilon \downarrow 0} \left[\oint_{\partial B_{\epsilon d}(a_1')} \int_{\Omega^p}\frac{ \mathfrak{A}_p(u,z)}{(z-a_1')^{\frac32}}d^{2p}u \,\frac{dz}{2\pi i} \right]da_1'.
$$
By \eqref{eq:ap-coeff-bound} and dominated convergence, we may pull out the limit outside of the $da_1'$ integral. So, let us study the triple integral, which is equal to (fix a straight path $\gamma$ from some $a_b$ to $a_1$ so that we may fix a uniform $\delta_0$, which is valid for $a_1'$ along $\gamma$, then write $\gamma_\epsilon: = \bigcup_{a_1'\in \gamma}B_{\epsilon d}(a_1')$ and $\Omega_\epsilon := \Omega \setminus \gamma_\epsilon$)

\begin{align*}
    \int_{a_b}^{a_1} \oint_{\partial B_{\epsilon d}(a_1')} \int_{\Omega^p}\frac{ \mathfrak{A}_p(u,z)\mathbf{1}_{\Omega_\epsilon}(u_p)}{(z-a_1')^{\frac32}}d^{2p}u \,\frac{dz}{2\pi i} \,da_1' + \int_{a_b}^{a_1} \oint_{\partial B_{\epsilon d}(a_1')} \int_{\Omega^p}\frac{ \mathfrak{A}_p(u,z)\mathbf{1}_{\gamma_\epsilon}(u_p)}{(z-a_1')^{\frac32}}d^{2p}u \,\frac{dz}{2\pi i} \,da_1'\\
    = \int_{a_b}^{a_1} \int_{\Omega^p}\frac{\lambda(a_1';\mathfrak{A}_p)\mathbf{1}_{\Omega_\epsilon}(u_p)}{-e^{i\pi/4}}{d^{2p}u \,da_1'} + \int_{a_b}^{a_1} \oint_{\partial B_{\epsilon d}(a_1')} \int_{\Omega^p}\frac{ \mathfrak{A}_p(u,z)\mathbf{1}_{\gamma_\epsilon}(u_p)}{(z-a_1')^{\frac32}}d^{2p}u \,\frac{dz}{2\pi i} \,da_1'\\
    =  \int_{\Omega^p} \int_{a_b}^{a_1} \frac{\lambda(a_1';\mathfrak{A}_p)\mathbf{1}_{\Omega_\epsilon}(u_p)}{-e^{i\pi/4}}{da_1'\,d^{2p}u} + \int_{a_b}^{a_1} \oint_{\partial B_{\epsilon d}(a_1')} \int_{\Omega^p}\frac{ \mathfrak{A}_p(u,z)\mathbf{1}_{\gamma_\epsilon}(u_p)}{(z-a_1')^{\frac32}}d^{2p}u \,\frac{dz}{2\pi i} \,da_1',
\end{align*}
Note that $\mathfrak{A}_p$ is meromorphic (on the double cover) unlike $A_p$, so we do not need to take the limit in order to use the residue formula, and Remark \ref{rem:up-analysis-1} justifies the multiple instances of Fubini for the first term.

For the remainder term (second term), note that
\begin{multline*}
\int_{\Omega^{p-1}}{\mathfrak{A}_p(u,z)}d^2u_1\cdots d^2u_{p-1} = \frac{1}{\pi}\left[i\alpha(u_p)\overline{A_{p-1}(u_p)}\right]\star \frac{\big\langle  \sigma_{a_1'}\sigma_{a_2}\cdots\sigma_{a_n}\psi_{z}\psi^{[\star]}_{u_p}\big\rangle_{0}}{\langle \sigma_{a_1'}\sigma_{a_2}\cdots \sigma_{a_n}\rangle_{0}} \\
=\frac{i\alpha(u_p)}{2\pi}\left[\overline{A_{p-1}(u_p)} \frac{\langle  \sigma_{a_1'}\sigma_{a_2}\cdots\sigma_{a_n}\psi_{z}\psi^{}_{u_p}\rangle_{0}}{\langle \sigma_{a_1'}\sigma_{a_2}\cdots \sigma_{a_n}\rangle_{0}} -{A_{p-1}(u_p)} \frac{\langle  \sigma_{a_1'}\sigma_{a_2}\cdots\sigma_{a_n}\psi_{z}\psi^{*}_{u_p}\rangle_{0}}{\langle \sigma_{a_1'}\sigma_{a_2}\cdots \sigma_{a_n}\rangle_{0}}\right].
\end{multline*}
From \eqref{eq:ap-coeff-bound} and using the same notation for $C$, we have for $u_p \in B_{d/16}(a_1')$ 
$$
\left| A_{p-1}^{\dagger\dagger} (u_p)\right| \leq {C^{p-1}d^{-\frac12}}{{|u_p-a_1'|}^{\frac12}}\text{ for }\beta_{(p-1)}\in i\mathbb R,\text{ where }A_{p-1}^{\dagger\dagger} (u_p):= A_{p-1} (u_p) + \frac{e^{-\frac{i\pi}{4}}\beta_{(p-1)}}{\sqrt{u_p-a_1'}}.
$$
In addition, thanks to (anti)meromorphicity, the critical OPE \eqref{eq:critical-ope}, and the bounds \eqref{eq:critical-ff-bound}, we have for $u_p \in B_{d/16}(a_1'), z\in \partial B_{\epsilon d}(a_1')$
\begin{align*}
    \left|\frac{\langle  \sigma_{a_1'}\sigma_{a_2}\cdots\sigma_{a_n}\psi_{z}\psi^{}_{u_p}\rangle_{0}}{\langle \sigma_{a_1'}\sigma_{a_2}\cdots \sigma_{a_n}\rangle_{0}} + \frac{e^{\frac{i\pi}{4}}}{\sqrt{u_p - a_1'}}\frac{\langle  \mu_{a_1'}\sigma_{a_2}\cdots\sigma_{a_n}\psi_{z}\rangle_{0}}{\langle \sigma_{a_1'}\sigma_{a_2}\cdots \sigma_{a_n}\rangle_{0}} \right| &\leq \frac{C|u_p-a_1'|^{\frac12}}{d^{\frac12}|z-u_p|},\\
    \left|\frac{\langle  \sigma_{a_1'}\sigma_{a_2}\cdots\sigma_{a_n}\psi_{z}\psi^{*}_{u_p}\rangle_{0}}{\langle \sigma_{a_1'}\sigma_{a_2}\cdots \sigma_{a_n}\rangle_{0}} + \frac{e^{-\frac{i\pi}{4}}}{\overline{\sqrt{u_p - a_1'}}}\frac{\langle  \mu_{a_1'}\sigma_{a_2}\cdots\sigma_{a_n}\psi_{z}\rangle_{0}}{\langle \sigma_{a_1'}\sigma_{a_2}\cdots \sigma_{a_n}\rangle_{0}}  \right| &\leq \frac{C|u_p-a_1'|^{\frac12}}{d^{\frac12}|z-u_p|}.
\end{align*}
Then it is straightforward to check that (the $1/|u_p-a_1'|$ term being in fact canceled)
$$
\Bigg|\int_{\Omega^{p-1}} { \mathfrak{A}_p(u,z)}d^{2}u_1\cdots d^2u_{p-1} - \frac{\alpha(u_p)}{\pi}\Im \Bigg[ \frac{e^{\frac{i\pi}{4}}\overline{A_{p-1}^{\dagger\dagger}(u_p)}}{{\sqrt{u_p - a_1'}}} \Bigg]\frac{\langle  \mu_{a_1'}\sigma_{a_2}\cdots\sigma_{a_n}\psi_{z}\rangle_{0}}{\langle \sigma_{a_1'}\sigma_{a_2}\cdots \sigma_{a_n}\rangle_{0}}\Bigg| \leq C^p\frac{1}{|z-u_p|}.
$$
We have, now for $C = C(\Omega,\kappa_\infty,n,\delta_0, \kappa_b, \gamma)$,
\begin{align*}
&\Bigg|\oint_{\partial B_{\epsilon d}(a_1')} \int_{\Omega}\int_{\Omega^{p-1}} { \mathfrak{A}_p(u,z)}d^{2}u_1\cdots d^2u_{p-1}\frac{\mathbf{1}_{\{\gamma_\epsilon \cap B_{d/16}(a_1')\}}(u_p)}{(z-a_1')^{\frac32}}d^{2}u_p \,\frac{dz}{2\pi i}\Bigg|\\
&\leq\! \Bigg| \int_{\Omega}\!\frac{\alpha(u_p)\mathbf{1}_{\{\gamma_\epsilon \cap B_{d/16}(a_1')\}}(u_p)}{\pi}\Im \!\Bigg[ \frac{e^{\frac{i\pi}{4}}\overline{A_{p-1}^{\dagger\dagger}(u_p)}}{{\sqrt{u_p - a_1'}}} \Bigg] \!d^2u_p \!\cdot\! \oint_{\partial B_{\epsilon d}(a_1')} \!\!\frac{\langle  \mu_{a_1'}\sigma_{a_2}\cdots\sigma_{a_n}\psi_{z}\rangle_{0}}{\langle \sigma_{a_1'}\sigma_{a_2}\cdots \sigma_{a_n}\rangle_{0}}\frac{dz}{2\pi i(z-a_1')^{\frac32}} \Bigg|\\ 
&\quad+ C^p\int_{\partial B_{\epsilon d}(a_1')} \int_{\Omega}\frac{\mathbf{1}_{\{\gamma_\epsilon \cap B_{d/16}(a_1')\}}(u_p)}{{|z-u_p|}|z-a_1'|^{\frac32}}d^{2}u_p \,{|dz|}\leq C^p\Bigg(\frac{\epsilon d}{d^{\frac12+1}}+ \frac{\epsilon d|\log(\frac{1}{\epsilon})|}{(\epsilon d)^{\frac12}}\Bigg)\leq  C^p\frac{\epsilon^{\frac12}|\log\epsilon|}{d^{\frac12}},
\end{align*}
using the above bound for $A_{p-1}^{\dagger\dagger}$ and since
$$
\int_{\{\gamma_\epsilon \cap B_{d/16}(a_1')\}}\!\frac{d^{2}u_p}{{|z-u_p|}}  \leq C \epsilon d\left|\log \frac{1}{\epsilon}\right|\enspace\text{and}\enspace\left|\oint_{\partial B_{\epsilon d}(a_1')} \!\frac{\langle  \mu_{a_1'}\sigma_{a_2}\cdots\sigma_{a_n}\psi_{z}\rangle_{0}}{\langle \sigma_{a_1'}\sigma_{a_2}\cdots \sigma_{a_n}\rangle_{0}}\frac{dz}{2\pi i(z-a_1')^{\frac32}} \right|
\leq Cd^{-1},
$$
by, e.g., a scaling argument and \eqref{eq:uniform-bound} respectively.

Similarly,
\begin{align*}
&\Bigg|\oint_{\partial B_{\epsilon d}(a_1')} \int_{\Omega}\int_{\Omega^{p-1}} { \mathfrak{A}_p(u,z)}d^{2}u_1\cdots d^2u_{p-1}\frac{\mathbf{1}_{\{\gamma_\epsilon \setminus B_{d/16}(a_1')\}}(u_p)}{(z-a_1')^{\frac32}}d^{2}u_p \,\frac{dz}{2\pi i}\Bigg|\\
&=\Bigg|\int_{\gamma_\epsilon \setminus B_{d/16}(a_1')}\frac{4e^{\frac{i\pi}{4}}}{\pi}\left[i\alpha(u_p)\overline{A_{p-1}(u_p)}\right]\star \frac{\langle  \partial_{a_1'}\mu_{a_1'}\sigma_{a_2}\cdots\sigma_{a_n}\psi^{[\star]}_{u_p}\rangle_{0}}{\langle \sigma_{a_1'}\sigma_{a_2}\cdots \sigma_{a_n}\rangle_{0}}d^{2}u_p\Bigg|\\
&\leq  \,C^p \Bigg|\int_{\gamma_\epsilon \setminus B_{d/16}(a_1')}\Bigg( \frac{d^\frac12}{\sqrt{|u_p-a_1'|}}+1\Bigg)\cdot \frac{d^{-1/2}}{|u_p-a_1'|^{2}} d^2u_p\Bigg|,
\end{align*}
where in the last line we used \eqref{eq:summand-bound} and Lemma \ref{lem:halforderexpansion-zeroth} respectively on the two factors, since $\frac{\langle  \partial_{a_1'}\mu_{a_1'}\sigma_{a_2}\cdots\sigma_{a_n}\psi^{[\star]}_{u_p}\rangle_{0}}{\langle \sigma_{a_1'}\sigma_{a_2}\cdots \sigma_{a_n}\rangle_{0}}$ may be recovered from the coefficient of $(z-a_1')^{1/2}$ of $\frac{\langle  \sigma_{a_1'}\sigma_{a_2}\cdots\sigma_{a_n}\psi_z\psi^{[\star]}_{u_p}\rangle_{0}}{\langle \sigma_{a_1'}\sigma_{a_2}\cdots \sigma_{a_n}\rangle_{0}}$ by the critical OPE \eqref{eq:critical-ope}. Continuing on,

\begin{align*}
&\leq C^p \left( \int_{\gamma_\epsilon \setminus B_{d/16}(a_1')} \frac{1}{|u_p-a_1'|^{\frac52}} d^2u_p + d^{-\frac12}\int_{\gamma_\epsilon \setminus B_{d/16}(a_1')} \frac{1}{|u_p-a_1'|^2} d^2u_p\right)\\
&\leq C^p\left(\epsilon\cdot d^{-\frac12} +  d^{-\frac12}\cdot \epsilon\cdot (1+\log|d|)\right) \\
&\leq C^p\epsilon d^{-\frac12} (1+\log|d|),
\end{align*}
using the fact that, e.g., $\gamma_\epsilon \setminus B_{d/16}(a_1')$ is contained in two wedges cornered at $a_1'$ whose angles are bounded by $C\cdot \epsilon$.

Line integrating (in $a_1'$) the above two estimates, we have
$$
\left|\int_{a_b}^{a_1} \lambda(a_1';A_p) da_1' - \int_{\Omega^p} \int_{a_b}^{a_1} {\lambda(a_1';\mathfrak{A}_p)\mathbf{1}_{\Omega_\epsilon}(u_p)} da_1'\,d^{2p}u\right| \leq  C^p \epsilon^{\frac12}|\log \epsilon|,
$$
where now $C = C(\Omega,\kappa_\infty,n,\delta_0, \gamma)$.

In addition, note that \eqref{eq:uniform-bound} (with preceding bounds, Lemma~\ref{lem: df-critical-bound} and \eqref{eq:critical-ff-bound}) implies that
\begin{align}\label{eq:lambda-frak-bound}|\lambda(a_1';\mathfrak{A}_p)| \leq |{\mathfrak{A}_{p-1}(u)}| \cdot \left(\frac{d^{-1}}{|u_p - a_1'||u_p-a_2|\cdots|u_p-a_n|} \right)\cdot P_{a_1'}(u_p).\end{align}
Since
$$\left| \mathbf{1}_{\Omega_\epsilon}(u_p) - \mathbf{1}_{\Omega_\epsilon^p}(u)\right| \leq \sum_{k=1}^{p-1} \mathbf{1}_{\gamma_\epsilon}(u_k)\mathbf{1}_{\Omega_\epsilon}(u_p), $$
and by \eqref{eq:summand-bound}, for any $k<p$, we have
\begin{align*}
    \int_{\Omega^{p-1}} &\left|\lambda(a_1';\mathfrak{A}_p) \mathbf{1}_{\gamma_\epsilon}(u_k)\mathbf{1}_{\Omega_\epsilon}(u_p)\right| d^{2}u_1\cdots d^2u_{p-1}\leq  
        \frac{C^p \epsilon^{\frac12} d^{-\frac12} \mathbf{1}_{\Omega_\epsilon}(u_p)}{|u_p - a_1'|^2|u_p-a_2|\cdots|u_p-a_n|}.
\end{align*}
Then we do the remaining integrals, first with respect to $u_p$:
$$
\int_\Omega\frac{C^p \epsilon^{\frac12} d^{-\frac12} \mathbf{1}_{\Omega_\epsilon}(u_p)}{|u_p - a_1'|^2|u_p-a_2|\cdots|u_p-a_n|}d^2u_p \leq C^p \epsilon^{\frac12} d^{-\frac12}|\log (\epsilon d)|,
$$
then doing the line integral finally gives
$$
\left|\int_{a_b}^{a_1} \lambda(a_1';A_p) da_1' - \int_{\Omega_\epsilon^p} \int_{a_b}^{a_1} {\lambda(a_1';\mathfrak{A}_p)da_1'\,d^{2p}u}\right| \leq {pC^p\epsilon^{\frac12}}|\log\epsilon|.
$$
The identical bound holds when we put real parts in front of the line integrals. By domain symmetry, it is easy to see that the identical bound holds for ${\lambda(a_1';\mathfrak{A}_p^{\text{sym}})}$.

Then, note that, if $\Re\big[\int_{a_b}^{a_1} {\lambda(a_1';\mathfrak{A}_p^{\text{sym}})}{da_1'}\big]$ is (absolutely) integrable in $\Omega^p$, dominated convergence yields \eqref{eq:line-integral-under} as desired. We show this in the following lemma.

\begin{lemma}\label{le:lambdaint}
    $\Re\big[\int_{a_b}^{a_1} {\lambda(a_1';\mathfrak{A}_p^{\text{sym}})}{da_1'}\big]$ is (absolutely) integrable.
\end{lemma}
\begin{proof}[Proof of Lemma \ref{le:lambdaint}]
    Note that, since $\lambda(a_1';\mathfrak{A}_p^{\text{sym}})$ is a $\partial_{a'}$-derivative of a real quantity (i.e. the integrand on the RHS of \eqref{eq:lambda-inside}, see also the note below about boundary limits), we may choose $a_b$ and the integration contour conveniently. For fixed $a_1$, we choose the contour $\gamma$ of length $l\leq C(\Omega, n, \delta_0, p)$ and its unit-speed parametrization $\gamma:[-l,0]\to \Omega$ satisfies, for each $j=2,\ldots, n$ and $k=1,\ldots,p$, we have, with $\gamma(t) = a_1'$,
    $$
    |\gamma(t)- a_j| \geq C(\Omega, n, \delta_0, p)\quad\text{and}\quad|\gamma(t)- u_k| \geq C(\Omega, n, \delta_0, p)(|a_1-u_k| + |t|).
    $$
    This can be done by identifying a wedge cornered at $a_1$, not containing any other $a_2, \ldots, a_n$ and $u_1, \ldots, u_p$ and having an angle bounded below by $C(\Omega, n, \delta_0, p)$.
    
    Having fixed $\gamma$, we first bound the non-symmetric $|\lambda(\gamma(t);\mathfrak{A}_p)|$, since symmetrization will preserve this bound.
    Let us carefully study the ($C(n,p)$ many) terms of the expansion in the bound \eqref{eq:lambda-frak-bound}, where we substitute in \eqref{eq:frak-estimate} for $p-1$ and $a_1=\gamma(t)$. Each term is a product of ($C(n,p)$ many) factors of type (at the highest order) $|u_k-u_{k+1}|^{-1}$ for $k<n$, $|a_j-u_k|^{-3/2}$ for $j>1$, $|\gamma(t)-u_k|^{-3/2}$ for $k<p$, and $|\gamma(t)-u_p|^{-2}$. We need to show that after the line integral we get small but strict improvements in the exponents. However, that is immediate from the following general computation: for $r_1, \ldots, r_N\in (1,2)$ and $e_1, \ldots, e_N>0$, we have by H\"older (set $R:= r_1 + \cdots + r_N$)
    $$
    \int_{-l}^0 \prod_{q=1}^N\frac{1}{|e_q + |t||^{r_q} }dt \leq \prod_{q=1}^N\left(\int_{-l}^0 \frac{1}{|e_q + |t||^{R} }dt \right)^{r_q/R}\leq C(N,l)\prod_{q=1}^N {e_q }^{-r_q+\frac{r_q}{R}}.
    $$
\end{proof}
This concludes the proof of Proposition~\ref{pr:spinexp}.
\end{proof}

\subsection{Fermion and disorder correlations}
In this section, we study the series expansion of spin-weighted correlations of fermions and disorders; since disorders are defined as the renormalized limit of the merger of spins and fermions, we study spin-weighted fermion correlations in significant detail first. Recall that we write $\Psi$ for either of $\psi,\psi^*$, or equivalently a real fermion $\psi^{[\eta]}$ with $|\eta|=1$.

\subsubsection{Fermion-fermion correlations} We will first treat the two-point correlation of fermions weighted by spin, in order to use it to expand general $2k$-point correlations.

\begin{proposition}\label{pr:spinferm2pt}
For distinct points $a_1,\dots,a_n$, $z_1,z_2$, the correlation $\frac{\langle \sigma_{A}\Psi_{z_1}\Psi_{z_2}\rangle_{m\alpha}}{\langle \sigma_A\rangle_{m\alpha}}$ has a series expansion around any $m=m_0 \in \R$ with the usual conditions. In particular, around $m_0=0$, we have
\begin{align}\label{eq:spinferm2pt}
\frac{\langle \sigma_{A}\Psi_{z_1}\Psi_{z_2}\rangle_{m\alpha}}{\langle \sigma_A\rangle_{m\alpha}}&= \int_{\Omega^p} \mathfrak{B}^{\text{sym}}_p(z_2;u,z_1) d^{2p}u\\
\nonumber&=\sum_{p=0}^\infty \left(-\frac{m}{\pi}\right)^p\frac{1}{p!}\int_{\Omega^p}d^{2p}u\prod_{k=1}^p\alpha(u_k)\sum_{S\subset \{ 1, \ldots, p\}}\frac{\langle \sigma_A \Psi_{z_1}\Psi_{z_2}\prod_{j\in S}\epsilon_{u_j}\rangle_0}{\langle \sigma_A\rangle_0}\\
\nonumber&\quad\times\sum_{\Lambda \in \Pi_{ \{ 1, \ldots, p\}\setminus S}}(-1)^{|\Lambda|}|\Lambda|!\prod_{B\in \Lambda}\frac{\langle \sigma_A \prod_{j\in B}\epsilon_{u_j}\rangle_0}{\langle \sigma_A\rangle_0},
\end{align}
where we again used the notation $\sigma_{a_1}\cdots \sigma_{a_n}=\sigma_A$ and the set $\Pi_U$ of partitions of $U$, with the convention that $\Pi_\emptyset = \{\emptyset \}$.
\end{proposition}
\begin{proof}
    Due to real (bi)linearity, it suffices to show this result for $\Psi_{z_1}=\psi_{z_1}$ and $\Psi_{z_2} = \psi_{z_2}^{[\eta]}$ (see also \eqref{eq:fermion-complex}), when the coefficients are given by $B_p$ in \eqref{eq:series-expansion}. Then we have series expansion under usual conditions by Remark \ref{re:continuity}, \eqref{eq:massive-ff} (for $p=0$), and \eqref{eq:summand-bound} (for $p \geq 1$).

    Around $m_0=0$, as in Proposition \ref{pr:Ap-induction}, we will only use integrand-level identities and symmetrization on components of $u$. Therefore, we do automatically have a local exponential integral bound for $|\mathfrak{B}^{\text{sym}}_p(z_2;\cdot, z_1)|$, obtained by symmetrization of the bounds \eqref{eq:summand-bound} as in Remark \ref{rem:up-analysis-2} (note again that the integrand in \eqref{eq:spinferm2pt} is invariant under any transposition).
    
    The inductive strategy as in the proof of Proposition \ref{pr:Ap-induction} applies here almost identically, so we leave out many details elaborated there. Noting that the $p=0$ case holds trivially, let us assume \eqref{eq:spinferm2pt} for $p$. Then, fixing $S$ and $\Lambda$ then omitting common parts as in \eqref{eq:spin-inductive-step}, we see that the $S$-specific part for the $p+1$ case would be written (again writing $S'=S\cup \{p+1\}$ and symmetrizing $p+1$ with $j\in S$, i.e. over $S'$)
    \begin{align*}
        \frac{i}{2}&\frac{\big\langle \sigma_A \psi^*_{u_{p+1}}\psi_{z_2}^{[\eta]}\prod_{j\in S}\epsilon_{u_j}\big\rangle_0}{\big\langle \sigma_A\big\rangle_0}\frac{\big\langle \sigma_{A}\psi_{z_1}\psi_{u_{p+1}}\big\rangle_0 }{\big\langle \sigma_{A}\big\rangle_0 }-\frac{i}{2}\frac{\big\langle \sigma_A \psi_{u_{p+1}}\psi_{z_2}^{[\eta]}\prod_{j\in S}\epsilon_{u_j}\big\rangle_0}{\big\langle \sigma_A\big\rangle_0}\frac{\big\langle \sigma_{A}\psi_{z_1}\psi_{u_{p+1}}^*\big\rangle_0 }{\big\langle \sigma_{A}\big\rangle_0 }\\
        \stackrel{sym}{=}&\frac{1}{|S|+1}\left(\frac{i}{2} \right)^{|S|+1}\sum_{j'\in S'}\Bigg[\frac{\big\langle \sigma_A \psi_{z_2}^{[\eta]}\psi^*_{u_{j'}}\prod_{j\in S'\setminus\{j'\}}\psi_{u_j}\psi_{u_j}^*\big\rangle_0}{\big\langle \sigma_A\big\rangle_0}\frac{\big\langle \sigma_{A}\psi_{u_{j'}}\psi_{z_1}\big\rangle_0 }{\big\langle \sigma_{A}\big\rangle_0 }\\ 
        &\qquad\qquad\qquad\qquad\qquad-\frac{\big\langle \sigma_A \psi_{z_2}^{[\eta]}\psi_{u_{j'}}\prod_{j\in S'\setminus\{j'\}}\psi_{u_j}\psi_{u_j}^*\big\rangle_0}{\big\langle \sigma_A\big\rangle_0}\frac{\big\langle \sigma_{A}\psi^*_{u_{j'}}\psi_{z_1}\big\rangle_0 }{\big\langle \sigma_{A}\big\rangle_0 } \Bigg]\\
        =&\frac{1}{|S'|}\Bigg[\frac{\big\langle \sigma_A \psi_{z_1}\psi_{z_2}^{[\eta]}\prod_{j\in S'}\epsilon_{u_j}\big\rangle_0}{\big\langle \sigma_A\big\rangle_0}-\frac{\big\langle \sigma_A\prod_{j\in S'}\epsilon_{u_j}\big\rangle_0}{\big\langle \sigma_A\big\rangle_0}\frac{\big\langle \sigma_{A}\psi_{z_1}\psi_{z_2}^{[\eta]}\big\rangle_0 }{\big\langle \sigma_{A}\big\rangle_0 } \Bigg].
    \end{align*}
    Similarly to the proof of Proposition \ref{pr:Ap-induction}, the first term gives rise to accurate $(\Lambda, S')$ terms in the $p+1$ case where $|S'|\geq1$, after one more symmetrization of $p+1$ and $1, 2, \ldots, p$. The second term, on the other hand, corresponds to the $(\Lambda', \emptyset)$ terms, where $\Lambda' = \Lambda \cup\{S'\}$. The minus sign corrects $(-1)^{|\Lambda|}$ pre-factor to $(-1)^{|\Lambda'|}$, and after symmetrization the above term will appear $|S'|$ times. Since there are  $|\Lambda'|$ choices of $S'\in \Lambda'$, we get the necessary pre-factor of $|\Lambda'|$ as well.
\end{proof}

Then, we state the following generalization for the $2k$-point correlations.
\begin{lemma}\label{le:spinferm2kpt}
For distinct points $a_1,\dots,a_n$, $z_1,\dots,z_{2k}$, the correlation $\frac{\langle \sigma_{A}\Psi_{z_1}\cdots \Psi_{z_{2k}}\rangle_{m\alpha}}{\langle \sigma_A\rangle_{m\alpha}}$ admits a series expansion around any $m=m_0 \in \R$ with the usual conditions. Around $m_0=0$, we have
\begin{multline}\label{eq:spinferm2kpt}
\frac{\langle \sigma_{A}\Psi_{z_1}\cdots \Psi_{z_{2k}}\rangle_{m\alpha}}{\langle \sigma_A\rangle_{m\alpha}}=\sum_{p=0}^\infty \left(-\frac{m}{\pi}\right)^p\frac{1}{p!}\int_{\Omega^p}d^2w_1\cdots d^2 w_p \alpha(w_1)\cdots \alpha(w_p) \\ 
\times\sum_{S\subset \{1, \ldots, p\}}\frac{\langle \sigma_A \Psi_{z_1}\cdots \Psi_{z_{2k}}\prod_{j\in S}\epsilon_{w_j}\rangle_0}{\langle \sigma_A\rangle_0} \sum_{\Lambda \in \Pi_{\{1, \ldots, p\} \setminus S}}(-1)^{|\Lambda|}|\Lambda|!\prod_{B\in \Lambda}\frac{\langle \sigma_A \prod_{j\in B}\epsilon_{w_j}\rangle_0}{\langle \sigma_A\rangle_0}.
\end{multline}
\end{lemma}

Our proof of Lemma \ref{le:spinferm2kpt} (specifically \eqref{eq:spinferm2kpt}), to be presented after introducing some tools, goes through an identity for critical spin-fermion-energy correlation functions, and this we find most convenient to prove through tools coming from Grassmann integrals -- though it is likely that there are more combinatorial proofs as well.

While analyticity and the usual conditions are simple consequences of the corresponding properties for the two-point correlations and analyticity of the Pfaffian (see Remark \ref{rem:usual-conditions}), we find it instructive to do the following manual counting.

\begin{remark}\label{rem:fermion-pfaffian-analysis}
Note the Pfaffian expansion \eqref{eq:pfaffian} may be written as a sum of $\frac{(2k)!}{2^kk!}$ terms, each having $k$ factors in the form of $2$-point correlations. Counting non-negative integer partitions of $p$ into $k$ terms, we see that the bound \eqref{eq:local-uniform-boundedness} for the $2$-point correlation coefficients (which came from \eqref{eq:uniform-compact-bound} and linearity) translate to the following local exponential bound for the coefficients of $(m-m_0)^p$ for the $2k$-point correlation:
\begin{equation}\label{eq:multiple-integral-combination}
\frac{(2k)!}{p!\cdot2^kk!}\binom{p+k-1}{k-1}\cdot C(M,\Omega,\kappa_\infty,n,\delta_0,d_z)^p \leq  C(k,M,\Omega,\kappa_\infty,n,\delta_0,d_z)^p,
\end{equation}
where $0<d_z \leq \min(\delta_{z_1}, \ldots, \delta_{z_{2k}},\min_{j<j'}|z_j-z_{j'}|)$. That is, in the following where we merge fermion insertions with other fields, in order to claim local exponential bounded coefficients (and thus usual conditions, since local continuity can then be shown using arguments akin to Remark \ref{re:continuity}), we need only check that the bound \eqref{eq:uniform-compact-bound} is kept.
\end{remark}

\subsubsection{Pfaffian expansion and Grassmann integrals}
Our proof of Lemma \ref{le:spinferm2kpt} will be by induction, and it is convenient to have a recursive definition of the Pfaffian of a skew symmetric matrix. Recall that for a $2k\times 2k$ skew symmetric matrix $M$, we have 
\begin{equation}\label{eq:pfrec}
\mathrm{Pf}(M)=\sum_{l=2}^{2k}(-1)^lM_{1,l}\mathrm{Pf}(M_{i,j})_{i,j\neq 1,l}.
\end{equation}
Another basic fact about Pfaffians we will use is that $(\mathrm{Pf}(M))^2=\det(M)$ (for a skew symmetric matrix $M$). In particular, if $\mathrm{Pf}(M)\neq 0$, then $M$ is invertible.

We now review the connection between Pfaffians and Grassmann integrals. For a more detailed treatment, see \cite[Section I]{Feldman}. 

Our starting point is that we consider an exterior algebra with an even number of generators $b_1,\dots,b_{2q}$. Then given any Grassmann monomial, namely an element of the exterior algebra of the form $b_{i_1}\cdots b_{i_j}$ with $i_1<i_2<\dots<i_j$, we define the Grassmann integral 
\begin{equation}\label{eq:gint}
\int Db\, b_{i_1}\cdots b_{i_j}=\begin{cases}
1, & j=2q \text{ (and }i_k=k \text{ for } k=1,\dots,2q)\\
0, & \text{else}
\end{cases}.
\end{equation}
Since Grassmann monomials form a (vector space) basis of the exterior algebra, we can extend the integral $\int Db$ into a linear functional on the exterior algebra.

Since the square of a generator of the exterior algebra vanishes, $b_i^2=0$ (by antisymmetry), an analytic function of the generators (defined by series expansion around the origin) produces an element of the exterior algebra and we can define its integral. In particular, for any skew symmetric $2q\times 2q$ complex valued matrix $M$, we have by \eqref{eq:gint}
\begin{equation}\label{eq:2q-form}
\int Db\, e^{\frac{1}{2}\sum_{i,j=1}^{2q}b_i M_{ij}b_j}=\frac{1}{2^qq!}\int Db\Bigg(\sum_{i,j=1}^{2q}b_i M_{i,j}b_j\Bigg)^q
\end{equation}
(since all other terms in the expansion of the exponential either vanish identically or integrate to zero due to their degree). Expanding the power, we see that the only terms that survive the $2q$-form sum \eqref{eq:2q-form} are of the form 
\begin{align*}
\int Db\, e^{\frac{1}{2}\sum_{i,j=1}^{2q}b_i M_{ij}b_j}&=\frac{1}{2^qq!}\sum_{\sigma\in S_{2q}}\prod_{i=1}^q M_{\sigma(2i-1),\sigma(2i)}\int Db\, b_{\sigma(1)}\cdots b_{\sigma(2q)}\\
&=\frac{1}{2^q q!}\sum_{\sigma\in S_{2q}}\mathrm{sgn}(\sigma)\prod_{i=1}^q M_{\sigma(2i-1),\sigma(2i)}=\mathrm{Pf}(M).
\end{align*}

For our purposes, we will actually find it more convenient to write suitable ratios of Grassmann integrals in terms of Pfaffians. More precisely, if $S$ is any $2q\times 2q$ skew symmetric complex matrix that is invertible, and $1\leq i_1<\dots<i_j\leq 2q$, then it is proven in \cite[Section I]{Feldman} that 
\begin{align}\label{eq:feld}
\frac{\int Db\, b_{i_1}\cdots b_{i_j}e^{-\frac{1}{2}\sum_{i,j=1}^{2q}b_i (S^{-1})_{i,j}b_j}}{\int Db\, e^{-\frac{1}{2}\sum_{i,j=1}^{2q}b_i (S^{-1})_{i,j}b_j}}=\mathrm{Pf}(S_{i_r,i_s})_{r,s=1}^{j}.
\end{align}

We now turn to expressing critical Ising correlation functions in terms of Grassmann integrals. We choose the matrix $S$ in \eqref{eq:feld} suitably so that we recover spin-fermion Ising correlation functions. We will also show that by allowing $S$ to depend on suitable parameters, we can obtain also energy correlation functions via differentiation with respect to these parameters.

We are interested in a correlation function of the form
\begin{align*}
\frac{\langle\sigma_A \Psi_{z_1}\cdots \Psi_{z_{2k}}\prod_{j\in C}\epsilon_{w_j}\rangle_0}{\langle \sigma_A\rangle_0}.
\end{align*}
Let us enumerate the points of $C$: $C=\{c_1,\dots,c_l\}$. Since $\epsilon = \frac{i}{2}\psi \psi^*$ at criticality, it is enough for us to understand correlation functions of the form 
\begin{align}\label{eq:spin-fermion-general}
\frac{\langle\sigma_A \Psi_{z_1}\cdots \Psi_{z_{2k}}\psi_{u_1}\psi_{w_{c_1}}^*\cdots \psi_{u_l}\psi_{w_{c_l}}^*\rangle_0}{\langle \sigma_A\rangle_0}
\end{align}
and then take $u_i\to w_{c_i}$ (and multiply by a suitable constant). Note that this is either analytic or anti-analytic (with branching) in the fermionic variables and does not vanish identically (as one can check from the OPEs). Thus the correlation function is non-zero at generic points.

To connect to Grassmann integrals, we introduce the $2(k+l)\times 2(k+l)$ skew symmetric complex matrix for which we have
\begin{equation}
S_{ij}=\begin{cases}
\frac{\big\langle \sigma_A\Psi_{z_i}\Psi_{z_j}\big\rangle_0}{\langle \sigma_A\rangle_0}, & i,j\leq 2k\\
\frac{\big\langle \sigma_A\Psi_{z_i}\psi_{u_{j-2k}}\big\rangle_0}{\langle \sigma_A\rangle_0}, & i\leq 2k<j\leq 2k+l\\
\frac{\big\langle \sigma_A\Psi_{z_i}\psi_{w_{c_{j-2k-l}}}^*\big\rangle_0}{\langle \sigma_A\rangle_0}, & i\leq 2k,j> 2k+l\\
\frac{\big\langle \sigma_A\psi_{u_{i-2k}}\psi_{u_{j-2k}}\big\rangle_0}{\langle \sigma_A\rangle_0}, & 2k<i,j\leq 2k+l\\
\frac{\big\langle \sigma_A\psi_{u_{i-2k}}\psi_{w_{c_{j-2k-l}}}^*\big\rangle_0}{\langle \sigma_A\rangle_0}, & 2k<i\leq 2k+l<j\\
\frac{\big\langle \sigma_A\psi_{w_{c_{i-2k-l}}}^*\psi_{w_{c_{j-2k-l}}}^*\big\rangle_0}{\langle \sigma_A\rangle_0}, &  2k+l<i,j
\end{cases}
\end{equation}
As the Pfaffian of this matrix is the correlation function \eqref{eq:spin-fermion-general} that we are considering, by \eqref{eq:pfaffian}, and the correlation function is non-vanishing generically, we see that $S$ is invertible generically. Let us go on to define for some small enough parameters $t_1,\dots,t_l$ a matrix
\begin{equation}
(S(t)^{-1})_{i,j}=(S^{-1})_{i,j}-\sum_{q=1}^l t_q(\delta_{2k+q,i}\delta_{2k+l+q,j}-\delta_{2k+q,j}\delta_{2k+l+q,i}).
\end{equation}
For small enough $t$, this matrix exists (generically).

Our point of introducing this matrix is the following fact. 
\begin{lemma}\label{le:diff}
We have 
\begin{align}\label{eq:cfdiff}
&\frac{\big\langle\sigma_A \Psi_{z_1}\cdots \Psi_{z_{2k}}\psi_{u_1}\psi_{w_{c_1}}^*\cdots \psi_{u_l}\psi_{w_{c_l}}^*\big\rangle_0}{\langle \sigma_A\rangle_0}\\
\nonumber&\qquad=\left.\frac{\partial}{\partial t_1}\cdots \frac{\partial}{\partial t_l}\right|_{t=0} \frac{\int Db\, b_1\cdots b_{2k}e^{-\frac{1}{2}\sum_{i,j=1}^{2(k+l)}b_i (S(t)^{-1})_{i,j}b_j}}{\int Db\, e^{-\frac{1}{2}\sum_{i,j=1}^{2(k+l)}b_i(S^{-1})_{i,j}b_j}}.
\end{align}
\end{lemma}
\begin{proof}
Since the generators anticommute and even elements of the exterior algebra (namely linear combinations of products of an even number of generators) commute, we can write 
\begin{align*}
e^{-\frac{1}{2}\sum_{i,j=1}^{2(k+l)}b_i (S(t)^{-1})_{i,j}b_j}&=e^{-\frac{1}{2}\sum_{i,j=1}^{2(k+l)}b_i (S^{-1})_{i,j}b_j+\sum_{q=1}^l t_q b_{2k+q}b_{2k+l+q}}\\
&=e^{-\frac{1}{2}\sum_{i,j=1}^{2(k+l)}b_i (S^{-1})_{i,j}b_j}\prod_{q=1}^l(1+t_qb_{2k+q}b_{2k+l+q}).
\end{align*}
We thus see that differentiation and setting $t=0$ picks out the term
\begin{align*}
\frac{\int Db\, b_{1}\cdots b_{2k}b_{2k+1}b_{2k+l+1}\cdots b_{2k+l}b_{2k+2l} e^{-\frac{1}{2}\sum_{i,j=1}^{2(k+l)}b_i (S^{-1})_{i,j}b_j}}{\int Db\, e^{-\frac{1}{2}\sum_{i,j=1}^{2(k+l)}b_i(S^{-1})_{i,j}b_j}}.
\end{align*}
By the definition of $S$ and \eqref{eq:feld}, we see that we can express this as the same Pfaffian as the correlation function we are after, so we have proven our claim.
\end{proof}
The point of this lemma is that we can express the right hand side of \eqref{eq:cfdiff} as a ratio of Pfaffians and use the recursive property of Pfaffians to get a useful recursion for spin-fermion-energy correlation functions. The precise claim is as follows.
\begin{lemma}\label{le:mainid}
We have the identity 
\begin{align}
&\frac{\langle \sigma_A\Psi_{z_1}\cdots \Psi_{z_{2k}}\prod_{i\in C}\epsilon_{w_i}\rangle_0}{\langle \sigma_A\rangle_0}\\
\nonumber&\quad=\sum_{l=2}^{2k}(-1)^l\sum_{U\subset C}\sum_{U_1\sqcup U_2=U} \frac{\langle \sigma_A\Psi_{z_1}\Psi_{z_l}\prod_{j_1\in U_1}\epsilon_{w_{j_1}}\rangle_0}{\langle \sigma_A\rangle_0}\frac{\langle \sigma_A \Psi_{z_2}\stackrel{\hat l}{\cdots}\Psi_{z_{2k}}\prod_{j_2\in U_2}\epsilon_{w_{j_2}}\rangle_0}{\langle \sigma_A\rangle_0}\\
\nonumber&\qquad\times \sum_{\pi \in \Pi_{C\setminus U}}(-1)^{|\pi|}|\pi|!\prod_{B\in \pi}\frac{\langle \sigma_A \prod_{j\in B}\epsilon_{w_{j}}\rangle_0}{\langle \sigma_A\rangle_0},
\end{align}
where if $U=C$, we interpret the sum over $\pi$ as $1$.
\end{lemma}
\begin{proof}
As mentioned before, we can replace energies by pairs of fermions. Also we can assume that we are first dealing with generic points where $S$ is invertible, then extend to all points by taking limits. 

We begin by using Lemma \ref{le:diff} to write 
\begin{align*}
\frac{\langle\sigma_A \Psi_{z_1}\cdots \Psi_{z_{2k}}\psi_{u_1}\psi_{w_{c_1}}^*\cdots \psi_{u_q}\psi_{w_{c_q}}^*\rangle_0}{\langle \sigma_A\rangle_0}&=\left.\frac{\partial}{\partial t_1}\cdots \frac{\partial}{\partial t_q}\right|_{t=0} \frac{\int Db\, b_1\cdots b_{2k}e^{-\frac{1}{2}\sum_{i,j=1}^{2(k+q)}b_i (S(t)^{-1})_{i,j}b_j}}{\int Db\, e^{-\frac{1}{2}\sum_{i,j=1}^{2(k+q)}b_i(S^{-1})_{i,j}b_j}}\\
&=\left.\frac{\partial}{\partial t_1}\cdots \frac{\partial}{\partial t_q}\right|_{t=0} \frac{\mathrm{Pf}(-S(t)^{-1})}{\mathrm{Pf}(-S^{-1})}\mathrm{Pf}(S(t))_{i,j=1}^{2k}.
\end{align*}
Next we use \eqref{eq:pfrec} for the last Pfaffian and find 
\begin{align*}
&\frac{\langle\sigma_A \Psi_{z_1}\cdots \Psi_{z_{2k}}\psi_{u_1}\psi_{w_{c_1}}^*\cdots \psi_{u_q}\psi_{w_{c_q}}^*\rangle_0}{\langle \sigma_A\rangle_0}\\
&\quad=\left.\frac{\partial}{\partial t_1}\cdots \frac{\partial}{\partial t_q}\right|_{t=0} \frac{\mathrm{Pf}(-S(t)^{-1})}{\mathrm{Pf}(-S^{-1})}\sum_{l=2}^{2k}(-1)^l S_{1,l}(t)\mathrm{Pf}(S_{i,j}(t))_{i,j\neq 1,l}\\
&\quad=\sum_{l=2}^{2k}(-1)^l\left.\frac{\partial}{\partial t_1}\cdots \frac{\partial}{\partial t_q}\right|_{t=0} \frac{\mathrm{Pf}(-S^{-1})}{\mathrm{Pf}(-S(t)^{-1})}\frac{\int Db\, b_1b_l e^{-\frac{1}{2}\sum_{i,j=1}^{2(k+q)}b_i (S(t)^{-1})_{i,j}b_j}}{\int Db\, e^{-\frac{1}{2}\sum_{i,j=1}^{2(k+q)}b_i (S^{-1})_{i,j}b_j}}\\
&\quad\qquad \times \frac{\int Db\, b_2\stackrel{\hat l}{\cdots}b_{2k} e^{-\frac{1}{2}\sum_{i,j=1}^{2(k+q)}b_i (S(t)^{-1})_{i,j}b_j}}{\int Db\, e^{-\frac{1}{2}\sum_{i,j=1}^{2(k+q)}b_i (S^{-1})_{i,j}b_j}}.
\end{align*}
We now split this into a sum over all the possible ways the derivatives can act on the different factors:
\begin{align*}
\frac{\langle\sigma_A \Psi_{z_1}\cdots \Psi_{z_{2k}}\psi_{u_1}\psi_{w_{c_1}}^*\cdots \psi_{u_q}\psi_{w_{c_q}}^*\rangle_0}{\langle \sigma_A\rangle_0}&=\sum_{l=2}^{2k}(-1)^l \sum_{U_1\sqcup U_2\sqcup U_3=C}\prod_{r_1\in U_1}\left.\frac{\partial}{\partial t_{r_1}}\right|_{t_{r_1}=0}\frac{\mathrm{Pf}(-S^{-1})}{\mathrm{Pf}(-S(t)^{-1})}\\
&\quad\times \prod_{r_2\in U_2}\frac{\partial}{\partial t_{r_2}}\bigg|_{t_{r_2}=0}\frac{\int Db\, b_1b_l e^{-\frac{1}{2}\sum_{i,j=1}^{2(k+q)}b_i (S(t)^{-1})_{i,j}b_j}}{\int Db\, e^{-\frac{1}{2}\sum_{i,j=1}^{2(k+q)}b_i (S^{-1})_{i,j}b_j}}\\
&\quad\times \prod_{r_3\in U_3}\frac{\partial}{\partial t_{r_3}}\bigg|_{t_{r_3}=0} \!\frac{\int Db\, b_2\stackrel{\hat l}{\cdots}b_{2k} e^{-\frac{1}{2}\sum_{i,j=1}^{2(k+q)}b_i (S(t)^{-1})_{i,j}b_j}}{\int Db\, e^{-\frac{1}{2}\sum_{i,j=1}^{2(k+q)}b_i (S^{-1})_{i,j}b_j}}.
\end{align*}
The last two rows are readily expressed as correlation functions using Lemma \ref{le:diff}. For the remaining one, we will make use of the fact that if $f$ is sufficiently smooth in a neighborhood of the origin and $f(0)=1$, we can then use Faà di Bruno's formula to write
\begin{align*}
\left.\frac{\partial}{\partial t_1}\cdots \frac{\partial}{\partial t_m}\right|_{t=0} \frac{1}{f(t)}=\sum_{\pi \in \Pi_m}(-1)^{|\pi|}|\pi|!\prod_{B\in \pi}\Bigg(\prod_{j\in B}\left.\frac{\partial}{\partial t_j}\right|_{t_j=0} f(t)\Bigg).
\end{align*}
Note that for our $f$, the derivatives generate spin-energy correlation functions. We thus find after using our genericity argument, merging $u_j$ and $w_{c_j}$ and multiplying by a suitable constant, 
\begin{align*}
&\frac{\langle \sigma_A\Psi_{z_1}\cdots \Psi_{z_{2k}}\prod_{j\in C}\epsilon_{w_j}\rangle_0}{\langle \sigma_A\rangle_0}\\
&\quad=\sum_{l=2}^{2k}(-1)^l\sum_{U_1\sqcup U_2 \sqcup U_3=C}\frac{\langle \sigma_A\Psi_{z_1}\Psi_{z_l}\prod_{j_1\in U_1}\epsilon_{w_{j_1}}\rangle_0}{\langle \sigma_A\rangle_0}\frac{\langle \sigma_A \Psi_{z_2}\stackrel{\hat l}{\cdots}\Psi_{z_{2k}}\prod_{j_2\in U_2}\epsilon_{w_{j_2}}\rangle_0}{\langle \sigma_A\rangle_0}\\
&\quad\qquad \times \sum_{\pi \in \Pi_{U_3}}(-1)^{|\pi|}|\pi|!\prod_{B\in \pi}\frac{\langle \sigma_A \prod_{j_3\in B}\epsilon_{w_{j_3}}\rangle_0}{\langle \sigma_A\rangle_0}.
\end{align*}
Slightly reorganizing the sum, we recover our claim.
\end{proof}

We are now in a position to prove Lemma \ref{le:spinferm2kpt}.
\begin{proof}[Proof of Lemma \ref{le:spinferm2kpt}]
As already pointed out, analyticity and existence of the series expansion (with usual conditions) follows from the case of two fermion insertions (Proposition \ref{pr:spinferm2pt}) and expressing Proposition \ref{prop:fermion-antisymmetry}. Thus it is sufficient to check that the expansion coefficients in the two series in the statement of the lemma match. We will do this by induction (the $p=0$ case being obvious).

Let us start with the right hand side of the claim in Lemma \ref{le:spinferm2kpt} and plug Lemma \ref{le:mainid} into it. We then have 
\begin{align*}
&\sum_{C\subset \{1, \ldots, p\}}\frac{\langle \sigma_A \Psi_{z_1}\cdots \Psi_{z_{2k}}\prod_{j\in C}\epsilon_{w_j}\rangle_0}{\langle \sigma_A\rangle_0} \sum_{\Lambda \in \Pi_{\{1, \ldots, p\} \setminus C}}(-1)^{|\Lambda|}|\Lambda|!\prod_{B\in \Lambda}\frac{\langle \sigma_A \prod_{j\in B}\epsilon_{w_j}\rangle_0}{\langle \sigma_A\rangle_0}\\
&=\sum_{l=2}^{2k}(-1)^l\!\sum_{C\subset \{1,\dots,p\}}\sum_{U\subset C}\sum_{U_1\sqcup U_2=U}\sum_{\Lambda\in \Pi_{\{1,\dots,p\}\setminus C}} \!(-1)^{|\Lambda|}|\Lambda|!\sum_{\nu \in \Pi_{C\setminus U}} \!(-1)^{|\nu|}|\nu|!\prod_{B'\in \nu}\frac{\langle \sigma_A\prod_{j'\in B'}\epsilon_{w_{j'}}\rangle_0}{\langle \sigma_A\rangle_0}\\
&\quad \times \frac{\langle \sigma_A\Psi_{z_1}\Psi_{z_l}\prod_{j_1\in U_1}\epsilon_{w_{j_1}}\rangle_0}{\langle \sigma_A\rangle_0}\frac{\langle \sigma_A\Psi_{z_2}\stackrel{\hat l}{\cdots}\Psi_{z_{2k}}\prod_{j_2\in U_2}\epsilon_{w_{j_2}}\rangle_0}{\langle \sigma_A\rangle_0} \prod_{B\in \Lambda}\frac{\langle \sigma_A\prod_{j\in B}\epsilon_{w_j}\rangle_0}{\langle \sigma_A\rangle_0}.
\end{align*}
Next we change the order of the $U$ and $C$ sums, and note that summing over $C$ is equivalent to summing over subsets of $\{1,\dots,p\}\setminus U$. This becomes  
\begin{align*}
&\sum_{l=2}^{2k}(-1)^l \sum_{U\subset \{1,\dots,p\}} \sum_{U_1\sqcup U_2=U}\frac{\langle \sigma_A\Psi_{z_1}\Psi_{z_l}\prod_{j_1\in U_1}\epsilon_{w_{j_1}}\rangle_0}{\langle \sigma_A\rangle_0}\frac{\langle \sigma_A\Psi_{z_2}\stackrel{\hat l}{\cdots}\Psi_{z_{2k}}\prod_{j_2\in U_2}\epsilon_{w_{j_2}}\rangle_0}{\langle \sigma_A\rangle_0}\\
&\times\hspace{-2pt}\sum_{D\subset \{1,\dots,p\}\setminus U} \sum_{\Lambda \in \Pi_{\{1,\dots,p\}\setminus (U\cup D)}} \sum_{\nu \in \Pi_{D}}(-1)^{|\Lambda|+|\nu|}|\Lambda|!|\nu|!\prod_{B\in \Lambda }\frac{\langle \sigma_A \prod_{j\in B}\epsilon_{w_j}\rangle_0}{\langle \sigma_A\rangle_0}\prod_{B'\in \nu}\frac{\langle \sigma_A\prod_{j'\in B'}\epsilon_{w_{j'}}\rangle_0}{\langle \sigma_A\rangle_0}.
\end{align*}

We now turn to the left hand side. By \eqref{eq:pfaffian}, \eqref{eq:pfrec}, \eqref{eq:spinferm2pt} and our induction assumption \eqref{eq:spinferm2kpt}, we can write 
\begin{align*}
&\frac{\langle \sigma_A \Psi_{z_1}\cdots \Psi_{z_{2k}}\rangle_{m\alpha}}{\langle \sigma_A\rangle_{m\alpha}}=\sum_{l=2}^{2k}(-1)^l \frac{\langle \sigma_A\Psi_{z_1}\Psi_{z_l}\rangle_{m\alpha}}{\langle \sigma_A\rangle_{m\alpha}}\frac{\langle \sigma_A \Psi_{z_2}\stackrel{\hat l}{\cdots} \Psi_{z_{2k}}\rangle_{m\alpha}}{\langle \sigma_A\rangle_{m\alpha}}\\
&=\sum_{l=2}^{2k}(-1)^l\sum_{p,p'=0}^\infty \frac{(-m)^{p+p'}}{\pi^{p+p'}\cdot p!(p')!}\int_{\Omega^{p+p'}}d^2 w_1\cdots d^2 w_p d^2 u_1\cdots d^2 u_{p'}\alpha(w_1)\cdots \alpha(w_p)\alpha(u_1)\cdots \alpha(u_{p'})\\
&\quad \times \sum_{C\subset \{1, \ldots,  p\},C'\subset \{1, \ldots, p'\}}\frac{\langle \sigma_A \Psi_{z_1}\Psi_{z_l}\prod_{i\in C}\epsilon_{w_i}\rangle_0}{\langle \sigma_A\rangle_0} \frac{\langle \sigma_A \Psi_{z_2}\stackrel{\hat l}{\cdots}\Psi_{z_{2k}}\prod_{i'\in C'}\epsilon_{u_{i'}}\rangle_0}{\langle \sigma_A\rangle_0}\\
&\quad\times\sum_{\Lambda \in \Pi_{\{1,\ldots,p\}\setminus C},\Lambda'\in \Pi_{\{1,\ldots,p'\}\setminus C'}}(-1)^{|\Lambda|+|\Lambda'|}|\Lambda|!|\Lambda'|!\prod_{B\in \Lambda}\frac{\langle \sigma_A\prod_{j\in B}\epsilon_{w_j}\rangle_0}{\langle \sigma_A\rangle_0}\prod_{B'\in \Lambda'}\frac{\langle \sigma_A \prod_{j'\in B'}\epsilon_{u_{j'}}\rangle_0}{\langle \sigma_A\rangle_0}.
\end{align*}
Using the symmetry in the $w,u$-variables, we can combine the $p,p'$ sums by noting that the combinatorial factor $\frac{(p+p')!}{p!(p')!}$ amounts to summing over subsets $D$ of $\{1,\dots,p+p'\}$ of size $p$. We will also form a partition of $\{1,\dots,p\}\setminus (C\cup C')$ from $\Lambda$ and $\Lambda'$. More precisely, we write ($p+p'$ having now been relabeled as $p$)
\begin{align*}&\frac{\langle \sigma_A \Psi_{z_1}\cdots \Psi_{z_{2k}}\rangle_{m\alpha}}{\langle \sigma_A\rangle_{m\alpha}}=\sum_{l=2}^{2k}(-1)^l\sum_{p=0}^\infty \frac{(-m)^{p}}{\pi^p\cdot p!}\int_{\Omega^{p}}d^{2} w_1\cdots d^{2} w_p\,\alpha(w_1)\cdots \alpha(w_p)\\
&\times\!\sum_{D\subset \{1,\dots,p\}}\sum_{C\sqcup C'=D}\frac{\langle \sigma_A\Psi_{z_1}\Psi_{z_l}\prod_{i\in C}\epsilon_{w_i}\rangle_0}{\langle \sigma_A\rangle_0}\frac{\langle \sigma_A\Psi_{z_2}\stackrel{\hat l}{\cdots}\Psi_{z_{2k}}\prod_{i'\in C'}\epsilon_{w_{i'}}\rangle_0}{\langle \sigma_A\rangle_0}\\
&\times \sum_{I\subset \{1,\dots,p\}\setminus D}\sum_{\Lambda\in \Pi_I, \Lambda'\in \Pi_{(\{1,\dots,p\}\setminus D)\setminus I}}\!(-1)^{|\Lambda|+|\Lambda'|}|\Lambda|!|\Lambda'|!\prod_{B\in \Lambda}\!\frac{\langle \sigma_A\prod_{j\in B}\epsilon_{w_j}\rangle_0}{\langle \sigma_A\rangle_0}\!\!\prod_{B'\in \Lambda'}\!\!\frac{\langle \sigma_A\prod_{j'\in B'}\epsilon_{w_{j'}}\rangle_0}{\langle\sigma_A\rangle_0}.
\end{align*}
Apart from relabeling the summation indices, this is exactly what we found for the right hand side as well, so we are done.
\end{proof}

\subsubsection{Disorder insertions}
For full spin-weighted fermion-disorder correlation functions, we recap the definition of such correlation functions using the OPE at criticality: for $k+l$ even, 
\begin{align} \label{eq:dismerger}
\langle \sigma_V\mu_U\Psi_Z\rangle_{m\alpha}&=\!e^{-il\frac{\pi}{4}}\!\lim_{u'_1\to u_1}(u'_1-u_1)^{\frac12}\cdots \!\lim_{u'_l\to u_l}(u'_l-u_l)^{\frac12}\langle \sigma_V \psi_{u'_1}\sigma_{u_1}\!\cdots \psi_{u'_l}\sigma_{u_l}\Psi_Z\rangle_{m\alpha}\\
\nonumber:&=e^{-il\frac{\pi}{4}}\lim_{u'_1\to u_1}(u'_1-u_1)^{\frac12}\cdots \lim_{u'_l\to u_l}(u'_l-u_l)^{\frac12}\langle \sigma_V \sigma_U \psi_{U'} \Psi_Z\rangle_{m\alpha}.
\end{align}

We note the following analog of Lemma \ref{le:spinferm2kpt}.
\begin{lemma}\label{le:dismain}
For distinct points $v_1,\dots,v_n,u_1,\dots,u_l,z_1,\dots,z_k$, $l+k$ even, the correlation $\frac{\langle \sigma_V\mu_U\Psi_Z\rangle_{m\alpha}}{\langle \sigma_V\sigma_U\rangle_{m\alpha}}$ admits a series expansion with the usual conditions around any $m=m_0 \in \R$. Around $m_0=0$, we have
\begin{multline}\label{eq:dismain}
\frac{\langle \sigma_V\mu_U\Psi_Z\rangle_{m\alpha}}{\langle \sigma_V\sigma_U\rangle_{m\alpha}}=\sum_{p=0}^\infty \left(-\frac{m}{\pi}\right)^p\frac{1}{p!}\int_{\Omega^p}d^{2p}w \prod_{k=1}^p\alpha(w_k)\sum_{S\subset \{1, \ldots, p\}}\frac{\langle \sigma_V\mu_U\Psi_Z\prod_{i\in S}\epsilon_{w_i}\rangle_0}{\langle \sigma_V\sigma_U\rangle_0} \\ 
\times \sum_{\Lambda \in \Pi_{ \{ 1, \ldots, p\}\setminus S}}(-1)^{|\Lambda|}|\Lambda|! \prod_{B\in \Lambda}\frac{\langle \sigma_V\sigma_U\prod_{j\in B}\epsilon_{w_j}\rangle_0}{\langle \sigma_V\sigma_U\rangle_0}.
\end{multline}
\end{lemma}
\begin{proof}
    Note that due to the OPE \eqref{eq:dismerger}, this is the same claim as saying that the order of the limits $u'_i\to u_i$ can be interchanged with the sum/integral in a spin-fermion correlation, i.e., in the expansion of Lemma \ref{le:spinferm2kpt}, while keeping the usual conditions.
    
    Given the decomposition of a general spin-fermion correlation using Pfaffian and the argument of Remark \ref{rem:fermion-pfaffian-analysis}, it suffices to note that the coefficients remain locally exponentially bounded under a spin-fermion merger.
    
    Given any $\mathfrak{B}_p(x_0;\cdot,x_{p+1})$ or $\mathfrak{B}_p^{sym}(x_0;\cdot,x_{p+1})$ coming from the expansion of any $2$-point correlation $\frac{\langle \sigma_A\psi_{x_{p+1}} \psi_{x_0}^{[\eta]} \rangle_{m\alpha}}{\langle \sigma_A\rangle_{m\alpha}}$, we have from \eqref{eq:summand-bound}, for $x = (x_1, \ldots, x_p)$ and $C = C(\kappa_\infty, n, \Omega, \delta_0)$,
    $$\int_{\Omega^p}\left|(x_{p+1}-a_{k'})^{\frac12}\mathfrak{B}_p(x_0;x,x_{p+1})\right|d^{2p}x\leq C^p\sum_{j=1}^n\frac{1}{\sqrt{|x_0-a_j|}},$$
    a bound that is clearly shared by $(x_{p+1}-a_{k'})^{\frac12}\mathfrak{B}^{sym}_p(x_0;x,x_{p+1})$ (recall we only symmetrize $x_1$ to $x_p$), valid under $x_{p+1}\to a_{k'}$. For fermion-fermion and disorder-disorder correlations, we have analogous bounds valid under $x_{p+1}\to a_{k'}$ and $x_{0}\to a_{k''}$
    \begin{align*}
        \int_{\Omega^p}\left|(x_{p+1}-a_{k'})^{\frac12}(x_{0}-a_{k''})^{\frac12}\mathfrak{B}_p(x_0;x,x_{p+1})\right|d^{2p}x \leq C^p.
    \end{align*}
    Given these bounds and \eqref{eq:multiple-integral-combination}, it is straightforward to check that, for fixed $V, U, Z$, there is a small radius for $m$ in which we may exchange the limits with the sum and the integral in the expansion \eqref{eq:spinferm2kpt}, all using dominated convergence.
\end{proof}

\subsection{Energy insertions}\label{subsec:energy}
We now finally introduce the primary field \emph{energy}, or \emph{energy density} in massive settings, defined by the limit (within correlations with mass $m\alpha$)
\begin{equation}\label{eq:energy-def}
\epsilon_w = \frac{i}{2}\lim_{z\to w}\left[\psi_z\psi_w^* +\frac{m}{\pi}c(z,w)\right],
\end{equation}
where we write
\begin{equation}\label{eq:energy-def-c}
c(z,w)=\int_\Omega d^2 y \alpha(y)\big(\langle \psi_{z}\psi^*_{w} \epsilon_y\rangle_0-\langle \psi_{z}\psi^*_{w}\rangle_0 \langle \epsilon_y\rangle_0\big)=:\int_\Omega d^2 y \alpha(y)\big\langle (\psi_z\psi_w^*);\epsilon_y\big\rangle_0^\mathsf T,
\end{equation}
and
\begin{equation}
\langle \mathcal O_1; \mathcal O_2\rangle_0^\mathsf T=\langle \mathcal O_1\mathcal O_2\rangle_0-\langle \mathcal O_1\rangle_0\langle \mathcal O_2\rangle_0.
\end{equation}
We need to show that the limit \eqref{eq:energy-def} exists. From $\psi = \psi^{[1]} + i\psi^{[i]}$, $\psi^* = \psi^{[1]}- i\psi^{[i]}$ and \eqref{eq:logarithmic-correction} (and implicitly \eqref{eq:pfaffian} in case there are more fermions), we know that
\begin{equation}\label{eq:complex-fermion-asymp}
\begin{split}
\psi_z\psi_w &=  \frac{2}{z-w} + (\text{continuous function}),\quad \text{and}\\ 
\psi_z\psi_w^* &= -4mi\cdot \alpha(w)\log|z-w| + (\text{continuous function}),
\end{split}
\end{equation}
where the continuous remainders satisfy a bound that conforms to local uniform boundedness (where we do not require $z$ and $w$ to be uniformly separated).

For the second term, from \eqref{eq:pfaffian}, we have
\begin{equation}\label{eq:fermionic-wicks}
\langle\psi_z\psi_w^*\psi_y\psi_y^*\rangle_0 = \langle\psi_z\psi_w^*\rangle_0\langle\psi_y\psi_y^*\rangle_0+ \langle\psi_z\psi_y^*\rangle_0\langle\psi_w^*\psi_y\rangle_0-\langle\psi_z\psi_y\rangle_0\langle\psi_w^*\psi_y^*\rangle_0.
\end{equation}
When we express the energy insertions in the integrand in \eqref{eq:energy-def} using the critical relation $\epsilon_w = \frac{i}{2}\psi_w\psi_w^*$ then use \eqref{eq:fermionic-wicks} to simplify the 4-point correlation of fermions, the first term gets canceled out. The second term is $O(1)$ when $y$ tends towards $z, w$. For the third term, from \eqref{eq:complex-fermion-asymp}, we have
$$
\langle\psi_z\psi_y\rangle_0 = \frac{2}{z-y}+O(1) \quad \text{and}\quad \langle\psi_w^*\psi_y^*\rangle_0 = \frac{2}{\overline{w-y}}+O(1).
$$
In addition, by Lemma \ref{lem:ff-planar-critical} and (anti)holomorphicity, these $O(1)$ terms are uniform only depending on the domain and the distance of $z,w$ to the boundary. Then, since the integral of the third term becomes essentially a Cauchy transform, using arguments similar to Remark \ref{rem:logarithmic-correction}, it is straightforward to show that
\begin{equation}\label{eq:c-logarithmic-correction}
    c(z,w)= 4\pi i\cdot \alpha(w)\log|z-w| + (\text{continuous function}),
\end{equation}
where again the continuous remainder is bounded by a uniform quantity only depending on the domain, the derivative of $\alpha$ and the distance of $z,w$ to the boundary.
 Comparing with \eqref{eq:complex-fermion-asymp}, we see that the singularity is precisely canceled and the limit \eqref{eq:energy-def} exists.

Therefore, we study the following limit
\begin{align*}
\left\langle \sigma_V \mu_U\Psi_Z \epsilon_{w_1}\cdots \epsilon_{w_r}\right\rangle_{m\alpha}
=\left(\frac{i}{2}\right)^r\lim_{w'_1\to w_1}\cdots \lim_{w'_r\to w_r}\left\langle \sigma_V \mu_U\Psi_Z \prod_{j=1}^r\left(\psi_{w'_j}\psi^*_{w_j}+\frac{m}{\pi}c(w'_j,w_j)\right)\right\rangle_{m\alpha}.
\end{align*}
From our work so far, using Lemma~\ref{le:dismain}, expanding the product and using \eqref{eq:energy-def-c}, we can write 
\begin{align*}
&\frac{\langle \sigma_V\mu_U\Psi_Z \prod_{j=1}^r (\psi_{w_j'}\psi_{w_j}^*+\frac{m}{\pi}c(w_j',w_j))\rangle_{m\alpha}}{\langle \sigma_V\sigma_U\rangle_{m\alpha}}\\
&=\sum_{p=0}^\infty \frac{(-m)^p}{p!\cdot \pi^p}\int_{\Omega^p}d^{2p}x \alpha(x_1)\cdots \alpha(x_p)\sum_{C\subset \{1,\ldots,p\}}\sum_{\Lambda\in \Pi_{\{1,\ldots,p\}\setminus C}}(-1)^{|\Lambda|}|\Lambda|!\\
&\quad \times \frac{\langle \sigma_V\mu_U\Psi_Z\prod_{j=1}^r (\psi_{w_j'}\psi_{w_j}^*+\frac{m}{\pi}c(w_j',w_j))\prod_{k\in C}\epsilon_{x_k}\rangle_0}{\langle \sigma_V\sigma_U\rangle_0}\prod_{B\in \Lambda}\frac{\langle \sigma_U\sigma_V\prod_{l\in B}\epsilon_{x_l}\rangle_0}{\langle \sigma_V\sigma_U\rangle_0}\\
&=\sum_{I\subset \{1,\dots,r\}} \sum_{p=0}^\infty \frac{(-m)^{p}}{p!\cdot \pi^{p+|I|}}\int_{\Omega^{p+|I|}}d^{2p}x \prod_{i\in I}dy_i \alpha(x_1)\cdots \alpha(x_p)\prod_{i\in I}\alpha(y_i)\sum_{C\subset \{1,\ldots,p\}}\sum_{\Lambda\in \Pi_{\{1,\ldots,p\}\setminus C}}\\
&\quad \times (-1)^{|\Lambda|}|\Lambda|! \prod_{i\in I}\langle \psi_{w_i'}\psi_{w_i}^*;\epsilon_{y_i}\rangle_0^\mathsf T\frac{\langle \sigma_V\mu_U\Psi_Z\prod_{j\notin I}\psi_{w_j'}\psi_{w_j}^*\prod_{k\in C}\epsilon_{x_k}\rangle_0}{\langle \sigma_V\sigma_U\rangle_0} \prod_{B\in \Lambda}\frac{\langle \sigma_V\sigma_U\prod_{l\in B}\epsilon_{x_l}\rangle_0}{\langle \sigma_V\sigma_U\rangle_0}\\
&=\sum_{q=0}^\infty \frac{(-m)^q}{q!\cdot \pi^q} \sum_{I\subset \{1,\dots,r\}}\sum_{p=0}^q\frac{q!}{p!}\delta_{q,p+|I|}\int_{\Omega^{q}} d^{2p}x \prod_{i\in I}dy_i \alpha(x_1)\cdots \alpha(x_p)\prod_{i\in I}\alpha(y_i)\!\sum_{C\subset \{1,\ldots,p\}}\sum_{\Lambda\in \Pi_{\{1,\ldots,p\}\setminus C}}\\
&\enspace \times (-1)^{|\Lambda|}|\Lambda|! \prod_{i\in I}\left( -\langle \psi_{w'_{i}}\psi_{w_{i}}^*;\epsilon_{y_i}\rangle_0^\mathsf T \right)\frac{\langle \sigma_V\mu_U\Psi_Z\prod_{j\notin I}\psi_{w_j'}\psi_{w_j}^*\prod_{k\in C}\epsilon_{x_k}\rangle_0}{\langle \sigma_V\sigma_U\rangle_0}\!\prod_{B\in \Lambda}\!\frac{\langle \sigma_V\sigma_U \prod_{l\in B}\epsilon_{x_l}\rangle_0}{\langle \sigma_V\sigma_U\rangle_0}.
\end{align*}
Next we note that due to symmetry, we can write 
\begin{align*}
&\sum_{I\subset \{1,\dots,r\}}\sum_{p=0}^q \frac{q!}{p!}\delta_{q,p+|I|}\int_{\Omega^{q}} d^{2p}x \prod_{i\in I}dy_i \alpha(x_1)\cdots \alpha(x_p)\prod_{i\in I}\alpha(y_i)\sum_{C\subset \{1,\ldots,p\}}\sum_{\Lambda\in \Pi_{\{1,\ldots,p\}\setminus C}}\\
&\enspace \times(-1)^{|\Lambda|}|\Lambda|!\prod_{i\in I}\left( -\langle \psi_{w'_{i}}\psi_{w_{i}}^*;\epsilon_{y_i}\rangle_0^\mathsf T \right)\frac{\langle \sigma_V\mu_U\Psi_Z\prod_{j\notin I}\psi_{w_j'}\psi_{w_j}^*\prod_{k\in C}\epsilon_{x_k}\rangle_0}{\langle \sigma_V\sigma_U\rangle_0}\prod_{B\in \Lambda}\frac{\langle \sigma_V\sigma_U \prod_{l\in B}\epsilon_{x_l}\rangle_0}{\langle \sigma_V\sigma_U\rangle_0}\\
&=\sum_{I\subset \{1,\dots,r\}} \sum_{\nu\in S_{I,[q]}}\int_{\Omega^q}d^{2q}x \alpha(x_1)\cdots \alpha(x_q) \sum_{C\subset \{1,\dots,q\}\setminus \nu(I)}\sum_{\Lambda\in \Pi_{\{1,\dots,q\}\setminus (\nu(I)\cup C)}}(-1)^{|\Lambda|}|\Lambda|!\\
&\enspace \times \prod_{i\in I}\left( -\langle \psi_{w'_{i}}\psi_{w_{i}}^*;\epsilon_{y_i}\rangle_0^\mathsf T \right)\frac{\langle \sigma_V\mu_U\Psi_Z\prod_{j\notin I}\psi_{w_j'}\psi_{w_j}^*\prod_{k\in C}\epsilon_{x_k}\rangle_0}{\langle \sigma_V\sigma_U\rangle_0}\prod_{B\in \Lambda}\frac{\langle \sigma_V\sigma_U\prod_{l\in B}\epsilon_{x_l}\rangle_0}{\langle \sigma_V\sigma_U\rangle_0},
\end{align*}
where $S_{I,[q]}$ denotes the set of injective mappings from $I$ into $\{1,\dots,q\}$ (the number of which is $\frac{q!}{(q-|I|)!}$ -- explaining where the factor $\frac{q!}{p!}$ goes). Note that moving the sums under the integral, the integrand has a limit as $w'\to w$ (i.e. $w_1'\to w_1, \ldots, w_r'\to w_r$), so we claim that: 
\begin{lemma}\label{le:enest}
For distinct points $v_1,\dots,v_n,u_1,\dots,u_l,z_1,\dots,z_k, w_1, \ldots, w_r$, $l+k$ even, the correlation $\frac{\langle \sigma_V\mu_U\Psi_Z\epsilon_W\rangle_{m\alpha}}{\langle \sigma_V\sigma_U\rangle_{m\alpha}}$ admits a series expansion around any $m=m_0$ with the usual conditions. For $m_0=0$, we have 
\begin{align}
\frac{\langle \sigma_V\mu_U\Psi_Z\epsilon_W\rangle_{m\alpha}}{\langle \sigma_V\sigma_U\rangle_{m\alpha}}&=\!\sum_{p=0}^\infty  \!\left(-\frac{m}{\pi}\right)^p \!\frac{1}{p!} \!\int_{\Omega^p} \!d^{2p}x \alpha(x_1)\cdots \alpha(x_p) \!\sum_{I\subset \{1,\dots,r\}}\sum_{\nu\in S_{I,[p]}}\sum_{C\subset \{1,\dots,p\}\setminus \nu(I)}\\
\nonumber&\quad\times\sum_{\Lambda\in \Pi_{\{1,\dots,p\}\setminus (\nu(I)\cup C)}}(-1)^{|\Lambda|}|\Lambda|!\prod_{i\in I}\left(-\langle \epsilon_{w_i};\epsilon_{x_{\nu(i)}}\rangle_0^\mathsf T\right)\\
\nonumber&\quad\times \frac{\langle \sigma_V \mu_U \Psi_Z\prod_{j\notin I}\epsilon_{w_j}\prod_{k\in C}\epsilon_{x_k}\rangle_0}{\langle \sigma_V\sigma_U\rangle_0}\prod_{B\in \Lambda}\frac{\langle \sigma_V\sigma_U\prod_{l\in B}\epsilon_{x_l}\rangle_0}{\langle \sigma_V\sigma_U\rangle_0}.
\end{align}
\end{lemma}
\begin{proof}
As in the proof of Lemma \ref{le:dismain} and Remark \ref{rem:fermion-pfaffian-analysis}, it suffices to note that as $w_s' \to w_s$, the local exponential bound is preserved. Note that in addition to $w' \to w$, we would be merging $u'\to u$ in the spin-fermion correlation as in \eqref{eq:dismerger}:
\begin{equation}\label{eq:ur-spin-fermion}
\frac{\langle \sigma_V\sigma_U \psi_{U'}\Psi_Z\prod_{j=1}^r (\psi_{w_j'}\psi_{w_j}^*+\frac{m}{\pi}c(w_j',w_j))\rangle_{m\alpha}}{\langle \sigma_V\sigma_U\rangle_{m\alpha}},
\end{equation}
which admits a Pfaffian decomposition in terms of the $2$-point correlations (after expanding $ \prod_{j=1}^r$) and series expansion as in Remarks \ref{rem:usual-conditions}.

Let $S, S'$ be disjoint subsets of $\{1, 2, \ldots, r\}$. Then, let us collect together all terms in the expansion of \eqref{eq:ur-spin-fermion} which share singular factors as all $w' \to w$:
\begin{equation}\label{eq:general-term}
(\text{regular as $w' \to w$})\cdot \prod_{s\in S}\frac{\langle \sigma_V\sigma_U \psi_{w_s'}\psi_{w_s}^*\rangle_{m\alpha}}{\langle \sigma_V\sigma_U\rangle_{m\alpha}}\cdot \prod_{s'\in S'}\left(\frac{m}{\pi}c(w_{s'}',w_{s'})\right).
\end{equation}
The first factor collects together factors coming from partitioning the following subset of the fermions in \eqref{eq:ur-spin-fermion}
\begin{equation}\label{eq:fermion-subset}
    \psi_{U'}\Psi_Z\prod_{j\in[r]\setminus (S\cup S')} (\psi_{w_j'}\psi_{w_j}^*+\frac{m}{\pi}c(w_j',w_j))
\end{equation} into $2$-point correlations, but only those without any factor of the form $\frac{\langle \sigma_V\sigma_U \psi_{w_j'}\psi_{w_j}^*\rangle_{m\alpha}}{\langle \sigma_V\sigma_U\rangle_{m\alpha}}$. Note that, for this reason, in any two terms of type \eqref{eq:general-term} which have the same $\mathfrak{S}=S\cup S'$, the first factor is exactly the same: it is clear that they sum over the same set of pairwise partitions of \eqref{eq:fermion-subset}, and the sign of the term in the decomposition is the same whether the nearest-neighbors $\psi_{w_j'}\psi_{w_j}^*$ were present in the original multi-point correlation and was paired off (i.e. $j\in S$), or the term $\frac{m}{\pi}c(w_j',w_j)$ was chosen ($j\in S'$), as seen directly from \eqref{eq:pfaffian} or the well-known correspondence to the parity of the number of crossings in the chord diagram. Let us write $R(S \cup S') = R(\mathfrak{S})$ for the first factor.

Then we argue that, for any fixed $\mathfrak{S}$, the sum
\begin{align*}
&\sum_{S\subset \mathfrak S}R(\mathfrak{S})\cdot \prod_{s\in S}\frac{\langle \sigma_V\sigma_U \psi_{w_s'}\psi_{w_s}^*\rangle_{m\alpha}}{\langle \sigma_V\sigma_U\rangle_{m\alpha}}\cdot \prod_{s'\in S'}\left(\frac{m}{\pi}c(w_{s'}',w_{s'})\right)\\ 
&\quad= R(\mathfrak{S})\cdot \prod_{s\in \mathfrak{S}}\left(\frac{\langle \sigma_V\sigma_U \psi_{w_s'}\psi_{w_s}^*\rangle_{m\alpha}}{\langle \sigma_V\sigma_U\rangle_{m\alpha}}+\frac{m}{\pi}c(w_{s}',w_{s})\right),
\end{align*}
has a series expansion with locally exponentially bounded coefficients that remain so under the spin-fermion merger $u'\to u$ and limit $w'\to w$. Indeed, the former only affects $R(\mathfrak{S})$, and we may estimate each factor in $R(\mathfrak{S})$ as in the proof of Lemma \ref{le:dismain}. For the latter limit, we know that each $\frac{\langle \sigma_V\sigma_U \psi_{w_s'}\psi_{w_s}^*\rangle_{m\alpha}}{\langle \sigma_V\sigma_U\rangle_{m\alpha}}+\frac{m}{\pi}c(w_{s}',w_{s})$ has a series expansion in $m-m_0$ with locally (we let $w_s'$ approach $w_s$, but away from other points and the boundary) exponentially bounded coefficients for $(m-m_0)^p$ for $p\geq 2$ thanks to \eqref{eq:summand-bound}. Since $\frac{\langle \sigma_V\sigma_U \psi_{w_s'}\psi_{w_s}^*\rangle_{m\alpha}}{\langle \sigma_V\sigma_U\rangle_{m\alpha}}+\frac{m}{\pi}c(w_{s}',w_{s})$ itself is locally uniformly bounded by \eqref{eq:complex-fermion-asymp} and \eqref{eq:c-logarithmic-correction}, we see that the coefficients of $(m-m_0)^0$ and $(m-m_0)^1$ must be locally bounded as well (again, letting $w_s' \to w_s$ away from other points and the boundary). The desired summability, integrability, and local continuity then follow from arguments similar to Remark \ref{rem:fermion-pfaffian-analysis} and the proof of Lemma \ref{le:dismain}.
\end{proof}

\section{Coefficients of Doubled Ising Correlations}\label{sec:ising-double}

In this section, we identify series expansions of {doubled} Ising correlations around criticality, i.e. $m_0=0$. Recall that we denote the fields in the second model by $\tilde\sigma, \tilde\mu$, etc., independent from $\sigma,\mu$, etc.

Our main goal is the proof of Theorem \ref{th:isingmain}, done in Section \ref{sec:proof-isingmain}, and in particular the doubled massive Ising primary field correlation near $m=0$ expressed as a series expansion in terms of critical correlations. Recall from Remark \ref{rem:usual-conditions} that the radius of convergence and other properties of the series expansions of single correlations are preserved in this case; we will therefore omit analytic statements, since results of this section are primarily about the identification of coefficients for the $m_0=0$ case.

\subsection{Pure spin correlations}
Recall the notation $\sigma_A=\sigma_{a_1}\cdots \sigma_{a_n}$ within correlation functions.
\begin{lemma}\label{le:spinbosoexp}
Around $m_0=0$, we have
\begin{align}
\log \frac{\langle\sigma_{A}\rangle_{m\alpha}^2}{\isingccf{\sigma_{A}}^2}
&=\sum_{p=0}^\infty \left(-\frac{m}{\pi}\right)^p\frac{1}{p!}\int_{\Omega^p}d^{2p}u\, \prod_{j=1}^p \alpha(u_j)\sum_{\Lambda\in \Pi_p}(-1)^{|\Lambda|-1}(|\Lambda|-1)!\\
\nonumber&\quad\times \Bigg(\prod_{B\in \Lambda}\frac{\isingccf{\sigma_{A}\widetilde \sigma_{A}\prod_{k\in B}(\epsilon_{u_k}+\widetilde \epsilon_{u_k})}}{\isingccf{\sigma_{A}\widetilde \sigma_{A}}}-\prod_{B\in \Lambda}\isingccf{\prod_{k\in B}(\epsilon_{u_k}+\widetilde \epsilon_{u_k})}\Bigg).
\end{align}
\end{lemma}
\begin{proof}
By Proposition \ref{pr:spinexp}, it is sufficient for us to prove that 
\begin{align*}
&\sum_{\Lambda\in \Pi_p}(-1)^{|\Lambda|-1}(|\Lambda|-1)! \prod_{B\in \Lambda}\frac{\isingccf{\sigma_{A}\widetilde \sigma_{A}\prod_{k\in B}(\epsilon_{u_k}+\widetilde \epsilon_{u_k})}}{\isingccf{\sigma_{A}\widetilde \sigma_{A}}}\\
&\quad=2\sum_{\Lambda\in \Pi_p}(-1)^{|\Lambda|-1}(|\Lambda|-1)! \prod_{B\in \Lambda}\frac{\isingccf{\sigma_{A}\prod_{k\in B}\epsilon_{u_k}}}{\isingccf{\sigma_{A}}}
\end{align*}
and 
\begin{align*}
&\sum_{\Lambda\in \Pi_p}(-1)^{|\Lambda|-1}(|\Lambda|-1)! \prod_{B\in \Lambda}\isingccf{\prod_{k\in B}(\epsilon_{u_k}+\widetilde \epsilon_{u_k})}=2\sum_{\Lambda\in \Pi_p}(-1)^{|\Lambda|-1}(|\Lambda|-1)! \prod_{B\in \Lambda}\isingccf{\prod_{k\in B}\epsilon_{u_k}}.
\end{align*}
We will prove the first claim. The second one is identical up to simplifying notation.

First of all, we note that for any set $B\subset \{1,\dots,p\}$ we have 
\begin{align}\label{eq:spin-nu-origin}
&\frac{\isingccf{\sigma_{A}\widetilde \sigma_{A}\prod_{k\in B}(\epsilon_{u_k}+\widetilde \epsilon_{u_k})}}{\isingccf{\sigma_{A}\widetilde \sigma_{A}}}=\sum_{U\subset B}\frac{\isingccf{\sigma_A\prod_{k\in U}\epsilon_{u_k}}}{\isingccf{\sigma_A}}\frac{\big\langle\widetilde\sigma_A\prod_{l\in B\setminus U}\widetilde\epsilon_{u_l}\big\rangle_0}{\isingccf{\widetilde\sigma_A}}.
\end{align}
Note that in the $U$ sum, $U$ and $U^\mathsf c$ produce identical terms. Thus if we expand the product over $B$ in the $\Lambda$ sum, we see that we get $2^{|\Lambda|}$ times a sum over all possible refinements $\nu$ of $\Lambda$ that are obtained by either splitting any given block in $\Lambda$ into two parts or keeping the block as is, since each factor originates, with or without tildes, from some term on the RHS of \eqref{eq:spin-nu-origin}. Let us write $\nu\leq_2\Lambda$ for this relation between $\Lambda$ and its refinement $\nu$. We see that interchanging the order of the $\Lambda$ and $\nu$ sums, we have 
\begin{align*}
&\sum_{\Lambda\in \Pi_p}(-1)^{|\Lambda|-1}(|\Lambda|-1)!\prod_{B\in \Lambda}\frac{\big\langle \sigma_A\widetilde \sigma_A\prod_{j\in B}(\epsilon_{w_j}+\widetilde\epsilon_{w_j})\big\rangle_0}{\langle \sigma_A\widetilde \sigma_A\rangle_0}\\
&\quad=\sum_{\nu \in \Pi_p}\prod_{B\in \nu}\frac{\big\langle \sigma_A\prod_{j\in B}\epsilon_{w_j}\big\rangle_0}{\langle \sigma_A\rangle_0}\sum_{\Lambda: \nu\leq_2\Lambda}(-1)^{|\Lambda|-1}(|\Lambda|-1)! 2^{|\Lambda|}.
\end{align*}
To evaluate this last sum, let us split it according to the number of parts of $|\Lambda|$. Since $\Lambda$ is obtained by pairwise merging some blocks of $\nu$, we must have $|\nu|/2\leq |\Lambda|\leq |\nu|$, so
\begin{equation*}
\sum_{\Lambda: \nu\leq_2\Lambda}(-1)^{|\Lambda|-1}(|\Lambda|-1)! 2^{|\Lambda|}=\sum_{|\nu|/2\leq k\leq |\nu|}(-1)^{k-1}(k-1)! 2^k \sum_{\substack{\Lambda\in \Pi_p:\\ |\Lambda|=k, \nu\leq_2\Lambda}}1.
\end{equation*}
To compute the $\Lambda$-sum, we note first that there must be $|\nu|-k$ parts of $\nu$ that get paired with another part of $\nu$. This means that there are $|\nu|-2(|\nu|-k)=2k-|\nu|$ parts of $\nu$ that are not paired. There are $\binom{|\nu|}{2k-|\nu|}$ ways we can choose these parts that are not paired with anything, and out of the remaining $2(|\nu|-k)$ parts, we can pair them in 
$
\frac{(2(|\nu|-k))!}{(|\nu|-k)!2^{|\nu|-k}}
$
ways. This means that 
\begin{align*}
\sum_{\Lambda: \nu\leq_2\Lambda}(-1)^{|\Lambda|-1}(|\Lambda|-1)! 2^{|\Lambda|}&=\sum_{\frac{|\nu|}{2}\leq k\leq |\nu|}(-1)^{k-1} (k-1)! 2^k \frac{|\nu|!}{(2(|\nu|-k))!(2k-|\nu|)!}\frac{(2(|\nu|-k))!}{(|\nu|-k)!2^{|\nu|-k}}\\
&=\sum_{\frac{|\nu|}{2}\leq k\leq |\nu|}(-1)^{k-1} (k-1)! 2^{2k-|\nu|}\frac{|\nu|!}{(2k-|\nu|)!(|\nu|-k)!}\\
&=\sum_{0\leq k\leq \frac{|\nu|}{2}}(-1)^{|\nu|-k-1}(|\nu|-k-1)!2^{|\nu|-2k}\frac{|\nu|!}{(|\nu|-2k)!k!}\\
&=(-1)^{|\nu|-1}(|\nu|-1)!\sum_{0\leq k\leq \frac{|\nu|}{2}}(-1)^{k}(|\nu|-k-1)!2^{|\nu|-2k}\frac{|\nu|-k+k}{(|\nu|-2k)!k!}\\
&=(-1)^{|\nu|-1}(|\nu|-1)!\sum_{0\leq k\leq \frac{|\nu|}{2}}(-1)^{k}2^{|\nu|-2k}\binom{|\nu|-k}{k}\\
&\quad +(-1)^{|\nu|-1}(|\nu|-1)!\sum_{0\leq k\leq \frac{|\nu|}{2}}(-1)^{k}2^{|\nu|-2k}\binom{|\nu|-k-1}{k-1}.
\end{align*}
Let us then introduce for $q\geq 0$ the quantities 
\begin{equation*}
S_q^{(1)}=\sum_{0\leq k\leq \frac{q}{2}}(-1)^k 2^{q-2k}\binom{q-k}{k} \qquad \text{and} \qquad S_q^{(2)}=\sum_{0\leq k\leq \frac{q}{2}}(-1)^k 2^{q-2k}\binom{q-k-1}{k-1}
\end{equation*}
and compute their generating functions:
\begin{align*}
\sum_{q=0}^\infty x^q S_q^{(1)}=\sum_{k=0}^\infty (-1)^k 2^{-2k} \sum_{q=2k}^\infty 2^q x^{q}\binom{q-k}{k}=\sum_{k=0}^\infty (-1)^k 2^{-k} x^k\sum_{q=k}^\infty 2^q x^{q}\binom{q}{k},
\end{align*}
now we use the fact that 
$
\sum_{n=k}^\infty t^n \binom{n}{k}=\frac{t^k}{(1-t)^{k+1}}
$
to find that (for $|x|<1$) 
\begin{align*}
\sum_{q=0}^\infty S_q^{(1)} x^q=\sum_{k=0}^\infty (-1)^k 2^{-k}x^k \frac{(2x)^k}{(1-2x)^{k+1}}&=\frac{1}{1-2x}\sum_{k=0}^\infty \left(-\frac{x^2}{1-2x}\right)^k\\
&=\frac{1}{1-2x}\frac{1}{1+\frac{x^2}{1-2x}}=\frac{1}{(1-x)^2}=\sum_{q=0}^\infty (q+1)x^q,
\end{align*}
implying that $S_q^{(1)}=q+1$. Similarly, 
\begin{align*}
\sum_{q=0}^\infty x^qS_q^{(2)}&=\sum_{k=0}^\infty (-1)^k 2^{-2k}\sum_{q=2k}^\infty 2^q x^q \binom{q-k-1}{k-1}\\
&=1+\sum_{k=0}^\infty (-1)^{k+1} 2^{-2(k+1)}\sum_{q=2(k+1)}^\infty 2^q x^q \binom{q-k-2}{k}\\
&=1+\sum_{k=0}^\infty (-1)^{k+1} 2^{-2(k+1)}2^{k+2}x^{k+2}\sum_{q=k}^\infty 2^q x^q \binom{q}{k}\\
&=1+\sum_{k=0}^\infty (-1)^{k+1} 2^{-2(k+1)}2^{k+2}x^{k+2}\frac{(2x)^k}{(1-2x)^{k+1}}\\
&=1-x^2 \frac{1}{1-2x}\frac{1}{1+\frac{x^2}{1-2x}}=1-\frac{x^2}{(1-x)^2}=1-\sum_{q=1}^\infty(q-1)x^q, 
\end{align*}
so for $q\geq 1$‚ $S_q^{(2)}=-(q-1)$. Since always $|\nu|\geq 1$, we conclude that 
\begin{equation*}
\sum_{\Lambda: \nu\leq_2\Lambda}(-1)^{|\Lambda|-1}(|\Lambda|-1)! 2^{|\Lambda|}=(-1)^{|\nu|-1}(|\nu|-1)!(S_{|\nu|}^{(1)}+S_{|\nu|}^{(2)})=2(-1)^{|\nu|-1}(|\nu|-1)!.
\end{equation*}
Thus 
\begin{align*}
&\sum_{\Lambda\in \Pi_p}(-1)^{|\Lambda|-1}(|\Lambda|-1)! \prod_{B\in \pi}\frac{\isingccf{\sigma_{A}\widetilde \sigma_{A}\prod_{k\in B}(\epsilon_{u_k}+\widetilde \epsilon_{u_k})}}{\isingccf{\sigma_{A}\widetilde \sigma_{A}}}\\
&\quad=\sum_{\nu \in \Pi_p}\prod_{B\in \nu}\frac{\langle \sigma_A\prod_{j\in B}\epsilon_{w_j}\rangle_0}{\langle \sigma_A\rangle_0}\sum_{\Lambda: \nu\leq_2\Lambda}(-1)^{|\Lambda|-1}(|\Lambda|-1)! 2^{|\Lambda|}\\
&\quad=2\sum_{\nu\in \Pi_p}(-1)^{|\nu|-1}(|\nu|-1)! \prod_{B\in \nu}\frac{\isingccf{\sigma_{A}\prod_{k\in B}\epsilon_{u_k}}}{\isingccf{\sigma_{A}}}
\end{align*}
which was our claim. 

Clearly the combinatorics of the claim for $A=\emptyset$ is identical. This concludes the proof of the lemma.
\end{proof}

\begin{theorem}\label{th:spinboso}
We have
\begin{equation}
\left\langle \sigma_{A}\right\rangle_{m\alpha}^2=\sum_{p=0}^\infty \left(-\frac{m}{\pi}\right)^p \frac{1}{p!} \int_{\Omega^p} d^{2p}u \prod_{j=1}^p \alpha(u_j)  \left\langle \sigma_{A}\widetilde{\sigma}_A; \epsilon_{u_1}+\tilde\epsilon_{u_1};\dots;\epsilon_{u_p}+\tilde\epsilon_{u_p}\right\rangle_0^\mathsf T.
\end{equation}
\end{theorem}

Before turning to the proof, we will need a simple identity for suitable correlation functions of bounded random variables. 
\begin{lemma}\label{le:rvid}
Let $X,Y_1,\dots,Y_p$ be bounded real-valued random variables with $\E(X)=1$. Then 
\begin{align}
&\sum_{\pi \in \Pi_p}\prod_{B\in \pi}\sum_{\sigma\in \Pi_B}(-1)^{|\sigma|-1}(|\sigma|-1)! \left(\prod_{C\in \sigma}\E\bigg[X\prod_{k\in C}Y_k\bigg]-\prod_{C\in \sigma}\E\bigg[\prod_{k\in C}Y_k\bigg]\right)\\
\nonumber&\quad=\sum_{\pi\in \Pi_{[p]\cup \{\infty\}}}(-1)^{|\pi|-1}(|\pi|-1)!\prod_{B\in \pi}\E\bigg[\prod_{j\in B}Z_j\bigg],
\end{align}
where $Z_\infty=X$ and $Z_j=Y_j$ for $j\in [p]=\{1,\dots,p\}$.
\end{lemma}
\begin{proof}
Let us start by looking at the function 
\begin{equation*}
(t_1,\dots,t_p)\mapsto \frac{\E(Xe^{\sum_{j=1}^p t_j Y_j})}{\E(e^{\sum_{j=1}^p t_j Y_j})}.
\end{equation*}
Since the random variables are bounded and $\E(X)\neq 0$, we see that this is an analytic function of $t_1,\dots,t_p$ in some neighborhood of the origin. Moreover, we know (say by Faà di Bruno) that 
\begin{equation*}
\partial_{t_1}\cdots \partial_{t_p}|_{t=0}\frac{\E(Xe^{\sum_{j=1}^p t_j Y_j})}{\E(e^{\sum_{j=1}^p t_j Y_j})}=\sum_{\pi\in \Pi_{[p]\cup \{\infty\}}}(-1)^{|\pi|-1}(|\pi|-1)!\prod_{B\in \pi}\E\bigg[\prod_{j\in B}Z_j\bigg].
\end{equation*}
Thus it remains to show that this derivative also equals the left hand side.

To verify this, let us define 
\begin{equation*}
L(t_1,\dots,t_p)=\log \frac{\E(Xe^{\sum_{j=1}^p t_j Y_j})}{\E(e^{\sum_{j=1}^p t_j Y_j})}
\end{equation*}
which is also analytic in some neighborhood of the origin since $\E(X)=1$. We can thus compute 
\begin{align*}
\partial_{t_1}\cdots \partial_{t_p}|_{t=0}\frac{\E(Xe^{\sum_{j=1}^p t_j Y_j})}{\E(e^{\sum_{j=1}^p t_j Y_j})}=\partial_{t_1}\cdots \partial_{t_p}|_{t=0}e^{L(t_1,\dots,t_p)}
=\sum_{\pi\in \Pi_p}\prod_{B\in \pi}\Bigg(\prod_{j\in B}\frac{\partial}{\partial t_j}\Bigg)\Bigg|_{t=0}L(t_1,\dots,t_p).
\end{align*}
It thus suffices to show that for each $B\subset [p]$
\begin{align*}
\Bigg(\prod_{j\in B}\frac{\partial}{\partial t_j}\Bigg)\Bigg|_{t=0}L(t_1,\dots,t_p)=\sum_{\sigma\in \Pi_{B}}(-1)^{|\sigma|-1}(|\sigma|-1)! \left(\prod_{C\in \sigma}\E\bigg[X\prod_{k\in C}Y_k\bigg]-\prod_{C\in \sigma}\E\bigg[\prod_{k\in C}Y_k\bigg]\right).
\end{align*}
In fact, it is even enough to prove that 
\begin{align*}
\Bigg(\prod_{j\in B}\frac{\partial}{\partial t_j}\Bigg)\Bigg|_{t=0}\log\E\left(X e^{\sum_{j=1}^p t_j Y_j}\right)=\sum_{\sigma\in \Pi_B}(-1)^{|\sigma|-1}(|\sigma|-1)!\prod_{C\in \sigma}\E\bigg[X\prod_{k\in C}Y_k\bigg]
\end{align*}
since the second term is obtained by setting $X=1$. Note that for each $n\geq 1$ and $x>0$
\begin{equation*}
\frac{d^n}{dx^n}\log x=(-1)^{n-1} (n-1)!x^{-n}.
\end{equation*}
Thus 
\begin{align*}
\prod_{j\in B}\left.\frac{\partial}{\partial t_j}\right|_{t=0}\log\E\left[X e^{\sum_{j=1}^p t_j Y_j}\right]
&=\sum_{\sigma\in \Pi_B}\hspace{-3pt}\bigg(\frac{\partial^{|\sigma|}}{\partial^{|\sigma|}x}\bigg|_{x=\E(X)}\log x\bigg)\bigg( \prod_{C\in \sigma}\prod_{j\in C}\left.\frac{\partial}{\partial t_j}\right|_{t=0}\bigg)\E\left[X e^{\sum_{j=1}^p t_j Y_j}\right]\\
&=\sum_{\sigma\in\Pi_B}(-1)^{|\sigma|-1}(|\sigma|-1)!\prod_{C\in \sigma}\E\bigg[X\prod_{j\in C}Y_j\bigg],
\end{align*}
which concludes the proof.
\end{proof}

\begin{proof}[Proof of Theorem \ref{th:spinboso}]
We begin by noting that combining Lemma \ref{le:spinbosoexp} with Proposition \ref{pr:massless} yields  
\begin{align*}
\left\langle \sigma_{A}\right\rangle_{m\alpha}^2 \hspace{-2pt}&=\hspace{-2pt}\bigg\langle \hspace{-2pt}\prod_{j=1}^n \hspace{-1pt}\big(\hspace{-1pt}\sqrt{2}\opcolon \cos\sqrt{\pi}\varphi(a_j)\opcolon \hspace{-1pt}\big)\hspace{-3pt}\bigg\rangle_{\hspace{-2pt}\gff}\hspace{-2pt}\exp\Bigg\{\hspace{-2pt}\sum_{p=0}^\infty \frac{(-m)^p}{\pi^p p!}\hspace{-2pt}\int_{\Omega^p}\hspace{-3pt}d^{2p}u\hspace{-2pt}\prod_{j=1}^p \hspace{-2pt}\alpha(u_j)\hspace{-5pt}\sum_{\Lambda\in \Pi_p}\hspace{-2pt}(-1)^{|\Lambda|-1}(|\Lambda|-1)!\\
&\qquad\times \Bigg[\prod_{B\in \Lambda} \frac{\big\langle \prod_{j=1}^n(\sqrt{2}\opcolon \cos\sqrt{\pi}\varphi(a_j)\opcolon  )\prod_{k\in B}(2 \opcolon \cos\sqrt{4\pi}\varphi(u_k)\opcolon )\big\rangle_\gff}{\big\langle \prod_{j=1}^n (\sqrt{2}\opcolon \cos\sqrt{\pi}\varphi(x_j)\opcolon )\big\rangle_\gff}\\
&\qquad\qquad-\prod_{B\in \Lambda}\bigg\langle \prod_{k\in B}\big(2\opcolon \cos\sqrt{4\pi}\varphi(u_k)\opcolon \big)\bigg\rangle_{\gff}\Bigg]\Bigg\}.
\end{align*}

We now expand the exponential as a series in $m$ (using Faà di Bruno), and find 
\begin{multline*}
\left\langle\sigma_{A}\right\rangle_{m\alpha}^2=\left\langle \prod_{j=1}^n \big(\sqrt{2}\opcolon \cos\sqrt{\pi}\varphi(a_j)\opcolon \big)\right\rangle_{\gff}\\
\shoveleft\qquad\qquad\times \sum_{p=0}^\infty \frac{(-m)^p}{\pi^p p!}\int_{\Omega^p}d^{2p}u\, \prod_{j=1}^p \alpha(u_j)\sum_{\pi\in \Pi_p}\prod_{B\in \pi}\sum_{\sigma\in \Pi_B}(-1)^{|\sigma|-1}(|\sigma|-1)!\\
\times \Bigg[\prod_{C\in \sigma}\frac{\big\langle\prod_{j=1}^n(\sqrt{2} \opcolon \cos\sqrt{\pi}\varphi(a_j)\opcolon )\prod_{k\in C}(2\opcolon \cos\sqrt{4\pi}\varphi(u_k)\opcolon )\big\rangle_\gff}{\big\langle\prod_{j=1}^n (\sqrt{2}\opcolon \cos \sqrt{\pi}\varphi(a_j)\opcolon )\big\rangle_\gff}\\
-\prod_{C\in \sigma}\bigg\langle\prod_{k\in C}(2\opcolon \cos\sqrt{4\pi}\varphi(u_k)\opcolon )\bigg\rangle_\gff\Bigg].
\end{multline*}
Note that the integrand is still integrable by construction and the series still converges absolutely (and uniformly when the points $a_j$ are in a suitable compact set).

Using Lemma \ref{le:rvid} with $X=\frac{\prod_{j=1}^n (\sqrt{2}\opcolon \cos \sqrt{\pi}\varphi(a_j)\opcolon )}{\langle \prod_{j=1}^n (\sqrt{2} \opcolon \cos \sqrt{\pi}\varphi(a_j)\opcolon )\rangle_{\gff}}$ and $Y_j=2\opcolon \cos \sqrt{4\pi}\varphi(u_j)\opcolon $ would be our next step, the issue however is that these are not bounded random variables (or even random variables). However, we do know that these correlation functions are limits of regularized ones by Lemma \ref{lem:gffcorr2}, and the regularized ones are bounded random variables so applying Lemma \ref{le:rvid} with a limiting argument, we conclude that 
\begin{align*}
&\sum_{\pi\in \Pi_p}\prod_{B\in \pi}\sum_{\sigma\in \Pi_B}(-1)^{|\sigma|-1}(|\sigma|-1)!\Bigg[\prod_{C\in \sigma}\frac{\big\langle\prod_{j=1}^n (\sqrt{2}\opcolon \cos\sqrt{\pi}\varphi(a_j)\opcolon )\prod_{k\in C}(2\opcolon \cos\sqrt{4\pi}\varphi(u_k)\opcolon )\big\rangle_\gff}{\big\langle\prod_{j=1}^n (\sqrt{2}\opcolon \cos \sqrt{\pi}\varphi(a_j)\opcolon )\big\rangle_\gff} \\
&\qquad\qquad\qquad- \prod_{C\in \sigma}\bigg\langle\prod_{k\in C}(2\opcolon \cos\sqrt{4\pi}\varphi(u_k)\opcolon )\bigg\rangle_\gff\Bigg]\\
&\quad=\sum_{\pi\in \Pi_{[p]}\cup \{\infty\}}(-1)^{|\pi|-1}(|\pi|-1)!\prod_{B\in \pi}\E\bigg[\prod_{k\in B}Z_k\bigg],
\end{align*}
with $Z_\infty=\frac{\prod_{j=1}^n (\sqrt{2}\opcolon \cos \sqrt{\pi}\varphi(a_j)\opcolon )}{\langle \prod_{j=1}^n(\sqrt{2} \opcolon \cos \sqrt{\pi}\varphi(a_j)\opcolon )\rangle_\gff}$ and $Z_k=2\opcolon \cos \sqrt{4\pi}\varphi(u_k)\opcolon $ for $k\in [p]$.

Note that by definition of cumulants, this is simply 
\begin{align*}
\frac{1}{\langle \prod_{j=1}^n (\sqrt{2}\opcolon \cos \sqrt{\pi}\varphi(a_j)\opcolon )\rangle_\gff}\left\langle \prod_{j=1}^n \big(\sqrt{2} \opcolon \cos \sqrt{\pi}\varphi(a_j)\opcolon \big); \prod_{k=1}^p(2\opcolon \cos \sqrt{4\pi}\varphi(u_k)\opcolon )\right\rangle_{\gff}^\mathsf T,
\end{align*}
the factor in front cancels and using Proposition \ref{pr:massless} again yields the desired expansion.
\end{proof}

\subsection{General spin weighted correlation functions}
In this section, we insert fermion, disorder and energy fields to doubled spin-weighted correlations. We start with a lemma we need for this. 

\begin{lemma}\label{le:comb}
For any $D\subset \{1,\dots,p\}$, we have
\begin{align}
&\sum_{\tau\in \Pi_{\{1,\dots,p\}\setminus D}}(-1)^{|\tau|}(|\tau|+1)!\prod_{B\in \tau}\frac{\langle \sigma_A\prod_{j\in B}\epsilon_{w_j}\rangle_0}{\langle \sigma_A\rangle_0}\\
\nonumber&\qquad=\sum_{\tau\in \Pi_{\{1,\dots,p\}\setminus D}}(-1)^{|\tau|}|\tau|!\prod_{B\in \tau}\frac{\langle \sigma_A\widetilde \sigma_A\prod_{j\in B}(\epsilon_{w_j}+\widetilde\epsilon_{w_j})\rangle_0}{\langle \sigma_A\widetilde \sigma_A\rangle_0}.
\end{align}
\end{lemma}
\begin{proof}
The proof is similar to an argument present in the proof of Lemma \ref{le:spinbosoexp}. Let us begin by noting that
\begin{align*}
\frac{\langle \sigma_A \widetilde \sigma_A\prod_{j\in B}(\epsilon_{w_j}+\widetilde \epsilon_{w_j})\rangle_0}{\langle \sigma_A\widetilde \sigma_A\rangle_0}=\sum_{U\subset B}\frac{\langle \sigma_A \prod_{i\in U}\epsilon_{w_i}\rangle_0\langle \sigma_A\prod_{j\in B\setminus U}\epsilon_{w_j}\rangle_0}{\langle \sigma_A\rangle_0 \langle \sigma_A\rangle_0}.
\end{align*}
Note also that $U$ and $U^\mathsf c$ produce the same term. As in the proof of Lemma \ref{le:spinbosoexp}, expanding the product over $B$, we see that we get $2^{|\tau|}$ times a sum over all possible refinements $\nu$ of $\tau$ that are either obtained by splitting a block into two parts; or keeping the block. We again write $\nu\leq_2\tau$ for this relation. We see that interchanging the order of the $\tau$ and $\nu$ sums, we have 
\begin{align*}
&\sum_{\tau\in \Pi_{\{1,\dots,p\}\setminus D}}(-1)^{|\tau|}|\tau|!\prod_{B\in \tau}\frac{\langle \sigma_A\widetilde \sigma_A\prod_{j\in B}(\epsilon_{w_j}+\widetilde\epsilon_{w_j})\rangle_0}{\langle \sigma_A\widetilde \sigma_A\rangle_0}\\
&\qquad=\sum_{\nu \in \Pi_{\{1,\dots,p\}\setminus D}}\prod_{B\in \nu}\frac{\langle \sigma_A\prod_{i\in B}\epsilon_{w_i}\rangle_0}{\langle \sigma_A\rangle_0}\sum_{\tau: \nu\leq_2\tau}(-1)^{|\tau|}|\tau|! 2^{|\tau|}.
\end{align*}
To compute this sum, let us split it according to the size of $\tau$. As in the proof of Lemma \ref{le:spinbosoexp},
\begin{align*}
\sum_{\tau: \nu\leq_2\tau}(-1)^{|\tau|}|\tau|! 2^{|\tau|}&=\sum_{\frac{|\nu|}{2}\leq k\leq |\nu|}(-1)^k k! 2^k \sum_{\substack{\tau: |\tau|=k, \nu\leq_2\tau}}1\\
&=\sum_{\frac{|\nu|}{2}\leq k\leq |\nu|}(-1)^k k! 2^k \frac{|\nu|!}{(2(|\nu|-k))!(2k-|\nu|)!}\frac{(2(|\nu|-k))!}{(|\nu|-k)!2^{|\nu|-k}}\\
&=(-2)^{|\nu|}|\nu|!\sum_{0\leq k\leq \frac{|\nu|}{2}}(-1)^k  2^{-2k}\frac{(|\nu|-k)!}{k!(|\nu|-2k)!}\\
&=(-1)^{|\nu|}|\nu|!\sum_{0\leq k\leq \frac{|\nu|}{2}}(-1)^k  2^{|\nu|-2k}\binom{|\nu|-k}{k}.
\end{align*}
Notice that this
is exactly the same quantity as $S_q^{(1)}$ from the proof of Lemma~\ref{le:spinbosoexp}, with $q=|\nu|$, which gives us $S_{|\nu|}^{(1)}=|\nu|+1$.
We conclude that 
$$
\sum_{\tau: \nu\leq_2\tau}(-1)^{|\tau|}|\tau|! 2^{|\tau|}=(-1)^{|\nu|}(|\nu|+1)!
$$
and
\begin{align*}
&\sum_{\tau\in \Pi_{\{1,\dots,p\}\setminus D}}(-1)^{|\tau|}|\tau|!\prod_{B\in \tau}\frac{\langle \sigma_A\widetilde \sigma_A\prod_{j\in B}(\epsilon_{w_j}+\widetilde\epsilon_{w_j})\rangle_0}{\langle \sigma_A\widetilde \sigma_A\rangle_0}\\
&\qquad=\sum_{\nu \in \Pi_{\{1,\dots,p\}\setminus D}}(-1)^{|\nu|}(|\nu|+1)!\prod_{B\in \nu}\frac{\langle \sigma_A\prod_{i\in B}\epsilon_{w_i}\rangle_0}{\langle \sigma_A\rangle_0},
\end{align*}
which was the claim.
\end{proof}

We now proceed to the main goal of this section.

\begin{lemma}\label{le:enboso}
Around $m_0=0$, we have
\begin{align*}
&\frac{\langle \sigma_V\widetilde \sigma_V \mu_U\widetilde \mu_U \Psi_Z\widetilde \Psi_Z\prod_{l=1}^r(\epsilon_{w_l}+\widetilde \epsilon_{w_l})\rangle_{m\alpha}}{\langle \sigma_V\widetilde \sigma_V\sigma_U\widetilde \sigma_U\rangle_{m\alpha}}=\sum_{p=0}^\infty \left(-\frac{m}{\pi}\right)^p\frac{1}{p!}\int_{\Omega^p}d^{2p}x \alpha(x_1)\cdots \alpha(x_p)\\
&\quad\times\sum_{I\subset \{1,\dots,r\}}\sum_{\nu\in S_{I,[p]}}\sum_{C\subset \{1,\dots,p\}\setminus \nu(I)}\sum_{\Lambda\in \Pi_{\{1,\dots,p\}\setminus (\nu(I)\cup C)}}(-1)^{|\Lambda|}|\Lambda|!\prod_{i\in I}\left(-\langle \epsilon_{w_i}+\widetilde \epsilon_{w_i};\epsilon_{x_{\nu(i)}}+\widetilde \epsilon_{x_{\nu(i)}}\rangle_0^\mathsf T\right)\\
&\quad \times \frac{\langle \sigma_V \widetilde \sigma_V\mu_U\widetilde \mu_U \Psi_Z\widetilde \Psi_Z \prod_{j\in\{1,\dots,r\}\setminus I}(\epsilon_{w_j}+\widetilde \epsilon_{w_j})\prod_{k\in C}(\epsilon_{x_k}+\widetilde \epsilon_{x_k})\rangle_0}{\langle \sigma_V\tilde\sigma_V\sigma_U\tilde\sigma_U\rangle_0}\\
&\quad \times \prod_{B\in \Lambda}\frac{\langle \sigma_V\widetilde \sigma_V \sigma_U\widetilde \sigma_U\prod_{l\in B}(\epsilon_{x_l}+\widetilde \epsilon_{x_l})\rangle_0}{\langle \sigma_V\widetilde \sigma_V\sigma_U\widetilde \sigma_U\rangle_0}.
\end{align*}
\end{lemma}
\begin{proof}
Using Lemma \ref{le:enest}, we have 
\begin{align*}
&\frac{\langle \sigma_V\widetilde \sigma_V \mu_U\widetilde \mu_U \Psi_Z\widetilde \Psi_Z\prod_{l=1}^r(\epsilon_{w_l}+\widetilde \epsilon_{w_l})\rangle_{m\alpha}}{\langle \sigma_V\widetilde \sigma_V\sigma_U\widetilde \sigma_U\rangle_{m\alpha}}\\
&=\sum_{L\subset \{1,\dots,r\}}\frac{\langle \sigma_V\mu_U\Psi_Z\prod_{l\in L}\epsilon_{w_l}\rangle_{m\alpha}}{\langle \sigma_V\sigma_U\rangle_{m\alpha}}\frac{\langle \sigma_V\mu_U\Psi_Z\prod_{l'\notin L}\epsilon_{w_{l'}}\rangle_{m\alpha}}{\langle \sigma_V\sigma_U\rangle_{m\alpha}}\\
&=\!\sum_{p,p'=0}^\infty \!\left(-\frac{m}{\pi}\right)^{p+p'}\!\frac{1}{p!(p')!}\int_{\Omega^{p+p'}}d^{2p}xd^{2p'}y\prod_{j=1}^p\alpha(x_j)\prod_{k=1}^{p'}\alpha(y_k)\!\sum_{L\subset \{1,\dots,r\}}\sum_{I\subset L,I'\subset \{1,\dots,r\}\setminus L}\sum_{\nu \in S_{I,[p]}}\\
&\enspace \times \!\!\sum_{\nu'\in S_{I',[p']}}\sum_{C\subset \{1,\dots,p\}\setminus \nu(I)}\sum_{C'\subset \{1,\dots,p'\}\setminus \nu'(I')} \sum_{\Lambda\in \Pi_{\{1,\dots,p\}\setminus (\nu(I)\cup C)}}\sum_{\Lambda'\in \Pi_{\{1,\dots,p'\}\setminus (\nu'(I')\cup C')}}\!\!(-1)^{|\Lambda|+|\Lambda'|}|\Lambda|!|\Lambda'|!\\
&\enspace \times \prod_{i\in I}\left(-\langle \epsilon_{w_i}; \epsilon_{x_{\nu(i)}}\rangle_0^\mathsf T\right)\prod_{i'\in I'}\left(-\langle \epsilon_{w_{i'}}; \epsilon_{y_{\nu'(i')}}\rangle_0^\mathsf T \right)\frac{\langle\sigma_V\mu_U\Psi_Z\prod_{j\in L\setminus I}\epsilon_{w_j}\prod_{k\in C}\epsilon_{x_k}\rangle_0}{\langle \sigma_V\sigma_U\rangle_0}\\
&\enspace \times\! \frac{\langle \sigma_V\mu_U\Psi_Z \prod_{j'\in (\{1,\dots,r\}\setminus L)\setminus  I'}\epsilon_{w_{j'}}\prod_{k'\in C'}\epsilon_{y_{k'}}\rangle_0}{\langle \sigma_V\sigma_U\rangle_0} \!\prod_{B\in \Lambda}\! \frac{\langle\sigma_V\sigma_U \prod_{l\in B}\epsilon_{x_l}\rangle_0}{\langle \sigma_V\sigma_U\rangle_0} \!\!\prod_{B'\in \Lambda'}\! \frac{\langle\sigma_V\sigma_U\prod_{l'\in B'}\epsilon_{x_{l'}}\rangle_0 }{\langle \sigma_V\sigma_U\rangle_0}.
\end{align*}
Again, using the symmetry of the integrands, we can combine $x_1,\ldots x_p, y_1, \ldots, y_{p'}$ into $x_1,\ldots,x_{q}$ for $q=p+p'$. First, we replace the sum $\sum_{p, p'=0}^\infty$ with $\sum_{q=0}^\infty \sum_{p=0}^q$ and rewrite $\frac{1}{p!(p')!}=\frac{1}{q!}\binom{q}{p}$. Then we replace $\sum_{p=0}^q \binom{q}{p}\sum_{\nu \in S_{I,p}}\sum_{\nu'\in S_{I',q-p}}$ with $\sum_{J\subset [q]}\sum_{\nu \in S_{I,J}}\sum_{\nu'\in S_{I',[q]\setminus J}}$, where $[q]=\{1,2,\ldots,q\}$ as usual and $S_{I,J}$ is the set of injections from $I$ to $J$, etc. Thus, letting $W=\{1,\dots,r\}$, we find 
\begin{align*}
&\frac{\langle \sigma_V\widetilde \sigma_V\mu_U \widetilde \mu_U\Psi_Z\widetilde \Psi_Z(\epsilon+\widetilde \epsilon)_W\rangle_{m\alpha}}{\langle (\sigma\widetilde \sigma)_{V}(\sigma\widetilde \sigma)_{U}\rangle_{m\alpha}}=\sum_{q=0}^\infty \left(-\frac{m}{\pi}\right)^{q}\frac{1}{q!}\int_{\Omega^{q}}d^{2q}x \prod_{j=1}^q \alpha(x_j)\\
&\qquad \times \sum_{L\subset W}\sum_{I\subset L}\sum_{I'\subset W\setminus L}\sum_{J\subset [q]}\sum_{\nu \in S_{I,J}}\sum_{\nu'\in S_{I',[q]\setminus J}}\prod_{i\in I}\left(-\langle \epsilon_{w_i};\epsilon_{x_{\nu(i)}}\rangle_0^\mathsf T \right)\prod_{i'\in I'}\left(-\langle \epsilon_{w_{i'}};\epsilon_{x_{\nu'(i')}}\rangle_0^\mathsf T\right)\\
&\qquad\times\sum_{C\subset J\setminus \nu(I)}\frac{\langle \sigma_V\mu_U \Psi_Z \prod_{j\in L\setminus I}\epsilon_{w_j}\prod_{k\in C}\epsilon_{x_k}\rangle_0}{\langle \sigma_{ V}\sigma_{ U}\rangle_0}\\
&\qquad \times\sum_{C'\subset [q]\setminus (J\cup \nu'(I'))}\frac{\langle \sigma_V\mu_U\Psi_Z \prod_{j'\in (W\setminus L)\setminus I'}\epsilon_{w_{j'}}\prod_{k'\in C'}\epsilon_{x_{k'}}\rangle_0}{\langle \sigma_{ V}\sigma_{ U}\rangle_0}\\
&\qquad \times \sum_{\Lambda \in \Pi_{J\setminus(\nu(I)\cup C)}}(-1)^{|\Lambda|}|\Lambda|!\prod_{B\in \Lambda}\frac{\langle\sigma_{ V}\sigma_{ U}\epsilon_B\rangle_0}{\langle \sigma_{ V}\sigma_{ U}\rangle_0}\sum_{\Lambda'\in \Pi_{[q]\setminus(J\cup \nu'(I')\cup C')}}(-1)^{|\Lambda'|}|\Lambda'|!\prod_{B'\in \Lambda'}\frac{\langle \sigma_{ V}\sigma_{ U}\epsilon_{B'}\rangle_0}{\langle \sigma_{ V}\sigma_{ U}\rangle_0}.
\end{align*}

Let us next combine the sum over $\nu, \nu'$ into a single sum over $\tau \in S_{I\cup I', [q]}$, subject to the condition that $\tau(I)\subset J$ and $\tau(I')\subset [q]\setminus J$. We can rewrite our correlation function as 
\begin{align*}
&\frac{\langle \sigma_V\widetilde \sigma_V\mu_U \widetilde \mu_U\Psi_Z\widetilde \Psi_Z(\epsilon+\widetilde \epsilon)_W\rangle_{m\alpha}}{\langle (\sigma\widetilde \sigma)_{V}(\sigma\widetilde \sigma)_{U}\rangle_{m\alpha}}=\sum_{q=0}^\infty \left(-\frac{m}{\pi}\right)^{q}\frac{1}{q!}\int_{\Omega^{q}}d^{2r}x \prod_{j=1}^q \alpha(x_j)\\
&\qquad\times \sum_{L\subset W}\sum_{I\subset L}\sum_{I'\subset W\setminus L}\sum_{\tau\in S_{I\cup I',[q]}}\prod_{i\in I\cup I'}\left(-\langle \epsilon_{w_i};\epsilon_{x_{\tau(i)}}\rangle_0^\mathsf T\right) \sum_{J\subset [q]} \1_{\{\tau(I)\subset J,\tau(I')\subset[q]\setminus J\}}\\
&\qquad\times \sum_{C\subset J\setminus \tau(I)}\frac{\langle \sigma_V\mu_U \Psi_Z \prod_{j\in L\setminus I}\epsilon_{w_j}\prod_{k\in C}\epsilon_{x_k}\rangle_0}{\langle\sigma_{ V}\sigma_{ U}\rangle_0}\\
&\qquad\times\sum_{C'\subset [q]\setminus (J\cup \tau(I'))}\frac{\langle \sigma_V\mu_U\Psi_Z \prod_{j'\in (W\setminus L)\setminus I'}\epsilon_{w_{j'}}\prod_{k'\in C'}\epsilon_{x_{k'}}\rangle_0}{\langle \sigma_{ V}\sigma_{ U}\rangle_0}\\
&\qquad\times \sum_{\Lambda \in \Pi_{J\setminus(\tau(I)\cup C)}}(-1)^{|\Lambda|}|\Lambda|!\prod_{B\in \Lambda}\frac{\langle \sigma_{ V}\sigma_{ U}\epsilon_B\rangle_0}{\langle \sigma_{ V}\sigma_{ U}\rangle_0}\sum_{\Lambda'\in \Pi_{[q]\setminus(J\cup \tau(I')\cup C')}}(-1)^{|\Lambda'|}|\Lambda'|!\prod_{B'\in \Lambda'}\frac{\langle \sigma_{ V}\sigma_{ U}\epsilon_{B'}\rangle_0}{\langle \sigma_{ V}\sigma_{ U}\rangle_0}.
\end{align*}
We will now rewrite the sums over $C, C'$ as sums over $D=C\cup C' \subset [q]\setminus \tau(I\cup I')$ and $C$. Then we may pull them out of the $J$ sum, as long as we enforce $C\subset J$ and $D\setminus C \subset [q]\setminus J$ later on:
\begin{align*}
&\frac{\langle \sigma_V\widetilde \sigma_V\mu_U \widetilde \mu_U\Psi_Z\widetilde \Psi_Z(\epsilon+\widetilde \epsilon)_W\rangle_{m\alpha}}{\langle (\sigma\widetilde \sigma)_{V}(\sigma\widetilde \sigma)_{U}\rangle_{m\alpha}}\\
&=\sum_{q=0}^\infty \left(-\frac{m}{\pi}\right)^{q}\frac{1}{q!}\int_{\Omega^{q}}d^{2q}x \prod_{j=1}^q \alpha(x_j)\sum_{L\subset W}\sum_{I\subset L}\sum_{I'\subset W\setminus L} \sum_{\tau\in S_{I\cup I',[q]}}\prod_{i\in I\cup I'}\left(-\langle \epsilon_{w_i};\epsilon_{x_{\tau(i)}}\rangle_0^\mathsf T\right) \\
&\enspace\times\sum_{D\subset [q]\setminus \tau(I\cup I')}\sum_{C\subset D} \frac{\langle \sigma_V\mu_U \Psi_Z\prod_{j\in L\setminus I}\epsilon_{w_j}\prod_{k\in C}\epsilon_{x_k}\rangle_0}{\langle \sigma_{ V}\sigma_{ U}\rangle_0}\frac{\langle \sigma_V\mu_U\Psi_Z\prod_{j'\in (W\setminus L)\setminus I'}\epsilon_{w_{j'}}\prod_{k'\in D \setminus C}\epsilon_{x_{k'}}\rangle_0}{\langle \sigma_{ V}\sigma_{ U}\rangle_0}\\
&\enspace\times \sum_{J\subset [q]} \1_{\{\tau(I)\subset J,\tau(I')\subset[q]\setminus J\}}\cdot\1_{\{C\subset J,D\setminus C\subset [q]\setminus J\}}\sum_{\Lambda \in \Pi_{J\setminus(\tau(I)\cup C)}}(-1)^{|\Lambda|}|\Lambda|!\prod_{B\in \Lambda}\frac{\langle \sigma_{ V}\sigma_{ U}\epsilon_B\rangle_0}{\langle\sigma_{ V}\sigma_{ U}\rangle_0}\\
&\enspace \times \sum_{\Lambda'\in \Pi_{[q]\setminus(J\cup \tau(I')\cup (D\setminus C))}}(-1)^{|\Lambda'|}|\Lambda'|!\prod_{B'\in \Lambda'}\frac{\langle \sigma_{ V}\sigma_{ U}\epsilon_{B'}\rangle_0}{\langle \sigma_{ V}\sigma_{ U}\rangle_0}.
\end{align*}

We will now try to merge $\Lambda,\Lambda'$ sums into a single sum, since $\Lambda\cup \Lambda'$ is a partition of $[q]\setminus (\tau (I\cup I')\cup D)$. 
We can thus write
\begin{align*}
&\sum_{\Lambda \in \Pi_{J\setminus(\tau(I)\cup C)}}(-1)^{|\Lambda|}|\Lambda|!\prod_{B\in \Lambda}\frac{\langle \sigma_{ V}\sigma_{ U}\epsilon_B\rangle_0}{\langle \sigma_{ V}\sigma_{ U}\rangle_0}\sum_{\Lambda'\in \Pi_{[q]\setminus(J\cup \tau(I')\cup (D\setminus C))}}(-1)^{|\Lambda'|}|\Lambda'|!\prod_{B'\in \Lambda'}\frac{\langle \sigma_{ V}\sigma_{ U}\epsilon_{B'}\rangle_0}{\langle \sigma_{ V}\sigma_{ U}\rangle_0}\\
&\quad= \sum_{\lambda\in \Pi_{[q]\setminus (\tau(I\cup I')\cup D)}}(-1)^{|\lambda|}\prod_{B\in \lambda} \frac{\langle \sigma_{ V}\sigma_{ U}\epsilon_B\rangle_0}{\langle \sigma_{ V}\sigma_{ U}\rangle_0}\sum_{\Lambda\in \Pi_{J\setminus (\tau(I)\cup C)}}\sum_{\Lambda'\in \Pi_{[q]\setminus(J\cup \tau(I')\cup (D\setminus C))}}|\Lambda|!|\Lambda'|!\1_{\{\lambda=\Lambda\cup \Lambda'\}}.
\end{align*}
We will now evaluate the remaining $J$ and $\Lambda,\Lambda'$ sums. First of all, we note that under the constraints on $J$, we can uniquely write $J=\tau(I)\cup C\cup E$ where $E$ ranges over subsets of $[q]\setminus (\tau(I\cup I')\cup D)$. Note that all of the sets are disjoint so we have $J\setminus (\tau(I)\cup C)=E$ and $[q]\setminus(J\cup \tau(I')\cup (D\setminus C))=([q]\setminus (\tau(I\cup I')\cup D))\setminus E$.

This means that the $J, \Lambda,\Lambda'$ sums can be combined into 
\begin{align*}
\sum_{E\subset [q]\setminus (\tau(I\cup I')\cup D)}\sum_{\Lambda \in \Pi_E}\sum_{\Lambda'\in \Pi_{([q]\setminus (\tau(I\cup I')\cup D))\setminus E}}|\Lambda|!|\Lambda'|!\1_{\{\lambda=\Lambda\cup\Lambda'\}}.
\end{align*}

To evaluate this sum, note that for any finite set $A$

\begin{align*}
\sum_{B\subset A}\sum_{\Lambda\in \Pi_{A\setminus B}}\sum_{\Lambda' \in \Pi_{B}}|\Lambda|!|\Lambda'|!\1_{\{\lambda=\Lambda\cup \Lambda'\}}=\sum_{\nu: \nu\subset \lambda}|\nu|!(|\lambda|-|\nu|)!.
\end{align*}
We can now split this according to the size of $\nu$:
\begin{align*}
\sum_{\nu: \nu\subset \lambda}|\nu|!(|\lambda|-|\nu|)!=\sum_{s=0}^{|\lambda|}\sum_{\nu\subset \lambda: |\nu|=s} s!(|\lambda|-s)!=\sum_{s=0}^{|\lambda|}s!(|\lambda|-s)!\sum_{\nu\subset \lambda: |\nu|=s}1. 
\end{align*}
Now, given $\lambda$, it has precisely $\binom{|\lambda|}{s}$ subsets of size $s$, so we have
\begin{align*}
\sum_{\nu: \nu\subset \lambda}|\nu|!(|\lambda|-|\nu|)!=\sum_{s=0}^{|\lambda|}s!(|\lambda|-s)!\frac{|\lambda|!}{s!(|\lambda|-s)!}=(|\lambda|+1)!.
\end{align*}

Relabeling $q,\lambda,\tau$ as $p, \Lambda,\nu$, we thus have
\begin{align*}
&\frac{\langle \sigma_V\widetilde \sigma_V \mu_U\widetilde \mu_U \Psi_Z\widetilde \Psi_Z\prod_{l=1}^r(\epsilon_{w_l}+\widetilde \epsilon_{w_l})\rangle_{m\alpha}}{\langle \sigma_V\widetilde \sigma_V\sigma_U\widetilde \sigma_U\rangle_{m\alpha}}=\sum_{p=0}^\infty \frac{(-m)^p}{p!\cdot \pi^p}\int_{\Omega^p}d^{2p}x \alpha(x_1)\cdots \alpha(x_p)\\
&\quad \sum_{L\subset \{1,\dots,r\}}\sum_{\substack{I\subset L, \\I'\subset \{1,\dots,r\}\setminus L}}\sum_{\nu \in S_{I\cup I',[p]}}\sum_{D\subset \{1,\dots,p\}\setminus \nu(I\cup I')}\sum_{C\sqcup C'=D}\prod_{i\in I\cup I'}\left(-\langle \epsilon_{w_i};\epsilon_{x_{\nu(i)}}\rangle_0^\mathsf T\right) \\
&\quad \times \frac{\langle\sigma_V\mu_U\Psi_Z\prod_{j\in L\setminus I}\epsilon_{w_j}\prod_{k\in C}\epsilon_{x_k}\rangle_0}{\langle\sigma_V\sigma_U\rangle_0}\frac{\langle \sigma_V\mu_U\Psi_Z \prod_{j'\in (\{1,\dots,r\}\setminus L)\setminus I'}\epsilon_{w_{j'}}\prod_{k'\in C'}\epsilon_{x_{k'}}\rangle_0}{\langle \sigma_V\sigma_U\rangle_0}\\
&\quad\times\sum_{\Lambda \in \Pi_{\{1,\dots,p\}\setminus (\nu(I\cup I')\cup D)}}(-1)^{|\Lambda|}(|\Lambda|+1)!\prod_{B\in \Lambda}\frac{\langle \sigma_V\sigma_U\prod_{l\in B}\epsilon_{x_l}\rangle_0}{\langle \sigma_V\sigma_U\rangle_0}.
\end{align*}
Next we use Lemma \ref{le:comb} on the last line, interchange the order of the $L$-sum and the $I,I'$-sum and simultaneously decompose the sum over $I,I'$ according to the sum over $E=I\cup I'$. We find 
\begin{align*}
&\frac{\langle \sigma_V\widetilde \sigma_V \mu_U\widetilde \mu_U \Psi_Z\widetilde \Psi_Z\prod_{l=1}^r(\epsilon_{w_l}+\widetilde \epsilon_{w_l})\rangle_{m\alpha}}{\langle \sigma_V\widetilde \sigma_V\sigma_U\widetilde \sigma_U\rangle_{m\alpha}}=\sum_{p=0}^\infty \frac{(-m)^p}{p!\cdot \pi^p}\int_{\Omega^p}d^{2p}x \alpha(x_1)\cdots \alpha(x_p)\\
&\quad\sum_{E\subset \{1,\dots,r\}}\sum_{I\sqcup I'=E}\sum_{\substack{L\subset \{1,\dots,r\}:\\ I\subset L, I'\subset \{1,\dots,r\}\setminus L}}\sum_{\nu \in S_{E,[p]}}\sum_{D\subset \{1,\dots,p\}\setminus \nu(E)}\sum_{C\sqcup C'=D}\prod_{i\in E}\left(-\langle \epsilon_{w_i};\epsilon_{x_{\nu(i)}}\rangle_0^\mathsf T\right) \\
&\quad \times \frac{\langle\sigma_V\mu_U\Psi_Z\prod_{j\in L\setminus I}\epsilon_{w_j}\prod_{k\in C}\epsilon_{x_k}\rangle_0}{\langle\sigma_V\sigma_U\rangle_0}\frac{\langle \sigma_V\mu_U\Psi_Z \prod_{j'\in (\{1,\dots,r\}\setminus L)\setminus I'}\epsilon_{w_{j'}}\prod_{k'\in C'}\epsilon_{x_{k'}}\rangle_0}{\langle \sigma_V\sigma_U\rangle_0}\\
&\quad \times \sum_{\Lambda \in \Pi_{\{1,\dots,p\}\setminus (\nu(E)\cup D)}}(-1)^{|\Lambda|}|\Lambda|!\prod_{B\in \Lambda}\frac{\langle \sigma_V\widetilde \sigma_V\sigma_U\widetilde \sigma_U\prod_{l\in B}(\epsilon_{x_l}+\widetilde \epsilon_{x_l})\rangle_0}{\langle \sigma_V\widetilde \sigma_V\sigma_U\widetilde \sigma_U\rangle_0}.
\end{align*}
Our next remark is that $L\setminus I\sqcup [(\{1,\dots,r\}\setminus L)\setminus I']=\{1,\dots,r\}\setminus E$, and summing over $L$ is equivalent to summing over the ways to partition $\{1,\dots,r\}\setminus E$ into two parts ($J=L\setminus I$ and $J'=(\{1,\dots,r\}\setminus L)\setminus I'$), so we can write this as
\begin{align*}
&\frac{\langle \sigma_V\widetilde \sigma_V \mu_U\widetilde \mu_U \Psi_Z\widetilde \Psi_Z\prod_{l=1}^r(\epsilon_{w_l}+\widetilde \epsilon_{w_l})\rangle_{m\alpha}}{\langle \sigma_V\widetilde \sigma_V\sigma_U\widetilde \sigma_U\rangle_{m\alpha}}\\
&=\!\sum_{p=0}^\infty \frac{(-m)^p}{p!\cdot \pi^p}\!\int_{\Omega^p} \!d^{2p}x \alpha(x_1)\cdots \alpha(x_p)\!\sum_{E\subset \{1,\dots,r\}}\sum_{I\sqcup I'=E}\sum_{J\sqcup J'=\{1,\dots,r\}\setminus E}\sum_{\nu \in S_{E,[p]}}\sum_{D\subset \{1,\dots,p\}\setminus \nu(E)}\sum_{C\sqcup C'=D}\\
&\quad \times \prod_{i\in E}\left(-\langle \epsilon_{w_i};\epsilon_{x_{\nu(i)}}\rangle_0^\mathsf T\right) \frac{\langle\sigma_V\mu_U\Psi_Z\prod_{j\in J}\epsilon_{w_j}\prod_{k\in C}\epsilon_{x_k}\rangle_0}{\langle\sigma_V\sigma_U\rangle_0}\frac{\langle \sigma_V\mu_U\Psi_Z \prod_{j'\in J'}\epsilon_{w_{j'}}\prod_{k'\in C'}\epsilon_{x_{k'}}\rangle_0}{\langle \sigma_V\sigma_U\rangle_0}\\
&\quad \times \sum_{\Lambda \in \Pi_{\{1,\dots,p\}\setminus (\nu(E)\cup D)}}(-1)^{|\Lambda|}|\Lambda|!\prod_{B\in \Lambda}\frac{\langle \sigma_V\widetilde \sigma_V\sigma_U\widetilde \sigma_U\prod_{l\in B}(\epsilon_{x_l}+\widetilde \epsilon_{x_l})\rangle_0}{\langle \sigma_V\widetilde \sigma_V\sigma_U\widetilde \sigma_U\rangle_0}\\
&=\sum_{p=0}^\infty \frac{(-m)^p}{p!\cdot \pi^p}\int_{\Omega^p}d^{2p}x \alpha(x_1)\cdots \alpha(x_p)\sum_{E\subset \{1,\dots,r\}}\sum_{I\sqcup I'=E}\sum_{\nu \in S_{E,p}}\sum_{D\subset \{1,\dots,p\}\setminus \nu(E)}\\
&\quad \times \prod_{i\in E}\left(-\langle \epsilon_{w_i};\epsilon_{x_{\nu(i)}}\rangle_0^\mathsf T\right) \frac{\langle\sigma_V\widetilde \sigma_V\mu_U\widetilde \mu_U\Psi_Z\widetilde \Psi_Z\prod_{j\in \{1,\dots,r\}\setminus E}(\epsilon_{w_j}+\widetilde \epsilon_{w_j})\prod_{k\in D}(\epsilon_{x_k}+\widetilde \epsilon_{x_k})\rangle_0}{\langle\sigma_V\widetilde \sigma_V\sigma_U\widetilde \sigma_U\rangle_0}\\
&\quad \times \sum_{\Lambda \in \Pi_{\{1,\dots,p\}\setminus (\nu(E)\cup D)}}(-1)^{|\Lambda|}|\Lambda|!\prod_{B\in \Lambda}\frac{\langle \sigma_V\widetilde \sigma_V\sigma_U\widetilde \sigma_U\prod_{l\in B}(\epsilon_{x_l}+\widetilde \epsilon_{x_l})\rangle_0}{\langle \sigma_V\widetilde \sigma_V\sigma_U\widetilde \sigma_U\rangle_0}.
\end{align*}
It remains to note that nothing in the sum depends on $I,I'$, so the sum over $I, I'$ produces a factor of $2^{|E|}$. To conclude, we note that by independence
\begin{equation*}
2^{|E|}\prod_{i\in E}\langle \epsilon_{w_i};\epsilon_{x_{\nu(i)}}\rangle_0^\mathsf T=\prod_{i\in E}\langle \epsilon_{w_i}+\widetilde \epsilon_{w_i}; \epsilon_{x_{\nu(i)}}+\widetilde \epsilon_{x_{\nu(i)}}\rangle_0^\mathsf T.
\end{equation*}
Relabeling $E, D$ as $I, C$ yields the claim.
\end{proof}

\subsection{Proof of Theorem \ref{th:isingmain}} \label{sec:proof-isingmain}
We can finally prove our main result about the Ising model.
\begin{proof}[Proof of Theorem \ref{th:isingmain}]
The points (1) and (2), i.e. existence, boundedness, analyticity, and local continuity, follow from our work in Sections \ref{sec:ising-analysis} and \ref{sec:ising-combinatorics}. Namely, pure spin correlations exist due to Proposition \ref{prop:spin}; spin-weighted correlations of other fields may then be constructed from \eqref{eq:dismain} and \eqref{eq:energy-def}. Both are analytic, with series expansion under usual conditions, by Proposition \ref{pr:spinexp} and Lemma \ref{le:enest}. Then, from Remark \ref{rem:usual-conditions}, it is easy to see that local uniform boundedness and continuity hold for $\left\langle \sigma_U\mu_V \psi_Z \psi^*_W\epsilon_A\right\rangle_{m\alpha;\Omega}$ within any single radius of convergence; then, since any radius of convergence is bounded below by $c = c(\Omega,M=2,\kappa_\infty,\kappa_b,\delta_{all},|U|+|V|+|Z|+|W|+|A|)$, we obtain the local uniform bound and continuity for $m=1$ by adding at most $2/c$ copies of the single bound as in the proof of Corollary \ref{cor:series-expansion-a}.

Multiplying the series expansions from Lemma \ref{le:enboso} and Theorem \ref{th:spinboso}, using the moment to cumulant map \eqref{eq:mtoc2}, we have,
\begin{align*}
&\langle \sigma_V\widetilde \sigma_V \mu_U\widetilde \mu_U \Psi_Z\widetilde \Psi_Z(\epsilon+\widetilde \epsilon)_W\rangle_{m\alpha}=\!\sum_{p,q=0}^\infty \!\left(-\frac{m}{\pi}\right)^{p+q}\!\frac{1}{p!q!}\int_{\Omega^{p+q}}\!d^{2p}x d^{2q} y \prod_{j=1}^p \alpha(x_j)\prod_{k=1}^q \alpha(y_k)\\
&\quad \times\sum_{I\subset W}\sum_{\nu\in S_{I,[p]}}\prod_{i\in I}\left(-\langle \epsilon_{w_i}+\widetilde \epsilon_{w_i};\epsilon_{x_{\nu(i)}}+\widetilde \epsilon_{x_{\nu(i)}}\rangle_0^\mathsf T\right)\\
&\quad \times \sum_{C\subset [p]\setminus \nu(I)}\frac{\langle \sigma_V \widetilde \sigma_V\mu_U\widetilde \mu_U \Psi_Z\widetilde \Psi_Z (\epsilon+\widetilde \epsilon)_{W\setminus I}\prod_{s\in C}(\epsilon+\widetilde \epsilon)_{x_s}\rangle_0}{\langle \sigma_V \widetilde \sigma_V\sigma_U\widetilde \sigma_U\rangle_0}\\
&\quad \times  \sum_{\Lambda\in \Pi_{[p]\setminus (\nu(I)\cup C)}}(-1)^{|\Lambda|}|\Lambda|!\prod_{B\in \Lambda}\frac{\langle \sigma_V\widetilde \sigma_V \sigma_U\widetilde \sigma_U\prod_{t\in B}(\epsilon+\widetilde \epsilon)_{x_t}\rangle_0}{\langle \sigma_V\widetilde \sigma_V\sigma_U\widetilde \sigma_U\rangle_0}\\
&\quad \times \sum_{C'\subset [q]}\bigg\langle (\sigma\widetilde \sigma)_{V}(\sigma\widetilde \sigma)_{U}\prod_{s'\in C'}(\epsilon+\widetilde \epsilon)_{y_{s'}}\bigg\rangle_0\sum_{\Lambda'\in \Pi_{[q]\setminus C'}}(-1)^{|\Lambda'|}|\Lambda'|!\prod_{B'\in \Lambda'}\bigg\langle \prod_{t'\in B'}(\epsilon+\widetilde \epsilon)_{y_{t'}}\bigg\rangle_0.
\end{align*}
We then call $p+q=r$ and define $x_{p+j}=y_j$, which allows us to write this as 
\begin{align*}
&\langle \sigma_V\widetilde \sigma_V \mu_U\widetilde \mu_U \Psi_Z\widetilde \Psi_Z(\epsilon+\widetilde \epsilon)_W\rangle_{m\alpha}=\sum_{r=0}^\infty \left(-\frac{m}{\pi}\right)^{r}\frac{1}{r!}\sum_{p=0}^r \binom{r}{p}\int_{\Omega^{r}}d^{2r}x  \prod_{j=1}^r \alpha(x_j)\\
&\quad\times\sum_{I\subset W}\sum_{\nu\in S_{I,[p]}}\prod_{i\in I}\left(-\langle \epsilon_{w_i}+\widetilde \epsilon_{w_i};\epsilon_{x_{\nu(i)}}+\widetilde \epsilon_{x_{\nu(i)}}\rangle_0^\mathsf T\right)\\
&\quad \times \sum_{C\subset [p]\setminus \nu(I)}\frac{\langle \sigma_V \widetilde \sigma_V\mu_U\widetilde \mu_U \Psi_Z\widetilde \Psi_Z (\epsilon+\widetilde \epsilon)_{W\setminus I}\prod_{s\in C}(\epsilon+\widetilde \epsilon)_{x_s}\rangle_0}{\langle \sigma_V \widetilde \sigma_V\sigma_U\widetilde \sigma_U\rangle_0}\\
&\quad \times  \sum_{\Lambda\in \Pi_{[p]\setminus (\nu(I)\cup C)}}(-1)^{|\Lambda|}|\Lambda|!\prod_{B\in \Lambda}\frac{\langle \sigma_V\widetilde \sigma_V \sigma_U\widetilde \sigma_U\prod_{t\in B}(\epsilon+\widetilde \epsilon)_{x_t}\rangle_0}{\langle \sigma_V\widetilde \sigma_V\sigma_U\widetilde \sigma_U\rangle_0}\\
&\quad \times \sum_{C'\subset (p,r]}\bigg\langle (\sigma\widetilde \sigma)_{V}(\sigma\widetilde \sigma)_{U}\prod_{s'\in C'}(\epsilon+\widetilde \epsilon)_{x_{s'}}\bigg\rangle_0\sum_{\Lambda'\in \Pi_{(p,r]\setminus C'}}(-1)^{|\Lambda'|}|\Lambda'|!\prod_{B'\in \Lambda'}\bigg\langle \prod_{t'\in B'}(\epsilon+\widetilde \epsilon)_{x_{t'}}\bigg\rangle_0.
\end{align*}
Due to the symmetry, we find from simply relabeling the integration variables that we can interpret the sum $\sum_{p=0}^r \binom{r}{p}$ as a sum over all possible subsets $J\subset [r]$, however we must replace e.g. $S_{I,[p]}$ with $S_{I,J}$ (injections from $I$ to $J$) and so on. This lets us write 
\begin{align*}
&\langle \sigma_V\widetilde \sigma_V \mu_U\widetilde \mu_U \Psi_Z\widetilde \Psi_Z(\epsilon+\widetilde \epsilon)_W\rangle_{m\alpha}=\sum_{r=0}^\infty \left(-\frac{m}{\pi}\right)^{r}\frac{1}{r!}\int_{\Omega^{r}}d^{2r}x  \prod_{j=1}^r \alpha(x_j)\sum_{I\subset W}\sum_{J\subset [r]}\sum_{\nu\in S_{I,J}}\\
&\quad \times \prod_{i\in I}\left(-\langle \epsilon_{w_i}+\widetilde \epsilon_{w_i};\epsilon_{x_{\nu(i)}}+\widetilde \epsilon_{x_{\nu(i)}}\rangle_0^\mathsf T\right)\sum_{C\subset J\setminus \nu(I)}\frac{\langle \sigma_V \widetilde \sigma_V\mu_U\widetilde \mu_U \Psi_Z\widetilde \Psi_Z (\epsilon+\widetilde \epsilon)_{W\setminus I}\prod_{s\in C}(\epsilon+\widetilde \epsilon)_{x_s}\rangle_0}{\langle \sigma_V \widetilde \sigma_V\sigma_U\widetilde \sigma_U\rangle_0}\\
&\quad \times  \sum_{\Lambda\in \Pi_{J\setminus (\nu(I)\cup C)}}(-1)^{|\Lambda|}|\Lambda|!\prod_{B\in \Lambda}\frac{\langle \sigma_V\widetilde \sigma_V \sigma_U\widetilde \sigma_U\prod_{t\in B}(\epsilon+\widetilde \epsilon)_{x_t}\rangle_0}{\langle \sigma_V\widetilde \sigma_V\sigma_U\widetilde \sigma_U\rangle_0}\\
&\quad \times \sum_{C'\subset [r]\setminus J}\bigg\langle (\sigma\widetilde \sigma)_{V}(\sigma\widetilde \sigma)_{U}\prod_{s'\in C'}(\epsilon+\widetilde \epsilon)_{x_{s'}}\bigg\rangle_0\sum_{\Lambda'\in \Pi_{[r]\setminus(J\cup  C')}}(-1)^{|\Lambda'|}|\Lambda'|!\prod_{B'\in \Lambda'}\bigg\langle \prod_{t'\in B'}(\epsilon+\widetilde \epsilon)_{x_{t'}}\bigg\rangle_0.
\end{align*}
We now write the sum $\sum_{J\subset [r]}\sum_{\nu\in S_{I,J}}$ in a slightly different way. Note that each pair $(J,\nu)$ can be (bijectively) identified with a triple $(K,E,\nu)$, where $\nu:I\to K\subset J$ is a bijection (we abuse notation as $\nu\in S_{I,J}$ maps to $J$ while it only takes values in $K$), and $E=J\setminus K$. Thus we can write 
\begin{align*}
\sum_{J\subset [r]}\sum_{\nu\in S_{I,J}}=\sum_{\substack{K\subset [r]:\\ |K|=|I|}}\sum_{\nu:I\leftrightarrow K}\sum_{E\subset [r]\setminus K}.
\end{align*}
With this representation, we have 
\begin{align*}
&\langle \sigma_V\widetilde \sigma_V \mu_U\widetilde \mu_U \Psi_Z\widetilde \Psi_Z(\epsilon+\widetilde \epsilon)_W\rangle_{m\alpha}=\sum_{r=0}^\infty \left(-\frac{m}{\pi}\right)^{r}\frac{1}{r!}\int_{\Omega^{r}}d^{2r}x  \prod_{j=1}^r \alpha(x_j)\sum_{I\subset W}\sum_{\substack{K\subset [r]:\\ |K|=|I|}}\sum_{\nu:I\leftrightarrow K}\\
&\times \prod_{i\in I}\left(-\langle \epsilon_{w_i}+\widetilde \epsilon_{w_i};\epsilon_{x_{\nu(i)}}+\widetilde \epsilon_{x_{\nu(i)}}\rangle_0^\mathsf T\right)\!\!\sum_{E\subset [r]\setminus K} \sum_{C\subset E}\frac{\langle \sigma_V \widetilde \sigma_V\mu_U\widetilde \mu_U \Psi_Z\widetilde \Psi_Z (\epsilon+\widetilde \epsilon)_{W\setminus I}\prod_{s\in C}(\epsilon+\widetilde \epsilon)_{x_s}\rangle_0}{\langle \sigma_V \widetilde \sigma_V\sigma_U\widetilde \sigma_U\rangle_0}\\
&\times \!\!\sum_{\Lambda\in \Pi_{E\setminus C}}(-1)^{|\Lambda|}|\Lambda|!\prod_{B\in \Lambda}\frac{\langle \sigma_V\widetilde \sigma_V \sigma_U\widetilde \sigma_U\prod_{t\in B}(\epsilon+\widetilde \epsilon)_{x_t}\rangle_0}{\langle \sigma_V\widetilde \sigma_V\sigma_U\widetilde \sigma_U\rangle_0}\!\!\sum_{C'\subset ([r]\setminus K)\setminus E}\!\!\bigg\langle (\sigma\widetilde \sigma)_{V}(\sigma\widetilde \sigma)_{U}\prod_{s'\in C'}(\epsilon+\widetilde \epsilon)_{x_{s'}}\bigg\rangle_0\\
&\times \sum_{\Lambda'\in \Pi_{([r]\setminus K)\setminus (E\cup  C')}}(-1)^{|\Lambda'|}|\Lambda'|!\prod_{B'\in \Lambda'}\bigg\langle \prod_{t'\in B'}(\epsilon+\widetilde \epsilon)_{x_{t'}}\bigg\rangle_0.
\end{align*}

We now swap the order of the $E$ and $C$ sums. This is convenient to do by writing $E=C\cup F$ with $F\subset [r]\setminus (K\cup C)$. Note that the constraint on $C'$ now becomes $C'\subset[r]\setminus (K\cup C\cup F)$ so we have 
\begin{align*}
&\langle \sigma_V\widetilde \sigma_V \mu_U\widetilde \mu_U \Psi_Z\widetilde \Psi_Z(\epsilon+\widetilde \epsilon)_W\rangle_{m\alpha}=\sum_{r=0}^\infty \left(-\frac{m}{\pi}\right)^{r}\frac{1}{r!}\int_{\Omega^{r}}d^{2r}x  \prod_{j=1}^r \alpha(x_j)\sum_{I\subset W}\sum_{\substack{K\subset [r]:\\ |K|=|I|}}\sum_{\nu:I\leftrightarrow K}\\
&\quad \times \prod_{i\in I}\left(-\langle \epsilon_{w_i}+\widetilde \epsilon_{w_i};\epsilon_{x_{\nu(i)}}+\widetilde \epsilon_{x_{\nu(i)}}\rangle_0^\mathsf T\right)\sum_{C\subset [r]\setminus K} \bigg\langle \sigma_V \widetilde \sigma_V\mu_U\widetilde \mu_U \Psi_Z\widetilde \Psi_Z (\epsilon+\widetilde \epsilon)_{W\setminus I}\prod_{s\in C}(\epsilon+\widetilde \epsilon)_{x_s}\bigg\rangle_0\\
&\quad \times \sum_{F\subset [r]\setminus(K\cup C)}  \sum_{\Lambda\in \Pi_{F}}(-1)^{|\Lambda|}|\Lambda|!\prod_{B\in \Lambda}\frac{\langle (\sigma\widetilde \sigma)_{V}(\sigma\widetilde \sigma)_{U}\prod_{t\in B}(\epsilon+\widetilde \epsilon)_{x_t}\rangle_0}{\langle (\sigma\widetilde \sigma)_{V}(\sigma\widetilde \sigma)_{U}\rangle_0}\\
&\quad \times \sum_{C'\subset [r]\setminus (K\cup C\cup F)}\frac{\left\langle (\sigma\widetilde \sigma)_{V}(\sigma\widetilde \sigma)_{U}\prod_{s'\in C'}(\epsilon+\widetilde \epsilon)_{x_{s'}}\right\rangle_0}{{\langle (\sigma\widetilde \sigma)_{V}(\sigma\widetilde \sigma)_{U}\rangle_0}}\\
&\quad\times\sum_{\Lambda'\in \Pi_{[r]\setminus (K\cup C\cup C'\cup F)}}(-1)^{|\Lambda'|}|\Lambda'|!\prod_{B'\in \Lambda'}\bigg\langle \prod_{t'\in B'}(\epsilon+\widetilde \epsilon)_{x_{t'}}\bigg\rangle_0.
\end{align*}
Let us try to perform this sum. First of all, there are two cases we will consider separately: either $C'=\emptyset$ or $C'\neq \emptyset$. 

\subsubsection*{Case one}
In the first case, we end up with the sum 
\begin{align*}
& \sum_{F\subset [r]\setminus(K\cup C)} \sum_{\Lambda\in \Pi_{F}} \!(-1)^{|\Lambda|}|\Lambda|! \!\prod_{B\in \Lambda} \!\frac{\langle (\sigma\widetilde \sigma)_{V}(\sigma\widetilde \sigma)_{U}\prod_{t\in B}(\epsilon+\widetilde \epsilon)_{x_t}\rangle_0}{\langle(\sigma\widetilde \sigma)_{V}(\sigma\widetilde \sigma)_{U}\rangle_0} \!\\
&\quad \times\sum_{\Lambda'\in \Pi_{[r]\setminus (K\cup C\cup F)}} \!(-1)^{|\Lambda'|}|\Lambda'|! \!\prod_{B'\in \Lambda'} \!\bigg\langle \prod_{t'\in B'}(\epsilon+\widetilde \epsilon)_{x_{t'}}\bigg\rangle_0.
\end{align*}

We insert within these sums $1=\sum_{\lambda\in \Pi_{[r]\setminus(K\cup C)}}\1_{\{\lambda=\Lambda\cup \Lambda'\}}$. We then pull out the $\lambda$ sum and note given $\lambda$ and $F$, $\Lambda$ and $\Lambda'$ are uniquely determined, and summing over $F$ is equivalent to just summing over subpartitions (subsets) of $\lambda$. So this sum becomes
\begin{align*}
\sum_{\lambda\in \Pi_{[r]\setminus (K\cup C)}}\!\!(-1)^{|\lambda|}\!\sum_{\tau: \tau\subset \lambda}\!|\tau|!(|\lambda|-|\tau|)!\prod_{B\in \tau}\frac{\langle (\sigma\widetilde \sigma)_{V}(\sigma\widetilde \sigma)_{U}  \prod_{t\in B}(\epsilon+\widetilde \epsilon)_{x_t}\rangle_0}{\langle (\sigma\widetilde \sigma)_{V}(\sigma\widetilde \sigma)_{U} \rangle_0}\!\prod_{B'\in \lambda\setminus \tau}\!\bigg\langle \prod_{t'\in B'}(\epsilon+\widetilde \epsilon)_{x_{t'}}\bigg\rangle_0.
\end{align*}

Before trying to further manipulate this quantity, let us look at the second case.

\subsubsection*{Case two} In this case, we insert $1=\sum_{\lambda\in \Pi_{[r]\setminus (K\cup C)}}\1_{\{\lambda=\Lambda\cup \Lambda'\cup\{C'\}\}}$ and again pull out the $\lambda$ sum. The idea is now to identify $\Lambda\cup \{C'\}$ as a (non-empty) subpartition of $\lambda$. There is a minor difference to earlier cases in that given a subpartition $\tau\subset \lambda$, there are $|\tau|$ different ways we can choose the block $C'$ ($\Lambda=\tau \setminus \{C'\}$ and $\Lambda'=\lambda\setminus \tau$ are then fixed). To be more precise, we have
\begin{align*}
&\sum_{F\subset [r]\setminus(K\cup C)}  \sum_{\Lambda\in \Pi_{F}}(-1)^{|\Lambda|}|\Lambda|!\prod_{B\in \Lambda}\frac{\langle (\sigma\widetilde \sigma)_{V}(\sigma\widetilde \sigma)_{U} \prod_{t\in B}(\epsilon+\widetilde \epsilon)_{x_t}\rangle_0}{\langle (\sigma\widetilde \sigma)_{V}(\sigma\widetilde \sigma)_{U}\rangle_0}\\
&\qquad \times \sum_{C'\subset [r]\setminus (K\cup C\cup F)}\frac{\left\langle (\sigma\widetilde \sigma)_{V}(\sigma\widetilde \sigma)_{U}\prod_{s'\in C'}(\epsilon+\widetilde \epsilon)_{x_{s'}}\right\rangle_0}{{\langle (\sigma\widetilde \sigma)_{V}(\sigma\widetilde \sigma)_{U}\rangle_0}}\\
&\qquad \times \sum_{\Lambda'\in \Pi_{[r]\setminus (K\cup C\cup C'\cup F)}}(-1)^{|\Lambda'|}|\Lambda'|!\prod_{B'\in \Lambda'}\bigg\langle \prod_{t'\in B'}(\epsilon+\widetilde \epsilon)_{x_{t'}}\bigg\rangle_0\\
&=\sum_{\lambda\in \Pi_{[r]\setminus(K\cup C)}}(-1)^{|\lambda|-1}\sum_{\tau: \tau \subset\lambda}\sum_{F\subset [r]\setminus(K\cup C)}  \sum_{\Lambda\in \Pi_{F}}(-1)^{|\Lambda|}|\Lambda|!\prod_{B\in \Lambda}\frac{\langle (\sigma\widetilde \sigma)_{V}(\sigma\widetilde \sigma)_{U}\prod_{t\in B}(\epsilon+\widetilde \epsilon)_{x_t}\rangle_0}{\langle (\sigma\widetilde \sigma)_{V}(\sigma\widetilde \sigma)_{U}\rangle_0}\\
&\qquad \times \sum_{\emptyset\neq C'\subset [r]\setminus (K\cup C\cup F)}\frac{\left\langle (\sigma\widetilde \sigma)_{V}(\sigma\widetilde \sigma)_{U}\prod_{s'\in C'}(\epsilon+\widetilde \epsilon)_{x_{s'}}\right\rangle_0}{{\langle (\sigma\widetilde \sigma)_{V}(\sigma\widetilde \sigma)_{U}\rangle_0}}\\
&\qquad \times \sum_{\Lambda'\in \Pi_{[r]\setminus (K\cup C\cup C'\cup F)}}(-1)^{|\Lambda'|}|\Lambda'|!\prod_{B'\in \Lambda'}\bigg\langle \prod_{t'\in B'}(\epsilon+\widetilde \epsilon)_{x_{t'}}\bigg\rangle_0\1_{\{\tau=\Lambda\cup C', \lambda\setminus \tau=\Lambda'\}}\\
&=\sum_{\lambda\in \Pi_{[r]\setminus(K\cup C)}}(-1)^{|\lambda|-1}\sum_{\tau: \emptyset\neq \tau \subset\lambda}\prod_{B\in \tau}\frac{\langle(\sigma\widetilde \sigma)_{V}(\sigma\widetilde \sigma)_{U}\prod_{t\in B}(\epsilon+\widetilde \epsilon)_{x_t}\rangle_0}{\langle (\sigma\widetilde \sigma)_{V}(\sigma\widetilde \sigma)_{U}\rangle_0}\prod_{B'\in \lambda\setminus \tau}\bigg\langle \prod_{t'\in B'}(\epsilon+\widetilde\epsilon)_{x_{t'}}\bigg\rangle_0\\
&\qquad \times \sum_{F\subset [r]\setminus(K\cup C)}\sum_{\Lambda\in\Pi_F}|\Lambda|!\sum_{\emptyset\neq C'\subset[r]\setminus(K\cup C\cup F)}\sum_{\Lambda'\in \Pi_{[r]\setminus(K\cup C\cup C'\cup F)}}|\Lambda'|!\1_{\{\tau=\Lambda\cup C',\lambda\setminus \tau=\Lambda'\}}
\end{align*}
Now note that $|\Lambda|=|\tau|-1$ and $|\Lambda'|=|\lambda|-|\tau|$ so we can pull these terms outside of the $F,\Lambda,C',\Lambda'$ sums. After this, we are simply counting how many ways we can choose $F,\Lambda,C',\Lambda'$ to satisfy the constraint. There are precisely $|\tau|$ different ways: $C'$ is any block of $\tau$, $F$ is the union of the remaining blocks, $\Lambda$ is chosen to be $\tau\setminus \{C'\}$, and $\Lambda'=\lambda\setminus \tau$. Thus this sum is simply  
\begin{multline*}
\sum_{\lambda\in \Pi_{[r]\setminus(K\cup C)}}(-1)^{|\lambda|-1}\sum_{\tau: \emptyset\neq \tau \subset\lambda}|\tau| (|\tau|-1)!(|\lambda|-|\tau|)!\\
\shoveright{\times\prod_{B\in \tau}\frac{\langle(\sigma\widetilde \sigma)_{V}(\sigma\widetilde \sigma)_{U}\prod_{t\in B}(\epsilon+\widetilde \epsilon)_{x_t}\rangle_0}{\langle (\sigma\widetilde \sigma)_{V}(\sigma\widetilde \sigma)_{U}\rangle_0}\prod_{B'\in \lambda\setminus \tau}\bigg\langle \prod_{t'\in B'}(\epsilon+\widetilde\epsilon)_{x_{t'}}\bigg\rangle_0}\\
\shoveleft=-\sum_{\lambda\in \Pi_{[r]\setminus(K\cup C)}}(-1)^{|\lambda|}\sum_{\tau: \emptyset\neq \tau \subset\lambda}|\tau|!(|\lambda|-|\tau|)!\\
\times\prod_{B\in \tau}\frac{\langle(\sigma\widetilde \sigma)_{V}(\sigma\widetilde \sigma)_{U}\prod_{t\in B}(\epsilon+\widetilde \epsilon)_{x_t}\rangle_0}{\langle (\sigma\widetilde \sigma)_{V}(\sigma\widetilde \sigma)_{U}\rangle_0}\prod_{B'\in \lambda\setminus \tau}\bigg\langle \prod_{t'\in B'}(\epsilon+\widetilde\epsilon)_{x_{t'}}\bigg\rangle_0.
\end{multline*}

\subsubsection*{Combining the two cases}
We thus find that adding the contribution from both of the cases gives
\begin{align*}
&\sum_{\lambda\in \Pi_{[r]\setminus (K\cup C)}}\!(-1)^{|\lambda|}\!\sum_{\tau: \tau\subset \lambda}\!|\tau|!(|\lambda|-|\tau|)!\prod_{B\in \tau}\frac{\langle (\sigma\widetilde \sigma)_{V}(\sigma\widetilde \sigma)_{U}\prod_{t\in B}(\epsilon+\widetilde \epsilon)_{x_t}\rangle_0}{\langle (\sigma\widetilde \sigma)_{V}(\sigma\widetilde \sigma)_{U}\rangle_0}\!\prod_{B'\in \lambda\setminus \tau}\!\bigg\langle \prod_{t'\in B'}(\epsilon+\widetilde \epsilon)_{x_{t'}}\bigg\rangle_0\\
&\quad -\sum_{\lambda\in \Pi_{[r]\setminus(K\cup C)}}(-1)^{|\lambda|}\sum_{\tau: \emptyset\neq \tau \subset\lambda}|\tau|!(|\lambda|-|\tau|)!\\
&\qquad\times\prod_{B\in \tau}\frac{\langle(\sigma\widetilde \sigma)_{V}(\sigma\widetilde \sigma)_{U}\prod_{t\in B}(\epsilon+\widetilde \epsilon)_{x_t}\rangle_0}{\langle (\sigma\widetilde \sigma)_{V}(\sigma\widetilde \sigma)_{U}\rangle_0}\prod_{B'\in \lambda\setminus \tau}\bigg\langle \prod_{t'\in B'}(\epsilon+\widetilde\epsilon)_{x_{t'}}\bigg\rangle_0.
\end{align*}
The two terms are identical apart from the second one having the constraint $\tau\neq \emptyset$. Thus all the other terms in the $\tau$ sums cancel and the only thing that is left is the $\tau=\emptyset$ term yielding 
\begin{align*}
\sum_{\lambda\in \Pi_{[r]\setminus (K\cup C)}}(-1)^{|\lambda|}|\lambda|!\prod_{B\in \lambda}\bigg\langle \prod_{t\in B}(\epsilon+\widetilde \epsilon)_{x_t}\bigg\rangle_0.
\end{align*}
This means that for our correlation function, we have
\begin{align*}
&\langle \sigma_V\widetilde \sigma_V \mu_U\widetilde \mu_U \Psi_Z\widetilde \Psi_Z(\epsilon+\widetilde \epsilon)_W\rangle_{m\alpha}\\
&=\sum_{r=0}^\infty \left(-\frac{m}{\pi}\right)^{r}\frac{1}{r!}\int_{\Omega^{r}}d^{2r}x  \prod_{j=1}^r \alpha(x_j)\sum_{I\subset W}\sum_{\substack{K\subset [r]:\\ |K|=|I|}}\sum_{\nu:I\leftrightarrow K}\prod_{i\in I}\left(-\langle \epsilon_{w_i}+\widetilde \epsilon_{w_i};\epsilon_{x_{\nu(i)}}+\widetilde \epsilon_{x_{\nu(i)}}\rangle_0^\mathsf T\right)\\
&\quad \times \sum_{C\subset [r]\setminus K}\bigg\langle \sigma_V \widetilde \sigma_V\mu_U\widetilde \mu_U \Psi_Z\widetilde \Psi_Z (\epsilon+\widetilde \epsilon)_{W\setminus I}\prod_{s\in C}(\epsilon+\widetilde \epsilon)_{x_s}\bigg\rangle_0\\
&\quad\times\sum_{\lambda\in \Pi_{[r]\setminus (K\cup C)}}(-1)^{|\lambda|}|\lambda|!\prod_{B\in \lambda}\bigg\langle \prod_{t\in B}(\epsilon+\widetilde \epsilon)_{x_t}\bigg\rangle_0,
\end{align*}
using the moment to cumulant map \eqref{eq:mtoc2} again yields the desired expansion.
\end{proof}

\bibliographystyle{abbrvnat}
\bibliography{sgising.bib}

\end{document}